September 12, 2017

# Multivariate Density Modeling for Retirement Finance

CHRISTOPHER J. ROOK

## ABSTRACT

Prior to the financial crisis mortgage securitization models increased in sophistication as did products built to insure against losses. Layers of complexity formed upon a foundation that could not support it and as the foundation crumbled the housing market followed. That foundation was the Gaussian copula which failed to correctly model failure-time correlations of derivative securities in duress. In retirement, surveys suggest the greatest fear is running out of money and as retirement decumulation models become increasingly sophisticated, large financial firms and robo-advisors may guarantee their success. Similar to an investment bank failure the event of retirement ruin is driven by outliers and correlations in times of stress. It would be desirable to have a foundation able to support the increased complexity before it forms however the industry currently relies upon similar Gaussian (or lognormal) dependence structures. We propose a multivariate density model having fixed marginals that is tractable and fits data which are skewed, heavy-tailed, multimodal, i.e., of arbitrary complexity allowing for a rich correlation structure. It is also ideal for stress-testing a retirement plan by fitting historical data seeded with black swan events. A preliminary section reviews all concepts before they are used and fully documented C/C++ source code is attached making the research self-contained. Lastly, we take the opportunity to challenge existing retirement finance dogma and also review some recent criticisms of retirement ruin probabilities and their suggested replacement metrics.

## TABLE OF CONTENTS

**Contact:** cjr5@njit.edu

A financial security that is purchased for $\$p_{t-1}$ at time t-1 with all distributions reinvested yields a value at time t called the *adjusted price*, say $\$P_t$, for t=1,2,...,T. The total return at time t is $R_t=(P_t-p_{t-1})/p_{t-1}$ and the total compounding return is $1+R_t=P_t/p_{t-1}$ so that $\$p_{t-1}(1+R_t)=\$P_t$. If the inflation rate between times t-1 and t is $I_t$ then $1+R_t=(1+I_t)(1+r_t)$, where $r_t=(1+R_t)/(1+I_t)-1$ is the real return at time t. The real price at time t is the value $\$p_t$ such that $(p_t-p_{t-1})/p_{t-1}=r_t$, which upon solving yields $\$p_t=\$P_t/(1+I_t)$. In an *efficient market* real prices are governed by a *geometric random walk* (GRW), that is, $\ln(\$p_t)=\ln(\$p_{t-1})+S_t$, where $S_t\sim N(\mu,\sigma^2)$. A value of $\mu>0$ represents a drift and is the expected log-scale price increase sufficient to compensate the investor for risk ($\sigma^2$) between times t-1 and t. In a random walk, the next value is the current value plus a random normal step, $S_t$, and the best predictor of it is the current value + $\mu$. Exponentiating both sides of the GRW model yields the alternative form $\$p_t=(\$p_{t-1})e^{S_t}$, where $e^{S_t}=(1+r_t)\sim\text{lognormal}(\mu,\sigma^2)$. Under strict conditions, the normally distributed step, $S_t$, can be justified. Decompose the time between t-1 and t into a series of smaller segments, say, d=1,2,...,D and let $1+r_d$ be independent and identically distributed (*iid*) random variables (RVs) for the compounding real return between times d-1 and d, so that the $\ln(1+r_d)$ are also *iid* RVs. The compounding real return at time t is $(1+r_t)=\prod_{d=1}^{D}(1+r_d)=e^{\sum_{d=1}^{D}\ln(1+r_d)}=e^{S_t}$ where $S_t=\sum_{d=1}^{D}\ln(1+r_d)\dot\sim N(\mu,\sigma^2)$ when $d\geq30$ by the Central Limit Theorem (CLT). Here, t can represent years and d days so that the compounding yearly return is the product of compounding daily returns. The lognormal assumption for $(1+r_t)$ breaks down when the $(1+r_d)$ are not *iid* for d=1,2,...,D and there is ample research to indicate that the correlation between returns increases as the time length decreases. We also find that compounding returns on liquid securities used in retirement finance are often better fit by the normal probability density function (PDF) than the lognormal, suggesting that short term real compounding returns may not be *iid* (see §**II.C**). A further complication is that the normal PDF is generally considered tractable, whereas the lognormal PDF is not. For example, a diversified portfolio of equities and bonds with real returns $e_t$ and $b_t$, respectively, has compounding real return $\alpha(1+e_t)+(1-\alpha)(1+b_t)$ where $\alpha$ is the *equity ratio*. Unfortunately, no known PDF exists for the sum of correlated lognormal RVs and we are left to approximate it for a given $\alpha$, see Rook & Kerman (2015) for an implementation. Despite the benefits of using normal RVs to model compounding real returns in finance many practitioners and researchers will not, due primarily to the lack of skewness and heavy tail, but also because the normal PDF can generate negative prices. The spectacular failure of Gaussian copulas during the financial crisis reinforces the skepticism. Unfortunately, those who reject the normal PDF do not benefit from discrete-time finance models optimized using it. This research is motivated by the dilemma, particularly the desire for skewed, heavy-tailed, multimodal, tractable PDFs to model the compounding real return on a diversified portfolio in discrete-time finance applications. Of particular interest is the 1895 claim by Karl Pearson that the moments of a lognormal PDF are virtually indistinguishable from a mixture of normals (McLachlan & Peel, 2000).

## I. Literature Review

During the housing boom residential mortgages were packaged and sold as securities. The price of a security is the present value of future cash flows, which here are the mortgage payments. The products were partitioned into tranches so that as borrowers defaulted low-level holders suffered first, followed by mid-level and then top-level (MacKenzie & Spears, 2014). Cash flows and timings are needed to price a tranche which is a function of which loans have defaulted by each time point. Default times can be modeled using an exponential PDF with the probability of default before some time returned by its cumulative distribution function (CDF). The probability of simultaneous defaults before given times is computed from the copula or multivariate CDF and depends on the correlation between default times. There is no way to estimate the true correlation between default times of residential borrowers due to lack of data. Li (2000) suggested translating the copula on simultaneous defaults to an equivalent expression using normal RVs. The correlation between these RVs is pulled from a measure on the underlying debt instrument for which the normal assumption is reasonable and sufficient data exists. Using these correlations the Gaussian copula can return the probability of simultaneous defaults before specific times. Samples on the correlated exponential failure times can then be simulated from the Gaussian copula and used to value the security.

Loan pools held ≅5000 mortgages with equity tranches acting like a stock and senior tranches like a safe bond. Low interest rates led to excess liquidity and produced an insatiable appetite from pension and sovereign wealth funds for AAA-rated senior tranches which yielded more than U.S. Treasurys (Kachani, 2012). A fatal flaw in the system was that economists have assumed for decades that financial data originates from regimes and correlations change during crises (Hamilton, 1989). Since housing busts follow housing booms it was unwise, in hindsight, to measure correlation with one value. As witnessed, default-time correlations increase in a crisis and senior tranches sold as safe bonds behaved more like a stock which devastated the insurers, who by 2007 had underwritten \$62.2 trillion of credit default contracts (up from \$1.6 trillion in 2002). Blame for the crisis has focused on the Gaussian copula (Salmon, 2009), with a takeaway being that 'normal returns are not appropriate in finance' (Nocera, 2009). Researchers and practitioners who warned against using the normal distribution were vindicated. Paolella (2013) subsequently declared 'the race is on' to find more suitable multivariate PDFs for financial applications and provides an overview of mixture densities, which are often used to model economic regimes and form the basis for this research.

The PDF we develop is a multivariate normal mixture having fixed normal mixture marginals. It is tractable when used in discrete time retirement decumulation models, and intuitive to understand. In §**II** we detail the needed theory/techniques, and in §**III** we fit generic univariate normal mixtures to sets of returns. In §**IV** we form the multivariate PDF and add correlations in §**V**. Finally in §**VI** and §**VII** we derive the expense-adjusted real compounding return on a diversified portfolio and use it within optimal decumulation models. Supporting proofs, derivations, and a full C/C++ implementation are included in the Appendix.

## II. Preliminaries

Foundational concepts needed for the density model developed in §**III** thru §**V** are presented here.

### *A. Probability Density & Cumulative Distribution Functions*

Let X be a continuous RV and $f(x)$ a function such that $f(x)\geq 0\ \forall\, x \in \mathbb{R}$ with $\int_{-\infty}^{\infty} f(t)dt=1$. The function $f(x)$ is a valid PDF for X (Casella & Berger, 1990). The CDF for X is defined as $F(x)=P(X\leq x)=\int_{-\infty}^{x} f(t)dt$. By the *2nd Fundamental Theorem of Calculus* (Anton, 1988), $F'(x)=f(x)-f(-\infty)=f(x)$, that is, the PDF of an RV X is the 1st derivative of its CDF. Note that X may be defined on a subset of $\mathbb{R}$, and $f(\cdot)$ usually depends on a vector of parameters, say $\boldsymbol{\theta}$, which may represent the mean and variance of X. Other common expressions for the PDF include $f(x,\boldsymbol{\theta})$, $f(x;\boldsymbol{\theta})$, $f(x|\boldsymbol{\theta})$, and it may also be denoted by $f_X(\cdot)$ to indicate the RV governed, written as $X\sim f_X(\cdot)$. For a single RV X, $f(x)$ is a univariate PDF, but the above also applies to an n-dimensional vector of RVs, $\mathbf{X}=(X_1,\ldots,X_n)'$, defined on $\mathbb{R}^n$. Here, $f(\boldsymbol{x})$ is a multivariate PDF with $f(\boldsymbol{x})\geq 0\ \forall\, \boldsymbol{x} \in \mathbb{R}^n$, and $\int_{\mathbb{R}^n} f(\boldsymbol{t})d\boldsymbol{t}=1$. The multivariate CDF of $\mathbf{X}$ is defined as $F(\boldsymbol{x})=P(\mathbf{X} \leq \boldsymbol{x})=\int_{\Omega} f(\boldsymbol{t})d\boldsymbol{t}$, where $\Omega = \{\, \boldsymbol{t}: \bigcap_{i=1}^{n} t_i \leq x_i \,\}$. Similar to the univariate case, differentiating a multivariate CDF yields the multivariate PDF, that is, $\frac{\partial^n}{\partial x_1 \cdots \partial x_n}[F(\boldsymbol{x})]=f(\boldsymbol{x})$, and the marginal PDF of one RV, say, $X_1$ is obtained by integrating out all other RVs. That is, $f(x_1)=\int_{\mathbb{R}^{n-1}} f(\boldsymbol{x})dx_2 \cdots dx_n$.

### *B. Finite Mixture Densities*

Let X be a continuous RV and let $f^1(x),\ldots,f^g(x)$ be g functions that satisfy the univariate PDF conditions in §**II.A**. Also, let $\pi_1,\ldots,\pi_g$ be probabilities ($0<\pi_i\leq 1$, i=1,...,g) such that $\pi_1+\ldots+\pi_g=1$. Then $f(x)=\pi_1 f^1(x)+\ldots+\pi_g f^g(x)$ also satisfies the PDF conditions in §**II.A** and is called a *finite mixture density* (Titterington, et al., 1985). If $X\sim f(x)$ then $F(x)=P(X\leq x)=\int_{-\infty}^{x}[\sum_{i=1}^{g} \pi_i f^i(t)]dt=\pi_1 F^1(x)+\ldots+\pi_g F^g(x)$ is the CDF of X. Let $\mu_i^{(r)}$ be the $r^{th}$ moment for $f^i(x)$. Then $E(X^r)=\mu^{(r)}=\int x^r f(x)dx=\int x^r[\sum_{i=1}^{g} \pi_i f^i(x)]dx=\sum_{i=1}^{g}[\pi_i \int x^r f^i(x)dx]=\sum_{i=1}^{g} \pi_i\, \mu_i^{(r)}$. Thus, $E(X)=\sum_{i=1}^{g} \pi_i\, \mu_i$ and $V(X)=E(X^2)-E(X)^2=\sum_{i=1}^{g} \pi_i\, \mu_i^{(2)} - (\sum_{i=1}^{g} \pi_i\, \mu_i)^2$. When $\mathbf{X}=(X_1,\ldots,X_n)'$ is an n-dimensional vector of RVs on $\mathbb{R}^n$, $f(\boldsymbol{x})=\pi_1 f^1(\boldsymbol{x})+\ldots+\pi_g f^g(\boldsymbol{x})$ is a multivariate mixture PDF and satisfies the multivariate PDF conditions set forth in §**II.A**.

A mixture PDF, $f(x)$, has two distinct interpretations: (1) $f(x)$ is a function that accurately models the PDF's shape/form for an RV X, or (2) the RV X originates from component density $f^i(x)$ with probability $\pi_i$, i=1,2,...,g, and the components have labels. While parameter estimation is unaffected by the interpretation, the underlying math is. During parameter estimation we adopt the interpretation that simplifies the math. Each component density $f^i(x)$ may depend on a parameter vector $\boldsymbol{\theta}^i$, and let $\boldsymbol{\pi}=(\pi_1,\ldots,\pi_g)'$ be the vector of component probabilities. When *iid* observations from $f(x)$ are drawn, say $x_t$, t=1,2,...,T, the objective is to estimate the parameters, $\boldsymbol{\theta}^i$, i=1,2,...,g and $\boldsymbol{\pi}$. Once estimated the PDF is fully specified and can be used.

Under interpretation (2) above, components have meaning. Mixture PDFs model serially correlated data via the components. Each component is considered a *state* and there are g states at each time t. If we assume that observation $x_{t+1}$ depends only on the prior observation $x_t$, and that the long-run probabilities of transitioning between states are stationary, then state transitions evolve over time as a *Markov chain* (Hillier & Lieberman, 2010). Define the gxg matrix $\{\pi_{ij}\}$ i,j=1,2,...,g as the conditional probability of being in state j at time t+1 given that we are in state i at time t. Serially correlated data originating from a mixture PDF thus requires estimation of the transition probabilities $\{\pi_{ij}\}$ in addition to $\boldsymbol{\theta}^i$ and $\boldsymbol{\pi}$, where $\pi_i$ is now interpreted as the unconditional probability of being in state i at time t (used at time t=0) (McLachlan & Peel, 2000). In finance/economics, the g states of a mixture PDF are called *regimes* and the process by which dependent observations transition between states over time is termed *regime switching* (Hamilton, 1989)[1].

As noted, the underlying math differs between interpretations (1) and (2) above. Under interpretation (2) components have labels, thus observations from a mixture PDF can be viewed as coming in pairs ($x_t$, $z_t$), where $x_t$ is the actual value and $z_t$ is the component that generated it, t=1,2,...,T. It is common to replace $z_t$ by a vector $\mathbf{z}_t=\{z_{tj}\}$, for t=1,2,...,T & j=1,2,...,g which has a 1 in the component slot and 0's elsewhere, e.g., component=2 at time t=1 can be expressed as $z_1 = 2$ or $\mathbf{z}_1 = (0,1,0,...,0)'$ (see Figure I).

***Figure I**. Mixture Data Collection when Components have Labels*

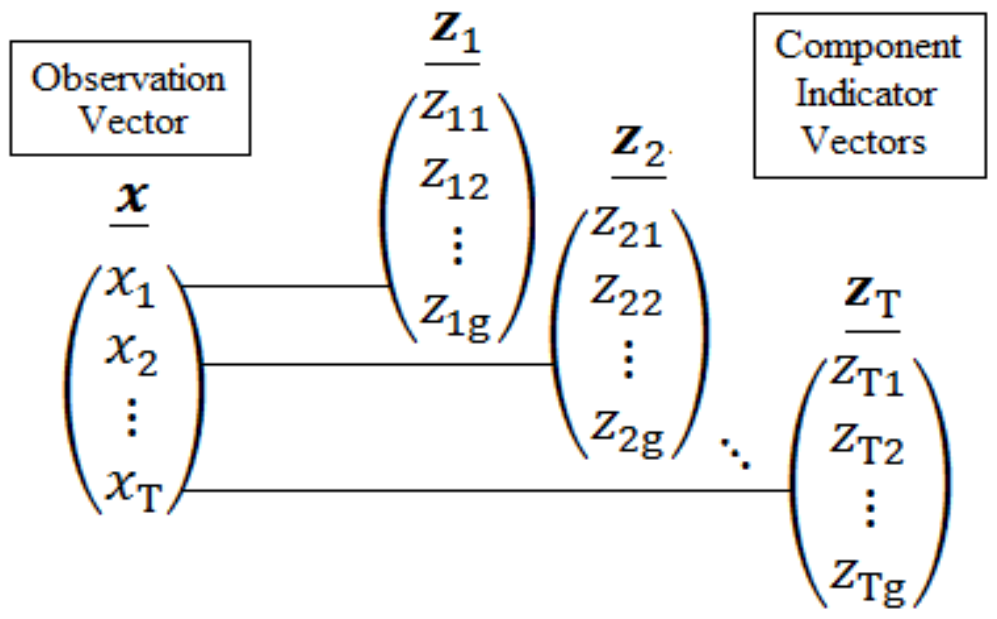

**Note:** Each $z_{tj} = 0$ or 1 and $\sum_{j=1}^{g} z_{tj}=1$, for t=1,2,...,T.

This representation applies to dependent and independent data. Mixture PDFs for dependent data are termed *hidden Markov models* (HMM) (Rabiner & Juang, 1986) because the state vector $\mathbf{z}_t$ generally cannot be observed (it is hidden). Thus a critical task in HMM model building is determining which state generated each observation, $x_t$, t=1,2,...,T. Under interpretation (2), the PDF incorporates the RV $\mathbf{Z}_t$, observed as $\mathbf{z}_t$, as $f(x_t,\mathbf{z}_t)=f(x_t|\mathbf{z}_t)\cdot p[\mathbf{Z}_t=\mathbf{z}_t]$ or $f(x_t, z_{t1}, ..., z_{tg})=f(x_t|z_{t1}, ..., z_{tg})\cdot p[\mathbf{Z}_t=(z_{t1}, ..., z_{tg})']$, which is given by:

$$f(x_t, z_{t1}, \dots, z_{tg}) = \begin{cases} f^1(x_t)\cdot\pi_1\ , & \text{for } z_{t1}=1 \\ f^2(x_t)\cdot\pi_2\ , & \text{for } z_{t2}=1 \\ \vdots & \\ f^g(x_t)\cdot\pi_g\ , & \text{for } z_{tg}=1 \end{cases}\ , \qquad (2.B.1)$$

[1] The regimes in Hamilton (1989) were time series models. In this research they are PDFs.

or, more compactly:

$$f(x_t, z_{t1}, \ldots, z_{tg}) = [\pi_1 \cdot f^1(x_t)]^{z_{t1}} \cdot \ldots \cdot [\pi_g \cdot f^g(x_t)]^{z_{tg}} = \prod_{i=1}^{g} [\pi_i \cdot f^i(x_t)]^{z_{ti}} . \tag{2.B.2}$$

For example, suppose N returns on a financial security are observed over time and appear symmetric around some overall mean, do not exhibit serial correlation, but do include *black swan* events[2] at a frequency greater than their corresponding tail probabilities under either a normal or lognormal PDF (Taleb, 2010).[3] An intuitive tractable PDF for such returns is Tukey's *contaminated normal* PDF (Huber, 2002), which is a mixture of two normals with equal means but unequal variances. It can be used to thicken the tail of a normal PDF. The density with larger variance generates outliers and has a smaller probability.[4] We can proceed intuitively by partitioning the returns into two sets with one holding the $n_1$ non-outliers, and the other holding the $n_2$ outliers. A normal PDF can be fit to each set using MLEs, for example with the mixture weights set to $\pi_i = n_i/N$, i=1,2.[5] If X is an RV representing these returns, then $X \sim f(x) = \pi_1 f^1(x) + \pi_2 f^2(x)$, where $f^i(x) = N(\mu, \sigma_i)$, i=1,2 and $\sigma_2 > \sigma_1$. After replacing all parameters by their estimates, suppose 75% of the returns originate from a "Common" PDF (non-outliers) which is N(0.08,0.10), and 25% from a "Gray Swan" PDF (outliers) which is N(0.08,0.30), where $\mu$=0.08 is the overall mean, estimated by $\bar{\mathbf{x}}$ (see Figure II).

***Figure II**. Example of Tukey's Contaminated Normal PDF*

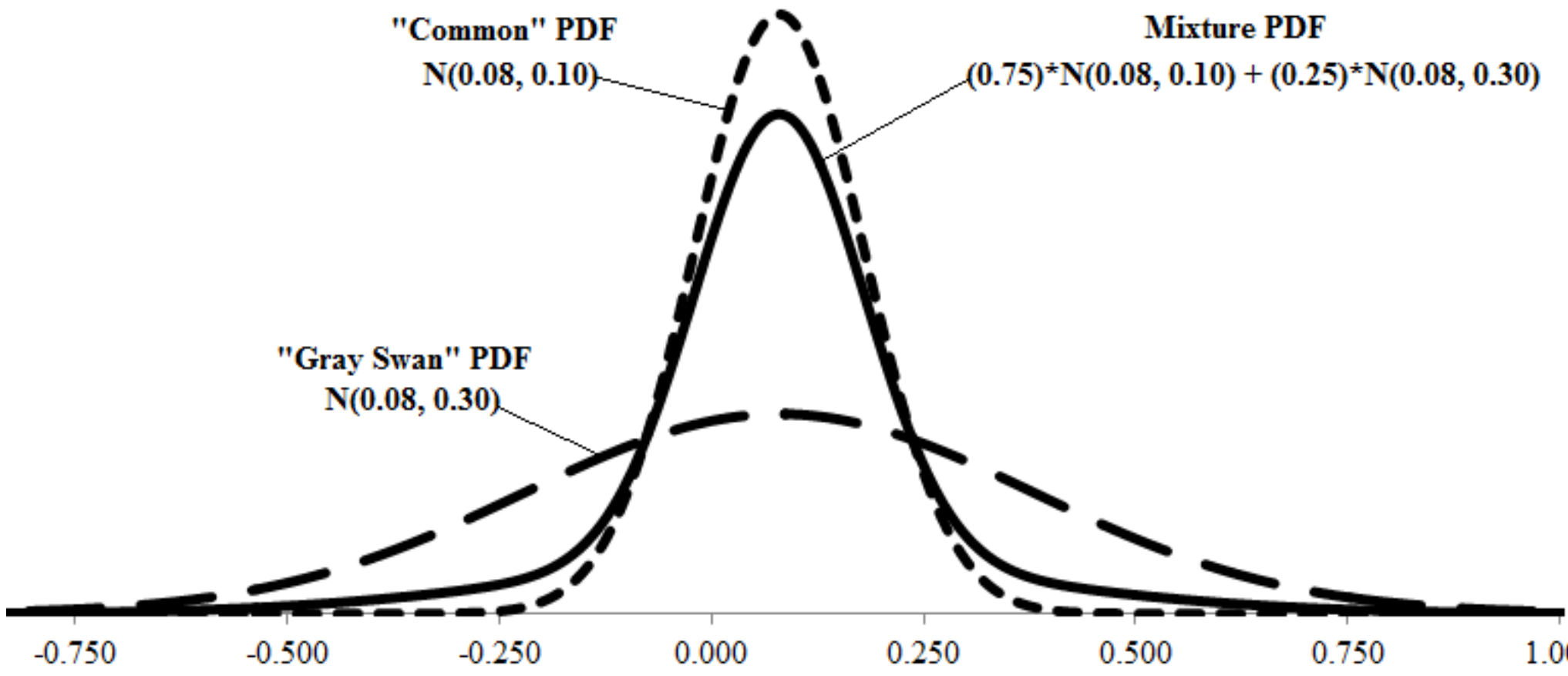

By labeling components we are using interpretation (2). Note that a mixture of normal PDFs is generally not normally distributed, and not symmetric (this being an exception). A note of caution is not to mistake the mixture PDF in Figure II with the RV $Z = 0.75 * X_1 + 0.25 * X_2$, where $X_1 \sim N(0.08,0.10)$ and $X_2 \sim N(0.08,0.30)$, as clearly Z *is* normally distributed. In practice, a g-component normal mixture can model any PDF and is tractable (McLachlan & Peel, 2000). For example, a mixture of two normals can closely approximate the lognormal PDF (Titterington, et al., 1985). Lastly, since the mixture PDF in Figure II does not model black swan events it may be deemed unsatisfactory. A solution could be to add a 3rd component labeled "Black Swan" as, say N(0.08,0.50), with small probability such as $\pi_3$=0.05, then adjust $\pi_1$ and $\pi_2$ so that $\sum_{i=1}^{3} \pi_i = 1$.[6]

[2] Observed outliers are called *gray swan* events. *Black swans* are extreme events that have not occurred before.
[3] Taleb (2010) discourages use of the normal PDF in finance/economics and refers to the lognormal PDF as a "bad compromise".
[4] Fama (1965) discussed using mixture PDFs to explain non-normality in stock prices, and Taleb (2010) also used mixtures in practice to add a heavy tail to the normal PDF as an alternative to the lognormal PDF.
[5] This ad-hoc procedure is for illustration only. Such methods were common prior to the advent of the EM algorithm (see, Johnson, et al., 1994). As will be seen, mixture PDFs are almost always calibrated today using either the EM Algorithm or a gradient procedure.

### *C. The Central Limit Theorem (CLT)*

Let $X_t \overset{iid}{\sim} f(x)$, t=1,2,...,T, with $E(X_t)=\mu$ and $V(X_t)=\sigma^2$. For large T, $\sum_{t=1}^{T} X_t \dot{\sim} N(T\mu, \sqrt{T}\sigma)$ per the *central limit theorem* (Freund, 1992). In words, the sum of *iid* RVs from any PDF, $f(x)$, is approximately normal (denoted $\dot{\sim}$) for large samples and the rule-of-thumb is T≥30. Unfortunately, this rule does not always apply to mixture PDFs. Consider $f(x)=\pi_1 f^1(x)+\pi_2 f^2(x)$, where $\pi_1=(1\text{-}10^{-5})$ and $\pi_2=10^{-5}$, and let $f^1(x)$= N(0,1) and $f^2(x)=N(10^{10},1)$. If $X\sim f(x)$, then $\mu=E(X)=10^5$ and $\sigma=\sqrt{V(X)} \cong 3.16E^7$ (from §**II.B**). In an *iid* sample of size T=30 from $f(x)$ we are unlikely to draw an observation from $f^2(x)$ leaving $\sum_{t=1}^{30} X_t \dot{\sim} N(0,\sqrt{30})$ in violation of the CLT, which ensures $\sum_{t=1}^{T} X_t \dot{\sim} N(30\cdot 10^5, \sqrt{30}\cdot 3.16E^7)$. Such a sample was generated from $f(x)$ and $\sum_{t=1}^{30} X_t=6.77$ with all observations originating from $f^1(x)$. This value is within a σ of the CLT PDF mean thus is a valid value. Repeating the process 9 times should produce an *iid* sample of size N=10 from the CLT PDF. It does not as all 10 values are < the CLT PDF mean.[7]

Caution is therefore advised when invoking the CLT on RVs from mixture PDFs. For example, let $r_d \sim f(r)$ be an RV for the daily real return[8] on the S&P 500 Index, d=1,2,...,D, where D=# of trading days/year. The amount \$B invested on January 1st, will grow to $\$E = \$B \cdot \prod_{d=1}^{D}(1 + r_d)$ on December 31st, in real dollars. The annual real compounding return is $A = \prod_{d=1}^{D}(1 + r_d)$ and $\ln(A) = \sum_{d=1}^{D} \ln(1 + r_d)$, which is $\dot{\sim}$ normal per the CLT when the $r_d$ are *iid* ($D \gg 30$). By definition, the RV $Y=e^{\ln(A)}=A$ would then be $\dot{\sim}$ lognormal, making the historical collection of annual real S&P 500 Index returns a lognormal random sample. This hypothesis was tested and rejected using the Anderson-Darling test (Rook & Kerman, 2015). One explanation is that daily returns are not independent, a claim supported by academic studies of short-term index returns (Baltussen et al., 2016).[9] Another is that daily returns are not identically distributed, or that daily returns originate from a mixture PDF with D not large enough for the CLT approximation.

### *D. The Density of a Future Observation*

Under certain assumptions the PDF of a future value can be derived before it is observed. Let $X_t \overset{iid}{\sim} N(\mu,\sigma)$ be compounding returns on a financial security at time t=1,2,...,T+1 (μ and σ unknown). Suppose $x_t$ is the observed value of $X_t$ for t=1,2,...,T, with $X_{T+1}$ reflecting the unobserved next value. Note that $\overline{X} \sim$ $N(\mu,\sigma/\sqrt{T})$, where $\overline{X} = \frac{1}{T}\sum_{t=1}^{T} X_t$ is the sample mean and $(T\text{-}1)S^2/\sigma^2 \sim \chi^2_{T-1}$, where $S^2 = \frac{1}{T-1}\sum_{t=1}^{T}(X_t - \overline{X})^2$ is the sample variance, and $\chi^2_{T-1}$ is a chi-square RV with T-1 degrees of freedom.[10] It has been established that $\overline{X}$ and $S^2$ are independent RVs (Ross, 2009). Since $\overline{X}$ and $X_{T+1}$ are also independent, $X_{T+1} - \overline{X} \sim N(0,\sigma\sqrt{1 + 1/T})$ and $(X_{T+1} - \overline{X})/\sigma\sqrt{1 + 1/T} \sim N(0,1)$. Thus (2.D.1) has form $N(0,1)/\sqrt{\chi^2_{T-1}/(T-1)}$ which follows a Student's *t*-distribution with T-1 degrees of freedom (Ross, 2009), denoted by $\mathrm{Ŧ}_{(T-1)}$:[11]

---

[6] Such an approach is considered for stress-testing a retirement plan.
[7] Full sample (N=10): 6.77, 11.71, 3.04, -6.75, 5.65, -0.49, -3.53, 4.58, 7.95, 5.55 (mean=3.45, standard deviation=5.56).
[8] The daily (real) compounding return can be approximated by $1+r_d = (1 + R_d)/\sqrt[D]{(1 + I)}$, where $R_d$ is the total daily return calculated as (end value – start value)/start value, and I is the annual inflation rate.
[9] While short term (i.e., daily) index returns have historically exhibited positive serial correlation, Baltussen, et al., (2016) suggest the ubiquity of index products may have eliminated the signal or even turned it negative.

$$\frac{\left(\frac{X_{T+1} - \overline{X}}{\sigma\sqrt{1 + \frac{1}{T}}}\right)}{\sqrt{\frac{S^2}{\sigma^2}}} = \left(\frac{X_{T+1} - \overline{X}}{\sigma\sqrt{1 + \frac{1}{T}}}\right) * \frac{\sigma}{S} = \left(\frac{X_{T+1} - \overline{X}}{S\sqrt{1 + \frac{1}{T}}}\right) \sim \mathrm{Ŧ}_{(T-1)}. \quad (2.D.1)$$

Consequently, $X_{T+1} \sim \overline{X} + \left(S\sqrt{1 + 1/T}\right)\mathrm{Ŧ}_{(T-1)}$, or a future observation has the PDF of a scaled Student's *t*-distribution centered at the sample mean. In a simulation study, this would be the preferred PDF for evaluating a financial plan. Simulating from a Student's t-distribution with T-1 degrees of freedom is straight-forward if we can simulate from the normal distribution. First, generate a N(0,1) random value for the numerator then generate, square, and sum T-1 additional N(0,1) values to construct a $\chi^2_{T-1}$ random value for the denominator.[12] Finally, use the ratio from the definition of a $\mathrm{Ŧ}_{(T-1)}$ RV given above (Law & Kelton, 2000). The PDF of a future value is $f_{X_{(T+1)}}(x)$ which can be derived by differentiating the CDF (see §**II.A)**. The CDF for a future observation $X_{T+1}$ is $F(x)=P(X_{T+1} \le x)=P\left(\mathrm{Ŧ}_{(T-1)} \le (x - \overline{X})/\left(S\sqrt{1 + 1/T}\right)\right)$, which is given by (Freund, 1992):

$$F_{\mathrm{Ŧ}_{(T-1)}}\left(\frac{(x - \overline{X})}{S\sqrt{1 + \frac{1}{T}}}\right) = \int_{-\infty}^{\left(\frac{x-\overline{X}}{S\sqrt{1+1/T}}\right)} \frac{\Gamma\left(\frac{T}{2}\right)}{\sqrt{\pi(T-1)}\,\Gamma\left(\frac{T-1}{2}\right)}\left(1 + \frac{t^2}{T-1}\right)^{-\frac{T}{2}} dt\,, \quad (2.D.2)$$

where $\Gamma(\cdot)$ is the gamma function. Finally the PDF for a future observation $X_{T+1}$ is $f_{X_{(T+1)}}(x) = \frac{d}{dx} F_{\mathrm{Ŧ}_{(T-1)}}\left((x - \overline{X})/\left(S\sqrt{1 + 1/T}\right)\right)$ which is derived using §**II.A** and the chain rule as:

$$f_{X_{(T+1)}}(x) = \frac{\Gamma\left(\frac{T}{2}\right)}{\sqrt{\pi(T-1)}\,\Gamma\left(\frac{T-1}{2}\right)}\left(1 + \frac{\left(\frac{x - \overline{X}}{S\sqrt{1 + 1/T}}\right)^2}{T-1}\right)^{-\frac{T}{2}}\left(\frac{1}{S\sqrt{1 + \frac{1}{T}}}\right), \text{ for } -\infty < x < \infty. \quad (2.D.3)$$

The PDF for future values $X_{T+2}, \ldots, X_{T+N}$ can be derived similarly. The multivariate PDF of N future values is the product of the univariate PDFs, assuming independence. Note that the technique just described breaks down when a 2nd asset is added. Let $X_t \overset{iid}{\sim} N(\mu_x,\sigma_x)$ and $Y_t \overset{iid}{\sim} N(\mu_y,\sigma_y)$ be compounding returns on two uncorrelated financial securities at times t=1,2,...,T+1. Interest is in modeling a future unobserved value on a diversified portfolio using these securities, say, $R_{T+1}=\alpha X_{T+1}+(1-\alpha)Y_{T+1}$, where $0\leq\alpha\leq 1$. It follows that $\alpha(X_{T+1}-\bar{X}) + (1-\alpha)(Y_{T+1}-\bar{Y}) \sim N(0,\sqrt{(1+1/T)(\alpha^2\sigma_x^2 + (1-\alpha)^2\sigma_y^2)})$, thus:

$$\frac{\alpha(X_{T+1}-\bar{X}) + (1-\alpha)(Y_{T+1}-\bar{Y})}{\sqrt{\left(1+\frac{1}{T}\right)(\alpha^2\sigma_x^2 + (1-\alpha)^2\sigma_y^2)}} \sim N(0,1). \tag{2.D.4}$$

[10] The distributions of $\overline{X}$ and $S^2$ here are not approximations (i.e., the CLT is not involved), therefore are valid for any sample size T.

[11] The normal numerator and chi-square denominator must be independent RVs.

[12] If $X_1, X_2, \ldots, X_n \overset{iid}{\sim} N(0,1)$, then $\sum_{i=1}^{n} X_i^2 \sim \chi^2_n$, (Ross, 2009).

As with 1 asset, suppose a quantity $Q(\sigma_x,\sigma_y)$ exists such that a function of it and (2.D.4) has a known PDF that is not a function of $(\sigma_x,\sigma_y)$, and it can be solved for the numerator of (2.D.4). This would suggest a solution to the *Behren's-Fisher Problem* which is a famous unsolved problem in statistics (Casella & Berger, 1990).[13]

### *E. Maximum Likelihood (ML) Estimation*

Let $X_t \sim f(x,\boldsymbol{\theta})$, t=1,2,...,T, be continuous RVs and $x_t$ be the observed value of $X_t$. The likelihood of $x_t$ is $f(x_t,\boldsymbol{\theta})$, which is the PDF of the observed value. The vector $\boldsymbol{\theta}$ holds unknown parameters such as $\boldsymbol{\theta}=(\mu,\sigma)'$, where $E(X_t)=\mu$ and $V(X_t)=\sigma^2$. A likelihood value is not a probability and can be >1, however, it is a similar measure since values with higher likelihoods are more likely to be observed. The *likelihood function*, $\mathfrak{L}(\cdot)$, is the PDF written as a function of $\boldsymbol{\theta}$, i.e., $\mathfrak{L}(\boldsymbol{\theta}|x_t)=f(x_t,\boldsymbol{\theta})$. Extending this to the entire sample, the multivariate PDF of $\mathbf{X}=(X_1,...,X_T)'$ evaluated at $\boldsymbol{x}=(x_1,...,x_T)'$ is $f(\boldsymbol{x},\boldsymbol{\theta})=f(x_1,...,x_T,\boldsymbol{\theta})$ which can be written as $\mathfrak{L}(\boldsymbol{\theta}|\boldsymbol{x})=\mathfrak{L}(\boldsymbol{\theta}|x_1,...,x_T)$, the likelihood of the entire sample. An appealing estimate for $\boldsymbol{\theta}$ is that which maximizes $\mathfrak{L}(\boldsymbol{\theta}|\boldsymbol{x})$ and its value, denoted $\hat{\boldsymbol{\theta}}$, is called the *maximum likelihood estimator* (MLE). Since the natural log function is increasing, maximizing $\mathfrak{L}(\boldsymbol{\theta}|\boldsymbol{x})$ and $\ln(\mathfrak{L}(\boldsymbol{\theta}|\boldsymbol{x}))$ are equivalent problems and it is often easier to deal with the latter. If the RVs $X_t$ are independent, t=1,2,...,T, then $\mathfrak{L}(\boldsymbol{\theta}|\boldsymbol{x})=f(x_1,...,x_T,\boldsymbol{\theta})=\prod_{t=1}^{T} f(x_t,\boldsymbol{\theta})$, and $\ln(\mathfrak{L}(\boldsymbol{\theta}|\boldsymbol{x}))=\sum_{t=1}^{T}\ln(f(x_t,\boldsymbol{\theta}))$. MLEs possess many desirable statistical qualities, such as *consistency*, *efficiency*, *asymptotic normality*, and *invariance* (to functional transformations) thus are often considered the gold standard for parameter estimation. Such qualities, however, depend on certain *regularity conditions* being satisfied (Hogg, et al., 2005), see §**II.G**. Finding parameter estimates in statistics therefore enters the purview of engineering disciplines that specialize in constrained optimization techniques, see §**II.K**. The likelihood function for an *iid* sample originating from a g-component mixture PDF depends on the interpretation, see §**II.B**. Let $\mathfrak{L}^{(1)}(\boldsymbol{\theta}|\boldsymbol{x})=f(\boldsymbol{x},\boldsymbol{\theta})$ and $\mathfrak{L}^{(2)}(\boldsymbol{\theta}|\boldsymbol{x},\boldsymbol{z}_1,...,\boldsymbol{z}_T)=f(\boldsymbol{x},\boldsymbol{z}_1,...,\boldsymbol{z}_T,\boldsymbol{\theta})$ be the likelihoods of $\boldsymbol{\theta}=(\boldsymbol{\pi},\boldsymbol{\theta}^1,\boldsymbol{\theta}^2,...,\boldsymbol{\theta}^g)$ under interpretations (1) and (2), respectively. Then,

[13] The *Behren's-Fisher Problem* tests for equal means in two normal populations with unequal (and unknown) variances. Let $X_1,...,X_{n1} \overset{iid}{\sim} N(\mu_x,\sigma_x)$ and $Y_1,...,Y_{n2} \overset{iid}{\sim} N(\mu_y,\sigma_y)$ be independent samples. Under Ho: $\mu_x = \mu_y$, $(\bar{X}-\bar{Y})/\sqrt{\sigma_x^2/n_1 + \sigma_y^2/n_2} \sim N(0,1)$. A $Q(\sigma_x,\sigma_y)$ to make this $\mathrm{T}_\nu$ or any other fully specified PDF is unknown.

$$\mathfrak{L}^{(1)}(\boldsymbol{\theta}|\boldsymbol{x}) = \prod_{t=1}^{T}\left[\pi_1 f^1(x_t, \boldsymbol{\theta}^1) + \cdots + \pi_g f^g(x_t, \boldsymbol{\theta}^g)\right] = \prod_{t=1}^{T}\left[\sum_{i=1}^{g} \pi_i f^i(x_t, \boldsymbol{\theta}^i)\right] \tag{2.E.1}$$

$$\rightarrow \ln\left(\mathfrak{L}^{(1)}(\boldsymbol{\theta}|\boldsymbol{x})\right) = \sum_{t=1}^{T} \ln\left[\pi_1 f^1(x_t, \boldsymbol{\theta}^1) + \cdots + \pi_g f^g(x_t, \boldsymbol{\theta}^g)\right] = \sum_{t=1}^{T} \ln\left[\sum_{i=1}^{g} \pi_i f^i(x_t, \boldsymbol{\theta}^i)\right], \tag{2.E.2}$$

and, using (2.B.2),

$$\mathfrak{L}^{(2)}(\boldsymbol{\theta}|\boldsymbol{x}, \boldsymbol{z}_1, \ldots, \boldsymbol{z}_T) = \prod_{t=1}^{T}\left[\left(\pi_1 f^1(x_t, \boldsymbol{\theta}^1)\right)^{z_{t1}} \cdot \ldots \cdot \left(\pi_g f^g(x_t, \boldsymbol{\theta}^g)\right)^{z_{tg}}\right] = \prod_{t=1}^{T}\prod_{i=1}^{g}\left(\pi_i f^i(x_t, \boldsymbol{\theta}^i)\right)^{z_{ti}} \tag{2.E.3}$$

$$\rightarrow \ln\left(\mathfrak{L}^{(2)}(\boldsymbol{\theta}|\boldsymbol{x}, \boldsymbol{z}_1, \ldots, \boldsymbol{z}_T)\right) = \sum_{t=1}^{T}\sum_{i=1}^{g} z_{ti} \cdot \left[\ln(\pi_i) + \ln\left(f^i(x_t, \boldsymbol{\theta}^i)\right)\right]. \tag{2.E.4}$$

The goal is to collect data and maximize the log-likelihood from either (2.E.2) or (2.E.4) with respect to $\boldsymbol{\theta}$, obtaining the MLE $\widehat{\boldsymbol{\theta}}$. Unfortunately, the log-likelihood for interpretation (2) in (2.E.4) cannot be maximized directly because the component indicator RVs $\mathbf{Z}_t=(z_{t1}, z_{t2}, \ldots, z_{tg})'$ are missing (i.e., not observable). The log-likelihood from interpretation (1) in (2.E.2) can be maximized directly, however this function is unpleasant for a variety of reasons. For example, it can have multiple local maximums thus finding stationary points does not guarantee an MLE with its desirable properties. It also is unbounded for normal components thus given any value $\psi$, no matter how large, we can always find a setting for $\boldsymbol{\theta}$ such that $\ln(\mathfrak{L}^{(1)}(\boldsymbol{\theta}|\boldsymbol{x}))>\psi$, see Appendix A. When maximizing the mixture log-likelihood for normal components we will therefore restrict the parameter space for $\boldsymbol{\theta}$ to a region where $\ln(\mathfrak{L}^{(1)}(\boldsymbol{\theta}|\boldsymbol{x}))$ is finite and search for all local maximums, declaring $\widehat{\boldsymbol{\theta}}$ as the argmax. To restrict the parameter space for a g-component mixture PDF, let $\sigma_i^2$ be the variance for component i, i=1,...,g. The variance ratio constraint is $\max(\sigma_i^2)/\min(\sigma_i^2)<C$, where C is a given constant (McLachlan & Peel, 2000). A good choice for C will eliminate spurious maximizers which are optimal values that lack a meaningful interpretation and can occur when one component fits a small # of observations.

### *F. The Expectation-Maximization (EM) Algorithm*

Several researchers had been using a 2-step process to obtain MLEs in studies with missing data. The process was observed to possess many interesting properties and became formalized with proofs and a name by Dempster et al. (1977) in what has become one of the most influential statistics papers ever written. The procedure, termed the *Expectation-Maximization* (EM) algorithm, generates MLEs as follows: Let $(X_t,Y_t) \sim f(x_t,y_t,\boldsymbol{\theta})$ and suppose $X_t$ is observed as $x_t$ but $Y_t$ is missing, for t=1,...,T. When $(X_t,Y_t)$ are *iid* over time the joint PDF of $\mathbf{X}=(X_1,\ldots,X_T)'$ and $\mathbf{Y}=(Y_1,\ldots,Y_T)'$ is $f(\boldsymbol{x},\boldsymbol{y},\boldsymbol{\theta})=\prod_{t=1}^{T} f(x_t, y_t, \boldsymbol{\theta})$.[14] Depending on $\boldsymbol{\theta}$, the marginal PDF of $X_t$ may be obtained by integrating $y_t$ out of $f(x_t,y_t,\boldsymbol{\theta})$ when $Y_t$ is continuous, or summing it out when $Y_t$ is discrete, see §**II.A**. That is, $f(x_t,\boldsymbol{\theta})=\int f(x_t, y_t, \boldsymbol{\theta})\, dy_t$ or $f(x_t,\boldsymbol{\theta})=\sum_{y_t} f(x_t, y_t, \boldsymbol{\theta})$ for t=1,...,T, then

[14] The EM algorithm does not require *iid* observations and can also be used to estimate parameters in HMM models (see §**II.B**).

$f(\boldsymbol{x},\boldsymbol{\theta})=\prod_{t=1}^{T} f(x_t,\boldsymbol{\theta})$. This results in 2 likelihood functions for $\boldsymbol{\theta}$, one that includes the missing data and one that does not, i.e., $\mathfrak{L}(\boldsymbol{\theta}|\boldsymbol{x},\mathbf{y})=f(\boldsymbol{x},\mathbf{y},\boldsymbol{\theta})$ and $\mathfrak{L}(\boldsymbol{\theta}|\boldsymbol{x})=f(\boldsymbol{x},\boldsymbol{\theta})$. The EM-algorithm computes $\widehat{\boldsymbol{\theta}}$, the MLE of $\boldsymbol{\theta}$ as:

**Begin:** Initialize $\boldsymbol{\theta}$ to starting values $\boldsymbol{\theta}^0$.

**E-Step:** Compute $E_Y[\ln(\mathfrak{L}(\boldsymbol{\theta}|\boldsymbol{x},\mathbf{y}))]$. Use $\boldsymbol{\theta}^0$ only when taking expectations wrt Y.

**M-Step:** Maximize $E_Y[\ln(\mathfrak{L}(\boldsymbol{\theta}|\boldsymbol{x},\mathbf{y}))]$ with respect to $\boldsymbol{\theta}$. Go to E-Step using $\boldsymbol{\theta}=\widehat{\boldsymbol{\theta}}$ in the expectations.

**End:** Terminate when $\mathfrak{L}(\widehat{\boldsymbol{\theta}}|\boldsymbol{x})$ stops increasing (use %-change below some threshold).

The E-Step replaces the missing $y_t$, t=1,...,T, with constants as a result of taking expectations and the M-Step therefore only has $\boldsymbol{\theta}$ unknown. The value of $\mathfrak{L}(\boldsymbol{\theta}|\boldsymbol{x})$ will not decrease while iterating and it will end at the local maximum nearest to the starting values for $\boldsymbol{\theta}$, given $\mathfrak{L}(\boldsymbol{\theta}|\boldsymbol{x})$ is bounded in this region. If $\mathfrak{L}(\boldsymbol{\theta}|\boldsymbol{x})$ has multiple local maximums we use a variety of starting $\boldsymbol{\theta}$ values and take $\widehat{\boldsymbol{\theta}}$ as the argmax. This value will exhibit the desirable qualities noted in §**II.E** (McLachlan & Peel, 2000). The EM-algorithm can be used to find MLEs for a wide variety of models. The trick is to reformulate so that some RVs appear missing. Applications to mixture PDFs is straightforward. Under interpretation (2) the component indicator RVs $\mathbf{Z}_t=(z_{t1},...,z_{tg})'$ are missing (see §**II.B**, §**II.E**). The multivariate PDF of $X_t$ and $Z_t$ is given in (2.B.2) and the corresponding marginal of $X_t$ is $f(x_t,\boldsymbol{\theta})=\sum_{z_{t1}=1}^{z_{tg}=1} f(x_t,z_{t1},\dots,z_{tg}) = \sum_{z_{t1}=1}^{z_{tg}=1}([\pi_1\cdot f^1(x_t)]^{z_{t1}}\cdot\ldots\cdot[\pi_g\cdot f^g(x_t)]^{z_{tg}}) = \pi_1 f^1(x_t,\boldsymbol{\theta}^1)+\cdots+\pi_g f^g(x_t,\boldsymbol{\theta}^g)$, which is the PDF used in interpretation (1), see (2.E.1). (The $\mathbf{Z}_t$ are discrete RVs.) The likelihood function for an *iid* sample $(x_1,...,x_T)$ from a g-component mixture PDF including the missing data RVs is $\mathfrak{L}^{(2)}(\boldsymbol{\theta}|\boldsymbol{x},\boldsymbol{z}_1,\dots,\boldsymbol{z}_T)$ from (2.E.3), and the corresponding likelihood without missing data is $\mathfrak{L}^{(1)}(\boldsymbol{\theta}|\boldsymbol{x})$ from (2.E.1). The EM-algorithm E-Step uses $E_{\mathbf{Z}}\left[\ln\left(\mathfrak{L}^{(2)}(\boldsymbol{\theta}|\boldsymbol{x},\boldsymbol{z}_1,\dots,\boldsymbol{z}_T)\right)\right]$ which replaces the $\boldsymbol{z}_t$ by their expected value. Notice that $\ln\left(\mathfrak{L}^{(2)}(\boldsymbol{\theta}|\boldsymbol{x},\boldsymbol{z}_1,\dots,\boldsymbol{z}_T)\right)$ is linear in the $\boldsymbol{z}_t=(z_{t1},...,z_{tg})'$ (see 2.E.4), thus:

$$E_{\mathbf{Z}}\left[\ln\left(\mathfrak{L}^{(2)}(\boldsymbol{\theta}|\boldsymbol{x},\boldsymbol{z}_1,\dots,\boldsymbol{z}_T)\right)\right] = \sum_{t=1}^{T}\sum_{i=1}^{g} E(z_{ti})\cdot\left[\ln(\pi_i)+\ln\left(f^i(x_t,\boldsymbol{\theta}^i)\right)\right], \tag{2.F.1}$$

where $E(z_{ti})=E(z_{ti}|x_t,\boldsymbol{\theta})$, i.e., computed using all available non-missing data along with the current settings for $\boldsymbol{\theta}=(\boldsymbol{\pi},\boldsymbol{\theta}^1,\dots,\boldsymbol{\theta}^g)$. Since $z_{ti}$ is a discrete RV that equals 1 when $x_t$ originates from component i, and 0 otherwise, $E(z_{ti}|x_t,\boldsymbol{\theta}) = 1\cdot P[z_{ti}=1|x_t,\boldsymbol{\theta}] + 0\cdot P[z_{ti}=0|x_t,\boldsymbol{\theta}] = P[z_{ti}=1|x_t,\boldsymbol{\theta}]$, which is:

$$E(z_{ti}|x_t,\boldsymbol{\theta}) = P[z_{ti}=1|x_t,\boldsymbol{\theta}] = \frac{f(z_{ti}=1,x_t,\boldsymbol{\theta})}{f(x_t,\boldsymbol{\theta})} = \frac{f(x_t,\boldsymbol{\theta}|z_{ti}=1)\cdot P[z_{ti}=1]}{f(x_t,\boldsymbol{\theta})} = \frac{\pi_i f^i(x_t,\boldsymbol{\theta}^i)}{f(x_t,\boldsymbol{\theta})}, \tag{2.F.2}$$

for t=1,...,T and i=1,...,g. When $\boldsymbol{\theta}$ is given, this value is completely known and replaces $E(z_{ti})$ in (2.F.1). The resulting function with only $\boldsymbol{\theta}$ unknown is optimized in the M-Step. Initial values for $\boldsymbol{\theta}$ can be set randomly using simulation (McLachlan & Peel, 2000) and, since $\mathfrak{L}^{(1)}(\boldsymbol{\theta}|\boldsymbol{x})$ may have many local optimums, the strategy is to apply the EM-algorithm to a variety of starting $\boldsymbol{\theta}^0$ values and select $\widehat{\boldsymbol{\theta}}$ as the argmax.

### *G. Regularity Conditions*

Statistical tests, models and theorems, are built upon sets of assumptions and in *statistical inference* these assumptions are called the *regularity conditions*. Hogg, et al., (2005) Appendix A describe 9 such conditions and it is usually the case that only a subset need be satisfied for a given result. The 1st regularity condition applies to PDFs and deals with uniqueness, namely for PDF $f(\cdot)$, if $\boldsymbol{\theta}\neq\boldsymbol{\phi}$ then $f(x,\boldsymbol{\theta})\neq f(x,\boldsymbol{\phi})$. This condition clearly holds for $N(\mu,\sigma)$ PDFs since changing the mean or variance changes the distribution. However, consider the mixture PDF $f(x,\boldsymbol{\theta})=\pi_1 f^1(x)+\pi_2 f^2(x)$ where $f^i(x)\sim N(\mu_i,\sigma_i)$, i=1,2. The vector of unknown parameters is $\boldsymbol{\theta}=(\pi_1,\mu_1,\sigma_1,\pi_2,\mu_2,\sigma_2)'$. Define $\boldsymbol{\phi}=(\pi_2,\mu_2,\sigma_2,\pi_1,\mu_1,\sigma_1)'$ and note that $\boldsymbol{\theta}\neq\boldsymbol{\phi}$ but $f(x,\boldsymbol{\theta})=f(x,\boldsymbol{\phi})$ violating *regularity condition #1*. In general, mixture PDFs do not satisfy all regularity conditions and caution is advised when using results that requires them.

### *H. The Likelihood Ratio Test (LRT)*

Let $\psi(\boldsymbol{x},\boldsymbol{\theta})$ be an arbitrary statistical model and $\psi(\boldsymbol{x},\boldsymbol{\theta}_R)$ be the same model with some parameters dropped. For example, $\psi(\boldsymbol{x},\boldsymbol{\theta})$ may be a linear regression model with $\boldsymbol{\theta}$ holding the coefficients and $\psi(\boldsymbol{x},\boldsymbol{\theta}_R)$ is the corresponding reduced model that excludes some predictor variables. The principle of *parsimony* favors statistical models with fewer parameters and the *likelihood ratio test* (LRT) checks for a significant difference between the likelihood value of a full model vs. its reduced version. If the likelihoods are not significantly different the reduced model $\psi(\boldsymbol{x},\boldsymbol{\theta}_R)$, having fewer parameters, is preferred. The LRT tests for equivalence of the likelihoods, namely, $H_o$: $\mathfrak{L}(\widehat{\boldsymbol{\theta}}_R|\boldsymbol{x})=\mathfrak{L}(\widehat{\boldsymbol{\theta}}|\boldsymbol{x})$ vs $H_a$: $\mathfrak{L}(\widehat{\boldsymbol{\theta}}_R|\boldsymbol{x})<\mathfrak{L}(\widehat{\boldsymbol{\theta}}|\boldsymbol{x})$, where $\widehat{\boldsymbol{\theta}}$ and $\widehat{\boldsymbol{\theta}}_R$ are the MLEs for $\boldsymbol{\theta}$ and $\boldsymbol{\theta}_R$, respectively. The test statistic is given by (Hogg, et al., 2005):

$$\lambda = -2\ln\left(\frac{\mathfrak{L}(\widehat{\boldsymbol{\theta}}_R|\boldsymbol{x})}{\mathfrak{L}(\widehat{\boldsymbol{\theta}}|\boldsymbol{x})}\right) \qquad (2.H.1)$$

Note that $\mathfrak{L}(\widehat{\boldsymbol{\theta}}_R|\boldsymbol{x})\leq\mathfrak{L}(\widehat{\boldsymbol{\theta}}|\boldsymbol{x})$ since adding parameters to a model does not reduce its likelihood and under $H_o$ the test statistic $\lambda$ is close to zero since ln(1)=0. When $\mathfrak{L}(\widehat{\boldsymbol{\theta}}|\boldsymbol{x})\gg\mathfrak{L}(\widehat{\boldsymbol{\theta}}_R|\boldsymbol{x})$, $\lambda$ takes a large positive value and $H_o$ is therefore rejected when $\lambda > C$, for some critical value C. Under $H_o$ and select regularity conditions (including the 1st, see §**II.G**), $\lambda \dot{\sim} \chi^2_v$, where $v$ is the number of parameters dropped from $\boldsymbol{\theta}$ to create $\boldsymbol{\theta}_R$. A test of $H_o$ with *Type I Error* probability $\alpha$ will define C such that $P[\chi^2_v > C]=\alpha$, and a *Type I Error* means $H_o$ is rejected when true. We are interested in using the LRT to test for the optimal # of components in a mixture PDF however, $\lambda$ is not $\chi^2_v$ under $H_o$ since the applicable regularity conditions are violated. McLachlan (1987) suggests approximating the null distribution of $\lambda$ with a bootstrap. The hypothesis test is:

$H_o$: Data vector $\boldsymbol{x}$ originates from a g-component mixture PDF, say $f(x,\boldsymbol{\theta})$
vs. $H_a$: Data vector $\boldsymbol{x}$ originates from a (g+k)-component mixture PDF, say $f(x,\boldsymbol{\phi})$

Under $H_o$ we estimate the PDF of $\boldsymbol{x}$ as $f(x,\widehat{\boldsymbol{\theta}})$, where $\widehat{\boldsymbol{\theta}}$ is the MLE of $\boldsymbol{\theta}$ and under $H_a$ we estimate the PDF as $f(x,\widehat{\boldsymbol{\phi}})$, where $\widehat{\boldsymbol{\phi}}$ is the MLE of $\boldsymbol{\phi}$. A value for the test statistic $\lambda$ from (2.H.1) is computed using the corresponding likelihood functions of these PDFs. If we implicitly assume that $f(x,\hat{\boldsymbol{\theta}})$ generated our sample vector $\boldsymbol{x}$, a value from the distribution of λ under $H_o$ can be simulated by generating a random sample from $f(x,\hat{\boldsymbol{\theta}})$ and fitting both a g-component and a (g+k)-component mixture PDF using MLEs. The sample should be of the same size as our data vector $\boldsymbol{x}$. Repeating this process K times will simulate values $\lambda_1,...,\lambda_K$ which estimate of the distribution of λ under $H_o$ and the p-value for this test is approximated by (# of $\lambda_j \geq \lambda$)/K.

### *I. Variance-Covariance Matrices*

Let $\mathbf{X}=(X_1,X_2,...,X_N)'$ be RVs for the compounding return on N financial assets at a given time point. If $V(X_i)=\sigma_i^2$, $Cov(X_i,X_j)=\sigma_{ij}$, and $Corr(X_i,X_j)=\sigma_{ij}/(\sigma_i\sigma_j)=\rho_{ij}$ are the variances, covariances, and correlations for i,j=1,2,...,N then $V(\mathbf{X})=E[(\mathbf{X}-E(\mathbf{X}))(\mathbf{X}-E(\mathbf{X}))']$ is the *variance-covariance* (VC) matrix of **X**, written as:

$$V(\mathbf{X}) = \begin{bmatrix} \sigma_1^2 & \sigma_{12} & \cdots & \sigma_{1N} \\ \sigma_{12} & \sigma_2^2 & \cdots & \sigma_{2N} \\ \vdots & \vdots & \ddots & \vdots \\ \sigma_{1N} & \sigma_{2N} & \cdots & \sigma_N^2 \end{bmatrix} = \begin{bmatrix} \sigma_1^2 & \rho_{12}\sigma_1\sigma_2 & \cdots & \rho_{1N}\sigma_1\sigma_N \\ \rho_{12}\sigma_1\sigma_2 & \sigma_2^2 & \cdots & \rho_{2N}\sigma_2\sigma_N \\ \vdots & \vdots & \ddots & \vdots \\ \rho_{1N}\sigma_1\sigma_N & \rho_{2N}\sigma_2\sigma_N & \cdots & \sigma_N^2 \end{bmatrix} \quad (2.I.1)$$

The diagonals are the variances $\sigma_i^2$ and the off-diagonals are the covariances $\sigma_{ij}=\rho_{ij}\sigma_i\sigma_j$. Note that $V(\mathbf{X})$ is square and symmetric so that $V(\mathbf{X})=V(\mathbf{X})'$, but not every square symmetric matrix is a VC matrix. To qualify, all variances and correlations must satisfy $\sigma_i^2>0$ and $-1<\rho_{ij}<1$. The values must also make statistical sense, for example the following do not: $\rho_{12}\cong 0.95$, $\rho_{13}\cong 0.95$, and $\rho_{23}\cong -0.95$. The strong positive correlation of $X_1$ with both $X_2$ and $X_3$ implies that $X_2$ and $X_3$ should also have a strong positive correlation. It turns out that all conditions for a square symmetric matrix to be a valid VC matrix are met if it is *positive-definite* (+definite), that is, $\mathbf{a}'V(\mathbf{X})\mathbf{a}>0$ for any constant vector $\mathbf{a}'=(a_1,a_2,...,a_N)\neq\mathbf{0}$ (Wothke, 1993). Since $V(\mathbf{a}'\mathbf{X})=\mathbf{a}'V(\mathbf{X})\mathbf{a}$, the condition simply means that any non-zero linear combination of the RVs $X_i$, i=1,2,...,N is an RV with variance >0. A matrix is +definite if all *eigenvalues* are >0. The eigenvalues of $V(\mathbf{X})$ are the constants $\lambda_i$ that satisfy $|V(\mathbf{X}) - \lambda_i\mathbf{I}|=0$, i=1,2,...,N (Meyer, 2000). Since $V(\mathbf{X}) - \lambda_i\mathbf{I}$ has determinant=0 it is singular which implies the equation $(V(\mathbf{X}) - \lambda_i\mathbf{I})\cdot\mathbf{u}_i=\mathbf{0}$ can be solved by a $\mathbf{u}_i\neq\mathbf{0}$, thus $\lambda_i$ and $\mathbf{u}_i$ come in pairs. The $\mathbf{u}_i$ with length=1 is referred to as the *eigenvector* for $\lambda_i$. Note that $V(\mathbf{X})\cdot\mathbf{u}_i=\lambda_i\mathbf{u}_i$ so that $\mathbf{u}_i'V(\mathbf{X})\mathbf{u}_i=\lambda_i\mathbf{u}_i'\mathbf{u}_i=\lambda_i$, which reveals why each $\lambda_i$ must be >0 for $V(\mathbf{X})$ to be +definite (otherwise $Z=\mathbf{u}_i'\mathbf{X}$ has $V(Z)<0$). Note that the determinant of a matrix is the product of its eigenvalues, $|V(\mathbf{X})|=\prod_{i=1}^{N}\lambda_i$, and since each $\lambda_i>0$, $|V(\mathbf{X})|>0$ which ensures $V(\mathbf{X})^{-1}$ exists (Guttman, 1982). A matrix with determinant≤0 thus cannot be +definite. Also, a correlation of $\rho_{ij}=1$ (or $\rho_{ij}=-1$) is not allowed in a VC matrix. If two RVs $X_i$ and $X_j$ are perfectly correlated we can question why both are needed but beyond that $V(\mathbf{X})$ cannot be +definite. Set $a_i=1$, $a_j=-\sigma_i/\sigma_j$ and all other elements in $\mathbf{a}=(a_1,...,a_N)'$ to 0 and note that $V(Z)=\mathbf{a}'V(\mathbf{X})\mathbf{a}=\mathbf{0}$ where $Z=\mathbf{a}'\mathbf{X}$, since $V(Z)=1^2\sigma_i^2 + (\sigma_i/\sigma_j)^2\sigma_j^2 - 2(\sigma_i/\sigma_j)\rho_{ij}\sigma_i\sigma_j = 2\sigma_i^2 - 2\rho_{ij}\sigma_i^2 = 0$, with $\rho_{ij}=1$. (When $\rho_{ij}=-1$ use constants $a_i=1$ and $a_j=\sigma_i/\sigma_j$.)

*I.1 Repairing a Broken VC Matrix*

A VC matrix that is not +definite is said to be *broken* and can occur for a variety of reasons including missing data, ad-hoc estimation procedures, and iterative optimization methods. If encountered we can end the analysis with an error or repair the broken VC matrix and continue. We take the latter approach and perform a *ridge repair* (Wothke, 1993). A *ridge* is added to V(**X**) by multiplying the diagonal by a constant K>1. Start with K=1+ε and increase ε until the modified matrix, say $V_R(\mathbf{X})$ with diagonals $K\sigma_i^2$, is +definite. The entire matrix is then divided by K, which revert the diagonals back to $\sigma_i^2$ and forces the covariances to $(\rho_{ij}/K)\sigma_i\sigma_j$, which approach 0 (along with the correlations) as K increases (a diagonal matrix with elements >0 is +definite). The scaled matrix will be +definite since $\mathbf{a}'(1/K)V_R(\mathbf{X})\mathbf{a}>0$ when $\mathbf{a}'V_R(\mathbf{X})\mathbf{a}>0$, and K>0.

*I.2 Useful Derivatives for VC Matrices*

Let $\mathbf{X}_t=(X_{1t},...,X_{Nt})'$ be compounding returns on N financial assets at times t=1,...,T where $\mathbf{X}_t \overset{iid}{\sim} f(\boldsymbol{x}_t,\boldsymbol{\theta})$ $\sim N(\boldsymbol{\mu},V(\mathbf{X}))$, with $\boldsymbol{\theta}=(\boldsymbol{\mu},V(\mathbf{X}))$, $\boldsymbol{\mu}=(\mu_1,...,\mu_N)'$ and $\mu_i=E(X_{it})$. The terms in V(**X**) from (2.I.1) are unknown parameters that can be estimated as MLEs after collecting data (see §**II.E**). MLEs maximize $\mathfrak{L}(\boldsymbol{\theta}|\boldsymbol{x}_1,...,\boldsymbol{x}_T)$, which is the multivariate PDF of the data. The multivariate normal PDF for $\mathbf{X}_t$ is (Guttman, 1982):

$$f(\boldsymbol{x}_t,\boldsymbol{\theta}) = f\big(\boldsymbol{x}_t,\boldsymbol{\mu},V(\mathbf{X})\big) = \frac{1}{(2\pi)^{N/2}\,|V(\mathbf{X})|^{1/2}}\,e^{-\frac{1}{2}(\boldsymbol{x}_t-\boldsymbol{\mu})'\,V(\mathbf{X})^{-1}(\boldsymbol{x}_t-\boldsymbol{\mu})},\quad \boldsymbol{x}_t\in\mathbb{R}^N \tag{2.I.2}$$

The multivariate PDF for the entire *iid* RV sample ($\mathbf{X}_1$, ... $\mathbf{X}_T$) is:

$$f(\boldsymbol{x}_1,\dots,\boldsymbol{x}_T,\boldsymbol{\theta}) = f\big(\boldsymbol{x}_1,\dots,\boldsymbol{x}_T,\boldsymbol{\mu},V(\mathbf{X})\big) = \prod_{t=1}^{T}\frac{1}{(2\pi)^{N/2}\,|V(\mathbf{X})|^{1/2}}\,e^{-\frac{1}{2}(\boldsymbol{x}_t-\boldsymbol{\mu})'\,V(\mathbf{X})^{-1}(\boldsymbol{x}_t-\boldsymbol{\mu})} \tag{2.I.3}$$

The log-likelihood function for the unknown parameters $\boldsymbol{\theta}=(\boldsymbol{\mu},V(\mathbf{X}))$ is given by:

$$\ln\big(\mathfrak{L}(\boldsymbol{\mu},V(\mathbf{X})|\boldsymbol{x}_1,\dots,\boldsymbol{x}_T)\big) = -\frac{T\cdot N}{2}\ln(2\pi) - \frac{T}{2}\cdot\ln(|V(\mathbf{X})|) - \frac{1}{2}\sum_{t=1}^{T}(\boldsymbol{x}_t-\boldsymbol{\mu})'\,V(\mathbf{X})^{-1}(\boldsymbol{x}_t-\boldsymbol{\mu}) \tag{2.I.4}$$

The MLE for $\boldsymbol{\theta}=(\boldsymbol{\mu},V(\mathbf{X}))$ is found by maximizing (2.I.4) wrt $\boldsymbol{\theta}$. A critical point is where the vector of 1st derivatives equals $\mathbf{0}$ which would reveal the maximum or it could be found iteratively using a gradient method also requiring said derivatives.[15] Let $\mathbf{A}^{ij}$ be an NxN matrix having a 1 in the i-j and j-i positions and a 0 in all other positions, and let $V^{ij}(\mathbf{X})$ be the VC matrix from (2.I.1) with a 0 in the i-j and j-i positions. It follows that $V(\mathbf{X})=\sigma_{ij}\mathbf{A}^{ij}+V^{ij}(\mathbf{X})$ and $\frac{\partial}{\partial\sigma_{ij}}[V(\mathbf{X})]=\mathbf{A}^{ij}$. By definition, $V(\mathbf{X})V(\mathbf{X})^{-1}=\mathbf{I}$, thus $\frac{\partial}{\partial\sigma_{ij}}[V(\mathbf{X})V(\mathbf{X})^{-1}]=\mathbf{0}$ and via the product rule $V(\mathbf{X})\frac{\partial}{\partial\sigma_{ij}}[V(\mathbf{X})^{-1}]+\frac{\partial}{\partial\sigma_{ij}}[V(\mathbf{X})]V(\mathbf{X})^{-1}=\mathbf{0}$ so that $\frac{\partial}{\partial\sigma_{ij}}[V(\mathbf{X})^{-1}]=-V(\mathbf{X})^{-1}\mathbf{A}^{ij}\,V(\mathbf{X})^{-1}$ (Searle, et al., 1992). Denote $\sigma_{ii}=\sigma_i^2$ and the above holds for all elements. The determinant of $V(\mathbf{X})$ can be expressed using cofactor expansion with respect to the i-th row, i=1,...,N, as (Meyer, 2000):

[15] As noted in §**II.I.1**, iterative optimization methods are a source of broken VC matrices as we may step into an infeasible region.

$$|V(\mathbf{X})| = \sigma_i^2|V_{ii}(\mathbf{X})| + \sum_{j\neq i=1}^{N}(-1)^{i+j}\sigma_{ij}|V_{ij}(\mathbf{X})|, \quad \text{for } i = 1, \ldots, N, \tag{2.I.5}$$

where $V_{ij}(\mathbf{X})$ is $V(\mathbf{X})$ with the i-th row and j-th column removed, thus $\frac{\partial}{\partial\sigma_i^2}|V(\mathbf{X})| = |V_{ii}(\mathbf{X})|$. Let $\mathbf{A}_{NxN}=\{a_{ij}\}$ be a matrix with $a_{ij}=a_{ij}(t)$. The determinant is a function of the $a_{ij}$, say, $|\mathbf{A}| = d(a_{11}, \ldots, a_{1N}, \ldots, a_{N1}, \ldots, a_{NN})$. Via the chain rule (N×N dimensions), $\frac{\partial}{\partial t}|\mathbf{A}| = \sum_{i=1}^{N}\sum_{j=1}^{N}\frac{\partial}{\partial a_{ij}}d(\cdot)\frac{\partial}{\partial t}a_{ij}(t)$ (Anton, 1988). If $\mathbf{A}=V(\mathbf{X})$ and $t=\sigma_{ij}$, $i\neq j$, then $\frac{\partial}{\partial\sigma_{ij}}|V(\mathbf{X})| = \frac{\partial}{\partial a_{ij}}|V(\mathbf{X})| + \frac{\partial}{\partial a_{ji}}|V(\mathbf{X})|$, with $a_{ij}=a_{ji}=a_{ij}(\sigma_{ij})=a_{ji}(\sigma_{ij})=\sigma_{ij}$. Using (2.I.5), $\frac{\partial}{\partial a_{ij}}|V(\mathbf{X})| = (-1)^{i+j}|V_{ij}(\mathbf{X})|$ and $\frac{\partial}{\partial a_{ji}}|V(\mathbf{X})| = (-1)^{j+i}|V_{ji}(\mathbf{X})|$. Since $V(\mathbf{X})$ is symmetric, $V_{ij}(\mathbf{X})'=V_{ji}(\mathbf{X})$, so that $|V_{ij}(\mathbf{X})|=|V_{ij}(\mathbf{X})'|=|V_{ji}(\mathbf{X})|$ and for $i\neq j$, $\frac{\partial}{\partial\sigma_{ij}}|V(\mathbf{X})| = 2(-1)^{i+j}|V_{ij}(\mathbf{X})| = 2|V(\mathbf{X})^{(ij)}|$, where $V(\mathbf{X})^{(ij)}$ is $V(\mathbf{X})$ with a 1 in position i-j and a 0 in all other row i and column j positions (Searle, et al., 1992).[16]

## *J. Serial Correlation*

Let $X_t$, t=1,...,T be RVs for the compounding return on a financial asset at time t. If the unconditional mean and variance are constants let $E(X_t)=\mu$ and $V(X_t)=\sigma^2$ ∀ t, respectively. When $\sigma_{ij}=Cov(X_i,X_j)\neq 0$ for $i\neq j$, the returns are *serially correlated* and cannot be assumed *iid*. Serially correlated data are modeled using a time series model such as the *autoregressive* (AR) or *moving average* (MA) process. The AR model of order p is $Z_t=\phi_1 Z_{t-1}+\ldots+\phi_p Z_{t-p}+\varepsilon_t$, and the MA model of order q is $Z_t=\varepsilon_t-\theta_1\varepsilon_{t-1}-\ldots-\theta_q\varepsilon_{t-q}$, where $Z_t=X_t-\mu$ and $\varepsilon_t \overset{iid}{\sim} N(0,\sigma_\varepsilon^2)$ ∀ t (Box, et al., 1994). The *autoregressive moving average* (ARMA) model includes AR(p) and MA(q) terms. The appropriate model for a data set is identified by the *signature* of the *autocorrelation* (ACF) and *partial autocorrelation* (PACF) functions. The ACF at lag k is $\rho_{t,t-k}=Corr(X_t,X_{t-k})$ and the PACF at lag k is the correlation remaining after accounting for all lag<k correlations, namely $\phi_k$ in an AR(p) model. The classic AR(p) signature has the PACF cut off abruptly and the ACF decay after lag p, with the 1st PACF being >0. The classic MA(q) signature has the ACF cut off abruptly but the PACF decay after lag q, with the 1st PACF being <0. One of these patterns often emerges (perhaps after log or power transforms) and usually max(p,q)≤3 (Nau, 2014). A series with fixed $\mu$, $\sigma^2$ and $\sigma_{ij}$ is said to be *stationary*. A time series that drifts either up or down over time has a *trend* and is not stationary since $E(X_t)$ is not fixed. A non-stationary series can often be made stationary by differencing (Box, et al., 1994). Usually 1 or 2 differences will suffice where $D_t=Z_t-Z_{t-1}$ for t=2,...,T and $D^2_t=D_t-D_{t-1}$ for t=3,...,T are the 1st & 2nd differences. When $D_t\sim AR(0)$, $Z_t=Z_{t-1}+\varepsilon_t$ is referred to as a *random walk* (RW). Each observation in a RW is the prior value plus a random *step* governed by $\varepsilon_t$. An alternative to differencing for a trend is to fit a regression with time as a predictor then model the stationary residuals. Define the *backshift* operator B as $B^k Z_t=Z_{t-k}$ so that the AR(p) model can be written as $Z_t=\phi_1 BZ_t+\ldots+\phi_p B^p Z_t+\varepsilon_t$ or $\phi(B)Z_t=\varepsilon_t$, where $\phi(B)=(1-\phi_1 B-\ldots-\phi_p B^p)$ is the *characteristic*

[16] The notation $V(\mathbf{X})^{(ij)}$ is ours and introduced since dealing only with NxN matrices simplifies the code. A new NxN matrix $V(\mathbf{X})^{(ij)}_{(rs)}$ will be introduced (§**V.A.2**) when taking 2nd derivatives which is $V(\mathbf{X})^{(ij)}$ with a 1 in position r-s and 0 in all other row r and column s positions.

*polynomial*. An AR(p) process is stationary if all roots of $\phi(B)$ are >1 in magnitude (Box, et al., 1994). The MA(q) model is stationary by design since $E(Z_t)=0 \rightarrow E(X_t)=\mu$, $\sigma^2=V(Z_t)=\sigma_\varepsilon^2(1+\theta_1^2+...+\theta_q^2)$, and $\sigma_{t,t-k}=Cov(Z_t,Z_{t-k})=\sigma_\varepsilon^2\left(-\theta_k+\sum_{i=k+1}^{q}\theta_i\theta_{i-k}\right)$ are the fixed $\mu$, $\sigma^2$ and $\sigma_{t,t-k}$ for $k\leq q$. The covariance was derived using the model's definition as $\sigma_{t,t-k}=Cov(-\theta_k\varepsilon_{t-k},\varepsilon_{t-k})+Cov(\theta_{k+1}\varepsilon_{t-(k+1)},\theta_1\varepsilon_{t-(k+1)})+...+Cov(\theta_q\varepsilon_{t-q},\theta_{q-k}\varepsilon_{t-q})=-\theta_kV(\varepsilon_{t-k})+\theta_{k+1}\theta_1V(\varepsilon_{t-(k+1)})+...+\theta_q\theta_{q-k}V(\varepsilon_{t-q})$ for $k\leq q$ and 0 for $k>q$. The correlation is $\rho_{t,t-k}=Corr(Z_t,Z_{t-k})=\sigma_{t,t-k}/\sigma^2=\left(-\theta_k+\sum_{i=k+1}^{q}\theta_i\theta_{i-k}\right)/\left(1+\sum_{i=1}^{q}\theta_i^2\right)$.

Consider the AR(1) process, $Z_t=\phi_1Z_{t-1}+\varepsilon_t$. Since $Z_{t-1}=\phi_1Z_{t-2}+\varepsilon_{t-1}$ and $Z_{t-2}=\phi_1Z_{t-3}+\varepsilon_{t-2}$, each centered observation can be rewritten as $Z_t=\phi_1(\phi_1(\phi_1Z_{t-3}+\varepsilon_{t-2})+\varepsilon_{t-1})+\varepsilon_t=\phi_1^3Z_{t-3}+\phi_1^2\varepsilon_{t-2}+\phi_1\varepsilon_{t-1}+\varepsilon_t$, and repeating without end yields $Z_t=\phi_1^{\infty}Z_{t-\infty}+\sum_{j=0}^{\infty}\phi_1^j\varepsilon_{t-j}$. The AR(1) model thus only works if $|\phi_1|\leq 1$, otherwise $\phi_1^{\infty}Z_{t-\infty}$ is infinite and $E(Z_t)$, $V(Z_t)$ are not defined. Further, if $|\phi_1|<1$, $\phi_1^{\infty}Z_{t-\infty}\rightarrow 0$ and an alternate form for the AR(1) model is $Z_t=\sum_{j=0}^{\infty}\phi_1^j\varepsilon_{t-j}$, with $E(Z_t)=0$, $E(X_t)=\mu$. The AR(1) model has characteristic polynomial $\phi(B)=(1-\phi_1B)$ with root $B=1/\phi_1$ and is hence stationary when $|1/\phi_1|>1$ or $|\phi_1|<1$, justifying the condition noted for a fixed $E(Z_t)$ and $V(Z_t)$. The conditional variance of a new observation, $X_t$, given all prior observations is $V(X_t|X_{t-1},X_{t-2},...)=V(Z_t|Z_{t-1},Z_{t-2},...)=V(\varepsilon_t)=\sigma_\varepsilon^2$, by definition of the AR(1) model. Using the alternative AR(1) model form, the unconditional variance of each $X_t$, given $|\phi_1|<1$, is $\sigma^2=V(Z_t)=\sum_{j=0}^{\infty}\phi_1^{2j}V(\varepsilon_{t-j})=\sigma_\varepsilon^2\sum_{j=0}^{\infty}\phi_1^{2j}=\sigma_\varepsilon^2/(1-\phi_1^2)$.[17] The AR(1) model does not imply that each new value $X_t$ depends only on the prior value $X_{t-1}$. In fact, it assumes a non-zero correlation between $X_t$ and all prior observations, $X_{t-1},X_{t-2},...$, which is a reason why $p>3$ is rarely needed in practice. To see this use the AR(1) alternative form and note that $\sigma_{t,t-k}=Cov(Z_t,Z_{t-k})=Cov(\phi_1^K\varepsilon_{t-k},\varepsilon_{t-k})+Cov(\phi_1^{K+1}\varepsilon_{t-(k+1)},\phi_1^1\varepsilon_{t-(k+1)})+...=\phi_1^KV(\varepsilon_{t-k})+\phi_1^{K+2}V(\varepsilon_{t-(k+1)})+...=\sigma_\varepsilon^2(\phi_1^K+\phi_1^{K+2}+...)=\sigma_\varepsilon^2\phi_1^K/(1-\phi_1^2)$, since the $\varepsilon_t$ are *iid*. The correlation between values k time points apart is $Corr(X_t,X_{t-k})=Cov(X_t,X_{t-k})/\sqrt{V(X_t)V(X_{t-k})}=\phi_1^K$ (exponential decline in k). The RW is an AR(1) model with *unit root*, i.e., $\phi_1=1$, having alternative form $Z_t=Z_{t-k}+\sum_{j=0}^{k-1}\varepsilon_{t-j}$. The unconditional variance is $V(Z_t)=V(Z_{t-k})+k\sigma_\varepsilon^2$ and since $V(Z_t)>V(Z_{t-k})$ the RW is not stationary ($V(Z_t)\rightarrow\infty$ as $t\rightarrow\infty$). Parameters in an AR(1) model can be estimated by least squares (with adjustments to account for serial correlation), maximum likelihood, or the Yule-Walker equations (method of moments) (Box, et al., 1994). The likelihood function of an observed sample is the multivariate PDF of the data, written as a function of the parameters $\phi_1$ and $\sigma_\varepsilon^2$. Using $Z_t$, this is the likelihood from (2.I.4) with $\boldsymbol{\mu}=\mathbf{0}$, and $V(\mathbf{Z})=\sigma_\varepsilon^2/(1-\phi_1^2)\boldsymbol{\Sigma}$ is the VC matrix with $\boldsymbol{\Sigma}$ having diagonals=1 and i-j off-diagonals $\phi_1^{|j-i|}$ for $i\neq j$.

A time series *mean reverts* if $|E(Z_{t+k}|Z_t,Z_{t-1},...)|\rightarrow 0$ and $V(Z_{t+k})<\infty$ as $k\rightarrow\infty$. In words, future centered values, $Z_{t+k}$, come from a PDF with fixed $\sigma^2$ and $\mu\rightarrow 0$ as k increases. Stationary time series thus mean revert due to their $E(Z_t)=0$ and fixed $V(Z_t)$, but RWs do not. Mean reversion strength is measured by speed or half-life. This is the k required for $E(Z_{t+k}|Z_t,Z_{t-1},...)=z_t/2$, i.e., the time before the process's conditional mean equals ½ of the last observed value $Z_t=z_t$. In an AR(1) process, $Z_{t+k}=\phi_1^k Z_t+\sum_{j=0}^{k-1}\phi_1^j \varepsilon_{t+k-j}$ so that $E(Z_{t+k}|Z_t=z_t)=\phi_1^k z_t$, and the half-life is k such that $\phi_1^k z_t=z_t/2$, or $k=\ln(0.5)/\ln(|\phi_1|)$ (Tsay, 2011). Mean reversion speed thus strengthens as $\phi_1\to 0$ and weakens as $|\phi_1|\to 1$, which is intuitive since $\phi_1$ is the damping factor applied to prior values. The fastest mean reverting AR(1) process has $\phi_1=0$, which is a random sample without serial correlation. The half-life of k=0 implies the process generates $Z_t=z_t$ then instantly reverts to a mean 0, finite $\sigma^2$ PDF. As $\phi_1\to 1$ the AR(1) process approaches a RW with half-life $k=\infty$, and does not mean revert. Thus in an AR(1) process as serial correlation strengthens mean reversion weakens, and vice versa. This is intuitive since as mean reversion strengthens (i.e., $\phi_1\to 0$) new values increasingly depend on the mean, but as serial correlation strengthens (i.e., $|\phi_1|\to 1$) new values increasingly depend on the prior value.

[17] Let $S=\phi_1^0+\phi_1^2+\phi_1^4+...$, so that $\phi_1^2S=\phi_1^2+\phi_1^4+\phi_1^6...$, and $S-\phi_1^2S=\phi_1^0$, resulting in $S=\phi_1^0/(1-\phi_1^2)=1/(1-\phi_1^2)$, *iff* $|\phi_1|<1$, otherwise $S=\infty$.

Data from 18 time series have been retrieved and tested for serial correlation (see Table I, Appendix B). Conclusions are draft and subject to relevant diagnostics (Box, et al., 1994). The correct significance level, $\alpha$=P(Type I Error), for testing multiple simultaneous hypotheses to control the family-wise error rate, $\alpha^*$=P(≥ 1 Type I Error), is discussed below. Market efficiency is the default and each test is formed as Ho: No Serial Correlation vs. Ha: Serial Correlation. Evidence is needed to reject market efficiency for liquid assets as it may suggest a profitable arbitrage trade for retirement advisors who can predict the path of future prices. A Type 1 Error occurs if we reject Ho when it is true, i.e., we falsely reject efficiency for a security.

***Table I**. Tests for Serial Correlation in 18 Finance/Economics Time Series*

| Data Returns (Annual) | P-Value[1] | Process | Data Returns (Annual) | P-Value[1] | Process |
|---|---|---|---|---|---|
| CPI-U (Inflation Rate) | <0.00001 | AR(3) | Cash (No Interest) | <0.00001 | AR(3) |
| Real S&P 500 | 0.90219 | RS | Real S&P 500-to-Bond RP | 0.71082 | RS |
| Total S&P 500 | 0.96352 | RS | Real S&P 500-to-Small Cap RP | 0.00359 | RS[3] |
| Real Small Cap Equity | 0.77708 | RS | Real Small Cap-to-Bond RP | 0.80516 | RS |
| Total Small Cap Equity | 0.58475 | RS | 10-Year Avg. Real S&P 500 | <0.00001 | AR(1)[4] |
| Real U.S. 10-Year T-Bond | 0.58194 | RS | 10-Year Avg. Total S&P 500 | <0.00001 | AR(1)[4] |
| Total U.S. 10-Year T-Bond | 0.47060 | RS[2] | 1st Diff. Shiller CAPE Ratio | 0.39993 | RW[5,6] |
| Real U.S. 3-Month T-Bill | <0.00001 | AR(3) | 1st Diff. Log Shiller CAPE Ratio | 0.96966 | GRW[5,7] |
| Real Gold Return | 0.00430 | RS[3] | S&P 10-Year Avg. Real Earnings | <0.00001 | ARMA(2,1)[4,8] |

Abbreviations: RS = random sample, (G)RW = (geometric) random walk w/Drift, RP = risk premium.

[1] P-Value is for test Ho: No Serial Correlation vs. Ha: Serial Correlation, not for test Ho: AR(p) vs Ha: AR(p-1).

[2] Non-normal. [3] Non-normal, possible serial correlation.

[4] Near unit roots by design. Let $X_t \overset{iid}{\sim} f(x)$, t=1,...,T, with $E(X_t)=\mu$, $V(X_t)=\sigma^2$. Avg. 10-year returns are $Y_t=(X_t+...+X_{t+9})/10$, for t=1,...,T-9. If $f(x)\sim N(\mu,\sigma^2)$, $Y_{t+1}=Y_t-\frac{1}{10}X_t+\frac{1}{10}X_{t+10}=Y_t+\frac{1}{10}(X_{t+10}-X_t)=Y_t+\varepsilon_t$, $\varepsilon_t\sim N(0,\frac{\sqrt{2}\sigma}{10})\neq$RW as $Cov(Y_t,\varepsilon_t)=-\frac{1}{100}\sigma^2$. If $f(x)\sim Ln(\mu,\sigma^2)$, $\ln(X_t)$ is similar.

[5] Possible non-normal "steps". [6,7] Dickey-Fuller unit root test statistics -2.067, -2.187, respectively, with critical value -2.900 cannot reject hypothesis of unit root. Note: Since CAPE Ratio ≥ 0, GRW may be more plausible. Non-zero drifts are 0.0885 (RW) & 0.0040 (GRW).

[8] Preliminary model would be ARMA(2,1) on detrended data: $Y_t-[\mu+\beta_t]=ARMA(2,1)\to Y_t=\mu+\beta_t+\phi_1 Y_{t-1}+\phi_2 Y_{t-2}-\theta_1\varepsilon_{t-1}+\varepsilon_t$.

The goal is to test these hypotheses such that the family of conclusions is replicable with $(1-\alpha^*)$% confidence, where $\alpha^*$=P(≥1 Type I Error). In general, with N independent tests and $\alpha$=P(Type I Error) for each test, the probability of making $k$ (≤N) Type I Errors is $P(B=k)=\binom{N}{k}\alpha^k(1-\alpha)^{N-k}$ with $B\sim Binomial(N,\alpha)$. Thus the probability of making ≥1 Type I Errors in N independent tests is $P(B\geq 1)=\sum_{k=1}^{N}\binom{N}{k}\alpha^k(1-\alpha)^{N-k}=1-$

$P(B=0)=1-(1-\alpha)^N$. For the N=18 tests conducted above with $\alpha=0.05$ the probability of making ≥1 Type I Errors is $\sum_{k=1}^{18}\binom{18}{k}0.05^k(0.95)^{18-k}=1-P(B=0)=1-0.95^{18}\cong 0.60$. Being 95% confident on each of 18 independent tests translates into a 40% chance of replicating the family of conclusions with new data. To replicate the family with 95% confidence, i.e., $\alpha^*=P(\geq 1 \text{ Type I Error})=0.05$, we adjust the α used for each test. The Bonferonni adjustment uses $\alpha=\alpha^*/N$ and does not require independent tests but does ensure $P(\geq 1$ Type I Error$) \leq \alpha^*$ (Westfall, et al., 1999). With $\alpha^*=0.05$, the adjusted p-value for each test in Table I is $\alpha= 0.002778$. The conclusions in Table I lead to the Table II concerns about current retirement finance dogma.

***Table II**. Questionable Claims in the Retirement Finance Literature*

| Claim | Concern[1] |
|---|---|
| Shiller's CAPE Ratio (annual) can be used to time markets. Investors/retirees should sell when the CAPE ratio is high and buy when it is low relative to its historical average. | The hypothesis that the annual CAPE ratio behaves as a (G)RW cannot be rejected. Random walks have unit roots and do not mean revert. The best predictor of any future value in a random walk is the current value + drift (use logs for GRW). |
| Annual S&P 500 Returns (real or total) mean revert therefore are serially correlated and should be fit using an autoregressive model. | Serial correlation and mean reversion are opposites (i.e., as one strengthens the other weakens). Annual S&P 500 returns exhibit no serial correlation, they are random samples and mean revert. |
| Shiller's CAPE Ratio (annual) can be used to predict average future long-term returns.[3] A (log-transformed) linear regression predicting 10-year average future S&P 500 returns using CAPE values has a highly significant $R^2$. | The 10-year average S&P 500 return is strongly serial correlated (see Table I and Appendix B).[2] Fitting a regression line through these points is inappropriate and will result in underestimated variances and inflated Type I Error rates which commonly lead to false claims that predictors are significant. |

[1] Findings are preliminary. The fitted (linear) models are subject to appropriate diagnostics (see, Box, et al., 1994).
[2] This is by design. Let $X_t \overset{iid}{\sim} f(x)$, t=1,...,T, with $E(X_t)=\mu$, $V(X_t)=\sigma^2$. Average 10-year returns are constructed as $Y_t=(X_t+...+X_{t+9})/10$, for t=1,...,T-9, so that $Cov(Y_t,Y_{t+k})=(1/10)^2[V(X_{t+k})+...+V(X_{t+9})]=[(9-k+1)/100]\cdot\sigma^2$, for $0\leq k\leq 9$ and $Cov(Y_t,Y_{t+k})=0$, for $k\geq 10$.
[3] A similar claim is made about safe withdrawal rates (SWR) in retirement via regression with CAPE, and the exact same concern arises.

## *K. Constrained Optimization*

### *K.1 Linear Programming*

A *linear program* (LP) is an optimization problem where the objective and all constraints are linear functions of the decision variables. LPs are solved using the *simplex algorithm* and the standard form for an LP with n decision variables and k linearly independent constraints is (Jensen & Bard, 2003):

$$\begin{array}{llll} \underline{\textbf{Maximize:}} & Z = c_1x_1 + c_2x_2 + \dots + c_nx_n & (\mathbf{c'x}) & \text{(objective function)} \\ \underline{\textbf{Subject to:}} & a_{11}x_1 + a_{12}x_2 + \dots + a_{1n}x_n \leq b_1 & & \\ & a_{21}x_1 + a_{22}x_2 + \dots + a_{2n}x_n \leq b_2 & & \\ & \vdots & (\mathbf{Ax} \leq \mathbf{b}) & \text{(feasible region)} \\ & a_{k1}x_1 + a_{k2}x_2 + \dots + a_{kn}x_n \leq b_k & & \\ \underline{\textbf{Where:}} & x_i \geq 0,\ i=1,2,\dots,n & (\mathbf{x} \geq \mathbf{0}) & \text{(non-negativity constraints)} \end{array} \qquad (2.K.1)$$

All LPs come in pairs with the *dual* being an equivalent minimization problem. When the primary LP is solved the dual is solved and vice-versa. Whereas the primary LP has n decision variables and k constraints, the dual has k decision variables and n constraints and since min(Z)≡max(-Z), any LP can be solved assuming $k \leq n$. The simplex algorithm recognizes that when Z is linear, global solutions must occur at corner points of the feasible region.[18] A constraint *binds* when '≤' becomes '=' for given values of the decision variables. When m constraints bind, m decision variables are fixed by the constraints and there are $\binom{k}{m}$ ways to select the these constraints, m=0,...,k. The remaining n-m decision variables must equal 0 at a corner point, and there are $\binom{n}{n-m}$ ways this can occur. The total # of corner points is thus $\sum_{m=0}^{k}\binom{k}{m}\binom{n}{n-m}=\binom{n+k}{k}$, with each a potential solution. The simplex algorithm partitions $\mathbf{A}$ as $\begin{bmatrix}\mathbf{A}_{11} & \mathbf{A}_{12}\\ \mathbf{A}_{21} & \mathbf{A}_{22}\end{bmatrix}$ such that $\mathbf{A}_{11}$ is m×m and of full rank and let $\mathbf{c}'=[\mathbf{c}_1|\mathbf{c}_2]$, $\mathbf{x}'=[\mathbf{x}_1|\mathbf{x}_2]$, $\mathbf{b}'=[\mathbf{b}_1|\mathbf{b}_2]$ be the corresponding vector partitions. It follows that $\mathbf{A}_{11}\cdot\mathbf{x}_1+\mathbf{A}_{12}\cdot\mathbf{x}_2=\mathbf{b}_1$ and $\mathbf{x}_1=\mathbf{A}_{11}^{-1}\cdot\mathbf{b}_1 - \mathbf{A}_{11}^{-1}\cdot\mathbf{A}_{12}\cdot\mathbf{x}_2$ reflects the m decision variables fixed by the binding constraints. The objective Z becomes $\mathbf{c}_1\cdot\mathbf{x}_1+\mathbf{c}_2\cdot\mathbf{x}_2=\mathbf{c}_1\cdot\mathbf{A}_{11}^{-1}\cdot\mathbf{b}_1-(\mathbf{c}_1\cdot\mathbf{A}_{11}^{-1}\cdot\mathbf{A}_{12}-\mathbf{c}_2)\cdot\mathbf{x}_2$ which is a constant less a linear combination of the decision variables in $\mathbf{x}_2$. If any coefficients in this linear combination are negative the problem is unbounded and has no solution. Simply increase the corresponding decision variable in $\mathbf{x}_2$ to increase the objective Z. To have a solution all coefficients for $\mathbf{x}_2$ in Z must be ≤0 and when this occurs, all decision variables in $\mathbf{x}_2$ must equal 0 to maximize Z. A constant less a quantity ≥0 is maximized when that quantity equals 0. Finally, when $\mathbf{x}_2=0$, $\mathbf{x}_1=\mathbf{A}_{11}^{-1}\cdot\mathbf{b}_1$ and if all constraints are satisfied then $\mathbf{x}'=[\mathbf{x}_1|\mathbf{x}_2]$ is a *basic feasible solution* (BFS) to the LP (Jensen & Bard, 2003). The simplex algorithm starts with a corner point of the feasible region and cycles through adjacent corner points (i.e., bases defined by $\mathbf{A}_{11}$) such that Z does not decrease. Thus the LP can be solved with a small number of evaluations and the algorithm ends at a global maximizer. Note that when k ≤ n both $\mathbf{A}_{21}$ and $\mathbf{A}_{22}$ vanish so that each basis is formed by setting n-m decision variables equal to 0 and solving for the remaining variables.

If any quantity is random then (2.K.1) is a *stochastic linear program* (SLP). While solving SLPs using theory is involved (Kall & Mayer, 2010), simulation can be a practical alternative. For example suppose $\mathbf{b}=(b_1,...,b_k)'$ is a set of RVs with $\mathbf{b}\sim f_B(\mathbf{b})$, $E(\mathbf{b})=\boldsymbol{\mu}_B$.[19] Since any solution is a function of $\mathbf{b}$ it must also be an RV, say $\mathbf{x}\sim f_X(\mathbf{x})$, $E(\mathbf{x})=\boldsymbol{\mu}_X$. When $f_B(\mathbf{b})$ is known/approximated a heuristic SLP solution is obtained by generating random values for $\mathbf{b}$, say $\mathbf{b}_i=(b_{1i},...,b_{ki})$, where sample i yields an LP BFS, say $\mathbf{x}_i=(x_{1i},...,x_{ni})'$, i=1,...,N. The solution is then taken as $\bar{\mathbf{x}}=\frac{1}{N}\sum_{i=1}^{N}\mathbf{x}_i$. Since each $\mathbf{x}_i$ satisfies $\mathbf{A}\mathbf{x}_i \leq \mathbf{b}_i$ it follows that $\mathbf{A}\bar{\mathbf{x}} \leq \bar{\mathbf{b}}$ and as N→∞, $\mathbf{A}\boldsymbol{\mu}_X \leq \boldsymbol{\mu}_B$ or $E(\mathbf{A}\mathbf{x}) \leq E(\mathbf{b})$. Simulated solutions therefore asymptotically satisfy the constraint set, in expectation. An exceedingly large number of real-life problems can be formulated and solved as LPs and some non-linear programs can also be closely approximated by an LP.

[18] The solution to an LP may or may not be unique, but it is global. The feasible region of an LP is referred to as a polyhedron which sits in the 1st quadrant since all decision variables are ≥ 0. A technically incorrect but useful visualization tool when n=2 is that four people of different heights are holding a flat board (objective function is a plane) over a stop sign laying flat on the ground (feasible region in 1st quadrant). The highest point on the board inside the sign will be directly above a corner of the sign, it cannot be above an interior point.

[19] Here, **b** may reflect any randomly occurring quantity such as supply, demand, temperature, sales revenue, profit, etc...

*K.2 Quadratic Programming*

When the objective function in (2.K.1) is of the form $Z=\sum_{i=1}^{n} a_i x_i - \sum_{i=1}^{n}\sum_{j>i}^{n} b_{ij}x_i x_j - \sum_{i=1}^{n} c_i x_i^2$, the surface being maximized is quadratic (not linear) and the resulting optimization is referred to as a *quadratic program* (QP). If $b_{ij}=0\ \forall\ i<j=1,\ldots,n$, Z is said to be *separable* as the objective separates into a sum of 1-decision variable functions, i.e., $Z=\sum_{i=1}^{n} g^i(x_i)$ (Hillier & Lieberman, 2010). A separable QP can be approximated by an LP. To begin, convert the QP to a minimization problem noting that max(Z)≡min(–Z), then write each function in –Z as $f^i(x_i) = -g^i(x_i) = c_i(x_i - a_i/2c_i)^2$ which adds the constant $\sum_{i=1}^{n} a_i^2/4c_i$ to -Z posing no issue since minimizing -Z and $-Z+\sum_{i=1}^{n} a_i^2/4c_i$ are equivalent problems. Proceed by partitioning each $x_i$-axis into equidistant constants $[x_{i1},\ldots,x_{iS+1}]$ which define S line segments that trace out $f^i(x_i)$. These S+1 values are chosen and allow $x_i$ to be replaced by a weight vector $\boldsymbol{\alpha}_i=(\alpha_{i1},\ldots,\alpha_{iS+1})'$, where $\sum_{j=1}^{S+1}\alpha_{ij}=1$. Any $x_{i1}\le x_i\le x_{iS+1}$ is reachable by a weighted sum of the constants using $\boldsymbol{\alpha}_i$, namely, $x_i=\sum_{j=1}^{S+1}\alpha_{ij}x_{ij}$, and each function $f^i(x_i) \cong \sum_{j=1}^{S+1}\alpha_{ij} f^i(x_{ij})$ where the approximation sharpens as S increases when adjacent weights are used (Jensen & Bard, 2003), see Figure III. Since the objective is min(–Z), adjacent weights must be used in an optimal solution, compare the blue dot objectives for the dashed and red lines in Figure III.

***Figure III**. Separable QP Approximation by an LP (S=4) (Minimization Objective)*

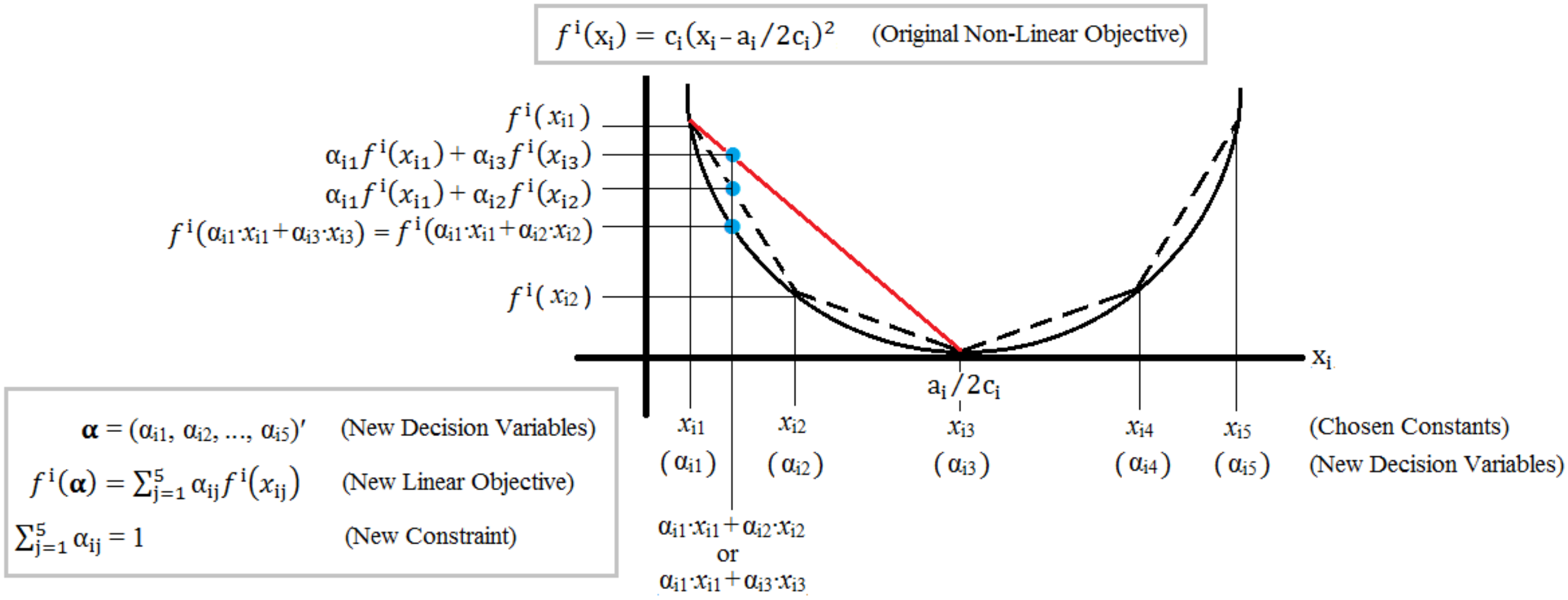

The QP max(Z), $Z = \sum_{i=1}^{n} a_i x_i - \sum_{i=1}^{n} c_i x_i^2$, is then approximated by the following LP (linear in $\boldsymbol{\alpha}$):[20]

| | | | | |
|---|---|---|---|---|
| **<u>Minimize:</u>** | $-Z \cong \sum_{i=1}^{n}\sum_{j=1}^{S+1}\alpha_{ij} f^i(x_{ij})$ | | $(\mathbf{c}'\boldsymbol{\alpha})$ | (objective function) |
| **<u>Subject to:</u>** | $a_{m1}\sum_{j=1}^{S+1}\alpha_{1j}x_{1j} + \ldots + a_{mn}\sum_{j=1}^{S+1}\alpha_{nj}x_{nj} \le b_m$, | m=1,2,...,k | $(\mathbf{A}\boldsymbol{\alpha} \le \mathbf{b})$ | (feasible region) |
| | $\sum_{j=1}^{S+1}\alpha_{ij} \le 1$, $-\sum_{j=1}^{S+1}\alpha_{ij} \le -1$ | i=1,2,...,n | | (2.K.2) |
| **<u>Where:</u>** | $\alpha_{ij}\ge 0$, i=1,2,...,n & j=1,2,...,S+1 | | $(\boldsymbol{\alpha} \ge \mathbf{0})$ | (non-neg. constraints) |

[20] The standard form LP in (2.K.1) requires all constraints to have the form $f_m(\mathbf{x}) \le b_m$, m=1,2,...,k where $f_m(\mathbf{x})$ is linear in **x**. Express an equality constraint $f_m(\mathbf{x}) = b_m$ using only "≤" as $(f_m(\mathbf{x}) \le b_m \cap f_m(\mathbf{x}) \ge b_m) \equiv (f_m(\mathbf{x}) \le b_m \cap -f_m(\mathbf{x}) \le -b_m)$.

*K.3 Classical Convex Programming*

A *classical convex program* (CCP) is a non-linear optimization having objective of either minimizing a convex, or maximizing a concave function over a set of linear equality constraints (Jensen & Bard, 2003). A function $f(\mathbf{x})$ is convex *iff* $f(\alpha\mathbf{x}_1+(1-\alpha)\mathbf{x}_2)\leq\alpha f(\mathbf{x}_1)+(1-\alpha)f(\mathbf{x}_2)\ \forall\ \mathbf{x}_1, \mathbf{x}_2 \in \mathbb{R}^n$, and if $f(\mathbf{x})$ is convex then $-f(\mathbf{x})$ is concave. Further, $\max[-f(\mathbf{x})]\equiv\max[e^{-f(\mathbf{x})}]$ thus maximizing a concave function and a log-concave function are equivalent problems as $e^{-f(\mathbf{x})}$ is log-concave since $\ln(e^{-f(\mathbf{x})})=-f(\mathbf{x})$ is concave (Lovasz & Vempala, 2006). A CCP has the desirable property that local optimums are global optimums, thus the problem reduces to finding any local optimum. The following formulation is of interest:

| | | | | |
|---|---|---|---|---|
| | **Minimize:** | $f(x_1,x_2,\ldots,x_n)$ | $[f(\mathbf{x})]$ | (convex objective function) |
| (or) | **Maximize:** | $-f(x_1,x_2,\ldots,x_n)$ | $[-f(\mathbf{x})]$ | (concave objective function) |
| | **Subject to:** | $a_{11}x_1 + a_{12}x_2 + \ldots + a_{1n}x_n = b_1$ | | (2.K.3) |
| | | $a_{21}x_1 + a_{22}x_2 + \ldots + a_{2n}x_n = b_2$ | | |
| | | $\vdots$ | $(\mathbf{Ax} = \mathbf{b})$ | (feasible region) |
| | | $a_{k1}x_1 + a_{k2}x_2 + \ldots + a_{kn}x_n = b_k$ | | |
| | **Where:** | $x_i \geq 0$, i=1,2,...,n | $(\mathbf{x} \geq \mathbf{0})$ | (non-negativity constraints) |

As with LPs, redundant constraints are removed. If k>n the feasible region is empty and there is no solution. If k=n the feasible region consists of the point $\mathbf{x}=\mathbf{A}^{-1}\mathbf{b}$, which is also the solution. If k=0 the solution is $\mathbf{x}^*$ such that $\nabla f(\mathbf{x}^*)=\mathbf{0}$. If $0<k<n$, $\nabla f(\mathbf{x}^*)=\mathbf{0}$ and $\mathbf{Ax}^*=\mathbf{b}$ then the unconstrained solution solves the constrained problem. Most likely none of the above will hold, and we are left to optimize $f(\mathbf{x})$ with $0<k<n$, $\mathbf{Ax}=\mathbf{b}$, and rank($\mathbf{A}$)=k. To solve this problem in 1-step we locate the critical point of the *Lagrangian*, $\mathcal{L}(\cdot)$, defined as, $\mathcal{L}(x_1,\ldots,x_n,\lambda_1,\ldots,\lambda_k)=\mathcal{L}(\mathbf{x},\boldsymbol{\lambda})=f(\mathbf{x})-\sum_{i=1}^{k}\lambda_i(a_{i1}x_1 + a_{i2}x_2 + \cdots + a_{in}x_n - b_i)$.[21] The Lagrangian, $\mathcal{L}(\mathbf{x},\boldsymbol{\lambda})$, incorporates all constraints into $f(\mathbf{x})$ and k new decision variables are introduced, $\lambda_i$, i=1,2,...,k, called the *Lagrange multipliers*. The solution occurs at $\nabla\mathcal{L}(\mathbf{x}, \boldsymbol{\lambda})=\mathbf{0}$, namely:

$$\nabla\mathcal{L}(\mathbf{x}, \boldsymbol{\lambda}) = \begin{pmatrix} \frac{\partial}{\partial x_1}f(\mathbf{x}) - \sum_{i=1}^{k}\lambda_i a_{i1} \\ \vdots \\ \frac{\partial}{\partial x_n}f(\mathbf{x}) - \sum_{i=1}^{k}\lambda_i a_{in} \\ -(a_{11}x_1 + a_{12}x_2 + \cdots + a_{1n}x_n - b_1) \\ \vdots \\ -(a_{k1}x_1 + a_{k2}x_2 + \cdots + a_{kn}x_n - b_k) \end{pmatrix} = \begin{pmatrix} 0 \\ \vdots \\ 0 \\ 0 \\ \vdots \\ 0 \end{pmatrix}. \quad (2.K.4)$$

This non-linear system can be solved using *Newton's method* which approximates $\nabla\mathcal{L}(\mathbf{x}, \boldsymbol{\lambda})$ linearly in the neighborhood of $\binom{\mathbf{x}^0}{\boldsymbol{\lambda}^0}$ as $\nabla\mathcal{L}(\mathbf{x},\boldsymbol{\lambda})\cong\nabla\mathcal{L}(\mathbf{x}^0,\boldsymbol{\lambda}^0)+\nabla^2\mathcal{L}(\mathbf{x}^0,\boldsymbol{\lambda}^0)\left[\binom{\mathbf{x}}{\boldsymbol{\lambda}}-\binom{\mathbf{x}^0}{\boldsymbol{\lambda}^0}\right]$. Setting the approximation=0 and solving yields $\binom{\mathbf{x}^1}{\boldsymbol{\lambda}^1}=\binom{\mathbf{x}^0}{\boldsymbol{\lambda}^0}-[\nabla^2\mathcal{L}(\mathbf{x}^0,\boldsymbol{\lambda}^0)]^{-1}\nabla\mathcal{L}(\mathbf{x}^0,\boldsymbol{\lambda}^0)$. Repeating the process generates the iterative solution

[21] A multi-step solution would write $\mathbf{Ax}=\mathbf{b}$ as $\mathbf{A}_1\mathbf{x}_1 + \mathbf{A}_2\mathbf{x}_2 = \mathbf{b}$ with $\mathbf{A}_1$ k×k and rank($\mathbf{A}_1$)=k then solve $\mathbf{x}_1 = \mathbf{A}_1^{-1}(\mathbf{b} - \mathbf{A}_2\mathbf{x}_2)$ and replace $\mathbf{x}_1$ in the objective, $f(\mathbf{x})$, making it an unconstrained function of $\mathbf{x}_2$ only. After solving $f'(\mathbf{x}_2^*) = \mathbf{0}$ we use the above to determine $\mathbf{x}_1^*$.

$\binom{\mathbf{x}^{i+1}}{\boldsymbol{\lambda}^{i+1}}$ from $\binom{\mathbf{x}^{i}}{\boldsymbol{\lambda}^{i}}$ and convergence occurs when $|\nabla\mathscr{L}(\mathbf{x}^*,\boldsymbol{\lambda}^*)|<\boldsymbol{\varepsilon}$. The (n+k)×(n+k) symmetric matrix of 2nd derivatives $\nabla^2\mathscr{L}(\mathbf{x},\boldsymbol{\lambda})$ is called the *bordered Hessian* which is given by:

$$\nabla^2\mathscr{L}(\mathbf{x},\boldsymbol{\lambda}) = \begin{bmatrix} \frac{\partial^2}{\partial x_1^2}f(\mathbf{x}) & \cdots & \frac{\partial^2}{\partial x_1 \partial x_n}f(\mathbf{x}) & -a_{11} & \cdots & -a_{k1} \\ \vdots & \ddots & \vdots & \vdots & \ddots & \vdots \\ \frac{\partial^2}{\partial x_n \partial x_1}f(\mathbf{x}) & \cdots & \frac{\partial^2}{\partial x_n^2}f(\mathbf{x}) & -a_{1n} & \cdots & -a_{kn} \\ -a_{11} & \cdots & -a_{1n} & 0 & \cdots & 0 \\ \vdots & \ddots & \vdots & \vdots & \ddots & \vdots \\ -a_{k1} & \cdots & -a_{kn} & 0 & \cdots & 0 \end{bmatrix} = \begin{bmatrix} \mathbf{H} & -\mathbf{A}' \\ -\mathbf{A} & \mathbf{0} \end{bmatrix}, \quad (2.K.5)$$

where $\mathbf{H}$ is the Hessian of $f(\mathbf{x})$. Newton's method uses $[\nabla^2\mathscr{L}(\mathbf{x},\boldsymbol{\lambda})]^{-1}$ thus (2.K.5) must be invertible. Note that $\nabla^2\mathscr{L}(\mathbf{x},\boldsymbol{\lambda})\binom{\mathbf{y}}{\mathbf{z}}=\mathbf{0}$ implies $\mathbf{Hy}-\mathbf{A}'\mathbf{z}=\mathbf{0}$ and $\mathbf{Ay}=\mathbf{0}$ thus $\mathbf{y}'\mathbf{Hy}=\mathbf{0}$. Since $f(\mathbf{x})$ is convex, $\mathbf{H}$ is +definite and $\mathbf{y}'\mathbf{Hy}=\mathbf{0}$ implies $\mathbf{y}=\mathbf{0}$, thus $\mathbf{A}'\mathbf{z}=\mathbf{0}$. (Same for $-f(\mathbf{x})$ concave, $\mathbf{H}$ –definite.) Since redundant constraints have been removed, $\mathbf{A}'$ is n×k with rank($\mathbf{A}'$)=k and $\mathbf{A}'\mathbf{z}=\mathbf{0}$ implies $\mathbf{z}=\mathbf{0}$. Therefore when $f(\mathbf{x})$ is convex [$-f(\mathbf{x})$ is concave], $\nabla^2\mathscr{L}(\mathbf{x},\boldsymbol{\lambda})\binom{\mathbf{y}}{\mathbf{z}}=\mathbf{0}$ implies $\binom{\mathbf{y}}{\mathbf{z}}=\mathbf{0}$ and the bordered Hessian is full-rank thus invertible (Border, 2013).

*K.4 General Non-Linear Programming*

A general non-linear program (NLP) seeks to minimize or maximize a smooth function $f(\mathbf{x})$, $\mathbf{x} \in \mathbb{R}^n$, subject to $g^i(\mathbf{x})\leq b_i$, i=1,2,...,k, where $f(\mathbf{x})$ is not necessarily convex or concave and $g^i(\mathbf{x})$ is generic. Such problems can have several local optimums and the goal is to find the best among these. Little can be said about NLP problems in general and the optimization strategy depends on the nature of the problem. In some cases the Lagrangian can be used to find local optimums. In others a *metaheuristic*, such as *tabu search*, *simulated annealing*, or a *genetic algorithm* can be used (Hillier & Lieberman, 2010). If all else fails we can generate random values and evaluate $f(\cdot)$ when the constraint set is satisfied, keeping a record of the optimal value. A better approach would generate random values that satisfy all constraints. Alternatively, we can take a random setting that is infeasible and project it to a point inside the feasible region then evaluate $f(\cdot)$ at that point.[22] Random starts can be effective when many local optimums exist and strategies for generating values have been developed for specific problems such as mixture likelihoods (see McLachlan & Peel, 2000).

## *L. Copula Modeling*

Let X be any continuous RV with PDF $f(x)$ and CDF $F(x)=P(X\leq x)=\int_{-\infty}^{x} f(t)dt$. Copula modeling is based on a fact that initially surprises, but is intuitive upon reflection. Namely that F(X)~uniform(0,1). The proof is straightforward, let U=F(X) then $F_U(u)=P(U\leq u)=P(F(X)\leq u)=P(X\leq F^{-1}(u))=F(F^{-1}(u))=u$, for $0\leq u\leq 1$. Random variables having the same CDF are identically distributed and the CDF for U is from a uniform(0,1) distribution. Let $\mathbf{X}=(X_1,...,X_N)'$ be RVs for the compounding return on N financial securities at a given time point. The marginal PDF and CDF of $X_i$ are $f_i(x_i)$ and $F_i(x_i)$ respectively, and the multivariate PDF and CDF of $\mathbf{X}$ are $f(\boldsymbol{x})$ and $F(\boldsymbol{x})=P(\mathbf{X}\leq\boldsymbol{x})=F(x_1,...,x_N)=P(X_1\leq x_1 \cap ... \cap X_N\leq x_N)$ respectively, see §**II.A**. As above, let $U_i=F_i(X_i)$ where $U_i\sim$uniform(0,1) and $X_i=F_i^{-1}(U_i)$. The CDF of $\mathbf{U}=(U_1,...,U_N)'$ is $G(\boldsymbol{u})=P(\mathbf{U}\leq\boldsymbol{u})=G(u_1,...,u_N)=P(U_1\leq u_1 \cap ... \cap U_N\leq u_N)$. Since $G(\boldsymbol{u})$ is a valid CDF, its derivative is the multivariate PDF of $\mathbf{U}$, namely $\frac{\partial^N}{\partial u_1 ... \partial u_N}[G(\boldsymbol{u})]=g(\boldsymbol{u})-g(\mathbf{0})=g(\boldsymbol{u})$, see §**II.A**. Note the relationship between $F(\boldsymbol{x})$ and $G(\boldsymbol{u})$ (Nelson, 2006):

$$F(\boldsymbol{x}) = F(x_1,...,x_N) = P(X_1\leq x_1 \cap ... \cap X_N\leq x_N) \tag{2.L.1}$$

$$= P(F_1^{-1}(U_1)\leq x_1 \cap ... \cap F_N^{-1}(U_N)\leq x_N) \tag{2.L.2}$$

$$= P(U_1\leq F_1(x_1) \cap ... \cap U_N\leq F_N(x_N)) \tag{2.L.3}$$

$$= G(F_1(x_1),...,F_N(x_N)). \tag{2.L.4}$$

[22] Consider maximizing a generic likelihood function that includes the VC matrix V($\mathbf{X}$) from (2.I.1). A constraint on the variances and covariances is that V($\mathbf{X}$) must be +definite. An alternative to discarding any point with a broken V($\mathbf{X}$) matrix is to repair it.

The multivariate PDF of $\mathbf{X}$ can then be derived by differentiating $F(x_1,...,x_N)$, using the chain rule on $F_i(x_i)$ as:

$$f(\boldsymbol{x}) = f(x_1,...,x_N) = \frac{\partial^n}{\partial x_1 ... \partial x_N}[F(\boldsymbol{x})] \tag{2.L.5}$$

$$= \frac{\partial^n}{\partial x_1 ... \partial x_N}[G(F_1(x_1),...,F_N(x_N))] \qquad \text{[from (2.L.4)]} \tag{2.L.6}$$

$$= \frac{\partial}{\partial x_1}\left[\int_0^{F_1(x_1)} ... \left[\frac{\partial}{\partial x_N}\int_0^{F_N(x_N)} g(t_1,...,t_N)\,dt_N\right]...\,dt_1\right] \tag{2.L.7}$$

$$= g\big(F_1(x_1),...,F_N(x_N)\big)f_1(x_1)\times ... \times f_N(x_N) \tag{2.L.8}$$

Multivariate PDF | Multivariate Copula PDF | Product of Marginal PDFs

The literature refers to $G(\cdot)$ as the *copula* and $g(\cdot)$ as the *copula density*. The term indicates a *coupling* between the multivariate and marginal PDFs for a set of RVs (Nelson, 2006). When $X_1,...,X_N$ are independent $U_1,...,U_N$ are also independent and $g(\boldsymbol{u})=1\times...\times1=1$ so that $f(x_1,...,x_N)=f_1(x_1)\times...\times f_N(x_N)$. The copula term therefore models the dependence between a set of RVs, with the breakthrough being that marginal PDFs can be modeled separately. This quality is appealing, particularly for the field of retirement finance. When modeling the multivariate PDF for a set of real compounding returns, the marginal return on a given security should not depend on which other securities are involved. For this reason copula modeling is a standard for multivariate PDF modeling in finance and many forms have been proposed. For example if $g(\cdot)$ is Gaussian-induced then the copula will model dependence after mapping to normal RVs. Since the unknown parameters in a copula exist in the likelihood they can be estimated as MLEs. When building a multivariate PDF we can choose between candidates copulas by taking the one with smallest *root-mean-square error* using the empirical copula, which is constructed via the empirical CDF. In this research we propose a generic fixed-marginal tractable alternative to copula modeling that is well suited to retirement finance.

Assuming marginal PDFs $f_1(x_1),...,f_n(x_n)$ have been arbitrarily fit to the data set, a distribution with corresponding CDF H(·) is selected to model the dependence structure of RVs $X_1,...,X_n$. For example, during the housing boom securitized mortgage products modeled default times of named residential borrowers as exponential RVs, see §**I**. Here, $f_i(x_i)$ would be an exponential PDF for i=1,...,n. Further, the dependence structure was chosen as Gaussian so that H(·) becomes Φ(·). The multivariate PDF in (2.L.8) requires g(·) for full specification, and we say that it is *induced* by the choice of H(·) using transformations $U_1=H_1(Y_1),...,U_n=H_n(Y_n)$, where $H_j(\cdot)$ is the corresponding marginal CDF for element j. Following the standard procedure for transforming RVs (Freund, 1992), the multivariate PDF g(·) has form:

$$\mathrm{g}(u_1,\dots,u_\mathrm{n}) = h\left(\mathrm{H}_1^{-1}(u_1),\dots,\mathrm{H}_\mathrm{n}^{-1}(u_\mathrm{n})\right)\cdot\begin{vmatrix}\frac{\partial}{\partial u_1}[y_1] & \cdots & \frac{\partial}{\partial u_\mathrm{n}}[y_1]\\ \vdots & \ddots & \vdots\\ \frac{\partial}{\partial u_1}[y_\mathrm{n}] & \cdots & \frac{\partial}{\partial u_\mathrm{n}}[y_\mathrm{n}]\end{vmatrix}\ ,\ for\ u_\mathrm{i}\in[0,1] \tag{2.L.9}$$

where $h(\cdot)=\mathrm{H}'(\cdot)$ is the corresponding PDF for our chosen dependence structure. All off-diagonals in the Jacobian term are zero since each transformation involves only one RV. The diagonal terms are derived for i=1,...,n as follows:

$$\frac{du_\mathrm{i}}{dy_\mathrm{i}} = H_\mathrm{i}'(y_\mathrm{i}) = h_\mathrm{i}(y_\mathrm{i}) \tag{2.L.10}$$

Using rules from calculus, we can treat $\frac{du_\mathrm{i}}{dy_\mathrm{i}}$ as a ratio when $H_i(\cdot)$ is invertible, thus:

$$\frac{dy_\mathrm{i}}{du_\mathrm{i}} = \frac{1}{h_\mathrm{i}(y_\mathrm{i})} = \frac{1}{h_\mathrm{i}\left(\mathrm{H}_\mathrm{i}^{-1}(u_\mathrm{i})\right)} \tag{2.L.11}$$

The copula density g(·) thus takes form of (2.L.12) when induced by the dependence structure of CDF H(·):

$$\mathrm{g}(u_1,\dots,u_\mathrm{n}) = h\left(\mathrm{H}_1^{-1}(u_1),\dots,\mathrm{H}_\mathrm{n}^{-1}(u_\mathrm{n})\right)\cdot\prod_{\mathrm{i}=1}^{\mathrm{n}}\frac{1}{h_\mathrm{i}\left(\mathrm{H}_\mathrm{i}^{-1}(u_\mathrm{i})\right)}\ ,\ for\ u_\mathrm{i}\in[0,1] \tag{2.L.12}$$

When H(·) is Gaussian, $h(\cdot)=\phi(\cdot)$ and $h_i(\cdot)=H_i'(\cdot)$ are univariate standard normal PDFs. The copula density g(·) from (2.L.12) is used in (2.L.8) to complete the multivariate PDF of our sample data, $f(x_1,...,x_N)$. Note that Sklar's Theorem guarantees the PDF in (2.L.8) has marginals $f_i(x_i)$ when using g(·) (Nelson, 2006). Copula parameters can then be estimated as MLEs with only covariance terms unknown, since it is assumed that the univariate marginals are fully specified. Tran, et al., (2013) caution that a straight forward optimization of (2.L.8) may work for basic copula forms, but can fail when using commercial software and a complicated dependence structure. Giordoni, et al., (2009) solve a problem similar to the one we address, but in different ways. Namely, using a normal mixture copula along with normal mixture marginals, and also with a non-copula adaptive estimator that attempts to smooth out the differences between normal mixture marginals and the implied marginals of a multivariate normal mixture fit directly to the data.

### *M. Information Criteria*

In addition to the LRT from §**II.H**, fitted models in statistics can be compared using a metric called the *information criteria* (IC). Such values quantify the information lost by a model vs. the true state of nature. Smaller IC values are preferred as they indicate less information loss. Models with too many parameters are said to lack *parsimony*. When choosing amongst candidate models, IC metrics attempt to strike a balance between the likelihood value and # of parameters. Among the most widely used IC metrics is *Akaike's information criteria* (AIC), calculated as (Akaike, 1974):

$$\mathbf{AIC} = 2\times(\text{\# Parameters}) - 2\times\ln(\mathfrak{L}(\widehat{\boldsymbol{\theta}}|\boldsymbol{x})) \quad (2.M.1)$$

The parameters counted must be free. A model with $\pi_1$ and $\pi_2$ constrained linearly by $\pi_1+\pi_2=1$ only counts as 1 parameter since estimating $\pi_1$ estimates $\pi_2$. Thus, (# parameters) = (total # parameters) – (# independent constraints). The term $\ln(\mathfrak{L}(\widehat{\boldsymbol{\theta}}|\boldsymbol{x}))$ is the log-likelihood using the MLE, see §**II.E**. The AIC works well when interest is in controlling Type I Errors in subsequent hypothesis tests (Tao, et al., 2002), but it tends to over fit in small samples leading to models that lack parsimony. Thus the *corrected AIC* (AICC) has been proposed and includes a penalty that increases with the # of parameters (Hurvich & Tsai, 1989):

$$\mathbf{AICC} = \mathbf{AIC} + \frac{2\times(\text{\# Parameters}+1)\times(\text{\# Parameters}+2)}{(\text{Sample Size})-(\text{\# Parameters})-2} \quad (2.M.2)$$

As the sample size increases the penalty decreases, thus AICC≈AIC. Whereas α=P(Type I Error)=P(Reject Ho|Ho is True), β=P(Type II Error)=P(Accept Ho|Ho is False). The *power* of a hypothesis test is 1−β= P(Reject Ho|Ho is False). If interest is in controlling the power of subsequent hypothesis tests then *Bayesian information criteria* (BIC) is recommended (Tao, et al., 2002), and calculated as (Schwarz, 1978):

$$\mathbf{BIC} = -2\times\ln(\mathfrak{L}(\widehat{\boldsymbol{\theta}}|\boldsymbol{x})) + (\text{\# Parameters})\times\ln(\text{Sample Size}) \quad (2.M.3)$$

While determining the optimal # of components in a finite mixture PDF (see §**II.B**) is an unsolved problem in statistics, IC are often used as a heuristic to compare mixture PDFs of various sizes. Titterington, et al., (1985) caution that, in theory, their validity often relies on the unmet regularity conditions of §**II.G.**

### *N. The Probability of Ruin in Retirement*

Let t=1,2,...,T be the time points of a retirement horizon where the 1st withdrawal is made at time t=1 and the last withdrawal at time t=T, which can be fixed ($T_F$) or random ($T_R$). The PMF for $T_R$ is defined as $P(T_R=t)$, for t=0,1,...,T, and can be derived using lifetables published at SSA.gov for an individual or a group (Rook, 2014). The safe withdrawal rate (SWR) is a heuristic that suggests retirees withdraw (100%)$W_R$ in real terms from their savings at each time point (Bengen, 1994). A retirement plan often couples the withdrawal rate ($W_R$) with an asset allocation. If N securities are involved let $R_{ti}$, $r_{ti}$, $E_i$, $\alpha_{ti}$, and $I_t$ be the total and real returns, expense ratio, proportion allocated to security i, and inflation rate, respectively, all at time t. The *total compounding return* for security i at time t is:

$$(1+R_{ti})=(1+r_{ti})(1+I_t) \qquad (2.N.1)$$

The *expense-adjusted total compounding return* for security i at time t is:

$$(1-E_i)(1+R_{ti})=(1-E_i)(1+r_{ti})(1+I_t) \qquad (2.N.2)$$

The *expense-adjusted real compounding return* for security i at time t, denoted $\hat{r}_{ti}$, is:

$$\hat{r}_{ti}=(1-E_i)(1+r_{ti})=(1-E_i)(1+R_{ti})/(1+I_t) \qquad (2.N.3)$$

Since $\hat{r}_{ti}$ is a continuous RV it is governed by a univariate PDF, say $f_{ti}(\hat{r}_{ti})$. When $Cov(r_{ki},r_{sj})=0$ for times $k \neq s=1,...,T$ and securities $i,j=1,...,N$, then $\hat{r}_{ti}$ and $f_{ti}(\hat{r}_{ti})$ are independent of time. If we drop the time index they become $\hat{r}_i$ and $f_i(\hat{r}_i)$, respectively. The marginal PDF for security i, $f_i(\hat{r}_i)$, is modeled using historical data.[23] A retirement plan succeeds or fails based on the return of a diversified portfolio, not that of a single security. The *expense-adjusted real compounding return* for the portfolio at time t is:

$$\hat{r}(t)=(\alpha_{t-1,1})\hat{r}_1+...+(\alpha_{t-1,N})\hat{r}_N, \qquad (2.N.4)$$

where $\alpha_{t-1,1}+...+\alpha_{t-1,N}=1$. Consequently, $\hat{r}(t)$ is a function of time via the portfolio weights (i.e., the asset allocation) set at time t-1 and is derived as a linear transform ($\mathbb{R}^N \rightarrow \mathbb{R}^1$) of $\hat{\mathbf{r}}=(\hat{r}_1,...,\hat{r}_N)'$. The univariate PDF for $\hat{r}(t)$, say $h_t(\hat{r})$, is used to evaluate/optimize a retirement plan. In some cases this PDF is easily derived (i.e., normal), and in others there is no solution (i.e., lognormal). Various methods exist to derive the PDF of a transformed RV and one uses the multivariate PDF of the random vector $\hat{\mathbf{r}}=(\hat{r}_1,...,\hat{r}_N)'$, $f(\hat{\mathbf{r}})$ (Freund, 1992). Our goal is to model $h_t(\hat{r})$ using $f(\hat{\mathbf{r}})$, while maintaining the individual security marginals, that is, subject to $f_i(\hat{r}_i)=\int_{\neq i}^{\cdots}\int f(\hat{r}_1,\ldots,\hat{r}_N)d_{\hat{r}_1}...d_{\hat{r}_N}$. Here, $h_t(\hat{r})$ may be skewed, heavy-tailed and multimodal (i.e., generally non-normal) allowing higher PDF moments to aid in determining a plan's success or failure.

*N.1 Retirement Surveys and Alternative Metrics*

Retiree surveys reveal that the #1 concern is running out of money.[24] A retiree who runs out of money experiences financial ruin. The probability of this event occurring can be computed and shared with the retiree for a given decumulation strategy. Probabilities are bounded by [0,1] and multiplying by 100% yields a percentage which is bounded by [0, 100]. Percentages are a ubiquitous metric and universally understood. For example, a probability of 0.5 translates to 50% which can be described in words as a "coin flip". Retirees make withdrawals at time t and they experience the event of ruin at time t, denoted Ruin(t), *iff* the time t-1 withdrawal is successful but the account does not support, or is completely emptied by the time t withdrawal. The compliment of this event is avoiding ruin at time t, denoted $Ruin^C(t)$, which occurs *iff* the withdrawal at time t is successful, leaving a >$0 balance. Define Ruin(≤t) as the event of ruin occurring on or before time t and let $Ruin^C(\leq t)$ be its compliment.

[23] The tools described in §**II.J** can be used to detect serial correlation within and between securities over time.
[24] A number are referenced at the conclusion of this research.

Despite being extensively researched, the *probability of ruin* as a metric is not universally accepted with criticisms leveled from all directions. The most common being that the retirement ruin event could occur as Ruin(1) or Ruin(30), and there is a substantial real-life difference for the retiree between these events. This criticism argues that ruin as a binary outcome is too simplistic and a more nuanced approach would consider varying degrees of failure. A separate criticism is that the ruin metric is overly complicated and better left to actuaries at insurance companies. Under this argument, the metric is misunderstood and being abused by financial planners who lack the ability to properly calibrate the computation and/or fail to understand its inherent flaws such as the impact of covariances and higher order PDF moments.

While a retirement strategy can have varying degrees of failure, we are primarily interested in the compliment of the ruin event, which is success. Unlike retirement ruin, the event of retirement success does not have varying time point-degrees attached to it. Further, a decumulation model that maximizes the probability of success will also minimize the probability of ruin as these are equivalent optimization problems. None-the-less, a model that maximizes the probability of success may in fact fail, and in this case it is reasonable to try and limit the damage. Harlow and Brown (2016) introduce two *downside risk* metrics which do precisely this. Their approach uses fully stochastic discounting to compute a retirement present value (RPV) for withdrawals (cash flows) from a decumulation plan. The RPV is an RV and its PDF can be estimated via simulation. Values of RPV below zero indicate the account did not support all withdrawals and retirement ruin has occurred. A strategy that minimizes downside risk recognizes that the ruin section of the RPV's PDF can be markedly different for retirement plans having similar failure probabilities. The goal is to make this section of the PDF as palatable as possible if ruin occurs. Both the mean and standard deviation of negative RPV values are used as minimization metrics in the optimization and corresponding asset allocations are found. Harlow and Brown (2016) report that far lower equity ratios are optimal in the context of minimizing downside risk. This finding has the benefit of being intuitive as we can generally think of a retiree's bequest distribution as having a spread (variance) that increases with the equity ratio and a negative bequest (i.e., RPV<0) indicates that the retiree has exhausted their savings while still alive.

Milevsky (2016) takes the opposing view that ruin probabilities are being routinely abused and/or misunderstood by retirement planners, and advocates for replacing it altogether by a different metric, namely the *portfolio longevity* ($P_L$). Since investments are volatile $P_L$ is an RV that measures the length of time a retirement portfolio lasts. It takes values $\ell=0,1,2,...,\infty$ in discrete time and has PMF defined by $P(P_L=\ell)$. Note that $(\ell=0)\leftrightarrow$(0 successful withdrawals are made)$\leftrightarrow$Ruin(1), $(\ell=1)\leftrightarrow$(only 1 successful withdrawal is made)$\leftrightarrow$Ruin(2), ..., $(\ell=T\text{-}1)\leftrightarrow$(exactly T-1 successful withdrawals are made)$\leftrightarrow$Ruin(T). Finally, $(\ell\geq T)\leftrightarrow$ (All T withdrawals are made successfully)$\leftrightarrow \text{Ruin}^C(\leq T)\equiv$ Retirement Success (given horizon length T). Therefore, $P(P_L=\ell)=P[\text{Ruin}(\ell+1)]$ for $\ell=0,1,2,...,T\text{-}1$ and $P(\ell\geq T)=P[\text{Ruin}^C(\leq T)]$, see Table III. The mean, median and mode of $P_L$ will thus be functions of ruin probabilities, and any flaws inherent in their construction will propagate through to these statistics.

***Table III**. The Portfolio Longevity ($P_L$) PMF and Corresponding Statistics*

| Portfolio Longevity ($P_L=\ell$) | $P(P_L=\ell)$ | Mean ($\mu=E[P_L]$) | Median[1] | Mode[2] |
|---|---|---|---|---|
| 0 | P[Ruin(1)] | 0 | | |
| 1 | P[Ruin(2)] | 1*P[Ruin(2)] | | |
| 2 | P[Ruin(3)] | 2*P[Ruin(3)] | | |
| ⋮ | ⋮ | ⋮ | | |
| i | P[Ruin(i+1)] | i*P[Ruin(i+1)] | | X |
| ⋮ | ⋮ | ⋮ | | |
| j | P[Ruin(j+1)] | j*P[Ruin(j+1)] | X | |
| ⋮ | ⋮ | ⋮ | | |
| k | P[Ruin(k+1)] | k*P[Ruin(k+1)] | | X |
| ⋮ | ⋮ | ⋮ | | |
| | $\sum = 1$ | $\mu = E[P_L] = \sum$ | $\sum \geq 0.5$ | $\text{argmax}[P(P_L=\ell)]$ |

[1] Sum probabilities in either direction and stop when ≥ 0.5 The corresponding $\ell$-value is the median, shown as j above. If the sum is exactly 0.5 at $\ell$=j then median($P_L$)=(2j+1)/2.

[2] Locate the maximum probability(s) $P(P_L=\ell)$. All corresponding $\ell$-values are the mode(s), shown above as mode($P_L$)={i,k}.

Table III applies to any non-negative withdrawal rate ($W_R$), and $\sum_{t=1}^{\infty} P[\text{Ruin}(t)]=1$ implies that no investment account lasts in perpetuity, including when $W_R=0$. It is suggested that financial advisors examine the event $P_L<T_R$ as it implies the retiree outlives their savings. The probability of this event, $P(P_L<T_R)$, is the probability of ruin (with $T=T_R$). We can compute $P(P_L<T_R)$ using conditional probabilities as follows:

$$P_L<T_R \equiv (P_L < T_R) \cap S \quad \text{[where S is any generic sample space, i.e., P(S)=1]} \tag{2.N.5}$$

$$\equiv (P_L < T_R) \cap (T_R = 0 \cup T_R = 1 \cup \cdots \cup T_R = T) \quad \text{[replace S by the sample space for } T_R\text{]}$$

$$\equiv [(P_L < T_R) \cap (T_R = 0)] \cup \cdots \cup [(P_L < T_R) \cap (T_R = T)] \quad \text{[distributive property for sets]}$$

$$\rightarrow P(P_L<T_R) = \sum_{t=0}^{T} P(P_L < T_R \cap T_R = t) \quad \text{[probabilities for mutually exclusive events are summed]}$$

$$= \sum_{t=0}^{T} P(P_L < T_R \mid T_R = t)\, P(T_R = t) \quad \text{[since P(A|B) = P(A} \cap \text{B)/P(B)]}$$

$$= \sum_{t=0}^{T} P(P_L < t \mid T_F = t)\, P(T_R = t) \quad \text{[conditional probability uses fixed horizon length, } T_F\text{]}$$

$$= \sum_{t=1}^{T} [1 - P(P_L \geq t \mid T_F = t)]\, P(T_R = t) \quad \text{[drop t=0 from prior step since } P_L \geq 0\text{]}$$

$$= \sum_{t=1}^{T} \left[1 - P[\text{Ruin}^C(\leq T_F) \mid T_F = t]\right] P(T_R = t) \quad \text{[compute using success probabilities]}$$

$$= \sum_{t=1}^{T} \left[1 - P[\text{Retirement Success} \mid T_F = t]\right] P(T_R = t) \tag{2.N.6}$$

As noted, $P(P_L<T_R)$ is, by definition, the probability of ruin using a random time horizon. With respect to the portfolio longevity ($P_L$), the probability said to be of most interest is derived entirely using ruin probabilities, as shown in (2.N.6) where probabilities for the RV $P_L$ do not appear.

*N.2 Computing Ruin Probabilities*

Assume the retiree has made successful withdrawals at times t=1,...,t-1. The event Ruin(t) occurs when $\hat{r}(t) \leq RF(t-1)$, where $RF(t)=RF(t-1)/[\hat{r}(t)-RF(t-1)]$ for t=1,...,T and $RF(0)=W_R$ (Rook, 2014). Thus, $Ruin^C(t)$ occurs when $\hat{r}(t)>RF(t-1)$. RF(t) is called the *ruin factor* and reflects the retiree's funded status at time t with 1/RF(t) equal to the # of real withdrawals remaining (Rook, 2014). The event of achieving retirement success using an SWR, is thus defined for fixed and random horizons, respectively, as:

$$Ruin^C(\leq T_F) \equiv (\text{Retirement Success} \mid T=T_F) \equiv \cap_{t=1}^{T_F}[\hat{r}(t) > RF(t-1)],\ \hat{r}(t) \sim h_t(\hat{r}) \quad (2.N.7)$$

$$Ruin^C(\leq T_R) \equiv (\text{Retirement Success} \mid T=T_R) \equiv \cap_{t=1}^{T_R}[\hat{r}(t) > RF(t-1)],\ \hat{r}(t) \sim h_t(\hat{r}) \text{ and } T_R \sim P(T_R=t) \quad (2.N.8)$$

Recall that $\hat{r}(t)$ is the expense-adjusted real compounding return on a diversified portfolio of N securities and a function of the asset allocation set at time t-1. It follows that the corresponding success probability for fixed $T=T_F$ in (2.N.7), assuming independence of $\hat{r}(t)$ across time, is ($x_t$ replaces $\hat{r}(t)$ as vbl of integration):

$$P[Ruin^C(\leq T_F)] = \int_{RF(0)}^{\infty} \cdots \int_{RF(T_F-1)}^{\infty} h_1(x_1) \ldots h_{T_F}(x_{T_F}) dx_{T_F} \ldots dx_1 \quad (2.N.9)$$

The success probability for random $T=T_R$ in (2.N.8) is $1-P[Ruin(\leq T_R)]$, where $P[Ruin(\leq T_R)]$ was derived in (2.N.6) above and also uses (2.N.9). For any events A and B, if $A \subset B$ then $P(B \cap A^C)=P(B)-P(A)$. Since $Ruin^C(\leq t) \subset Ruin^C(\leq t-1)$ and $Ruin(t) \equiv [Ruin^C(\leq t-1) \cap Ruin(\leq t)]$, it follows that $P[Ruin(t)]=P[Ruin^C(\leq t-1) \cap Ruin(\leq t)]=P[Ruin^C(\leq t-1)]-P[Ruin^C(\leq t)]$.[25] Consequently, for $\hat{r}(t) \sim h_t(\hat{r})$ and $t \leq T_F$:

$$\begin{aligned} P[Ruin(t)] &= P\left(\cap_{i=1}^{t-1}[\hat{r}(i) > RF(i-1)]\right) - P\left(\cap_{i=1}^{t}[\hat{r}(i) > RF(i-1)]\right) \\ &= \int_{RF(0)}^{\infty} \cdots \int_{RF(t-2)}^{\infty} h_1(x_1) \ldots h_{t-1}(x_{t-1})\, dx_{t-1} \ldots dx_1 - \int_{RF(0)}^{\infty} \cdots \int_{RF(t-1)}^{\infty} h_1(x_1) \ldots h_t(x_t)\, dx_t \ldots dx_1 \end{aligned} \quad (2.N.10)$$

Given values for the security weights ($\alpha_{t-1,1},\ldots,\alpha_{t-1,N}$) at time t (i.e., the asset allocation), the terms in (2.N.10) are estimated using simulation or approximated recursively with a dynamic program (DP), see Rook (2014) and Rook (2015). Subsequently, Table III can be populated with these probabilities. Usually we are not given the weights but are tasked with deriving them according to some optimality criteria. Rook (2014) derives the weights that minimize the probability of ruin for a stock and bond portfolio using a dynamic glidepath (for both $T_F$ and $T_R$), and Rook (2015) derives the corresponding weights to minimize the probability of ruin using a static glidepath. Both solutions assume normally distributed compounding returns and, as noted, many financial practitioners/researchers reject this assumption. The primary purpose of this research is to extend these and other models to non-normal compounding returns. It was noted above that investment accounts do not last forever regardless of the withdrawal rate ($W_R$). When $W_R=0$, $RF(t)=0\ \forall$ t=

[25] See Rook (2014) for corresponding Venn diagrams.

0,...,T. As T→∞, Ruin(t) occurs when $\hat{r}(t) \leq 0$ for any t. Consequently, if $P[\hat{r}(t) \leq 0] > 0$, the event of ruin will eventually occur under an infinite time horizon. Unfortunately, the lognormal PDF is defined for values ≥0 with $f(0)=0$ and it does not allow compounding returns of zero. The PDF we develop for $\hat{r}(t)$ assigns a >0 probability to the event $\hat{r}(t) \leq 0$ as compounding returns and prices of securities can and do take values of zero.

## III. Univariate Density Modeling

The expense-adjusted real compounding return on a diversified portfolio determines the success or failure of a retirement strategy (§**II.N**). We assume independence across time and use securities from Table I that are random samples.[26] The multivariate PDF for the expense-adjusted real compounding return on S&P 500 (L), Small Cap Equities (S), and U.S. 10-Year T-Bonds (B) will be developed in this research. The RVs representing these returns are $(L,S,B)'$ and the multivariate PDF is $f(l,s,b)$. A diversified portfolio using these securities generates time t expense-adjusted real compounding return $\hat{r}(t)=\alpha_{t-1L}(1-E_L)L+\alpha_{t-1S}(1-E_S)S+\alpha_{t-1B}(1-E_B)B$, where $\alpha_{t-1,L},\alpha_{t-1,S},\alpha_{t-1,B}$ are the portfolio weights (≥0) set at time t-1 with $\alpha_{t-1,L}+\alpha_{t-1,S}+\alpha_{t-1,B}=1$, and $E_L$, $E_S$, $E_B$ are the expenses. Similar to copula modeling (§**II.L**), we first build univariate PDFs $f_L(l)$, $f_S(s)$ and $f_B(b)$ for L, S and B, respectively. The multivariate PDF $f(l,s,b)$ (built in §**IV** and §**V**) will preserve the marginals, i.e., $f_L(l)=\int_{-\infty}^{\infty}\int_{-\infty}^{\infty} f(l,s,b)\,ds\,db$, $f_S(s)=\int_{-\infty}^{\infty}\int_{-\infty}^{\infty} f(l,s,b)\,dl\,db$ and $f_B(b)=\int_{-\infty}^{\infty}\int_{-\infty}^{\infty} f(l,s,b)\,dl\,ds$. Univariate PDFs for $(L,S,B)'$ are fit to finite normal mixtures using the EM algorithm with random starts and a variance ratio constraint to eliminate spurious maximizers (§**II.E**). A novel *forward-backward* procedure is introduced to find the optimal # of univariate components, generally considered an unsolved problem in statistics (§**II.M**). The *forward* portion tests 1 vs. 2 components using a bootstrapped LRT (§**II.H**), then (1 or 2) vs. 3 components, up to the maximum # of univariate components allowed. If the forward procedure ends with the last significant test being g components, the *backward* portion tests g vs. g-1 components, then g vs. g-2, etc..., until a significant difference is found ending the procedure. For example, if backward test g vs. g-k yields a significant difference then the optimal # of components is g-k+1. Note that Anderson-Darling normality tests for $(L,S,B)'$ yield p-values 0.7206, 0.0984 and 0.2607, respectively, indicating that the normality assumption is not rejected at α=0.05 for these securities. The fact that all 3 can be assumed to originate from univariate normal distributions should not be lost in the forthcoming analysis.

### *A. Univariate PDFs*

All univariate tests of g vs. g+k components will use 1,000 LRT bootstrap samples. Each sample fits data to both g and g+k component mixtures using 6,000*(g-1) random starts for each execution of the EM algorithm when g>1. Random starts use values generated from the nearest fitted mixture for the same data but with fewer components. Each LRT sample value thus requires 6,000*(2g+k-2) EM executions.

---

[26] Retirement research that uses serially correlated assets should account for the dependence in the multivariate PDF, otherwise valid doubts may be raised. This would include strategies that use certain short-term cash equivalents such as U.S. T-Bills (see Table I).

*A.1 Univariate PDF for S&P 500 (L): $f_L(l)$*

***Figure IV**. Annual Real S&P 500 Compounding Returns (L) Histogram*

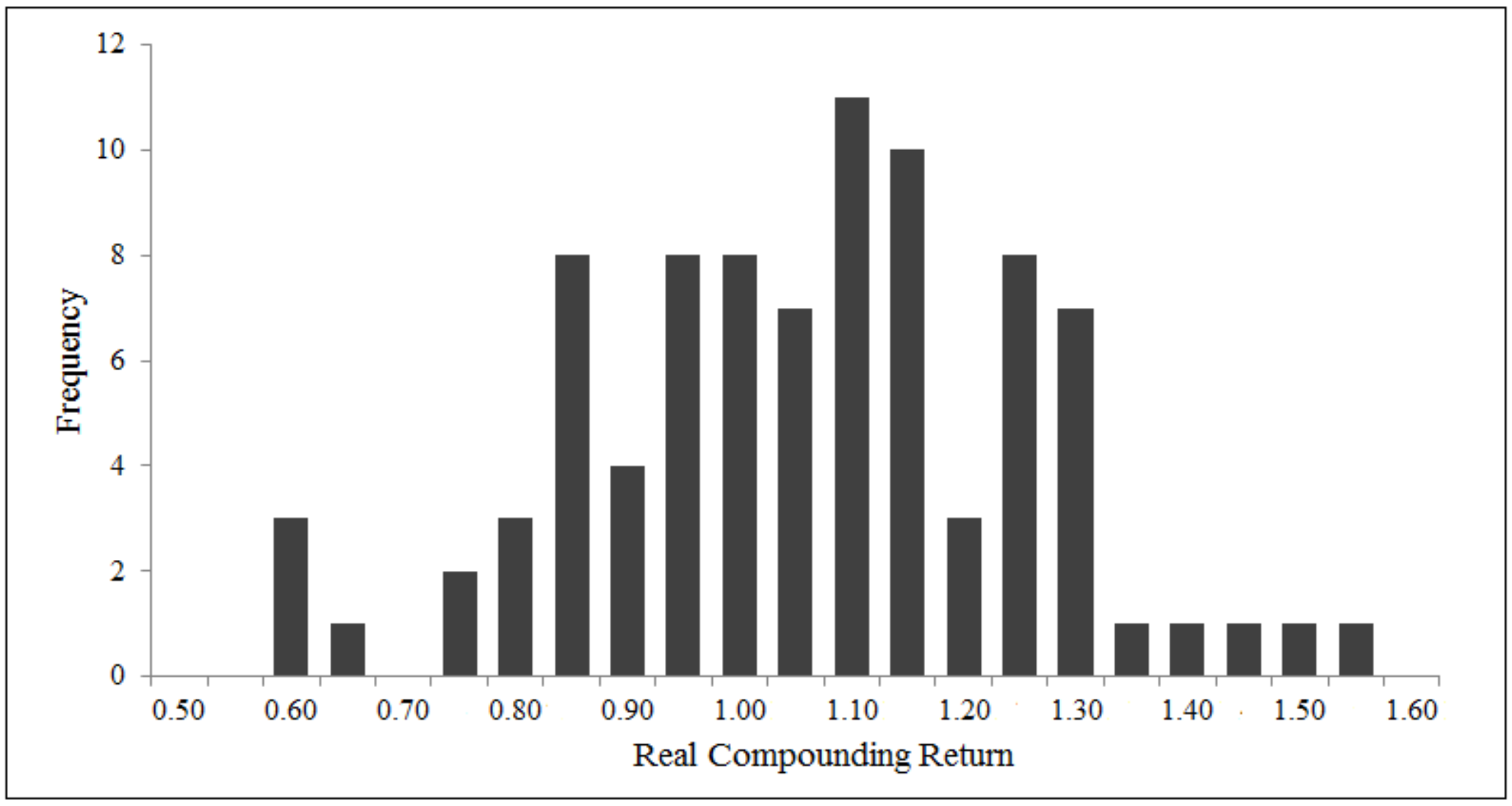

***Table IV**. Univariate Mixture PDFs for Annual Real S&P 500 Compounding Returns (L)*

| Number of Components | | | | | | | | | |
|---|---|---|---|---|---|---|---|---|---|
| 1 | | 2 | | 3 | | 4 | | 5 | |
| LL | 17.9019 | LL | 18.6383 | LL | 20.1144 | LL | 25.1860 | LL | 26.7351 |
| VR | 1.0000 | VR | 46.7606 | VR | 68.4102 | VR | 84.6646 | VR | 42.6613 |
| AIC | -31.8038 | AIC | -27.2766 | AIC | -24.2289 | AIC | -28.3721 | AIC | -25.4702 |
| AICC | -31.5181 | AICC | -26.2396 | AICC | -21.9211 | AICC | -24.2120 | AICC | -18.8035 |
| BIC | -26.8491 | BIC | -14.8899 | BIC | -4.4102 | BIC | -1.1214 | BIC | 9.2125 |
| μ(L) | 1.082 | μ(L) | 1.082 | μ(L) | 1.082 | μ(L) | 1.082 | μ(L) | 1.082 |
| σ(L) | 0.1974 | σ(L) | 0.1974 | σ(L) | 0.1974 | σ(L) | 0.1974 | σ(L) | 0.1974 |
| γ(L) | 0.00000 | γ(L) | -0.10701 | γ(L) | -0.26986 | γ(L) | -0.20207 | γ(L) | -0.14463 |
| κ(L) | 3.00000 | κ(L) | 2.87231 | κ(L) | 3.02576 | κ(L) | 2.76519 | κ(L) | 2.98090 |

| π | μ | σ | π | μ | σ | π | μ | σ | π | μ | σ | π | μ | σ |
|---|---|---|---|---|---|---|---|---|---|---|---|---|---|---|
| 1.000 | 1.082 | 0.197 | 0.056 | 1.287 | 0.029 | 0.833 | 1.055 | 0.204 | 0.092 | 0.998 | 0.016 | 0.114 | 0.998 | 0.017 |
| | | | 0.944 | 1.070 | 0.197 | 0.078 | 1.286 | 0.029 | 0.708 | 1.170 | 0.149 | 0.045 | 0.630 | 0.022 |
| | | | | | | 0.089 | 1.157 | 0.025 | 0.155 | 0.860 | 0.046 | 0.191 | 0.864 | 0.049 |
| | | | | | | | | | 0.045 | 0.629 | 0.022 | 0.037 | 1.505 | 0.064 |
| | | | | | | | | | | | | 0.613 | 1.174 | 0.110 |

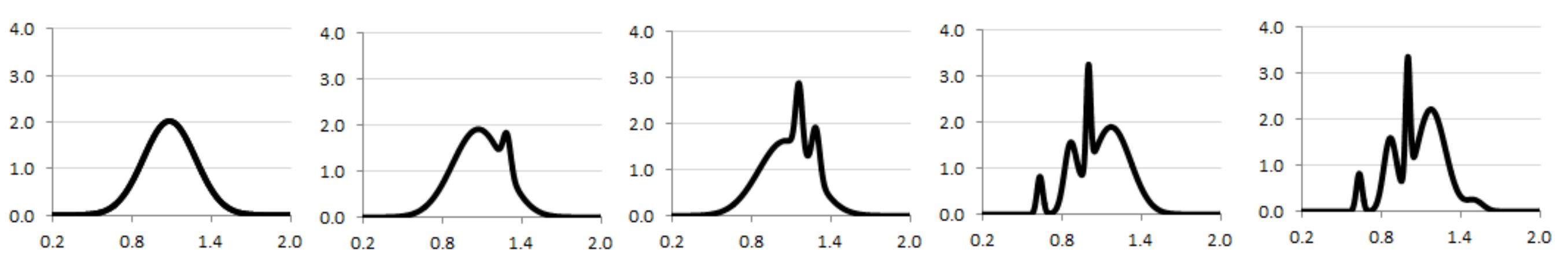

Abbreviations: LL=log-likelihood, VR=Variance Ratio, γ=skewness, κ=kurtosis.[27] See §**II.M** for AIC, AICC, BIC.

[27] Skewness(X) = $\gamma(X) = E[((X-\mu)/\sigma)^3] = [E(X^3) - 3\mu E(X^2) + 2\mu^3]/\sigma^3 = 0$ for the normal PDF, thus $E(X^3) = 3\mu E(X^2) - 2\mu^3$. Kurtosis(X) = $\kappa(X) = E[((X-\mu)/\sigma)^4] = [E(X^4) - 4\mu E(X^3) + 6\mu^2 E(X^2) - 3\mu^4]/\sigma^4 = 3$ for the normal PDF, thus $E(X^4) = 3\sigma^4 + 4\mu E(X^3) - 6\mu^2 E(X^2) + 3\mu^4$. (See, Hogg, et al., 2005 for definitions and Johnson, et al., 1994 for moment details.) Note that higher moments such as these can be difficult to interpret in multimodal distributions.

*__Figure V__. LRT Sampling Distribution for Testing the Optimal # of Univariate Mixture Components*
*Annual Real S&P 500 Compounding Returns (L)*

**Forward ↓** | **Backward ↑**

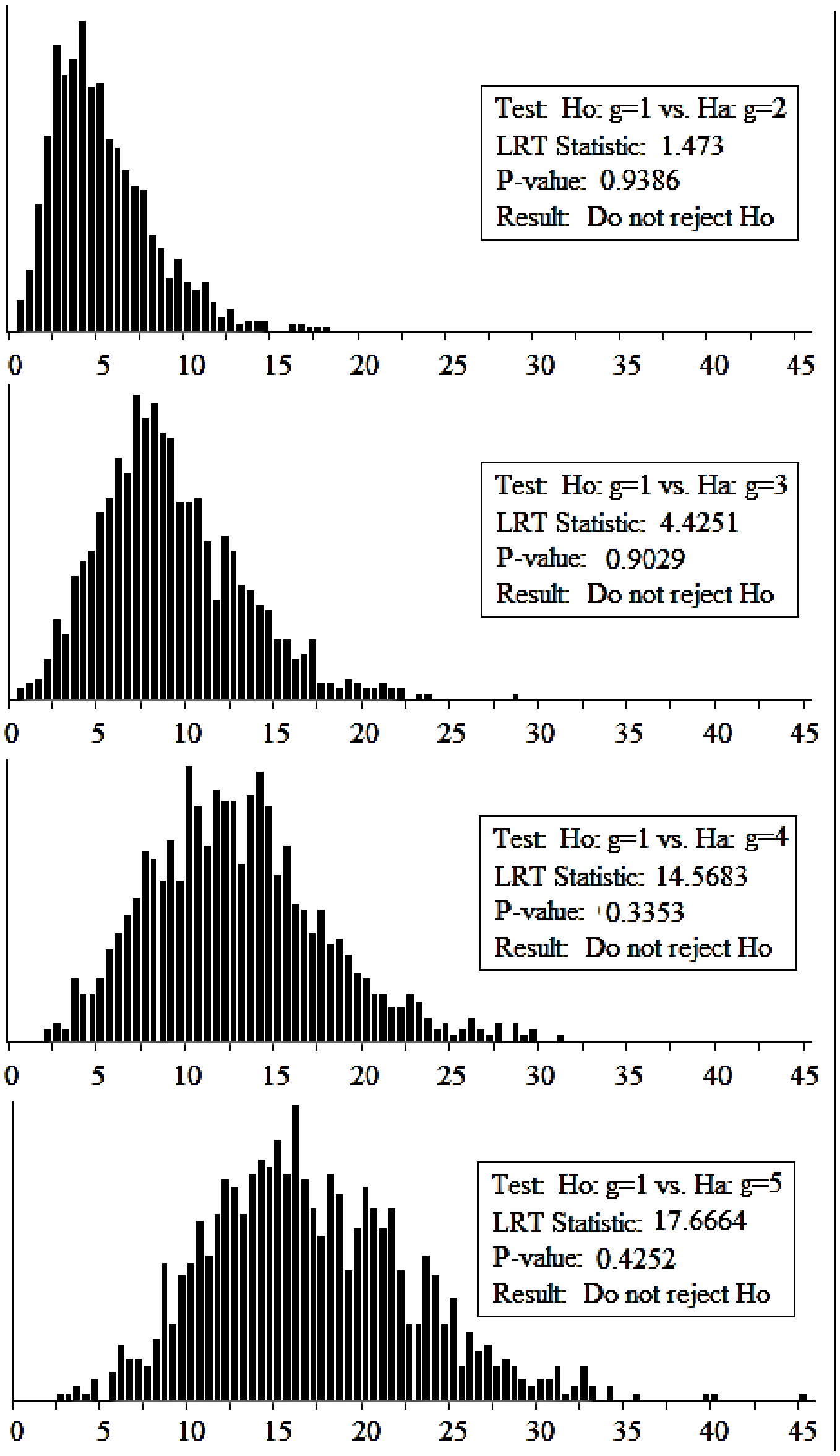

**NA**

Figure IV plots annual real compounding returns for the S&P 500 Index from 1928-2015 (n=88) in a histogram and Table IV fits these returns to univariate normal mixtures (≤5 components). Tests for the optimal # of components are done in Figure V using the forward-backward procedure described above. All displayed values are rounded throughout and unrounded values are used in calculations. The VR constraint is $16^2$ and a significance level of α=0.2500 is used for each test. Some evidence of non-normality will lead to rejection of Ho. As shown in Table IV, there is insufficient evidence to reject normality for annual real compounding S&P 500 Index returns. The forward-backward bootstrapped LRT procedure is in agreement with all 3 information criteria values (AIC, AICC, BIC) that a univariate normal PDF is appropriate for these returns, which also agrees with the AD test for normality (p=0.7206).

*A.2 Univariate PDF for Small Cap (S): $f_S(s)$*

***Figure VI**. Annual Real Small Cap Compounding Returns (S) Histogram*

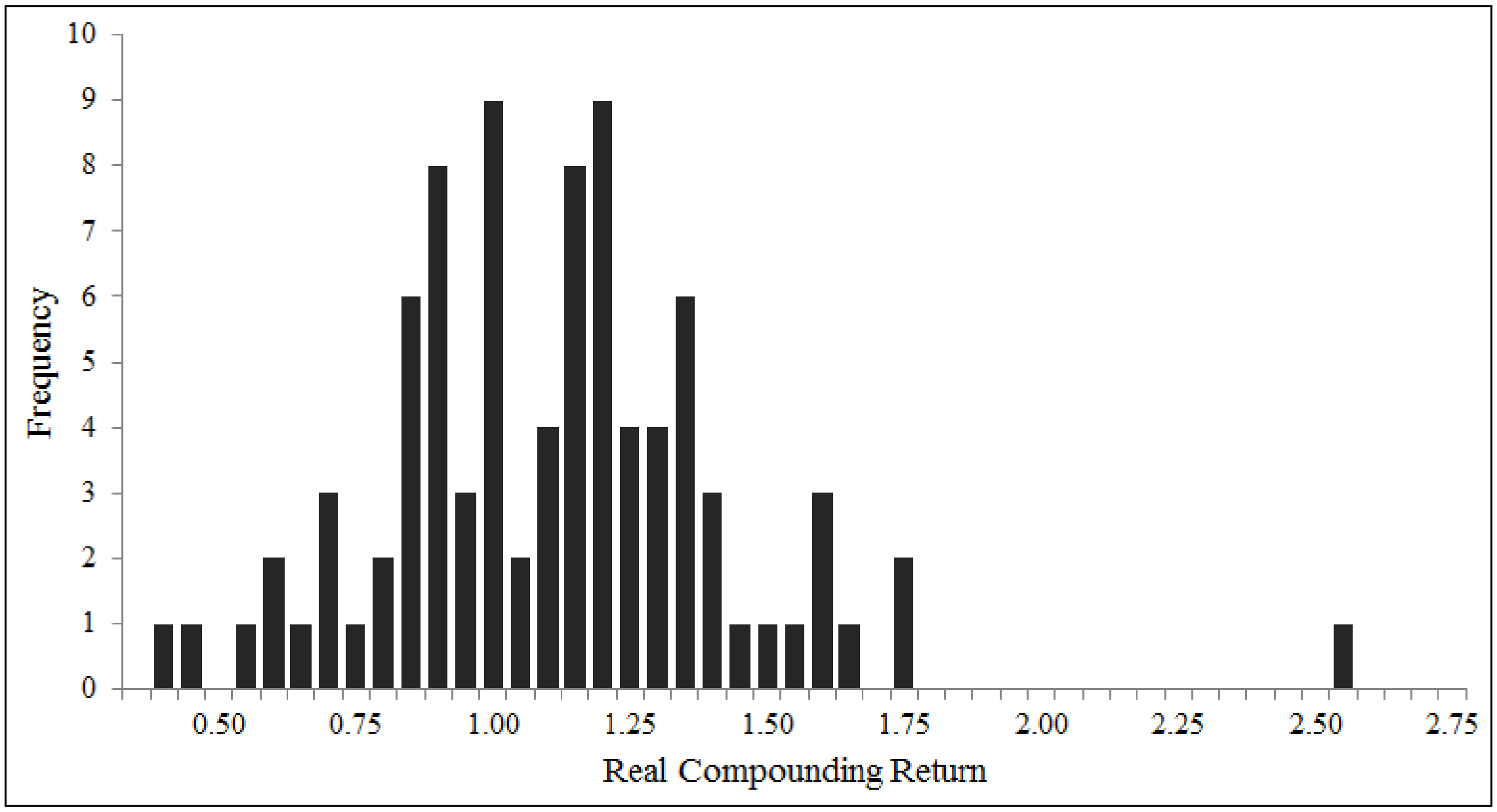

***Table V**. Univariate Mixture PDFs for Annual Real Small Cap Compounding Returns (S)*

| Number of Components | | | | | | | | | |
|---|---|---|---|---|---|---|---|---|---|
| 1 | | 2 | | 3 | | 4 | | 5 | |
| LL | -24.7605 | LL | -23.0510 | LL | -18.3915 | LL | -15.9514 | LL | -14.4531 |
| VR | 1.0000 | VR | 148.4732 | VR | 241.2802 | VR | 241.5788 | VR | 198.4935 |
| AIC | 53.5211 | AIC | 56.1021 | AIC | 52.7829 | AIC | 53.9028 | AIC | 56.9062 |
| AICC | 53.8067 | AICC | 57.1390 | AICC | 55.0907 | AICC | 58.0628 | AICC | 63.5729 |
| BIC | 58.4758 | BIC | 68.4888 | BIC | 72.6016 | BIC | 81.1535 | BIC | 91.5889 |
| μ(S) | 1.132 | μ(S) | 1.132 | μ(S) | 1.132 | μ(S) | 1.132 | μ(S) | 1.132 |
| σ(S) | 0.3206 | σ(S) | 0.3206 | σ(S) | 0.3206 | σ(S) | 0.3206 | σ(S) | 0.3206 |
| γ(S) | 0.00000 | γ(S) | 0.13061 | γ(S) | 0.23954 | γ(S) | 0.15342 | γ(S) | 0.61679 |
| κ(S) | 3.00000 | κ(S) | 3.02488 | κ(S) | 3.71740 | κ(S) | 4.60462 | κ(S) | 4.52729 |

| π | μ | σ | π | μ | σ | π | μ | σ | π | μ | σ | π | μ | σ |
|---|---|---|---|---|---|---|---|---|---|---|---|---|---|---|
| 1.000 | 1.132 | 0.321 | 0.928 | 1.150 | 0.326 | 0.164 | 0.945 | 0.065 | 0.483 | 1.155 | 0.426 | 0.194 | 1.185 | 0.029 |
| | | | 0.072 | 0.908 | 0.027 | 0.707 | 1.165 | 0.367 | 0.114 | 1.338 | 0.051 | 0.149 | 1.338 | 0.054 |
| | | | | | | 0.129 | 1.192 | 0.024 | 0.167 | 1.187 | 0.027 | 0.105 | 0.662 | 0.139 |
| | | | | | | | | | 0.237 | 0.948 | 0.074 | 0.258 | 1.370 | 0.415 |
| | | | | | | | | | | | | 0.295 | 0.954 | 0.076 |

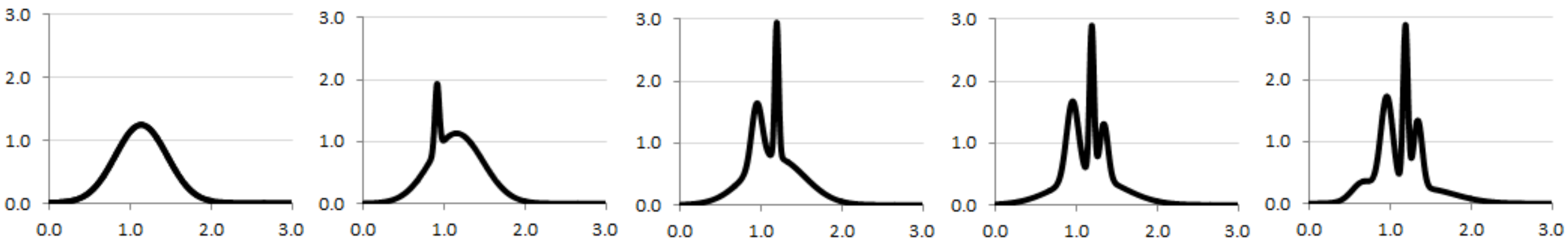

<u>Abbreviations:</u> LL=log-likelihood, VR=Variance Ratio, γ=skewness, κ=kurtosis. See §**II.M** for AIC, AICC, BIC.

***Figure VII**. LRT Sampling Distribution for Testing the Optimal # of Univariate Mixture Components*
*Annual Real Small Cap Compounding Returns (S)*

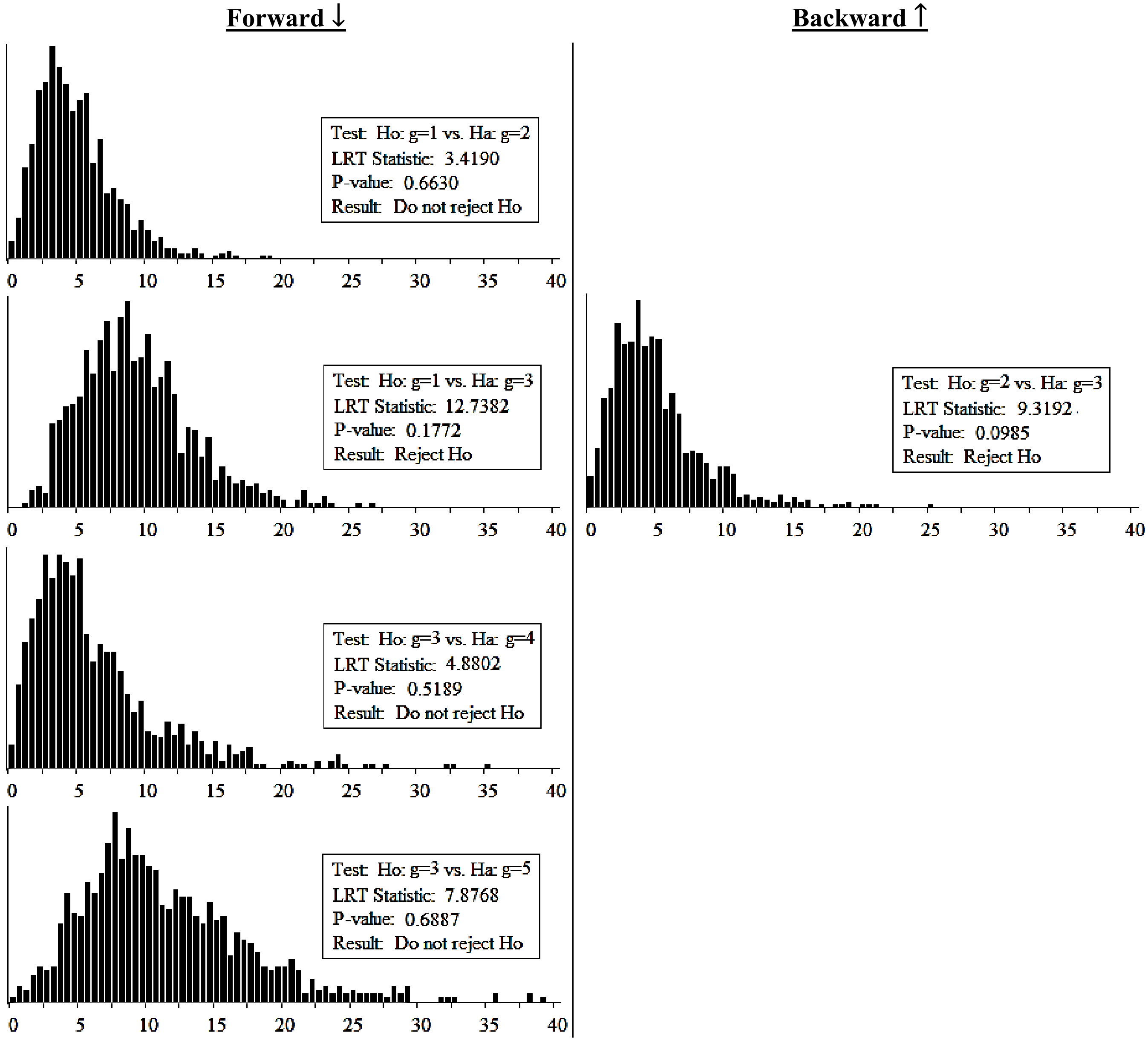

Annual real small cap compounding returns from 1928-2015 (n=88) are plotted in Figure VI and Table V fits these returns to univariate normal mixtures (≤5 components). Tests for the optimal # of components are shown in Figure VII using the forward-backward procedure described above with VR constraint of $16^2$ and significance level $\alpha$=0.2500. The test of 1 vs. 3 components yields a significant p-value and Ho is rejected. Backward processing begins by testing 2 vs. 3 components which is also significant, ending the procedure. A 3-component normal mixture PDF is therefore found appropriate for these returns. The coefficient $\gamma(S)$=0.23954>0 indicates the PDF has positive skew and $\kappa(S)$=3.71740>3 indicates a positive excess kurtosis which implies a heavier tail than the normal distribution. The fitted marginal PDF is evidently skewed, heavy-tailed, and multimodal (Table V).

*A.3 Univariate PDF for U.S. 10-Year T-Bonds (B): $f_B(b)$*

***Figure VIII**. Annual Real U.S. 10-Year T-Bond Total Compounding Returns (B) Histogram*

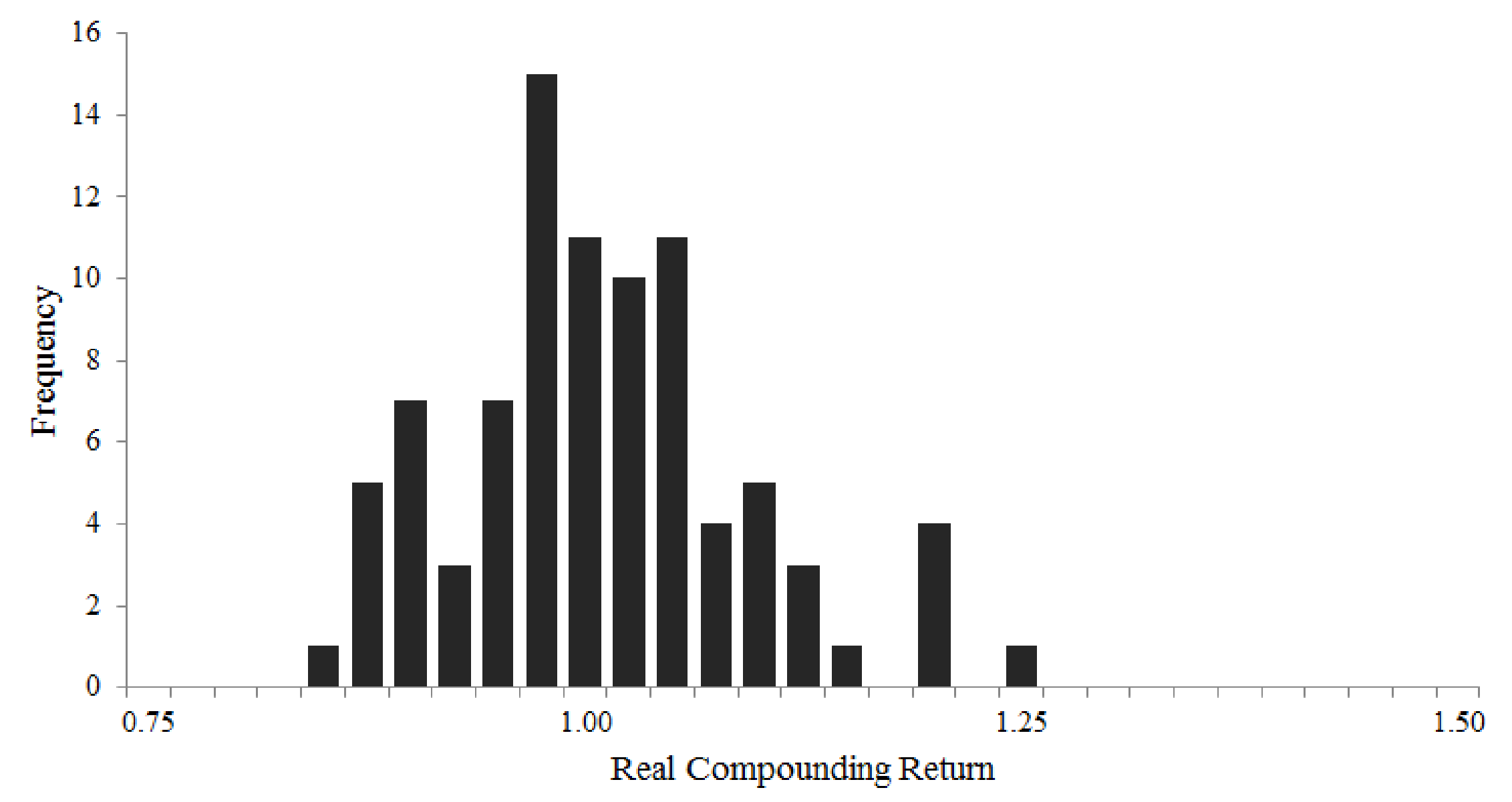

***Table VI**. Univariate Mixture PDFs for Annual Real U.S. 10-Year T-Bond Total Compounding Returns (B)*

**Number of Components**

| | 1 | 2 | 3 | 4 | 5 |
|---|---|---|---|---|---|
| LL | 94.9092 | 99.4731 | 100.2688 | 102.2525 | 104.1529 |
| VR | 1.0000 | 16.7844 | 20.2289 | 8.8448 | 125.3318 |
| AIC | -185.8184 | -188.9463 | -184.5377 | -182.5049 | -180.3057 |
| AICC | -185.5327 | -187.9092 | -182.2299 | -178.3450 | -173.6391 |
| BIC | -180.8637 | -176.5596 | -164.7190 | -155.2542 | -145.6230 |
| μ(B) | 1.022 | 1.022 | 1.022 | 1.022 | 1.022 |
| σ(B) | 0.0823 | 0.0823 | 0.0823 | 0.0823 | 0.0823 |
| γ(B) | 0.00000 | 0.47529 | 0.50035 | 0.49208 | 0.49047 |
| κ(B) | 3.00000 | 3.36595 | 3.26913 | 3.14599 | 3.14076 |

| π (1) | μ (1) | σ (1) | π (2) | μ (2) | σ (2) | π (3) | μ (3) | σ (3) | π (4) | μ (4) | σ (4) | π (5) | μ (5) | σ (5) |
|---|---|---|---|---|---|---|---|---|---|---|---|---|---|---|
| 1.000 | 1.022 | 0.082 | 0.948 | 1.011 | 0.070 | 0.893 | 1.004 | 0.065 | 0.089 | 1.118 | 0.020 | 0.350 | 0.991 | 0.029 |
| | | | 0.052 | 1.220 | 0.017 | 0.055 | 1.220 | 0.017 | 0.118 | 0.903 | 0.021 | 0.385 | 1.059 | 0.041 |
| | | | | | | 0.052 | 1.120 | 0.015 | 0.057 | 1.220 | 0.017 | 0.042 | 1.128 | 0.004 |
| | | | | | | | | | 0.736 | 1.014 | 0.051 | 0.167 | 0.908 | 0.024 |
| | | | | | | | | | | | | 0.057 | 1.220 | 0.017 |

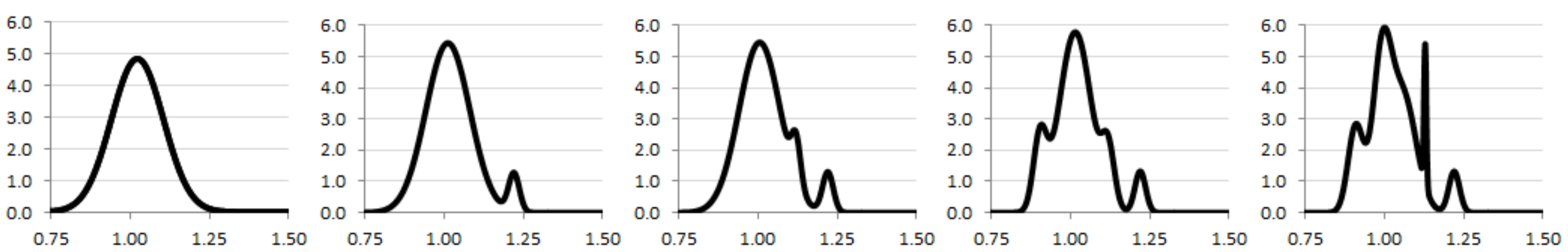

Abbreviations: LL=log-likelihood, VR=Variance Ratio, γ=skewness, κ=kurtosis. See §**II.M** for AIC, AICC, BIC.

***Figure IX**. LRT Sampling Distribution for Testing the Optimal # of Univariate Mixture Components*
*Annual Real U.S. 10-Year T-Bond Total Compounding Returns (B)*

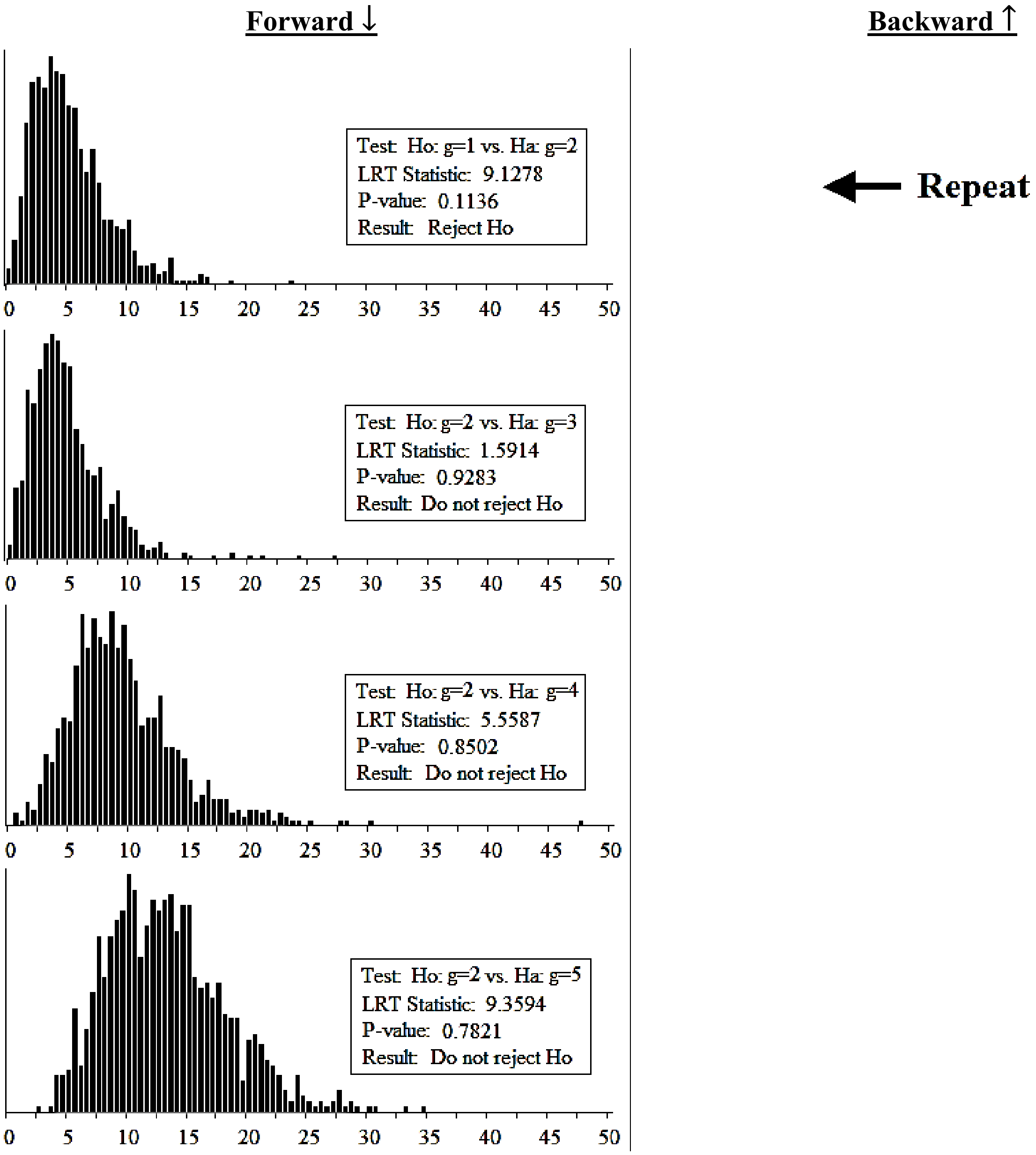

Annual real U.S. 10-Year T-Bond total compounding returns from 1928-2015 (n=88) are shown in Figure VIII and are fit to univariate normal mixtures in Table VI. The optimal # of components is found via the forward-backward procedure detailed in Figure IX with VR constraint of $16^2$ and significance level of α= 0.2500. The test of 1 vs. 2 components yields a significant p-value and Ho is rejected. Backward processing begins by testing 1 vs. 2 components which is a repeat test and not performed. A 2-component normal mixture PDF is thus appropriate for these returns. The coefficient γ(B)=0.47529>0 indicates this PDF has positive skew and κ(B)=3.36595>3 indicates a positive excess kurtosis which implies a heavier tail than the normal distribution. The fitted marginal PDF is evidently skewed, heavy-tailed, and multimodal (Table VI).

### B. *Univariate PDF Summary*

***Table VII**. Full Univariate PDF Parameterization*

| Security | Component | π | μ | σ |
|---|---|---|---|---|
| S&P 500 (L) | 1 | 1.000000000000000 | 1.082139318181818 | 0.197430382245555 |
| Small Cap (S) | 1 | 0.163796557010864 | 0.944667188140307 | 0.065233151408053 |
| | 2 | 0.707369571461765 | 1.165057494177362 | 0.366529325768043 |
| | 3 | 0.128833871527371 | 1.191903886074301 | 0.023596517545339 |
| U.S. 10-Year T-Bond (B) | 1 | 0.947744576049301 | 1.011164539967906 | 0.069579917666149 |
| | 2 | 0.052255423950700 | 1.220436091927283 | 0.016983666409906 |

Note: Use these estimates along with the historical data to reproduce the log-likelihood values.

Tests for the optimal # of components use α=0.2500. Large values are common defaults in forward-backward testing procedures. A variance ratio constraint of VR = 256 was used to eliminate spurious optimizers while finding MLEs (§**II.E**) with the EM algorithm (§**II.F**). Reducing this value or adding new constraints on either the probabilities or means will alter the PDF shapes, for example increasing or decreasing the # of modes/peaks (see Figure X). To add a constraint simply discard any random start that violates it. Skewed unimodal PDFs such as the lognormal were available but did not optimize the constrained objectives. Under interpretation (2) from §**II.B** components are assigned labels. The data suggest that annual real compounding S&P 500 returns originate from one regime, but that small cap equity returns originate from 3. Namely, a *dominant* N(1.165,0.367) PDF generates about 70% of returns (including outliers) and a *low* N(0.945,0.065) PDF generates >15%, with <15% originating from a *high* N(1.192,0.024) PDF. The high/low regimes add shoulders above and below the mean with the PDF evidently heavier tailed than a normal. Annual real U.S. 10-Year T-Bond total compounding returns originate from 2 regimes with the *dominant* N(1.011,0.070) PDF generating 95% of returns, and a high-outlier N(1.220,0.017) PDF regime generating the other 5%. Note that these returns averaged 1.033 from 2010 - 2015, which is above the dominant regime μ and overall historical mean of μ(B)=1.022. Consequently, widespread claims that current low yields invalidate retirement heuristics such as the "4% rule" should be met with skepticism.

***Figure X**. Univariate Mixture PDFs with Probability Weighted Component Regimes*

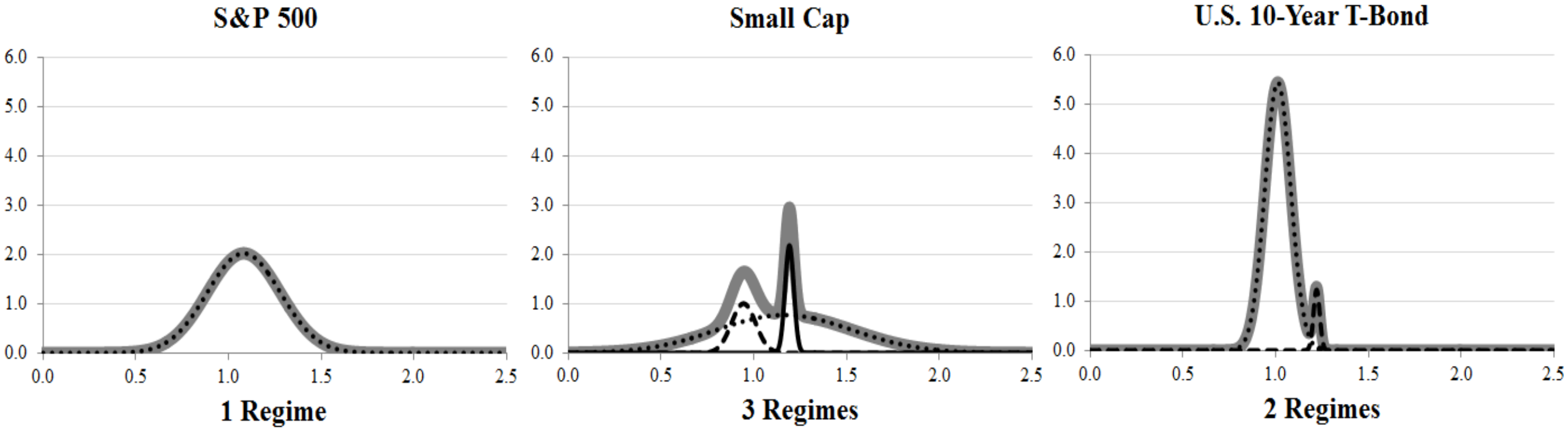

## IV. Multivariate Density Modeling w/out Covariances

The multivariate PDF for $(\mathrm{L,S,B})'$ is built in two steps. First, dependence is introduced without correlations and the result is the starting point for a final step of estimating correlations. Under interpretation (2) in §**II.B**, mixture observations come from labeled regimes as seen in §**III.B** for (L,S,B)′. An observation from the multivariate PDF can be viewed as originating from some combination of these regimes. There are 1,3,2 regimes governing L,S,B, respectively, thus at most 1·3·2=6 regimes will govern the multivariate PDF. A parsimonious multivariate PDF may call for eliminating combinations that have not produced data. The goal is to perform regime selection in an optimal manner, accounting for the sample size and total # of multivariate PDF parameters. We must also preserve the marginal PDFs that were derived in §**III.A**.

### *A. Multivariate Regimes*

Under mixture PDF interpretation (2), regimes produce observations. After estimating the PDF parameters in a mixture (2.B.2), we can estimate the probability that an observation is from a given regime. Let $X_t$ and $\mathbf{Z}_t = (z_{t1}, \ldots, z_{tg})'$ be the time t observation and component indicator RVs for a g-component univariate mixture PDF, for t=1,...,T. Assume all parameters have been estimated as in §**III.A** and that $x_t$ is the observed value at time t. The probability that $x_t$ is produced by component i, i=1,...,g is:

$$\mathrm{P}(z_{\mathrm{ti}} = 1|x_{\mathrm{t}}) = \frac{\mathrm{P}(z_{\mathrm{ti}} = 1, x_{\mathrm{t}})}{\mathrm{P}(x_{\mathrm{t}})} = \frac{\mathrm{P}(x_{\mathrm{t}}|z_{\mathrm{ti}} = 1) * \mathrm{P}(z_{\mathrm{ti}} = 1)}{\mathrm{P}(x_{\mathrm{t}})} = \frac{f^{\mathrm{i}}(x_{\mathrm{t}}) * \pi_{\mathrm{i}}}{\sum_{\mathrm{i}=1}^{\mathrm{g}} \pi_{\mathrm{i}} f^{\mathrm{i}}(x_{\mathrm{t}})} \qquad (4.A.1)$$

Since $\pi_{\mathrm{i}}$ has been estimated and $f^{\mathrm{i}}(x_{\mathrm{t}})=f^{\mathrm{i}}(x_{\mathrm{t}}|\boldsymbol{\theta}_{\mathrm{i}})$ with $\boldsymbol{\theta}_{\mathrm{i}}$ also estimated, this quantity can be computed for each observation $x_t$, t=1,...,T. *Bayes decision rule* assigns each observation to the component with largest probability and is considered an optimal allocation scheme (McLachlan & Peel, 2000). Let $n_{ijk}$ be the # of observations in regimes i, j, k of RVs L, S, B, respectively, using the assignment rule above. The probability that an observation on $(\mathrm{L, S, B})'$ originates from a given multivariate component can be estimated as $\hat{p}_{ijk}=n_{ijk}/T$, where $p_{ijk}$ will be the true (unknown) probability, see Figure XI.

***Figure XI**. Multivariate Regime Combinations for (L, S, B) with Estimated Probabilities*

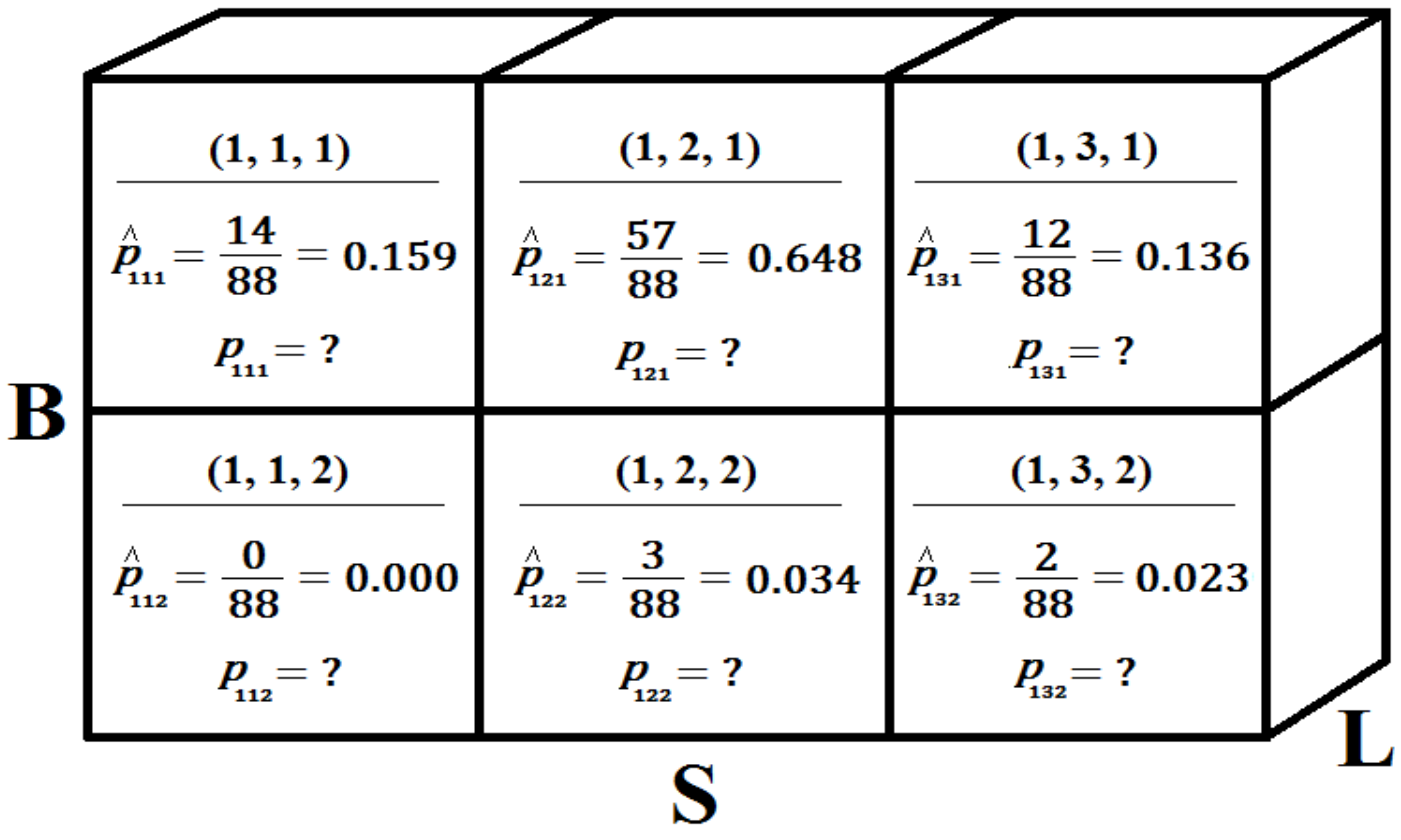

Note: Observations on the trivariate (L,S,B) exist in each cell and correlations may change across regimes. For example the correlation between S and B may be strongly positive in one cell and strongly negative in another (see §**V**).

If L,S and B are independent, the multivariate PDF is the product of the marginals, $f(l,s,b)=f_L(l)\cdot f_S(s)\cdot f_B(b)$. The marginals were fit to mixtures in §**III.A** and their product is a 6-component multivariate normal mixture that yields the fitted marginals. Under independence the probability that a multivariate observation $(l,s,b)$ is from a given regime is the product of the marginal probabilities. There is no basis to assume independence and Figure XI shows the multivariate component probabilities estimated from the data, namely $\hat{p}_{ijk}$. Dependence for multivariate mixture PDFs takes 2 forms: between component and within component. Between component dependence is modeled via the probabilities and within component dependence is modeled via the covariances. Both must preserve the marginals from §**III.A**, however the $\hat{p}_{ijk}$ estimates in Figure XI do not, thus they are infeasible. For example, using $f_S(s)$ the probability that S is from the "dominant" regime equals 0.707 (Table V) whereas using the data it is 60/88=0.682 (Figure XI). Conway (1979) gives conditions that guarantee both univariate and bivariate PMFs are enforced for probabilities in a 3-way contingency table. If the bivariate PDF is of interest we suggest deriving it from the trivariate PDF.

### *B. Multivariate Regime Selection via Linear Programs (LPs)*

Our goal is to parsimoniously model between component dependence while preserving the marginal PDFs from §**III.A**. Two approaches are presented and become initial solutions for a final step of estimating covariances and updating component probabilities. We limit the discussion to the problem at hand, however, the methods are completely generic and easily extendable to arbitrarily dimensional problems.

#### *B.1 Minimize Maximum Distance (Minimax) LP*

Define the maximum distance between true ($p_{ijk}$) and estimated ($\hat{p}_{ijk}$) component probabilities for L, S,B from Figure XI as $Z=\max\{|p_{111}-\hat{p}_{111}|,|p_{121}-\hat{p}_{121}|,|p_{131}-\hat{p}_{131}|,|p_{112}-\hat{p}_{112}|,|p_{122}-\hat{p}_{122}|,|p_{132}-\hat{p}_{132}|\}$. Values for $p_{ijk}$ that minimize Z and preserve the marginal PDFs from §**III.A** would be of interest. Namely, solving:

| | | |
|---|---|---|
| **Minimize:** | $Z=\max\{|p_{111}-\hat{p}_{111}|, |p_{121}-\hat{p}_{121}|, |p_{131}-\hat{p}_{131}|, |p_{112}-\hat{p}_{112}|, |p_{122}-\hat{p}_{122}|, |p_{132}-\hat{p}_{132}|\}$ | |
| **Subject to:** | $p_{111}+p_{121}+p_{131}+p_{112}+p_{122}+p_{132}=1.000$ | (L-marginals) |
| | $p_{111}+p_{112}=0.164$ | (S-marginals) |
| | $p_{121}+p_{122}=0.707$ | |
| | $p_{131}+p_{132}=0.129$ | (4.B.1) |
| | $p_{111}+p_{121}+p_{131}=0.948$ | (B-marginals) |
| | $p_{112}+p_{122}+p_{132}=0.052$ | |
| | $p_{111}+p_{121}+p_{131}+p_{112}+p_{122}+p_{132}=1.000$ | (Probabilities sum to 1) |
| **Where:** | $p_{ijk}\geq 0$, i=1, j=1,2,3, k=1,2 | (non-negativity constraints) |

Minimizing the maximum of a set is a *minimax* objective. While Z is not linear, the constrained optimization problem in (4.B.1) can be formulated and solved as an LP. Note that of the 7 constraints, the $4^{th}$ and $6^{th}$ are redundant since all probabilities sum to 1 as is the $7^{th}$ since L has 1 component. They are thus dropped, and since the maximum of a set must be greater-than or equal-to all set elements, (4.B.1) becomes:

| | | | |
|---|---|---|---|
| **Minimize:** | $Z$ | | |
| **Subject to:** | $Z \geq \mid p_{ijk} - \hat{p}_{ijk} \mid$ | i=1, j=1,2,3, k=1,2 | |
| | $p_{111} + p_{121} + p_{131} + p_{112} + p_{122} + p_{132} = 1.000$ | (L-marginals) | |
| | $p_{111} + p_{112} = 0.164$ | (S-marginals) | |
| | $p_{121} + p_{122} = 0.707$ | | (4.B.2) |
| | $p_{111} + p_{121} + p_{131} = 0.948$ | (B-marginals) | |
| **Where:** | $Z, p_{ijk} \geq 0$, i=1, j=1,2,3, k=1,2 | (non-negativity constraints) | |

Constraints in (4.B.2) include absolute values and are non-linear, however note that $X \geq |Y|$ *iff* $X \geq Y$ and $X \geq -Y$. To promote parsimony we penalize the objective when a cell from Figure XI containing 0 observations yields a >0 probability. This will ensure it only occurs when needed for feasibility. The constraints are $p_{ijk} \leq n_{ijk} + X_{ijk}$, and the penalty is $M \cdot \left(\sum_{i=1}^{1} \sum_{j=1}^{3} \sum_{k=1}^{2} X_{ijk}\right)$, where M is arbitrarily large. When $n_{ijk}=0$ and $p_{ijk}>0$, $X_{ijk}$ must be >0 to satisfy the constraint and Z suffers a penalty. The final LP formulation is:

| | | | |
|---|---|---|---|
| **Minimize:** | $Z + M \cdot \left(\sum_{i=1}^{1} \sum_{j=1}^{3} \sum_{k=1}^{2} X_{ijk}\right)$ | | |
| **Subject to:** | $Z \geq p_{ijk} - \hat{p}_{ijk}$ and $Z \geq \hat{p}_{ijk} - p_{ijk}$ | for i=1, j=1,2,3, k=1,2 | |
| | $p_{ijk} \leq n_{ijk} + X_{ijk}$ | for i=1, j=1,2,3, k=1,2 | |
| | $p_{111} + p_{121} + p_{131} + p_{112} + p_{122} + p_{132} = 1.000$ | (L-marginals) | |
| | $p_{111} + p_{112} = 0.164$ | (S-marginals) | (4.B.3) |
| | $p_{121} + p_{122} = 0.707$ | | |
| | $p_{111} + p_{121} + p_{131} = 0.948$ | (B-marginals) | |
| **Where:** | $Z, p_{ijk}, X_{ijk} \geq 0$, for i=1, j=1,2,3, k=1,2 | (non-negativity constraints) | |
| | M and $n_{ijk}$ for i=1, j=1,2,3, k=1,2 | (known constants) | |

This LP was solved using the techniques from §**II.K.1**. The solution is $p_{111}$=0.1638, $p_{121}$=0.6627, $p_{131}$=0.1212, $p_{112}$=0.0000, $p_{122}$=0.0447, $p_{132}$=0.0076, yielding the minimization objective Z=0.0151.

*B.2 Minimum Sum of Squared Distances LP*

Define the sum of squared distances between true ($p_{ijk}$) and estimated ($\hat{p}_{ijk}$) component probabilities for L,S,B from Figure XI as $Z=(p_{111}-\hat{p}_{111})^2+(p_{121}-\hat{p}_{121})^2+(p_{131}-\hat{p}_{131})^2+(p_{112}-\hat{p}_{112})^2+(p_{122}-\hat{p}_{122})^2+(p_{132}-\hat{p}_{132})^2$. Values of $p_{ijk}$ that minimize Z and preserve the marginal PDFs in §**III.A** are of interest. Namely, solving:

| | | | |
|---|---|---|---|
| **Minimize:** | $Z = \sum_{i=1}^{1} \sum_{j=1}^{3} \sum_{k=1}^{2} \left(p_{ijk} - \hat{p}_{ijk}\right)^2$ | | |
| **Subject to:** | $p_{111} + p_{121} + p_{131} + p_{112} + p_{122} + p_{132} = 1.000$ | (L-marginals) | |
| | $p_{111} + p_{112} = 0.164$ | (S-marginals) | |
| | $p_{121} + p_{122} = 0.707$ | | (4.B.4) |
| | $p_{111} + p_{121} + p_{131} = 0.948$ | (B-marginals) | |
| **Where:** | $p_{ijk} \geq 0$, i=1, j=1,2,3, k=1,2 | (non-negativity constraints) | |

The objective Z is quadratic and separable therefore (4.B.4) can be approximated by an LP as shown in §**II.K.2**. Each decision variable $p_{ijk}$ appears in one term of Z and has a convex shape similar to Figure III. Each horizontal axis range is $0 \leq p_{ijk} \leq 1$ and (4.B.4) is converted to an LP by replacing the quadratic terms with connected line segments. The horizontal axis is partitioned into S contiguous sections and any value $p_{\mathrm{ijk}}$ is reachable as $p_{\mathrm{ijk}}=\sum_{\mathrm{p=0}}^{\mathrm{S}}\alpha(\mathrm{p})_{\mathrm{ijk}}\,(\mathrm{p/S})$ with $\sum_{\mathrm{p=0}}^{\mathrm{S}}\alpha(\mathrm{p})_{\mathrm{ijk}}=1$. Each term $\left(p_{\mathrm{ijk}}-\hat{p}_{\mathrm{ijk}}\right)^2$ is then approximated linearly by $\sum_{\mathrm{p=0}}^{\mathrm{S}}\alpha(\mathrm{p})_{\mathrm{ijk}}\,(\mathrm{p/S}-\hat{p}_{\mathrm{ijk}})^2$, and the same penalty from (4.B.3) is applied thus (4.B.4) becomes:

| | | | |
|---|---|---|---|
| **Minimize:** | $\sum_{\mathrm{i=1}}^{1}\sum_{\mathrm{j=1}}^{3}\sum_{\mathrm{k=1}}^{2}\left[\sum_{\mathrm{p=0}}^{\mathrm{S}}\alpha(\mathrm{p})_{\mathrm{ijk}}\left(\mathrm{p/S}-\hat{p}_{\mathrm{ijk}}\right)^2\right]+\mathrm{M}\cdot\left(\sum_{\mathrm{i=1}}^{1}\sum_{\mathrm{j=1}}^{3}\sum_{\mathrm{k=1}}^{2}\mathrm{X}_{\mathrm{ijk}}\right)$ | | |
| **Subject to:** | $\sum_{\mathrm{p=0}}^{\mathrm{S}}\alpha(\mathrm{p})_{\mathrm{ijk}}\,(\mathrm{p/S})\leq \mathrm{n}_{\mathrm{ijk}}+\mathrm{X}_{\mathrm{ijk}}$ | for i=1, j=1,2,3, k=1,2 | |
| | $\sum_{\mathrm{i=1}}^{1}\sum_{\mathrm{j=1}}^{3}\sum_{\mathrm{k=1}}^{2}\left[\sum_{\mathrm{p=0}}^{\mathrm{S}}\alpha(\mathrm{p})_{\mathrm{ijk}}\,(\mathrm{p/S})\right]=1.000$ | (L-marginals) | |
| | $\sum_{\mathrm{i=1}}^{1}\sum_{\mathrm{j=1}}^{1}\sum_{\mathrm{k=1}}^{2}\left[\sum_{\mathrm{p=0}}^{\mathrm{S}}\alpha(\mathrm{p})_{\mathrm{ijk}}\,(\mathrm{p/S})\right]=0.164$ | (S-marginals) | |
| | $\sum_{\mathrm{i=1}}^{1}\sum_{\mathrm{j=2}}^{2}\sum_{\mathrm{k=1}}^{2}\left[\sum_{\mathrm{p=0}}^{\mathrm{S}}\alpha(\mathrm{p})_{\mathrm{ijk}}\,(\mathrm{p/S})\right]=0.707$ | | (4.B.5) |
| | $\sum_{\mathrm{i=1}}^{1}\sum_{\mathrm{j=1}}^{3}\sum_{\mathrm{k=1}}^{1}\left[\sum_{\mathrm{p=0}}^{\mathrm{S}}\alpha(\mathrm{p})_{\mathrm{ijk}}\,(\mathrm{p/S})\right]=0.948$ | (B-marginals) | |
| **Where:** | $\alpha(\mathrm{p})_{\mathrm{ijk}},\ \mathrm{X}_{\mathrm{ijk}}\geq 0$, for i=1, j=1,2,3, k=1,2, p=0,...,S | (non-negativity constraints) | |
| | M, S, $\hat{p}_{\mathrm{ijk}}$, $\mathrm{n}_{\mathrm{ijk}}$ for i=1, j=1,2,3, k=1,2 | (known constants) | |

The decision variables $p_{\mathrm{ijk}}$ in (4.B.4) are replaced by $\alpha(\mathrm{p})_{\mathrm{ijk}}$ in (4.B.5) and note that a minimization objective ensures only adjacent α's are $>0$ for each i,j,k. We use S=500 and solve the LP in (4.B.5) to obtain $p_{111}$= 0.1638, $p_{121}$=0.6614, $p_{131}$=0.1226, $p_{112}$=0.0000, $p_{122}$=0.0460, $p_{132}$=0.0063, yielding the objective Z=0.0008.

The LPs solved in §**IV.B.1** and §**IV.B.2** were customized for the current exercise of modeling the multivariate PDF of $(\mathrm{L,S,B})'$ and the concepts are easily extendable to any collection of securities. The code supplied in Appendix C models an arbitrary number of securities. Our purpose is twofold: (1) find a feasible solution to initialize the final step, and (2) eliminate as many unnecessary multivariate components as possible. In larger problems this step can eliminate >80% of multivariate cells. Note that dependence for $(\mathrm{L,S,B})'$ has been introduced without having estimated any covariances as the multivariate PDFs using these LP solutions are not the product of the marginals from §**III.A**. Specifically, between component dependence has been introduced using the data to guide which components occur together and at what frequency. Note that randomly reordering the returns on each of the 3 assets separately would not change the results of §**III.A**, however would change the probability estimates in §**IV.B.1** and §**IV.B.2**. The unknown $p_{\mathrm{ijk}}$ from Figure XI now have 2 sets of initial values that are feasible and preserve the marginals.

## V. Multivariate Density Modeling w/Covariances

Let $X_1,...,X_N$ be RVs for the real compounding return on N financial securities that are not serially correlated. Historical data on each $X_j$ will be a random sample over time, say, $x_{tj}$, t=1,...,T and j=1,...,N, where $X_{tj}=(1+r_{tj})$ with $r_{tj}$ being the real return from §**II.N**. The marginal density for each $X_j$ can be modeled as a normal mixture, $f_j(x_j)$, having $g_j$ components with $E(X_j|z_{ji}=1)=\mu_{ji}$, $V(X_j|z_{ji}=1)=\sigma_{ji}^2$ and $P(z_{ji}=1)=\pi_{ji}$ for j=1,...,N and i=1,...,$g_j$. The multivariate PDF for $(X_1,...,X_N)'$ will be modeled as a G-component normal mixture with $Cov(X_j,X_k|z_c=1)=\sigma_{cjk}$ for j≠k=1,...,N and $P[z_c=1]=p_c$, c=1,...,G. As seen in Figure XI, each $z_c$ defines a combination of univariate components. The multivariate PDF for $(X_1,...,X_N)'$ is:

$$f(x_1,\dots,x_N) = f(\boldsymbol{x}) = \sum_{c=1}^{G} \frac{p_c}{(\sqrt{2\pi})^N \sqrt{|\boldsymbol{\Sigma}_c|}} e^{-\frac{1}{2}(x-\boldsymbol{\mu}_c)'[\boldsymbol{\Sigma}_c^{-1}](x-\boldsymbol{\mu}_c)}, \quad \boldsymbol{x} \in \mathbb{R}^N \tag{5.1}$$

which, to maintain the fitted marginals, must satisfy:

$$f_j(x_j) = \int_{-\infty}^{\infty} \dots \int_{-\infty}^{\infty} f(x_1,\dots,x_N) dx_1 \dots dx_{j-1} dx_{j+1} \dots dx_N, \quad j = 1,\dots,N \tag{5.2}$$

The log-likelihood of the unknown parameters given the historical sample and known parameters is:

$$\ln[\mathfrak{L}(p_1,\dots,p_G,\boldsymbol{\Sigma}_1,\dots,\boldsymbol{\Sigma}_G|\boldsymbol{x}_1,\dots,\boldsymbol{x}_T,\boldsymbol{\mu}_1,\dots,\boldsymbol{\mu}_G,\boldsymbol{\sigma}_1^2,\dots,\boldsymbol{\sigma}_G^2)] = \sum_{t=1}^{T} \ln\left(\sum_{c=1}^{G} \frac{p_c}{(\sqrt{2\pi})^N \sqrt{|\boldsymbol{\Sigma}_c|}} e^{-\frac{1}{2}(x_t-\boldsymbol{\mu}_c)'[\boldsymbol{\Sigma}_c^{-1}](x_t-\boldsymbol{\mu}_c)}\right) \tag{5.3}$$

MLEs for the unknown parameters are found by solving the following general NLP (see, §**II.K.4**):

**Maximize:** $Z = \sum_{t=1}^{T} \ln[f(\boldsymbol{x}_t)]$ (5.4)

**Subject to:** $f_j(x_j) = \int_{-\infty}^{\infty} \dots \int_{-\infty}^{\infty} f(\boldsymbol{x}) dx_1 \dots dx_{j-1} dx_{j+1} \dots dx_N$, $j = 1,\dots,N$ (uphold marginals)

$\text{Eigen}_j(\boldsymbol{\Sigma}_c) > 0$, for j=1,...,N and c=1,...,G (VC matrices are +definite)

**Where:** $f(\boldsymbol{x}) = \sum_{c=1}^{G} \frac{p_c}{(\sqrt{2\pi})^N \sqrt{|\boldsymbol{\Sigma}_c|}} e^{-\frac{1}{2}(x-\boldsymbol{\mu}_c)'[\boldsymbol{\Sigma}_c^{-1}](x-\boldsymbol{\mu}_c)}$, $\boldsymbol{x} \in \mathbb{R}^N$ (multivariate PDF)

$0 \le p_c \le 1$, $\sigma_{cjk} \in \mathbb{R}$, for j≠k=1,...,N and c=1,...,G (decision variables)

$\boldsymbol{\mu}_c$, $\boldsymbol{\sigma}_c^2 = \text{Diag}(\boldsymbol{\Sigma}_c)$, c=1,...,G (known constants)

The log-likelihood for a G-component multivariate mixture is maximized in (5.4) with respect to the probabilities and off-diagonal VC elements (covariances). Linear probability constraints maintain marginals as in §**IV.B** and covariance constraints will ensure +definite VC matrices (see §**II.I**). Researchers found that mixture PDF parameters can be estimated conditionally by first deriving the probabilities holding other parameters constant, then estimating the remaining parameters holding probabilities constant (repeating until convergence). The ECME algorithm is one such approach that makes use of the actual or incomplete log-likelihood in (2.E.2) (see, Liu & Rubin, 1994; McLachlan & Krishnan, 2008) and is the technique we will employ in optimizing (5.4). We will define the indicator function I(c,j,i) as 1 when univariate component i of security j exists in multivariate component c, for j=1,...,N, i=1,...,$g_j$, c=1,...,G and 0 otherwise.

## *A. Multivariate PDF Optimization*

### *A.1 <u>Step 1:</u> Optimize wrt Probabilities ($p_c$) Holding Covariances Constant*

**Maximize:** $Z = \sum_{t=1}^{T} \ln(f(\boldsymbol{x}_t)) = \sum_{t=1}^{T} \ln(\sum_{c=1}^{G} f^c(\boldsymbol{x}_t) \cdot p_c)$ (5.A.1)

**Subject to:** $\sum_{c=1}^{G} I(c,j,i) \cdot p_c = \pi_{ji}$, for j=1,...,N and i=1,...,$g_j$ (uphold marginals)

**Where:** $p_c \ge 0$, for c=1,...,G (non-negativity constraints)

$f^c(\boldsymbol{x}_t) = \frac{1}{(\sqrt{2\pi})^N \sqrt{|\boldsymbol{\Sigma}_c|}} e^{-\frac{1}{2}(x_t-\boldsymbol{\mu}_c)'[\boldsymbol{\Sigma}_c^{-1}](x_t-\boldsymbol{\mu}_c)}$, for t=1,...,T & c=1,...,G (known constants)

Since, (1) linear functions are concave, (2) log of a concave function is concave, (3) sum of concave functions is concave, Z is concave in $p_c$ (Boyd & Vandenberghe, 2009). Marginals are enforced with independent linear constraints, thus (5.A.1) is a CCP from §**II.K.3** and local optimums are global optimums. A critical point of the Lagrangian for (5.A.1) yields the maximum where $\mathscr{L}(\boldsymbol{p},\boldsymbol{\lambda})=Z - \sum_{j=1}^{N}\sum_{i=1}^{g_j}\lambda_{ji}\left[\sum_{k=1}^{G} I(k,j,i)p_k - \pi_{ji}\right]$, (Jensen & Bard, 2003). The 1st derivatives are $\partial\mathscr{L}(\boldsymbol{p},\boldsymbol{\lambda})/\partial p_c=\sum_{t=1}^{T}\left[\frac{f^c(\boldsymbol{x}_t)}{f(\boldsymbol{x}_t)}\right]-\sum_{j=1}^{N}\sum_{i=1}^{g_j}\lambda_{ji}I(c,j,i)$, and $\partial\mathscr{L}(\boldsymbol{p},\boldsymbol{\lambda})/\partial\lambda_{ji}=\pi_{ji}-\sum_{k=1}^{G} I(k,j,i)p_k$, the 2nd derivatives are $\partial^2\mathscr{L}(\boldsymbol{p},\boldsymbol{\lambda})/\partial p_c^2=\sum_{t=1}^{T}\left[\frac{-f^c(\boldsymbol{x}_t)^2}{f(\boldsymbol{x}_t)^2}\right]$, $\partial^2\mathscr{L}(\boldsymbol{p},\boldsymbol{\lambda})/\partial p_c\partial p_d=\sum_{t=1}^{T}\left[\frac{-f^c(\boldsymbol{x}_t)f^d(\boldsymbol{x}_t)}{f(\boldsymbol{x}_t)^2}\right]$, and $\partial^2\mathscr{L}(\boldsymbol{p},\boldsymbol{\lambda})/\partial p_c\partial\lambda_{ji}=-I(c,j,i)$ for c≠d=1,...,G, j=1,...,N, i=1,...,$g_j$. All 2nd derivatives wrt the Lagrange multipliers are zero and non-negativity constraints on $p_c$ are enforced by dropping any components having $p_c < 0$.

*A.2 <u>Step 2:</u> Optimize wrt Covariances ($\sigma_{cjk}$) Holding Probabilities Constant*

| | | |
|---|---|---|
| **<u>Maximize:</u>** | $Z = \sum_{t=1}^{T}\ln\left(f(\boldsymbol{x}_t)\right) = \sum_{t=1}^{T}\ln\left(\sum_{c=1}^{G}\frac{p_c}{(\sqrt{2\pi})^N\sqrt{\lvert\boldsymbol{\Sigma}_c\rvert}}e^{-\frac{1}{2}(\boldsymbol{x}_t-\boldsymbol{\mu}_c)'[\boldsymbol{\Sigma}_c^{-1}](\boldsymbol{x}_t-\boldsymbol{\mu}_c)}\right)$ | (5.A.2) |
| **<u>Subject to:</u>** | $\text{Eigen}_j(\boldsymbol{\Sigma}_c) > 0$, for j=1,...,N and c=1,...,G | (VC matrices are +definite) |
| **<u>Where:</u>** | $\sigma_{cjk} \in \mathbb{R}$, for j≠k=1,...,N and c=1,...,G | (decision variables) |
| | $p_c \geq 0$, for c=1,...,G | (known constants) |

Maximizing a multivariate log-likelihood function with respect to variance components is a difficult general NLP (§**II.K.4**) as there may be multiple local optimums or saddle points with zero gradient as well as boundary optimums with non-zero gradient. Searle, et al., (1992) recommend a hill-climbing procedure based at a good starting point. The gradient (**g**) helps inform on direction and the Hessian (**H**) on step size. Levenberg (1944) suggests a modification to Newton's method that iterates as $\boldsymbol{\theta}^{i+1}=\boldsymbol{\theta}^{i}-[\mathbf{H}(\boldsymbol{\theta}^{i})+s^{i}\cdot\mathbf{I}]^{-1}\cdot\mathbf{g}(\boldsymbol{\theta}^{i})$ where $s^i$ adjusts the step size and climbing angle. Marquardt (1963) derived a similar modification iterating as $\boldsymbol{\theta}^{i+1}=\boldsymbol{\theta}^{i}-[\mathbf{H}(\boldsymbol{\theta}^{i})+s^{i}\cdot\text{Diag}(\mathbf{H}(\boldsymbol{\theta}^{i}))]^{-1}\cdot\mathbf{g}(\boldsymbol{\theta}^{i})$, which is considered an optimal compromise between Newton's method (which often diverges) and gradient ascent/descent (which converges too slowly). A class of Levenberg-Marquardt techniques based on these approaches has since been published, see Gavin (2017) for an overview. While designed to find least-squares estimates in non-linear models (a constrained minimization problem), Searle, et al. (1992) note that they are also useful in finding MLEs for variance components.

Both gradient ascent and Newton's method failed to optimize (5.A.2) for the reasons stated and a Levenberg-Marquardt approach was taken, namely iterations defined by $\boldsymbol{\theta}^{i+1}=\boldsymbol{\theta}^{i}-\lambda^{i}[\mathbf{H}(\boldsymbol{\theta}^{i})+s^{i}\cdot I]^{-1}\cdot\mathbf{g}(\boldsymbol{\theta}^{i})$ with $\lvert s^i\rvert \leq \text{Max}(\lvert\text{Diag}[\mathbf{H}(\boldsymbol{\theta}^{i})]\rvert)$ and line-search parameter $\lambda^i \in (0,1)$. A large # of random $s^i$ and $\lambda^i$ are generated at each iteration and we select randomly among the top performers. Varying $s^i$ and $\lambda^i$ prevents divergence with large $s^i$ mimicking gradient ascent and small $s^i$ Newton's method. This ensures the nearest maximum is found, relative to an informed start, while scanning nearby regions for better values. Iterating into an infeasible region is addressed by performing a ridge repair on the offending VC matrix (§**II.I.1**). Exact 1st and 2nd derivatives of the log-likelihood in (5.A.2) are derived below wrt covariance terms only (§**II.I.2**).

*A.2.1 First Derivatives for (5.A.2): (Gradient Terms)*

There are G·N·(N-1)/2 covariance terms in a G-component multivariate mixture PDF for N securities. Using the chain rule along with the results derived in §**II.I.2** and Z from (5.A.2), the 1st derivatives are:

$$\frac{\partial Z}{\partial \sigma_{cjk}} = \sum_{t=1}^{T} \frac{\left(\frac{p_c}{(\sqrt{2\pi})^N}\right)\left(|\boldsymbol{\Sigma}_c|^{-\frac{1}{2}} \frac{\partial}{\partial \sigma_{cjk}}\left[e^{-\frac{1}{2}(\boldsymbol{x}_t-\boldsymbol{\mu}_c)'[\boldsymbol{\Sigma}_c^{-1}](\boldsymbol{x}_t-\boldsymbol{\mu}_c)}\right] + e^{-\frac{1}{2}(\boldsymbol{x}_t-\boldsymbol{\mu}_c)'[\boldsymbol{\Sigma}_c^{-1}](\boldsymbol{x}_t-\boldsymbol{\mu}_c)} \frac{\partial}{\partial \sigma_{cjk}}\left[|\boldsymbol{\Sigma}_c|^{-\frac{1}{2}}\right]\right)}{\left(\sum_{k=1}^{G} \frac{p_k}{(\sqrt{2\pi})^N \sqrt{|\boldsymbol{\Sigma}_k|}} e^{-\frac{1}{2}(\boldsymbol{x}_t-\boldsymbol{\mu}_k)'[\boldsymbol{\Sigma}_k^{-1}](\boldsymbol{x}_t-\boldsymbol{\mu}_k)}\right)} \tag{5.A.3}$$

$$= \sum_{t=1}^{T} \frac{p_c \cdot f^c(\boldsymbol{x}_t) \cdot Q_{cjk}}{f(\boldsymbol{x}_t)}, \text{ where } Q_{cjk} = \frac{1}{2}(\boldsymbol{x}_t - \boldsymbol{\mu}_c)'[\boldsymbol{\Sigma}_c^{-1}]\mathbf{A}^{jk}[\boldsymbol{\Sigma}_c^{-1}](\boldsymbol{x}_t - \boldsymbol{\mu}_c) - \frac{\left|\boldsymbol{\Sigma}_c{}^{(jk)}\right|}{|\boldsymbol{\Sigma}_c|}, \tag{5.A.4}$$

which is a scalar for $c = 1, \ldots, G$ and $j \neq k = 1, \ldots, N$.

*A.2.2 Second Derivatives for (5.A.2): (Hessian Terms)*

The 2nd derivatives wrt terms $\sigma_{cjk}$, $\sigma_{prs}$ where $\binom{j}{r} < \binom{k}{s}$ and $\binom{c}{j} \leq \binom{p}{r}$ are as follows:

*A.2.2.1* Case 1: wrt $\sigma_{cjk}$, $\sigma_{prs}$ where $c \neq p$ (covariances are from different multivariate components)

$$\frac{\partial^2 Z}{\partial \sigma_{cjk} \partial \sigma_{prs}} = -\sum_{t=1}^{T} \frac{\left(p_c \cdot f^c(\boldsymbol{x}_t) \cdot Q_{cjk}\right) \cdot \left(p_p \cdot f^p(\boldsymbol{x}_t) \cdot Q_{prs}\right)}{f(\boldsymbol{x}_t)^2} \tag{5.A.5}$$

*A.2.2.2* Case 2: wrt $\sigma_{cjk}$, $\sigma_{prs}$ where $c = p$ (covariances are from the same multivariate component)

$$\frac{\partial^2 Z}{\partial \sigma_{cjk} \partial \sigma_{prs}} = \sum_{t=1}^{T} \frac{f(\boldsymbol{x}_t) \cdot \left(p_c \cdot f^c(\boldsymbol{x}_t) \cdot \frac{\partial}{\partial \sigma_{prs}}[Q_{cjk}] + Q_{cjk} \cdot \frac{\partial}{\partial \sigma_{prs}}[p_c \cdot f^c(\boldsymbol{x}_t)]\right) - p_c \cdot f^c(\boldsymbol{x}_t) \cdot Q_{cjk} \cdot \frac{\partial}{\partial \sigma_{prs}}[f(\boldsymbol{x}_t)]}{f(\boldsymbol{x}_t)^2}$$

$$= \sum_{t=1}^{T} \frac{p_c \cdot f^c(\boldsymbol{x}_t)\left[f(\boldsymbol{x}_t) \cdot \frac{\partial}{\partial \sigma_{prs}}[Q_{cjk}] + \left(f(\boldsymbol{x}_t) - p_c \cdot f^c(\boldsymbol{x}_t)\right) \cdot Q_{cjk} \cdot Q_{crs}\right]}{f(\boldsymbol{x}_t)^2}, \tag{5.A.6}$$

where,

$$\frac{\partial}{\partial \sigma_{prs}}[Q_{cjk}] = -\frac{1}{2}(\boldsymbol{x}_t - \boldsymbol{\mu}_c)'\left([\boldsymbol{\Sigma}_c^{-1}]\mathbf{A}^{jk}[\boldsymbol{\Sigma}_c^{-1}]\mathbf{A}^{rs}[\boldsymbol{\Sigma}_c^{-1}] + [\boldsymbol{\Sigma}_c^{-1}]\mathbf{A}^{rs}[\boldsymbol{\Sigma}_c^{-1}]\mathbf{A}^{jk}[\boldsymbol{\Sigma}_c^{-1}]\right)(\boldsymbol{x}_t - \boldsymbol{\mu}_c) - \frac{\partial}{\partial \sigma_{prs}}\left[\frac{\left|\boldsymbol{\Sigma}_c{}^{(jk)}\right|}{|\boldsymbol{\Sigma}_c|}\right]. \tag{5.A.7}$$

The last term in (5.A.7) depends on the location of $\sigma_{prs}$ in $\boldsymbol{\Sigma}_c{}^{(jk)}$ and is derived piecewise as:

$$\frac{\partial}{\partial \sigma_{prs}}\left[\frac{\left|\boldsymbol{\Sigma}_c{}^{(jk)}\right|}{|\boldsymbol{\Sigma}_c|}\right] = \begin{cases} \frac{\left|\boldsymbol{\Sigma}_{c(sr)}^{(jk)}\right| \cdot |\boldsymbol{\Sigma}_c| - 2\left|\boldsymbol{\Sigma}_c{}^{(jk)}\right| \cdot \left|\boldsymbol{\Sigma}_c{}^{(rs)}\right|}{|\boldsymbol{\Sigma}_c|^2}, & \text{for } j = r \text{ and/or } k = s \text{ (i.e., } \sigma_{prs} \text{ below diagonal in } \boldsymbol{\Sigma}_c{}^{(jk)}) \\ \frac{\left|\boldsymbol{\Sigma}_{c(rs)}^{(jk)}\right| \cdot |\boldsymbol{\Sigma}_c| - 2\left|\boldsymbol{\Sigma}_c{}^{(jk)}\right| \cdot \left|\boldsymbol{\Sigma}_c{}^{(rs)}\right|}{|\boldsymbol{\Sigma}_c|^2}, & \text{for } k = r \text{ (i.e., } \sigma_{prs} \text{ above diagonal in } \boldsymbol{\Sigma}_c{}^{(jk)}) \\ \frac{\left(\left|\boldsymbol{\Sigma}_{c(rs)}^{(jk)}\right| + \left|\boldsymbol{\Sigma}_{c(sr)}^{(jk)}\right|\right) \cdot |\boldsymbol{\Sigma}_c| - 2\left|\boldsymbol{\Sigma}_c{}^{(jk)}\right| \cdot \left|\boldsymbol{\Sigma}_c{}^{(rs)}\right|}{|\boldsymbol{\Sigma}_c|^2}, & \text{otherwise (i.e., } \sigma_{prs} \text{ above/below diagonal in } \boldsymbol{\Sigma}_c{}^{(jk)}) \end{cases} \tag{5.A.8}$$

Whereas quasi-Newton methods approximate the Hessian, this Levinson-Marquardt approach will use exact 2nd derivatives, supplied in (5.A.[5–8]). The multivariate PDF log-likelihood is maximized by iterating over (5.A.1) and (5.A.2).[28] The objective is non-decreasing across steps and iterations. We begin with the informed starts from §**IV.B.1** and §**IV.B.2** then iterate until the log-likelihood stops increasing. Non-border solutions are maximums if the region around (5.A.2) is concave, i.e., the Hessian is negative semi-definite (all eigenvalues ≤ 0). Border solutions may exist in non-concave regions. Note that eigenvalues computed from an ill-conditioned matrix are unstable and should not be used. Sparse matrices containing extremely large and small (diagonal) entries are often ill-conditioned. The Hessian derived above in §**V.A.2.2** may appear problematic via inspection, however, it is real and symmetric thus its eigenvectors are orthogonal and the condition # is 1 (in theory). Bauer-Fike give bounds for the accuracy of eigenvalues as $\pm\kappa(\mathbf{U})\cdot\|\mathbf{E}\|_2$, where the Hessian **H** is subject to error **E** and **U** is the matrix of column eigenvectors for **H** with $\|\cdot\|_2$ denoting the Euclidean norm (2-norm). The condition # is $\kappa(\mathbf{U})=\|\mathbf{U}\|_2$ and when $\kappa(\mathbf{U})\gg 1$ the matrix is ill-conditioned and calculated eigenvalues are suspect (Meyer, 2000). We will select the PDF having min(AIC) across the informed starts from §**IV.B** and conduct an in-depth analysis of the surface's properties at the maximizer, considering also whether or not it is spurious and/or ill-conditioned.

[28] The 2-step algorithm from (5.A.1) and (5.A.2) is an ECME approach, e.g. see Liu & Sun (1997) who apply Newton's method to the EM algorithm M-step for faster convergence. See also Liu & Rubin (1994) and McLachlan & Krishnan (2008) for ECME algorithm details.

### *B. Multivariate PDF for (L, S, B)′*

The approach from §**V.A** was used to estimate the probabilities and covariances for *f*(*l*,*s*,*b*), the multivariate PDF of (L,S,B)′. The result is a 5-component multivariate mixture PDF (see Table VIII) which supplements the univariate PDF estimates in Table VII. The univariate regime labels propagate through to the multivariate PDF. Without a doubt many disciplines would discard this solution as a spurious maximizer since components 3 & 5 model only a few observations and much of the log-likelihood improvement over a multivariate normal is via these components.[29] Finance, however, differs from other industries in that a primary focus is studying risk. Extreme events drive risk and instead of discarding outliers, finance assigns them labels such as 'gray' or 'black' swans (Taleb, 2010). Conventional wisdom suggests, for example, that such outliers can cause a bank to fail or a retiree to experience financial ruin thus they must be accounted for. It is also a reason why the normal distribution may be rejected in financial research. Our model accounts for risk by either explicitly modeling low outliers which adds density to the tail or by modeling high outliers which shifts the dominant regimes left (along with their tails). In this application the latter occurs. There is a tradeoff with either approach as the within regime variance shrinks when observations are separated. The kurtosis indicates whether or not the mixture PDF is heavier tailed than a normal PDF as described in §**III**. As an aside, the predictive modeler may accuse us of memorizing the training data and suggest that this is always possible with a model of sufficient complexity but that such models are poor at prediction. A best practice when predicting is to partition the data into 3 sets then train models on the 1st set, evaluate/select the best using the 2nd, and report results after applying the chosen model to the 3rd. Unfortunately, there is insufficient data to use this practice on annual historical returns in finance. Lastly, using information criteria such as AIC from **§II.M**, the 5-component multivariate mixture from Table VIII (LL=159.09 & 28 free parameters) is superior to a multivariate normal (LL=134.43 & 9 free parameters), since -262.18 < -250.86.

[29] The solution would be to tighten the marginal constraints (lower the variance ratio or add constraints on the means and/or probabilities).

***Table VIII****. Full Multivariate PDF Parameterization*

| | **Component** | | | | |
|---|---|---|---|---|---|
| | **1** | **2** | **3** | **4** | **5** |
| (L,S,B) | (1,1,1) | (1,2,1) | (1,2,2) | (1,3,1) | (1,3,2) |
| $\pi$ | 0.163796557010864 | 0.683298902828332 | 0.024070668633433 | 0.100649116210105 | 0.028184755317266 |
| $\rho_{LS}$ | 0.718841320548123 | 0.846153215066181 | 0.047531637731748 | 0.577235331054455 | 0.962441058378833 |
| $\rho_{LB}$ | 0.162032898398328 | -0.095463348011069 | -0.703796286452937 | 0.555733717419093 | 0.487911866790132 |
| $\rho_{SB}$ | 0.156156396700733 | -0.082959075055513 | -0.743051448755313 | 0.698818203652309 | 0.673278286734948 |
| Det($\boldsymbol{\Sigma}$) | 0.000000376625410 | 0.000007134870354 | 0.000000000000100 | 0.000000033408954 | 0.000000000091838 |

Note: Use these estimates with those in Table VII (and the historical data) to reproduce the multivariate log-likelihood values.

Table VIII estimates were generated using the 2-step procedure from §**V.A**, which converged in 2 iterations. Step 1 from (5.A.1) required 5 and 4 sub-iterations while Step 2 from (5.A.2) required 54 then 1, respectively. The optimization in (5.A.1) is convex and will converge to a global maximum, while that in (5.A.2) is not. Starting at the LP solution from §**IV.B.2**, (5.A.2) instantly requires 5 ridge repairs and lands in a non-concave region. The final VC matrix repair occurs at sub-iteration 6/54 and between sub-iterations 8/54 and 47/54 the procedure finds a concave region and methodically climbs it. The Hessian matrix condition # begins at 1 and slowly increases, ending at 1.001278, perhaps revealing some numerical instability. Between sub-iterations 47/54 and 54/54 a Hessian eigenvalue turns positive and Step 1 ends at a saddle point. Step 2 ends nearby and is a boundary solution since $\Sigma_3$ is borderline +definite, which is a constraint in (5.A.2). To enforce it we require that all VC matrix eigenvalues are >0 with determinants $>0.1^{13}$. Table VIII reveals that det($\boldsymbol{\Sigma}_3$) is at the threshold and note that the log-likelihood value can be driven higher at this solution by lowering the threshold, however the condition # becomes large and the result so unstable that minor rounding of $\boldsymbol{\Sigma}_3$ produces a non +definite matrix. Such is the nature of a border solution.

The covariance between RVs X and Y is $\sigma_{XY}$=E[(X-E(X))(Y-E(Y))]=E(XY) - $\mu_X\mu_Y$. Using the *law of total expectation*, E(X)=$E_Z$[E(X|Z)] for any RV Z, thus for a G-component multivariate mixture PDF:

$$\sigma_{XY} = E_Z[E(XY|Z)] - \mu_X\mu_Y \quad (5.B.1)$$

$$= \left(\sum_{c=1}^{G}[E(XY|Z_c = 1) \cdot P(Z_c = 1)]\right) - \mu_X\mu_Y \quad (5.B.2)$$

$$= \left(\sum_{c=1}^{G}[(\sigma_{cXY} + \mu_{cX}\mu_{cY}) \cdot p_c]\right) - \mu_X\mu_Y \quad (5.B.3)$$

The correlation between RVs X and Y is then derived as $\rho_{XY}=\sigma_{XY}/(\sigma_X\sigma_Y)$. Using the multivariate PDF, $\rho_{LS}$= 0.693, $\rho_{LB}$=0.014, and $\rho_{SB}$=-0.030. Similar values are common in retirement research, likely derived using the unbiased sample estimator, $\hat{\sigma}_{XY}=\sum_{i=1}^{N}(x_i - \overline{X})(y_i - \overline{Y})/\sqrt{\sum_{i=1}^{N}(x_i - \overline{X})^2 \sum_{i=1}^{N}(y_i - \overline{Y})^2}$. The corresponding sample correlations are $\hat{\rho}_{LS}$=0.801, $\hat{\rho}_{LB}$=0.046, and $\hat{\rho}_{SB}$=-0.056. The MLEs produced by iterating over (5.A.1) and (5.A.2) are subject to constraints that maintain the marginals from §**III**.[30] Further, the LPs solved in §**IV** set the initial PDF structure and thereby introduce between-regime dependence. For this reason, $\rho$ and $\hat{\rho}$ are unlikely to perfectly align. Since variances and covariances are defined as expectations (averages) they can be skewed by extreme values. The small positive estimate for $\rho_{\mathrm{LB}}$ (and $\hat{\rho}_{\mathrm{LB}}$) masks the fact that in over 70% of years the correlation between L and B is negative (regimes 2 & 3), and in another 12% it is strongly positive (regimes 4 & 5). Mixture modeling thus uncovers insights previously not known as the within component means/variances are used to derive these correlations. As witnessed during the financial crisis, extreme value correlations can invalidate models as simple RV dependence structures do not hold during times of stress. In hindsight, many blame the 2008 crisis on the Gaussian copula and its failure to accurately model failure time correlations of derivative securities under duress. We should not expect other simple structures, such as a single-parameter copula family, to perform any better. Retirement research increasingly advocates for use of complex instruments to improve outcomes. Coupled with the near universal use of a Gaussian (or lognormal) copula to model dependence reveals a situation that sounds all too familiar.[31]

[30] In general MLEs for variance components are biased because the degrees of freedom are not discounted for the estimated means. A procedure such as *restricted maximum likelihood* (REML) corrects this by estimating the variances/covariances after removing the means.

### *C. Simulating from a Multivariate Mixture PDF*

Let $\mathbf{X}=(X_1,...,X_N)'$ be RVs for the compounding return on N financial securities at a given time point where $\mathbf{X}\sim f(\boldsymbol{x})=p_1{\cdot}f^1(\boldsymbol{x})+...+p_G{\cdot}f^G(\boldsymbol{x})$ is a G-component multivariate normal mixture PDF. Simulating values from $f(\boldsymbol{x})$ is a 2-step process. First generate a uniform(0,1) random value, say $u$, to determine the component. If $u \le p_1$, the observation is from regime 1, else if $\sum_{c=1}^{k-1} p_c < u \le \sum_{c=1}^{k} p_c$, the observation is from regime k, k=2,...,G. Next, generate a value from the selected regime, say $f^k(\boldsymbol{x})$. If $\boldsymbol{\Sigma}_k$ has non-zero covariances we apply a decorrelating transformation. Recall that the N eigenvalues ($\lambda_i$) and eigenvectors ($\mathbf{u}_i$) of $\boldsymbol{\Sigma}_k$ satisfy $\boldsymbol{\Sigma}_k{\cdot}\mathbf{u}_i=\lambda_i\mathbf{u}_i$ so that $\mathbf{u}_i'\boldsymbol{\Sigma}_k\mathbf{u}_i=\lambda_i\mathbf{u}_i'\mathbf{u}_i=\lambda_i$ (see §**II.I**). Also from §**V.A.2**, the eigenvectors of a real/symmetric matrix are orthogonal, thus $\mathbf{u}_i'\mathbf{u}_j=0$ for $i\neq j=1,...,N$. Let $\mathbf{U}=[\mathbf{u}_1|...|\mathbf{u}_N]$ and make the linear transform $\mathbf{Z}=\mathbf{U}\mathbf{X}$, where $E(\mathbf{Z})=\mathbf{U}{\cdot}E(\mathbf{X})$ and $V(\mathbf{Z})=\mathbf{U}'{\cdot}\boldsymbol{\Sigma}_k{\cdot}\mathbf{U}=\boldsymbol{\Lambda}$ is a diagonal matrix with variances $\lambda_i$. If the independent and normally distributed $Z=(Z_1,...,Z_N)'$ are simulated individually as $\boldsymbol{z}$ then $\boldsymbol{x}=\mathbf{U}'\boldsymbol{z}$ is a sample on $\mathbf{X}\sim f(\boldsymbol{x})$.

### *D. Stress-Testing a Retirement Plan*

Let $\mathbf{X}=(X_1,...,X_N)'\sim f(\boldsymbol{x})$ be RVs for the compounding return on N financial securities at a given time point. If $f(\boldsymbol{x})$ has been developed using the historical sample then it accounts for "gray" swans (observed outliers). A *stress-test* will be defined as evaluating a retirement strategy using a multivariate PDF, $g(\boldsymbol{x})$, that can produce "black" swans (unobserved outliers), a subjective determination. This is accomplished by seeding the historical data $\mathbf{X}$ with extreme events. Note that the model proposed here can be fit to such data whereas the normal (or lognormal) PDF cannot. The proposed retirement strategy is then subjected to $g(\boldsymbol{x})$.

[31] The problem is not avoided by backtesting a given strategy as the success/failure at any retirement start year is a Bernoulli RV which is highly correlated with the success/failure in nearby years. This correlation is rarely, if ever, accounted for in retirement research.

## VI. Expense-Adjusted Real Compounding Return on a Diversified Portfolio

Assume a retiree holds N securities and let $p_{t-1,i}$ and $P_{t,i}$ be the price (value) of their security i holdings at times t-1 and t, respectively, for i=1,...,N. The time t total return on security i is $R_{t,i}=[P_{t,i}-p_{t-1,i}]/p_{t-1,i}$ and the compounding return is $1+R_{t,i}=P_{t,i}/p_{t-1,i}$, so that $(1+R_{t,i})\cdot p_{t-1,i}=P_{t,i}$. The real compounding return for security i is $1+r_{t,i}$, where $1+R_{t,i}=(1+r_{t,i})(1+I_t)$ and $I_t$ is the inflation rate between times t-1 and t. Solving yields $1+r_{t,i}=p_{t,i}/p_{t-1,i}$, where $p_{t,i}=P_{t,i}/(1+I_t)$ is the real price (value) of security i holdings at time t and $(1+r_{t,i})\cdot p_{t-1,i}=p_{t,i}$. If security i includes an expense ratio, $E_i$, then $E_i\cdot P_{i,t}$ is paid as a cost at time t and the expense-adjusted price (value) is $P_{t,i}-E_i\cdot P_{t,i}=(1-E_i)\cdot P_{i,t}$, the expense-adjusted compounding return is $(1-E_i)(1+R_{t,i})=(1-E_i)\cdot P_{t,i}/p_{t-1,i}$, and the expense-adjusted real compounding return is $(1-E_i)(1+r_{t,i})=(1-E_i)\cdot p_{t,i}/p_{t-1,i}$. Adding up all holdings, denote the expense-adjusted total account values at times t-1 and t by $v_{t-1}=\sum_{i=1}^{N} p_{t-1,i}$ and $V_t=\sum_{i=1}^{N}(1-E_i)\cdot P_{t,i}$, respectively. The expense-adjusted total return on the account between times t-1 and t is $R_t=(V_t-v_{t-1})/v_{t-1}$ and the expense-adjusted compounding return is $1+R_t=V_t/v_{t-1}$. The time t expense-adjusted real compounding return on the account must satisfy $1+R_t=(1+r_t)(1+I_t)$ so that $\hat{r}(t)=1+r_t=(1+R_t)/(1+I_t)=v_t/v_{t-1}$, where $v_t=V_t/(1+I_t)$ is the expense-adjusted real account value at time t. Combining it all,

$$\hat{r}(t)=1+r_t=\frac{V_t}{(1+I_t)\cdot v_{t-1}}=\frac{\sum_{i=1}^{N}(1-E_i)\cdot P_{t,i}}{(1+I_t)\cdot\sum_{i=1}^{N}p_{t-1,i}}=\frac{\sum_{i=1}^{N}(1-E_i)\cdot p_{t,i}}{\sum_{i=1}^{N}p_{t-1,i}}=\frac{\sum_{i=1}^{N}(1-E_i)(1+r_{t,i})\cdot p_{t-1,i}}{\sum_{i=1}^{N}p_{t-1,i}} \tag{6.1}$$

$$=\sum_{i=1}^{N}\left(\frac{p_{t-1,i}}{\sum_{i=1}^{N}p_{t-1,i}}\right)(1-E_i)(1+r_{t,i})=\sum_{i=1}^{N}\alpha_{t-1,i}\cdot(1-E_i)(1+r_{t,i}), \tag{6.2}$$

where $\alpha_{t-1,i}>0$ is the proportion invested in security i at time t-1 and $\{\alpha_{t-1,i},...,\alpha_{t-1,N}\}$ is the asset allocation with $\sum_{i=1}^{N}\alpha_{t-1,i}=1$, which proves (2.N.4). When modeling returns as RVs that are not serially correlated the time index on $(1+r_{t,i})$ can be dropped, however $\hat{r}(t)$ remains a function of time through the asset allocation.

### *A. Expense-Adjusted Real Compounding Return for Portfolio using (L, S, B)*

Let $E_L$, $E_S$, and $E_B$ be annual expenses for the S&P 500 (L), Small Cap Equities (S), and U.S. 10-Year T-Bonds (B). Define the asset allocation, set at time t-1, as $\{\alpha_{t-1L},\alpha_{t-1S},\alpha_{t-1B}\}$ where $\alpha_{t-1L}+\alpha_{t-1S}+\alpha_{t-1B}=1$. Applying (6.2), the expense-adjusted real compounding return, $\hat{r}(t)$, is a linear transform of (L,S,B)′, namely:

$$\hat{r}(t)=\alpha_{t-1L}(1-E_L)L+\alpha_{t-1S}(1-E_S)S+\alpha_{t-1B}(1-E_B)B \tag{6.A.1}$$

The expenses and asset allocations are constants or decision variables. Using means $(\mu_{cl},\mu_{cs},\mu_{cb})$ and variances $(\sigma_{cl}^2,\sigma_{cs}^2,\sigma_{cb}^2)$ from Table VII and probabilities $(p_c)$ and covariances $(\sigma_{cls},\sigma_{clb},\sigma_{csb})$ from Table VIII, $f(l,s,b)$ is:

$$f(l,s,b)=\sum_{c=1}^{5}p_c\cdot f^c(l,s,b)=\sum_{c=1}^{5}\frac{p_c}{\left(\sqrt{2\pi}\right)^3\sqrt{|\boldsymbol{\Sigma}_c|}}e^{-\frac{1}{2}\begin{pmatrix}l-\mu_{cl}\\ s-\mu_{cs}\\ b-\mu_{cb}\end{pmatrix}'[\boldsymbol{\Sigma}_c^{-1}]\begin{pmatrix}l-\mu_{cl}\\ s-\mu_{cs}\\ b-\mu_{cb}\end{pmatrix}},\ \boldsymbol{\Sigma}_c=\begin{bmatrix}\sigma_{cl}^2 & \sigma_{cls} & \sigma_{clb}\\ \sigma_{cls} & \sigma_{cs}^2 & \sigma_{csb}\\ \sigma_{clb} & \sigma_{csb} & \sigma_{cb}^2\end{bmatrix},\ (l,s,b)'\in\mathbb{R}^3 \tag{6.A.2}$$

As in §**II.N**, let $h_t(\hat{r})$ be the univariate PDF of $\hat{r}(t)$ at time t, then $h_t(\hat{r})$ satisfies $H_t(\hat{r})=P[\hat{r}(t)\le\hat{r}]=\int_{-\infty}^{\hat{r}}h_t(x)dx$, where $H_t(\hat{r})$ is the CDF of $\hat{r}(t)$ and $\frac{d}{d\hat{r}}[H_t(\hat{r})]=h_t(\hat{r})$. Define the sample space $S=\{S_1\cup S_2\cup S_3\cup S_4\cup S_5\}$, where

$S_c$ is the event that $Z_c = 1$ and c=1,...,5. These are mutually exclusive with P(S)=1 and,

$$H_t(\hat{r}) = P[\hat{r}(t) \le \hat{r}] = P[(\hat{r}(t) \le \hat{r}) \cap S] = P[(\hat{r}(t) \le \hat{r}) \cap \{S_1 \cup S_2 \cup S_3 \cup S_4 \cup S_5\}] \quad (6.A.3)$$

$$= P\left[\bigcup_{c=1}^{5}\{(\hat{r}(t) \le \hat{r}) \cap S_c\}\right] = \sum_{c=1}^{5} P[(\hat{r}(t) \le \hat{r}) \cap S_c] = \sum_{c=1}^{5} P[(\hat{r}(t) \le \hat{r}) \mid Z_c = 1] \cdot P(Z_c = 1) = \sum_{c=1}^{5}\left[p_c \int_{-\infty}^{\hat{r}} \phi(x, \mu_{tc}, \sigma_{tc}^2)\, dx\right],$$

where $\phi(x, \mu_{tc}, \sigma_{tc}^2)$ is a normal PDF having the following mean and variance for t=1,...,T, c=1,...,5:

$$\mu_{tc} = \alpha_{t\text{-}1L}(1\text{-}E_L)\mu_{cl} + \alpha_{t\text{-}1S}(1\text{-}E_S)\mu_{cs} + \alpha_{t\text{-}1B}(1\text{-}E_B)\mu_{cb} \quad (6.A.4)$$

$$\sigma_{tc}^2 = \alpha_{t\text{-}1L}^2(1\text{-}E_L)^2\sigma_{cl}^2 + \alpha_{t\text{-}1S}^2(1\text{-}E_S)^2\sigma_{cs}^2 + \alpha_{t\text{-}1B}^2(1\text{-}E_B)^2\sigma_{cb}^2 \quad (6.A.5)$$

$$+ 2\alpha_{t\text{-}1L}(1\text{-}E_L)\alpha_{t\text{-}1S}(1\text{-}E_S)\sigma_{cls} + 2\alpha_{t\text{-}1L}(1\text{-}E_L)\alpha_{t\text{-}1B}(1\text{-}E_B)\sigma_{clb} + 2\alpha_{t\text{-}1S}(1\text{-}E_S)\alpha_{t\text{-}1B}(1\text{-}E_B)\sigma_{csb}$$

The PDF for $\hat{r}(t)$ is a 5-component univariate Gaussian mixture PDF, derived as:

$$h_t(\hat{r}) = \frac{d}{d\hat{r}}[H_t(\hat{r})] = \sum_{c=1}^{5}[p_c \cdot \phi(\hat{r}, \mu_{tc}, \sigma_{tc}^2)], \quad \hat{r} \in \mathbb{R} \quad (6.A.6)$$

Various PDFs using (6.A.6) are shown in Figure XII and feature asset allocations that are dominant in one of the securities (L,S,B). They are non-normal, skewed ($\gamma>0$) and evidently (mostly) heavier-tailed ($\kappa>3$) than a normal PDF. Each approaches its univariate shape as the proportion for that security increases (see §**VI.B**).

***Figure XII**. PDF for $\hat{r}(t)$ (Expense-Adjusted Real Compounding Return on a Diversified Portfolio)*

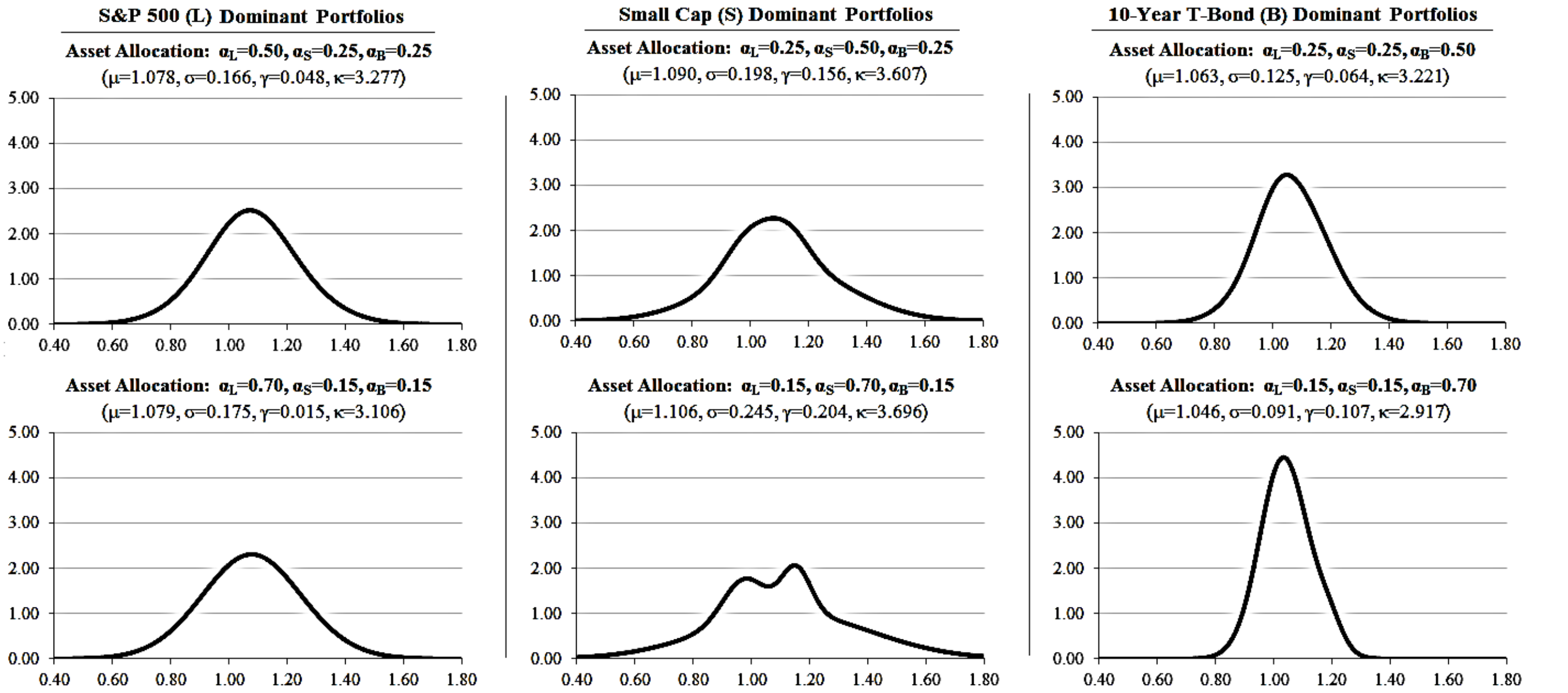

Notes: 1.) Expenses for all asset allocations are $E_L$=0.0015 (0.15%), $E_S$=0.0025 (0.25%), and $E_B$=0.0020 (0.20%).
2.) Minimum variance portfolio ($\sigma$=0.076) occurs at asset allocation $\alpha_L$=0.141 , $\alpha_S$=0.003, $\alpha_B$=0.856.

## *B. Marginal Distributions*

### *B.1: S&P 500 (L): $f_L(l)$*

Set $E_L$=0, $\alpha_{tL}$=1, $\alpha_{tS}$=0, $\alpha_{tB}$=0 in (6.A.1), (6.A.4) and (6.A.5), then $\hat{r}(t)=L\sim f_L(l)=\sum_{c=1}^{5} p_c \cdot \phi(\hat{r}, \mu_{cl}, \sigma_{cl}^2)$, where $\mu_{cl}$=1.082, $\sigma_{cl}^2=0.197^2$, c=1,...,5, and $\sum_{c=1}^{5} p_c=1$. Thus, matching §**III.A.1**:

$$L\sim f_L(l)=\phi(1.082, 0.197^2) \quad (6.A.7)$$

*B.2: Small Cap (S): $f_S(s)$*

Set $E_S$=0, $\alpha_{tL}$=0, $\alpha_{tS}$=1, $\alpha_{tB}$=0 in (6.A.1), (6.A.4) and (6.A.5), then $\hat{r}(t)=S\sim f_S(s)=\sum_{c=1}^{5} p_c \cdot \phi(\hat{r}, \mu_{cs}, \sigma_{cs}^2)$, where $\mu_{cs}$=0.945, $\sigma_{cs}^2=0.065^2$, $p_c$=0.164 for c=1, $\mu_{cs}$=1.165, $\sigma_{cs}^2=0.367^2$ for c=2,3 with $p_2+p_3$=0.683+0.024 =0.707, and $\mu_{cs}$=1.192, $\sigma_{cs}^2=0.024^2$ for c=4,5 with $p_4+p_5$=0.101+0.028=0.129. Thus, matching §**III.A.2**:

$$S\sim f_S(s)=(0.164)\cdot\phi(0.945, 0.065^2) + (0.707)\cdot\phi(1.165, 0.367^2) + (0.129)\cdot\phi(1.192, 0.024^2) \quad (6.A.8)$$

*B.3: U.S. 10-Year T-Bonds (B): $f_B(b)$*

Set $E_B$=0, $\alpha_{tL}$=0, $\alpha_{tS}$=0, $\alpha_{tB}$=1 in (6.A.1), (6.A.4) and (6.A.5), then $\hat{r}(t)=B\sim f_B(b)=\sum_{c=1}^{5} p_c \cdot \phi(\hat{r}, \mu_{cb}, \sigma_{cb}^2)$, where $\mu_{cb}$=1.011, $\sigma_{cb}^2=0.070^2$ for c=1,2,4 with $p_1+p_2+p_4$=0.164+0.683+0.101=0.948, and $\mu_{cb}$=1.220, $\sigma_{cb}^2$= $0.017^2$ for c=3,5 with $p_3+p_5$=0.024+0.028=0.052. Thus, matching §**III.A.3**:

$$B\sim f_B(b)=(0.948)\cdot\phi(1.011, 0.070^2) + (0.052)\cdot\phi(1.220, 0.017^2) \quad (6.A.9)$$

## VII. Retirement Portfolio Optimization

Dynamic retirement glidepaths are asset allocations that adapt over time to changes in either a retiree's funded status or market conditions, whereas static glidepaths are fixed allocations the retiree can 'set and forget'. Glidepaths are often considered in the context of safe withdrawal rates (SWR), see §**II.N**. The optimal glidepath outperforms all others with respect to some measure. Rook (2014) and (2015) derive the dynamic and static glidepaths that minimize the probability of ruin using an SWR, respectively. Both models use normally distributed returns which is an assumption many practitioners and researchers reject due to the lack of skewness and heavy tail. The purpose of this research is to extend those models to returns that are non-normal, skewed, heavy-tailed, multi-model, i.e., of generic complexity. The multivariate PDF we develop also allows for stress-testing a retirement plan using data seeded with extreme events (see §**V.D)**.

### *A. Optimal Dynamic Retirement Glidepaths*

The dynamic glidepath that minimizes the probability of ruin using an SWR is derived in Rook (2014) via a dynamic program (DP) for fixed and random time horizons. It applies to an individual or a group. The DP value function V(·) is defined on 3 dimensions: t, RF(t) and $\hat{r}(t)$. Both t and RF(t) are discretized (see §**II.N**) to construct the corresponding grid that manages V(·) and it can be solved when the CDF for $\hat{r}(t)$ is tractable. Under lognormal returns the CDF for $\hat{r}(t)$ is intractable although methods do exist to approximate it, see Rook & Kerman (2015) for one implementation. We have derived $\hat{r}(t)$, the expense-adjusted annual real compounding return for a diversified portfolio using S&P 500 (L), Small Cap Equity (S), and U.S. 10-Year T-Bonds (B) returns. The CDF for $\hat{r}(t)$ in (6.A.3) is a function of the normal CDF, Φ(·), which is considered tractable as near exact approximation routines are readily available. C/C++ source code to solve the 2-asset portfolio problem is supplied in Rook (2014). For example, an SWR plan using small cap stocks

(S) and bonds (B) can be optimized using $\hat{r}(t)$ from (6.A.3) with $\alpha_{tL}=0$, $\alpha_{tS}\geq 0$, $\alpha_{tB}\geq 0$ and $\alpha_{tS}+\alpha_{tB}=1$, so that $\alpha_{tB}=1-\alpha_{tS}$. Incorporating additional securities (such as L) into the optimal dynamic glidepath problem is straight-forward.

### *B. Optimal Static Retirement Glidepaths*

Expressions for the probability of ruin were supplied in §**II.N** for fixed (2.N.7) and random (2.N.8) time horizons. Minimizing the probability of ruin and maximizing the probability of success are equivalent optimization problems. Using $\hat{r}(t) \sim h_t(\hat{r})$ from (6.A.6) and (2.N.9), under a fixed time horizon $T_F$:

$$P[\text{Ruin}^C(\leq T_F)] = \int_{RF(0)}^{\infty} \dots \int_{RF(T_F-1)}^{\infty} \prod_{t=1}^{T_F} \left( \sum_{c=1}^{5} [p_c \cdot \phi(x_t, \mu_{tc}, \sigma_{tc}^2)] \right) dx_{T_F} \dots dx_1 \tag{7.B.1}$$

The optimal static glidepath is found by maximizing (7.B.1) with respect to the asset allocation, and is solved for a 2-asset portfolio in Rook (2015) using both gradient ascent and Newton's method for fixed and random time horizons. As in §**VII.A**, assume S and B are used so that $\alpha_{tL}=0$, $\alpha_{tS}\geq 0$, $\alpha_{tB}\geq 0$ and $\alpha_{tS}+\alpha_{tB}=1$. Since $\alpha_{tB}=1-\alpha_{tS}$, the probability of success is a function of $\{\alpha_{0S},\dots,\alpha_{T_F-1S}\}$. The 1st derivative wrt each $\alpha_{iS}$, $i=0,\dots,T_F-1$ is:

$$\frac{\partial}{\partial \alpha_{iS}} P[\text{Ruin}^C(\leq T_F)] = \int_{RF(0)}^{\infty} \dots \int_{RF(T_F-1)}^{\infty} \prod_{\substack{t=1 \\ t\neq i}}^{T_F} \left( \sum_{c=1}^{5} [p_c \cdot \phi(x_t, \mu_{tc}, \sigma_{tc}^2)] \right) \cdot \frac{\partial}{\partial \alpha_{iS}} \left( \sum_{g=1}^{5} [p_g \cdot \phi(x_i, \mu_{ig}, \sigma_{ig}^2)] \right) dx_{T_F} \dots dx_1 \tag{7.B.2}$$

$$= \sum_{g=1}^{5} p_g \cdot \left[ \int_{RF(0)}^{\infty} \dots \int_{RF(T_F-1)}^{\infty} \prod_{\substack{t=1 \\ t\neq i}}^{T_F} \left( \sum_{c=1}^{5} [p_c \cdot \phi(x_t, \mu_{tc}, \sigma_{tc}^2)] \right) \cdot \frac{\partial}{\partial \alpha_{iS}} \phi(x_i, \mu_{ig}, \sigma_{ig}^2)\, dx_{T_F} \dots dx_1 \right], \tag{7.B.3}$$

The 2nd derivatives wrt the same $\alpha_{iS}$, $i=0,\dots,T_F-1$ are:

$$\frac{\partial^2}{\partial \alpha_{iS}^2} P[\text{Ruin}^C(\leq T_F)] = \sum_{g=1}^{5} p_g \cdot \left[ \int_{RF(0)}^{\infty} \dots \int_{RF(T_F-1)}^{\infty} \prod_{\substack{t=1 \\ t\neq i}}^{T_F} \left( \sum_{c=1}^{5} [p_c \cdot \phi(x_t, \mu_{tc}, \sigma_{tc}^2)] \right) \cdot \frac{\partial^2}{\partial \alpha_{iS}^2} \phi(x_i, \mu_{ig}, \sigma_{ig}^2)\, dx_{T_F} \dots dx_1 \right] \tag{7.B.4}$$

The 2nd derivatives wrt $\alpha_{jS}$, $j\neq i=0,\dots,T_F-1$ are:

$$\frac{\partial^2}{\partial \alpha_{iS}\, \partial \alpha_{jS}} P[\text{Ruin}^C(\leq T_F)] \tag{7.B.5}$$
$$= \sum_{k=1}^{5} p_k \left( \sum_{g=1}^{5} p_g \left[ \int_{RF(0)}^{\infty} \dots \int_{RF(T_F-1)}^{\infty} \prod_{\substack{t=1 \\ t\neq i,j}}^{T_F} \left( \sum_{c=1}^{5} [p_c \cdot \phi(x_t, \mu_{tc}, \sigma_{tc}^2)] \right) \cdot \frac{\partial}{\partial \alpha_{iS}} \phi(x_i, \mu_{ig}, \sigma_{ig}^2) \cdot \frac{\partial}{\partial \alpha_{jS}} \phi(x_j, \mu_{jk}, \sigma_{jk}^2)\, dx_{T_F} \dots dx_1 \right] \right)$$

Each term in the sum of (7.B.3), (7.B.4), and (7.B.5) can be computed to an arbitrary level of precision, see Rook (2015) which includes the relevant C/C++ source code. As in §**VII.A**, the corresponding DPs would use the CDFs of univariate normal mixture PDFs for t other than i,j. Estimates for these expressions can also be generated using simulation.

## VIII. Conclusion

As retirement decumulation models increase in sophistication financial firms may guarantee their success. The retiree could pay for this as a percentage of funds remaining at death. Decumulation models are statistical and based on assumptions which if incorrect can render the model unsound. Quantitative mortgage products sold during the housing boom were priced using cash-flows generated by simulating default times with a Gaussian copula. In hindsight, the normal assumption was incorrect because default-time correlations change in a crisis. Since housing booms are followed by housing busts, model assumptions should have incorporated economic regimes. As our economy transitions from pensions to defined contribution plans, quantitative retirement products are proliferating. At present, the industry is built on a Gaussian (or lognormal) foundation which also fails to incorporate regimes or crises when modeling returns and their correlation.

The purpose of this research is to develop a multivariate PDF for asset returns that is suitable for quantitative retirement plans. The model fits any set of returns, however the curse of dimensionality will limit the number of securities. We propose a multivariate mixture having fixed mixture marginals using normal components. The model is motivated by the claim that a lognormal PDF is virtually indistinguishable from a mixture of normals. Whereas the lognormal PDF is intractable with regard to weighted sums, the normal mixture is not. The lognormal PDF is only justifiable when short-term returns are *iid* and the PDF is CLT-compatible for the given sample size. A typical retiree could endure several market crashes and we should not expect the historical sample to represent all possible extremes. We can stress test a retirement plan by subjecting it to a return PDF that has been fit on the historical sample seeded with black swan events. The normal or lognormal PDF are unhelpful in this regard as neither can accommodate such outliers.

The univariate and multivariate PDFs we have developed fit the historical returns closely and a valid criticism is that models which memorize the training data project poorly into the future. Adjusting the variance ratio constraint and/or p-value when bootstrapping the marginal LRTs can loosen the fit. We have used relatively high values for both. Larger variance ratio constraints and LRT p-values lead to more marginal peaks and more components. Since the multivariate PDF maintains the marginals, over fitting the marginals propagates through to the multivariate PDF. The user sets these values as desired. We fit the multivariate PDF in 3 steps. First generic mixture marginals are derived using the EM algorithm. Second, the multivariate PDF structure is set using LPs where the number of multivariate regimes is pruned by penalizing the objective when it includes components with no data. Lastly, covariances are added and probabilities updated using an ECME likelihood-based approach with the M-step split into convex and general NLP optimizations. For the NLP we use a Levenberg-Marquardt approach that simulates the step size and line search parameters while iterating. Lastly, a linear transform on the multivariate PDF forms the expense-adjusted real compounding return on a diversified portfolio and it is incorporated it into optimal discrete time retirement decumulation models using both static and dynamic asset allocation glidepaths.

## Data Sources

### CPI-U (Inflation Rate), Cash (No Interest)

### Real S&P 500, Total S&P 500, Real U.S. 10-Year T-Bond, Total U.S. 10-Year T-Bond, Real U.S. 3-Month T-Bill

### Real Small Cap Equity, Total Small Cap Equity

### Yearly Shiller CAPE Ratio (January), S&P 500 Earnings (January)

### Gold Returns

## Retiree Surveys

# IX. Appendix with Source Code

## Appendix A. Proof of Unbounded Likelihood for Normal Mixture PDF

Let $X_t \overset{iid}{\sim} f(x) = \pi_1 f^1(x) + \pi_2 f^2(x) + ... + \pi_g f^g(x)$, where $f(x)$ is a g-component mixture PDF with $f^i(x) \sim N(\mu_i,\sigma_i)$, i=1,2,...,g, t=1,2,...,T, and suppose that an *iid* sample of size T has been observed as $(x_1,x_2,...,x_T)$. The normal component PDFs are given by:

$$f^{\mathrm{i}}(x) = \frac{1}{\sqrt{2\pi}\sigma_{\mathrm{i}}} e^{-\frac{1}{2}\left(\frac{x-\mu_{\mathrm{i}}}{\sigma_{\mathrm{i}}}\right)^2}, \quad -\infty < x < \infty, \;\; \mathrm{i} = 1,2,\dots,\mathrm{g} \tag{9.A.1}$$

The vector of unknown parameters for $f(x)=f(x,\boldsymbol{\theta})$ is defined as $\boldsymbol{\theta}=(\boldsymbol{\pi},\boldsymbol{\mu},\boldsymbol{\sigma})'$ where $\boldsymbol{\pi}=(\pi_1,...,\pi_g)$, $\boldsymbol{\mu}=(\mu_1,...,\mu_g)$, and $\boldsymbol{\sigma}=(\sigma_1,...,\sigma_g)$. To obtain an arbitrarily large value for $\mathfrak{L}^{(1)}(\boldsymbol{\theta}|\boldsymbol{x})$ from (2.E.1), take a single component and dedicate it to one observation. For example, consider component k, $f^k(x)$, and observation j, $x_j$. Let ε be an arbitrarily small number and set $\pi_k=1/T$, $\mu_k = x_j-\varepsilon$, and $\sigma_k=\varepsilon$. The value of $f^k(x_j)$ is:

$$f^{\mathrm{k}}(x_{\mathrm{j}}) \propto \frac{1}{\varepsilon} e^{-\frac{1}{2}\left(\frac{\varepsilon}{\varepsilon}\right)^2} \propto \frac{1}{\varepsilon} \tag{9.A.2}$$

Recall that we initialize $\boldsymbol{\theta}$ to begin the EM-algorithm and $\mathfrak{L}^{(1)}(\boldsymbol{\theta}|\boldsymbol{x})$ is guaranteed to increase at each iteration. From (9.A.2) above, as $\varepsilon \to 0$, $f^k(x_t) \to \infty$. Therefore by choosing k, j, and ε we can initialize $f^k(x_t)$ to an arbitrarily large number making $\mathfrak{L}^{(1)}(\boldsymbol{\theta}|\boldsymbol{x})$ unbounded. Such a solution for $\boldsymbol{\theta}$, however, is not meaningful as it dedicates a near-degenerate component to a single observation. Similarly, a component that is trapped into fitting a small number of closely clustered observations also leads to a high likelihood value (due to the small variance) and is referred to as a spurious maximizer of $\mathfrak{L}^{(1)}(\boldsymbol{\theta}|\boldsymbol{x})$ when the solution is not meaningful. These should be identified and removed if they are not the MLE, $\widehat{\boldsymbol{\theta}}$. To avoid manually evaluating each $\mathfrak{L}^{(1)}(\boldsymbol{\theta}|\boldsymbol{x})$ for spuriousness, we impose a variance ratio constraint as noted in (§**II.E**). This prevents any single variance from becoming too small and eliminates both problems noted above (McLachlan & Peel, 2000).

## Appendix B. Diagnostic Plots for 18 Finance/Economics Time Series

Diagnostic plots for the 18 finance/economics time series analyzed in §**II.J** Table I are presented here. Each time series includes a plot of the uncentered raw observations, $X_t$, t=1,2,...,T as well as the ACF and PACF up to N/4 lags, where N = total # data points for the series (Box, et al., 1994). Annual values from 1928 – 2015 are used therefore N = 88 for all series that are not differenced or averaged. The sources for all data can be found in the *Data Sources* section located after *References* in the main paper. Each test is formed as Ho: No Serial Correlation vs Ha: Serial Correlation, with the p-value provided on the PACF plot. Preliminary conclusions about the behavior of each process are supplied in Table I followed by a discussion of the appropriate family-wise $\alpha^*$ to account for multiplicity. All security returns are annual compounding.

## 1.) *CPI-U (Inflation Rate)*

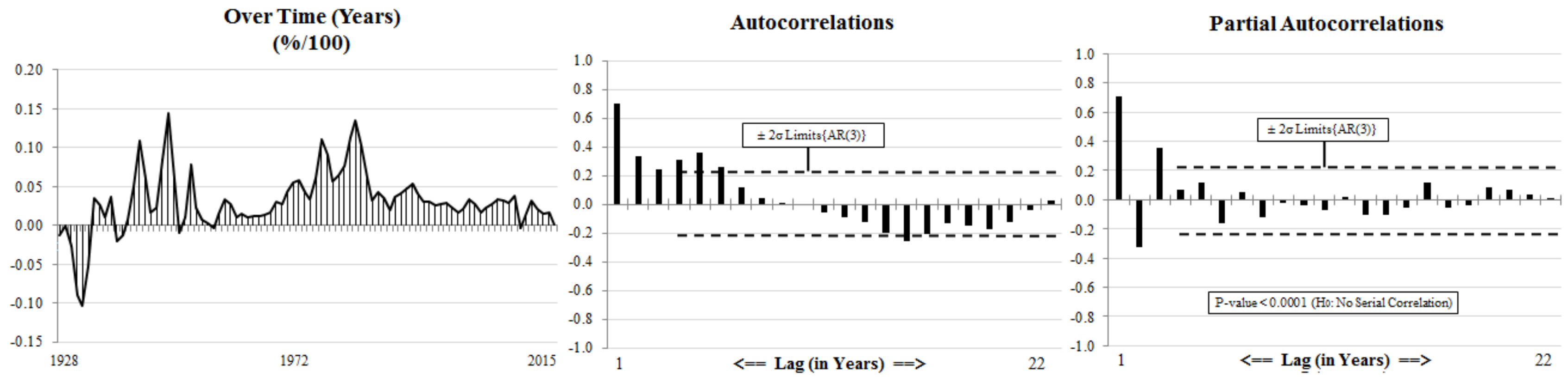

## 2.) *Real Compounding S&P 500*

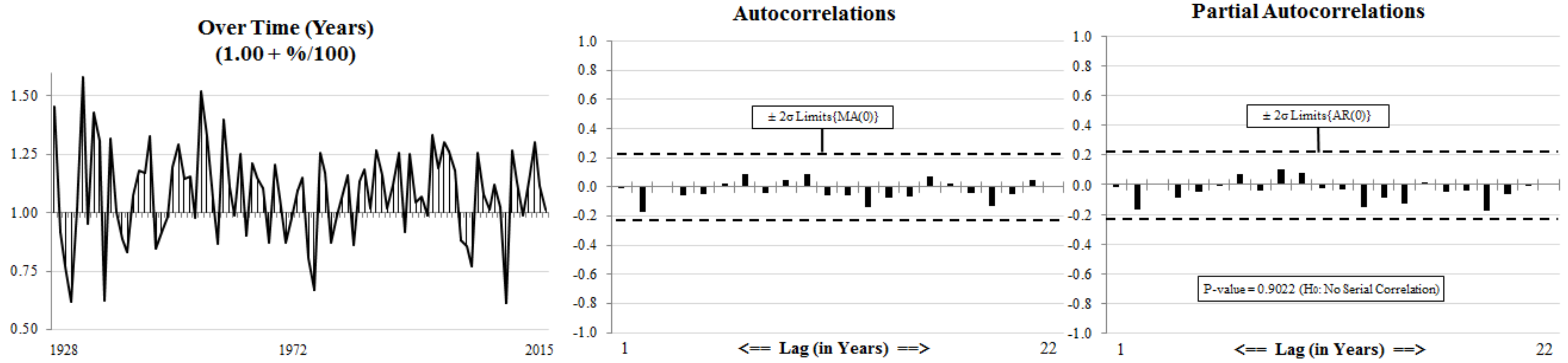

## 3.) *Total Compounding S&P 500*

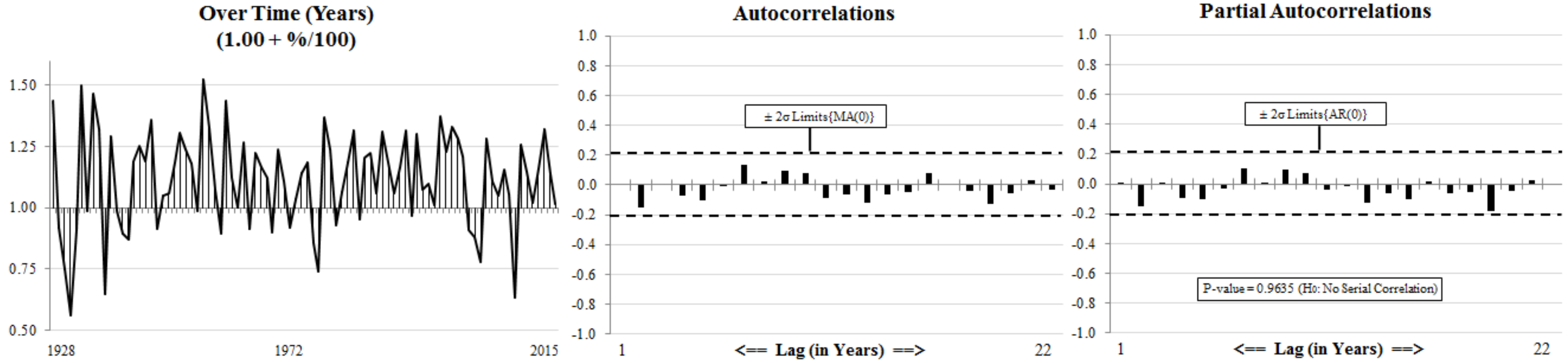

## *4.) Real Compounding Small Cap Equity*

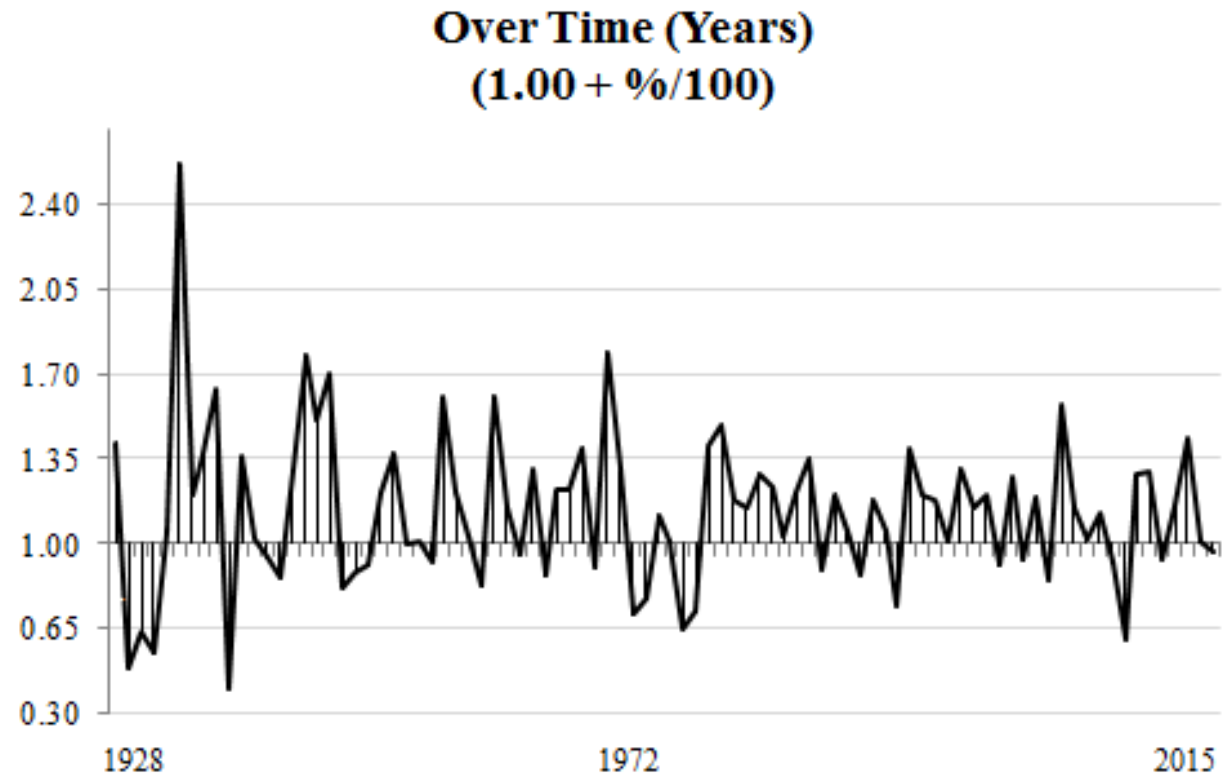

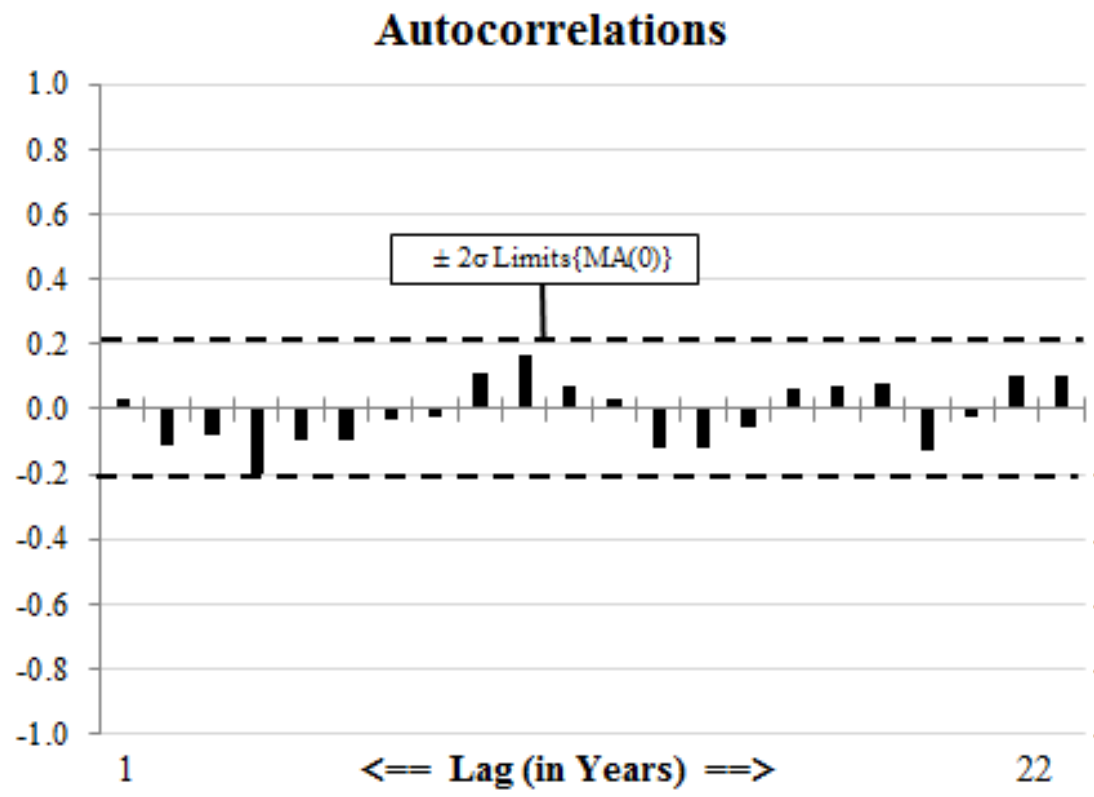

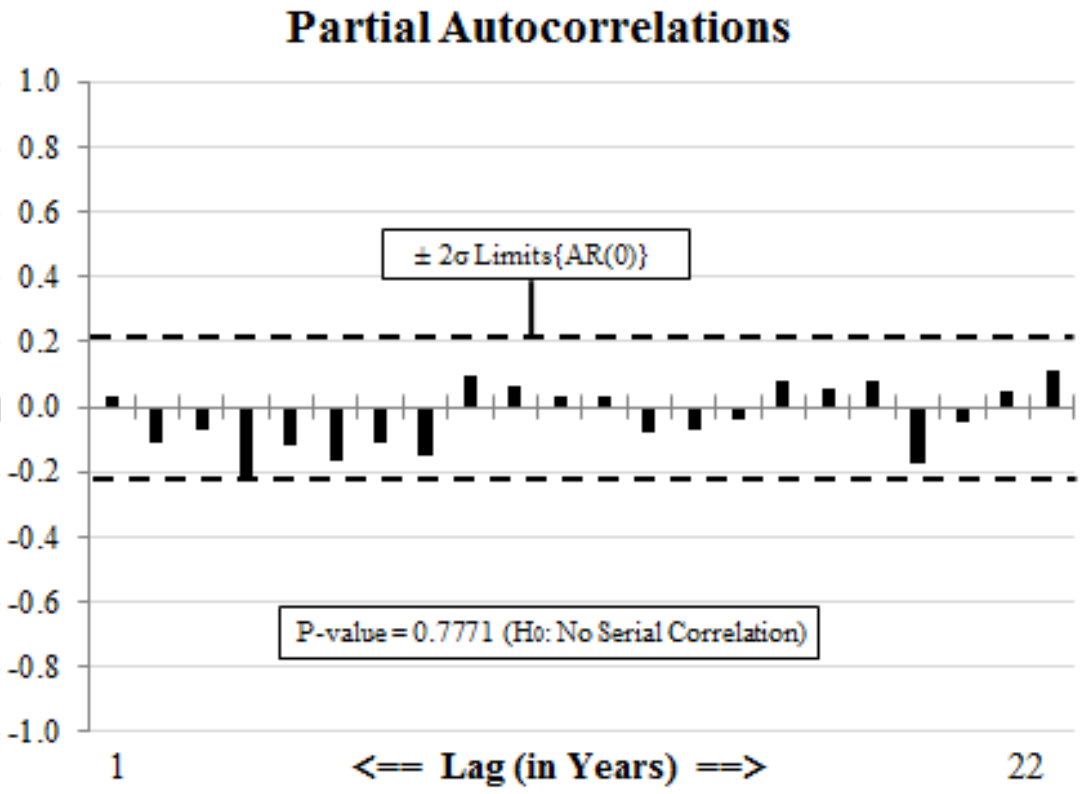

## *5.) Total Compounding Small Cap Equity*

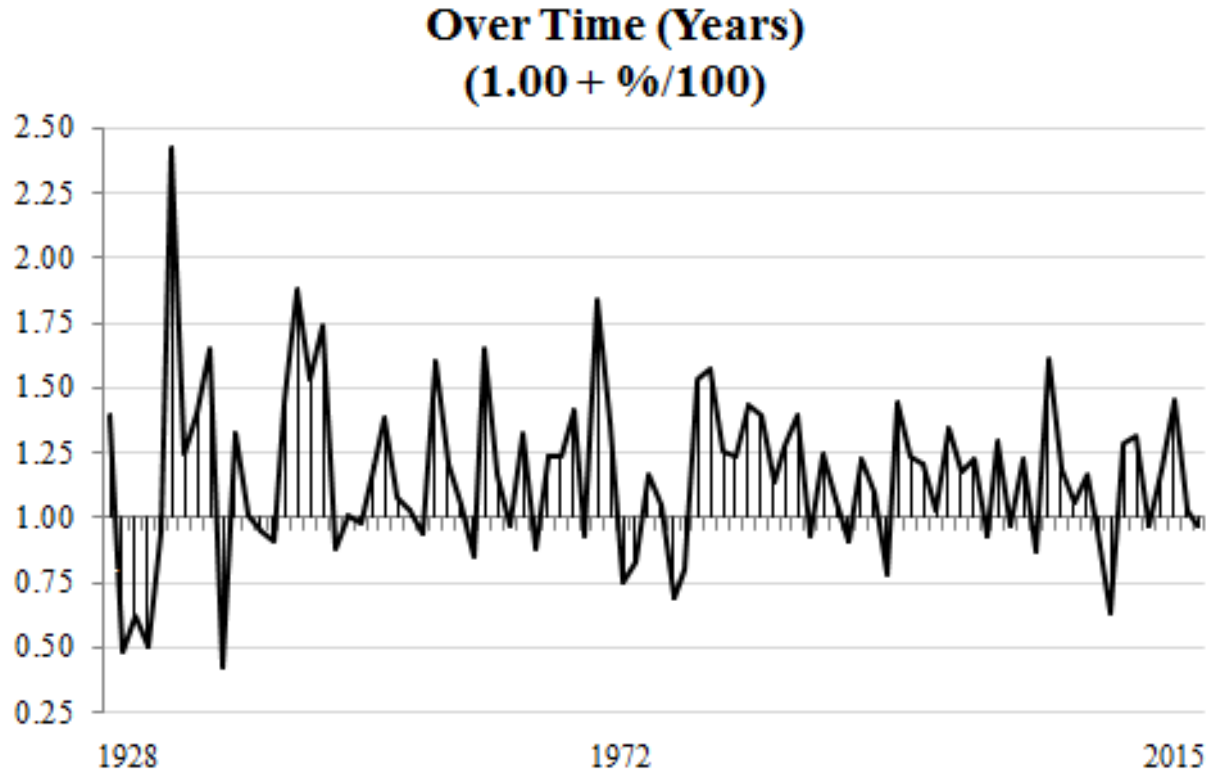

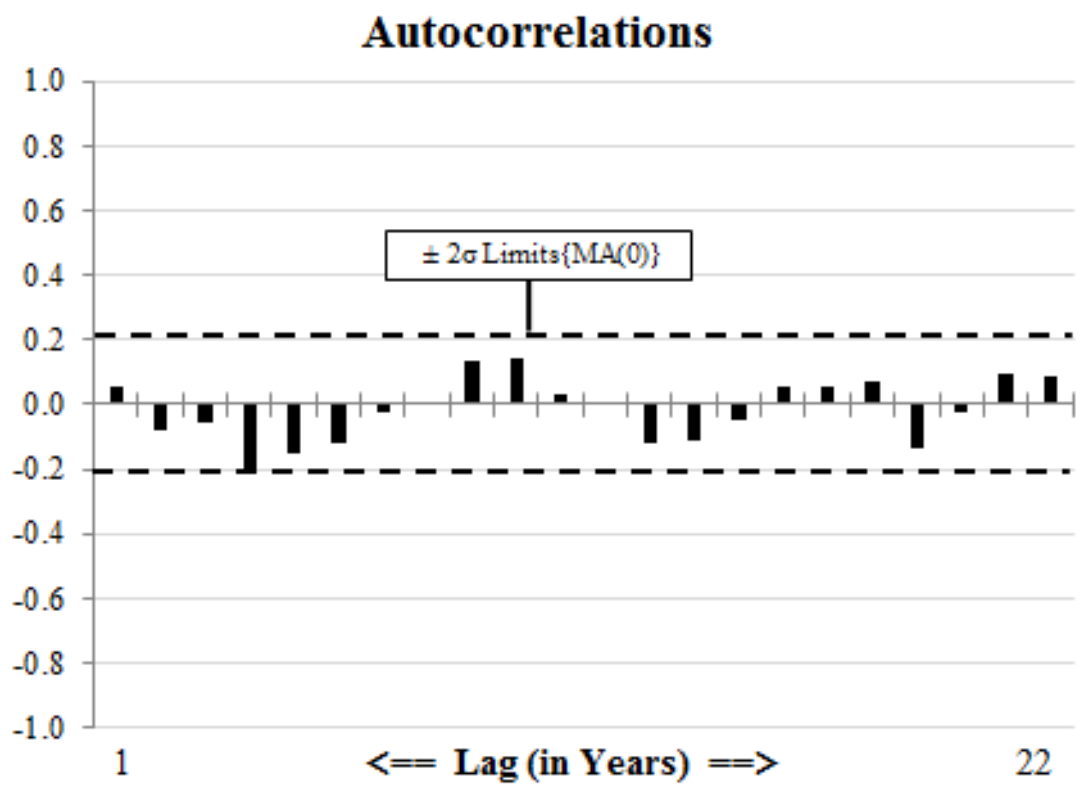

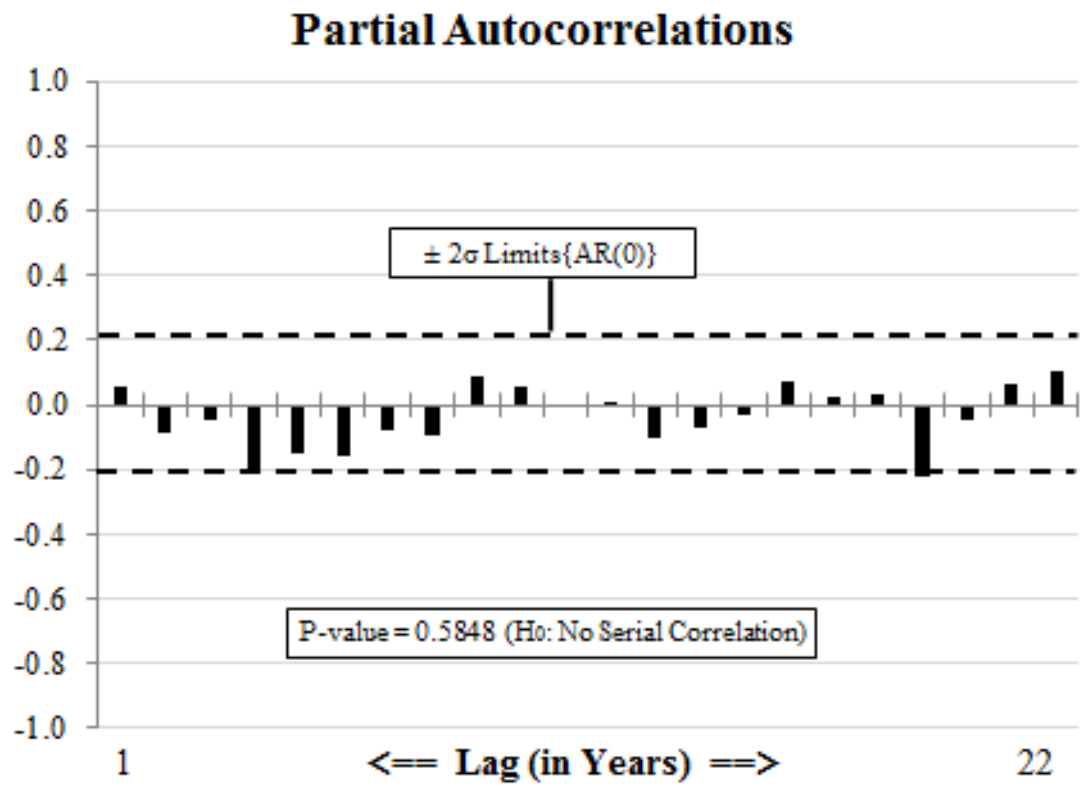

## *6.) Real Compounding U.S. 10-Year T-Bond*

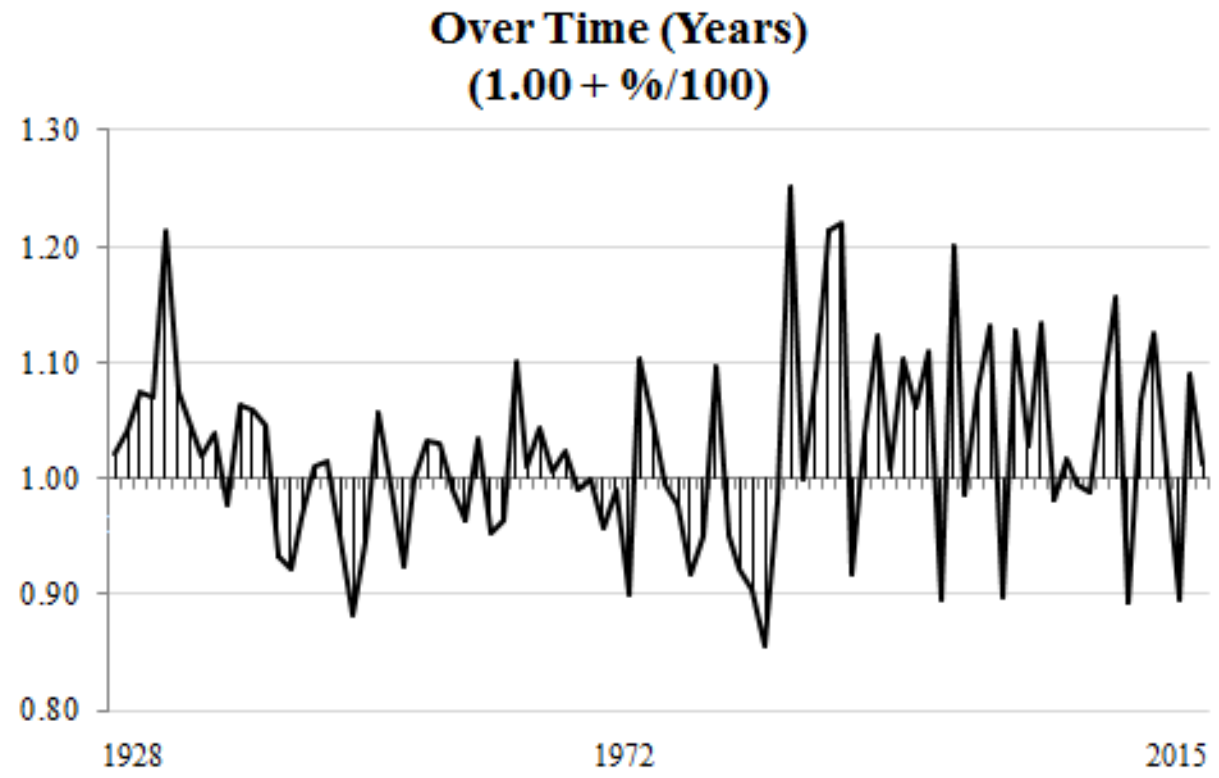

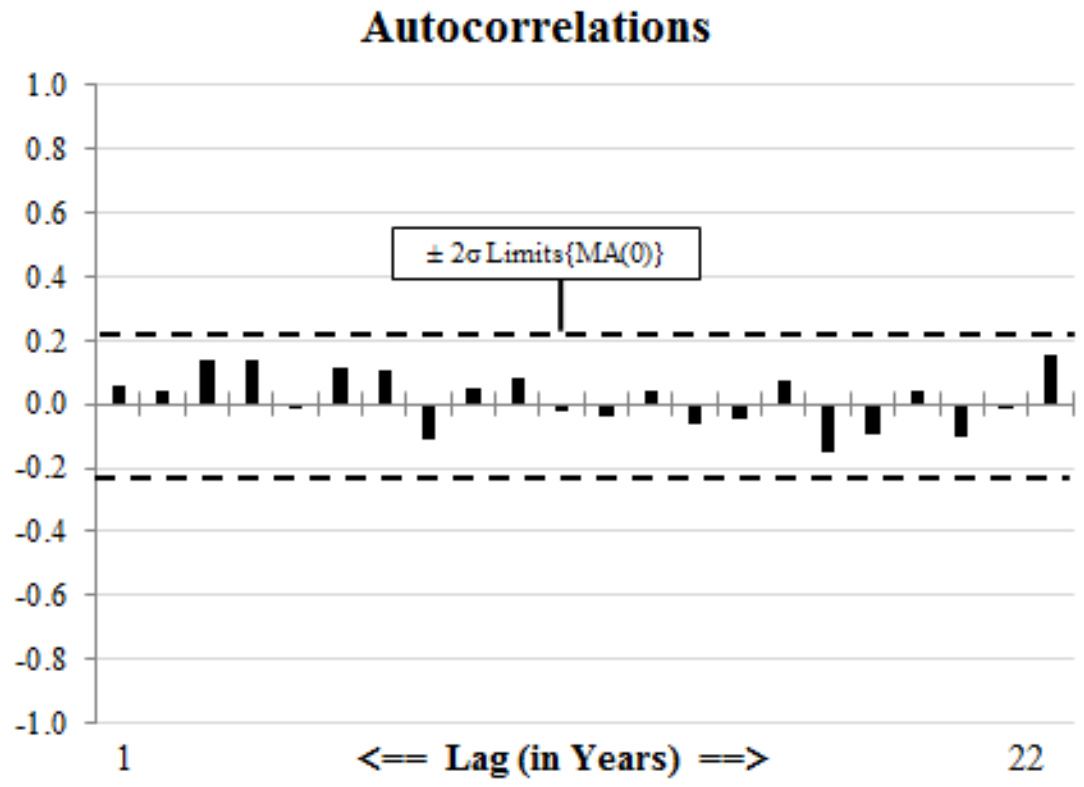

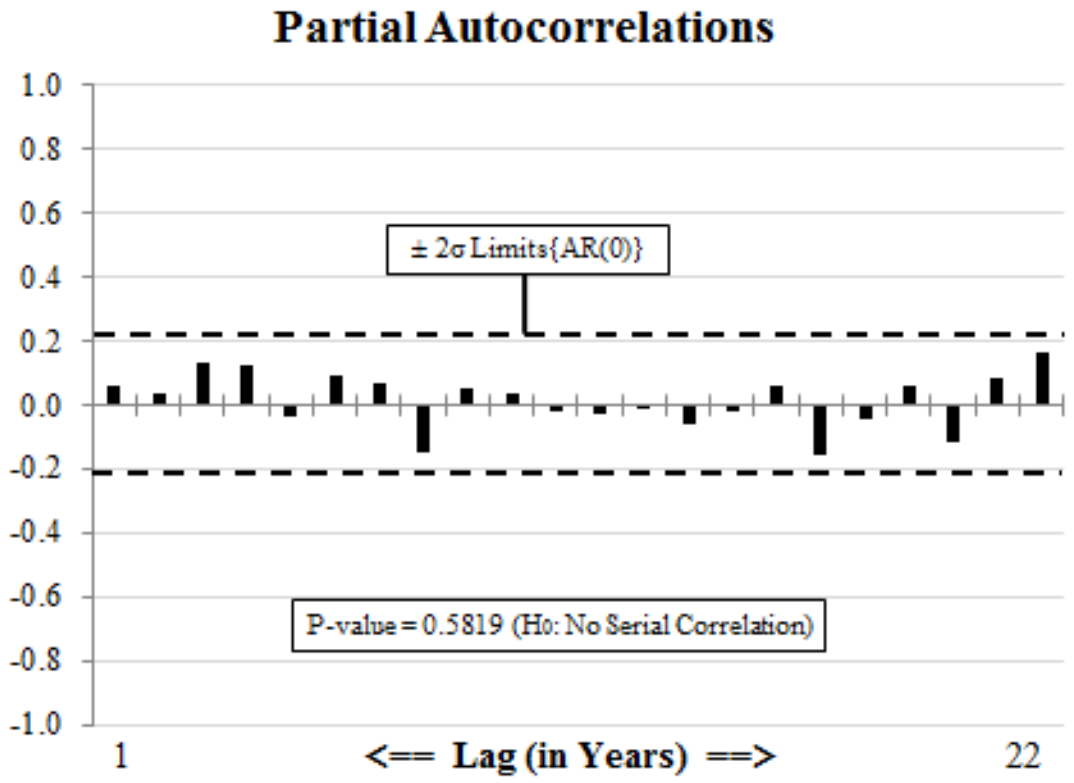

## 7.) *Total Compounding U.S. 10-Year T-Bond*

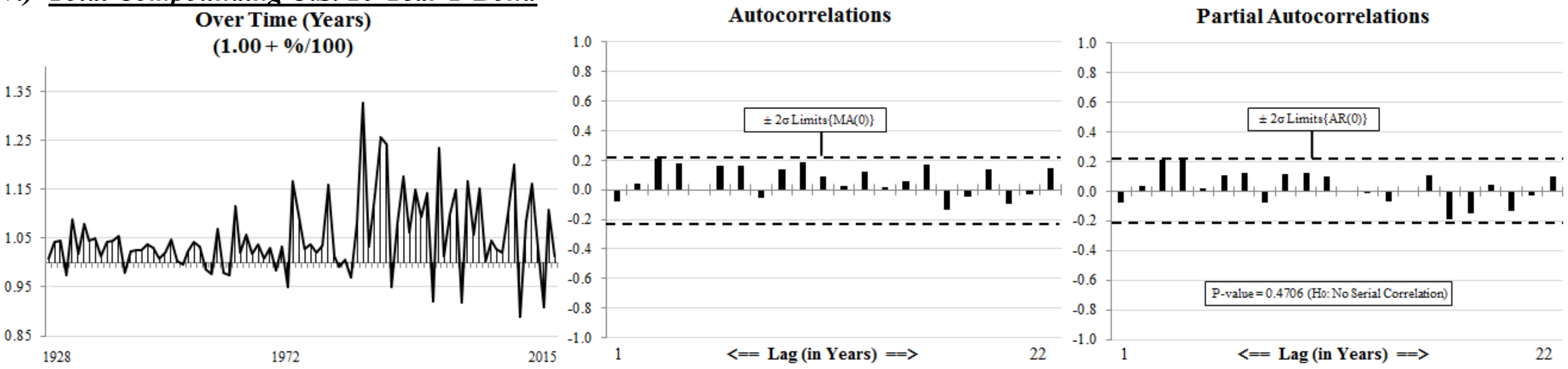

## 8.) *Real Compounding U.S. 3-Month T-Bill*

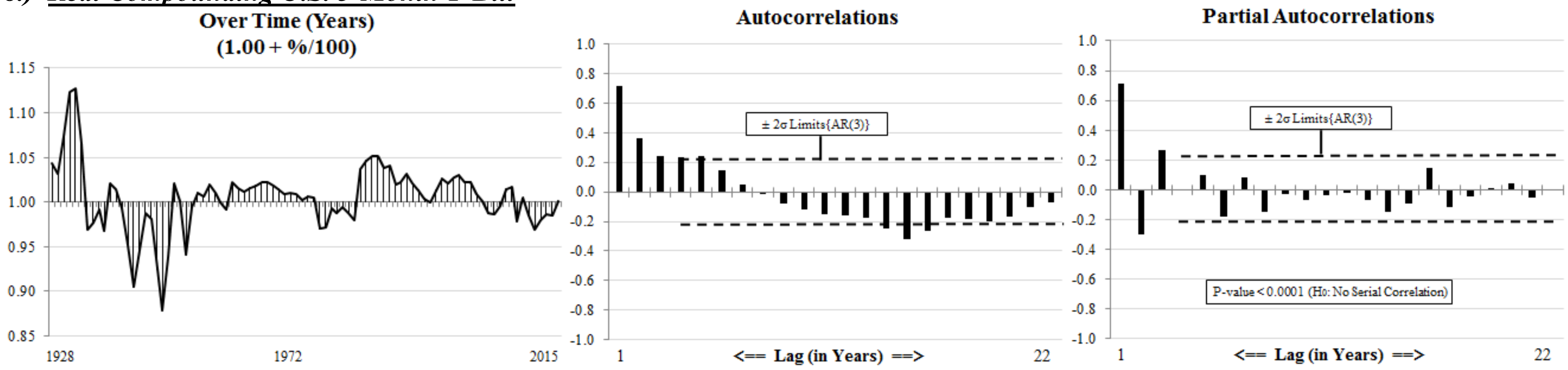

## 9.) *Real Compounding Gold Return*

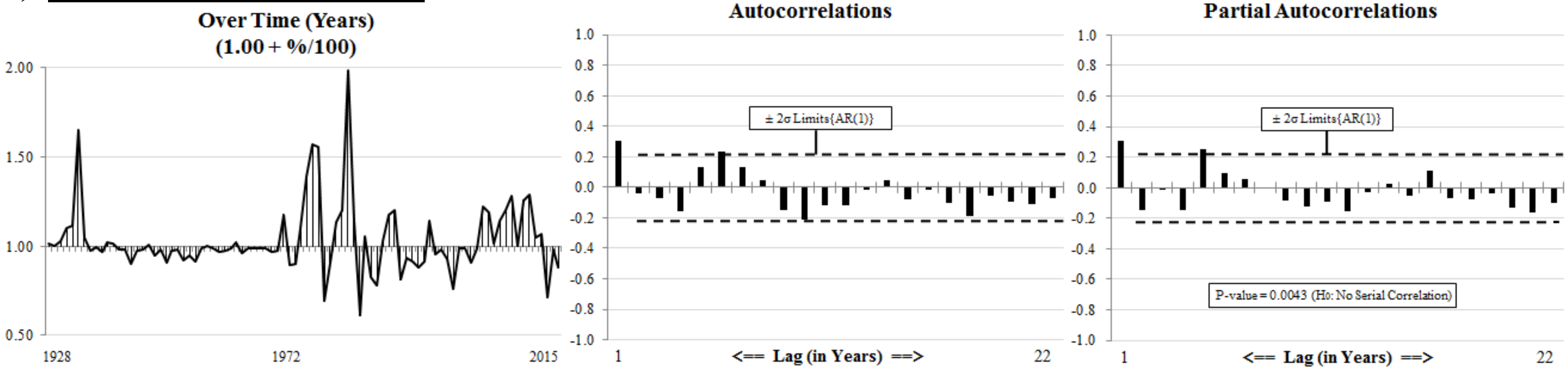

## 10.) *Cash Compounding (No Interest)*

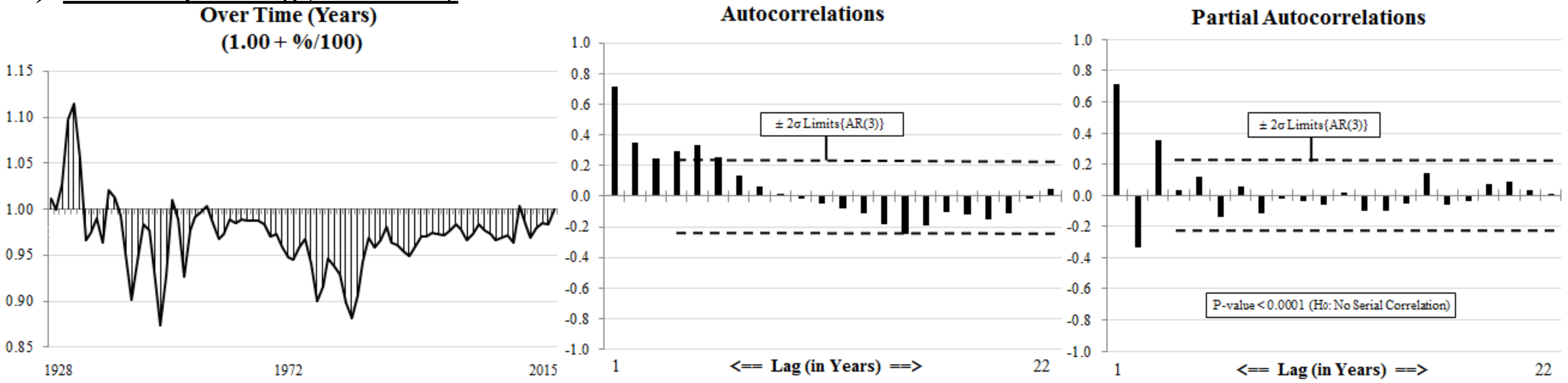

## 11.) *Real S&P 500-to-Bond Risk Premium*

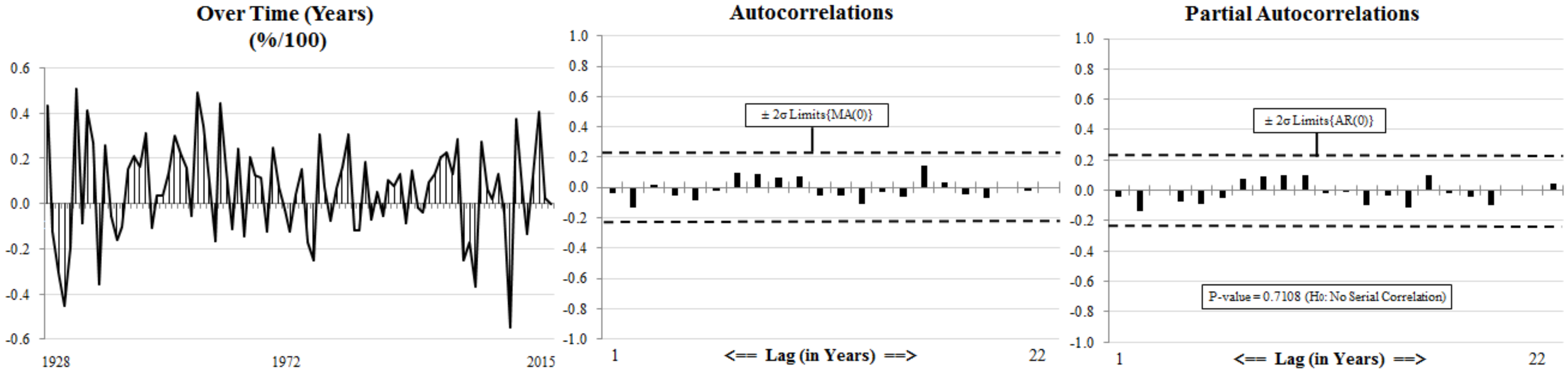

## 12.) *Real S&P 500-to-Small Cap Risk Premium*

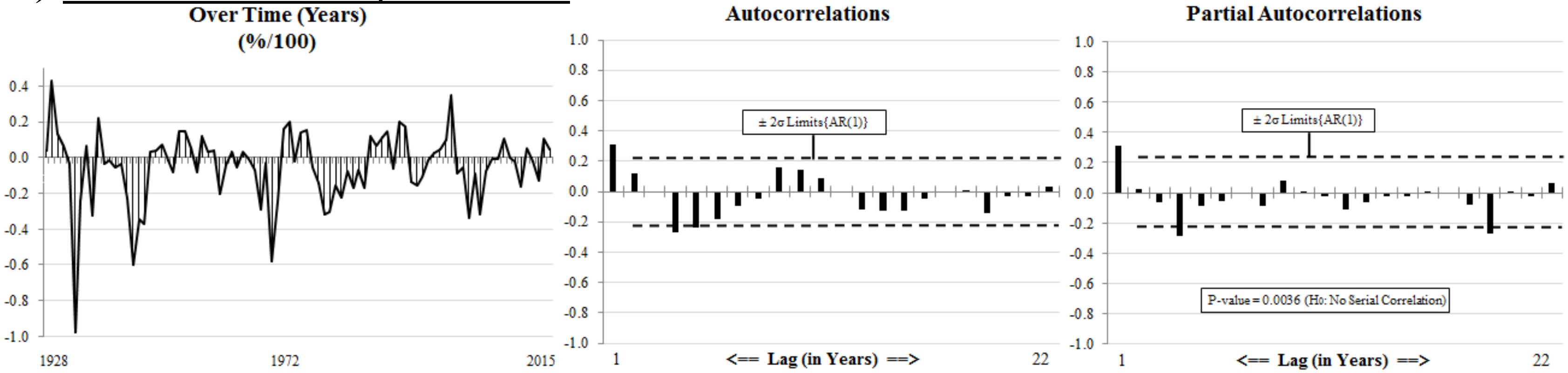

## 13.) *Real Small Cap-to-Bond Risk Premium*

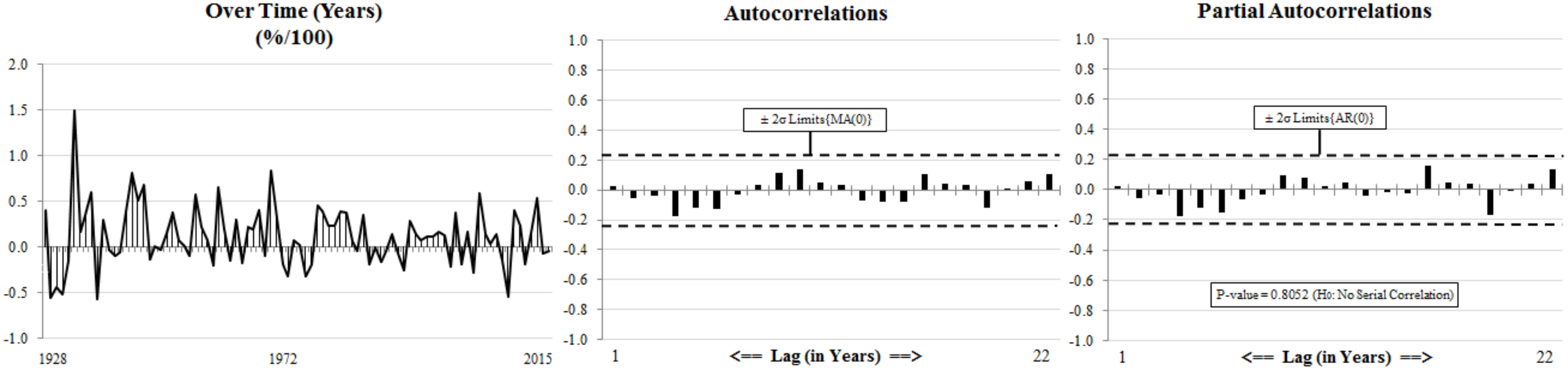

## 14.) *10-Year Avg. Real Compounding S&P 500*

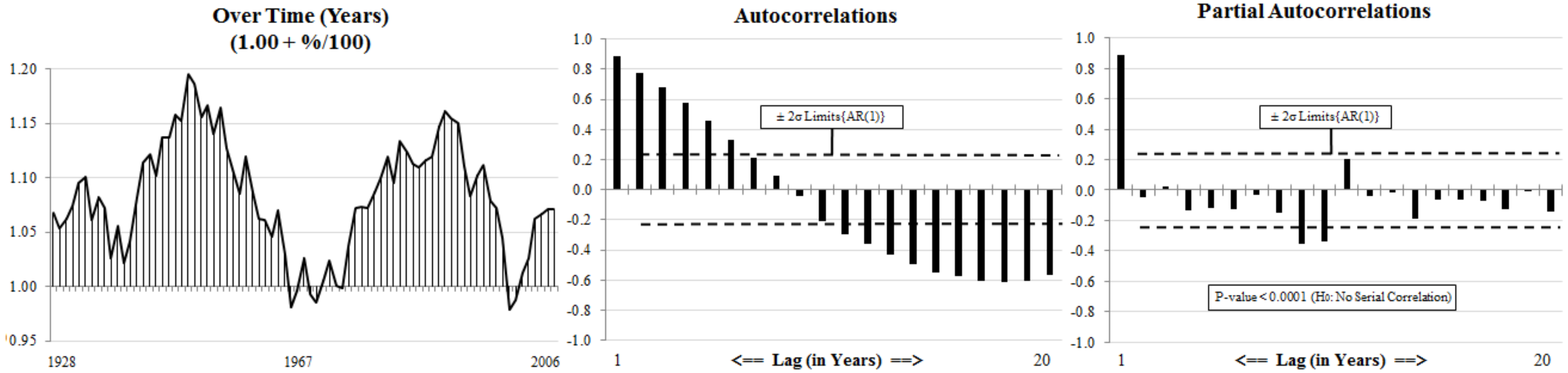

## 15.) *10-Year Avg. Total Compounding S&P 500*

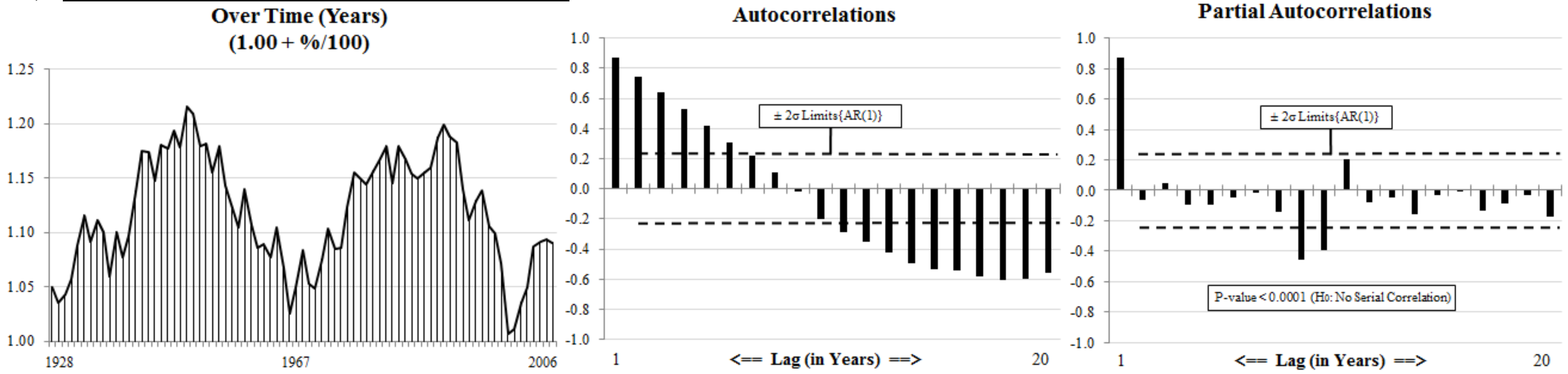

## *16.) 1st Diff. Shiller CAPE Ratio*

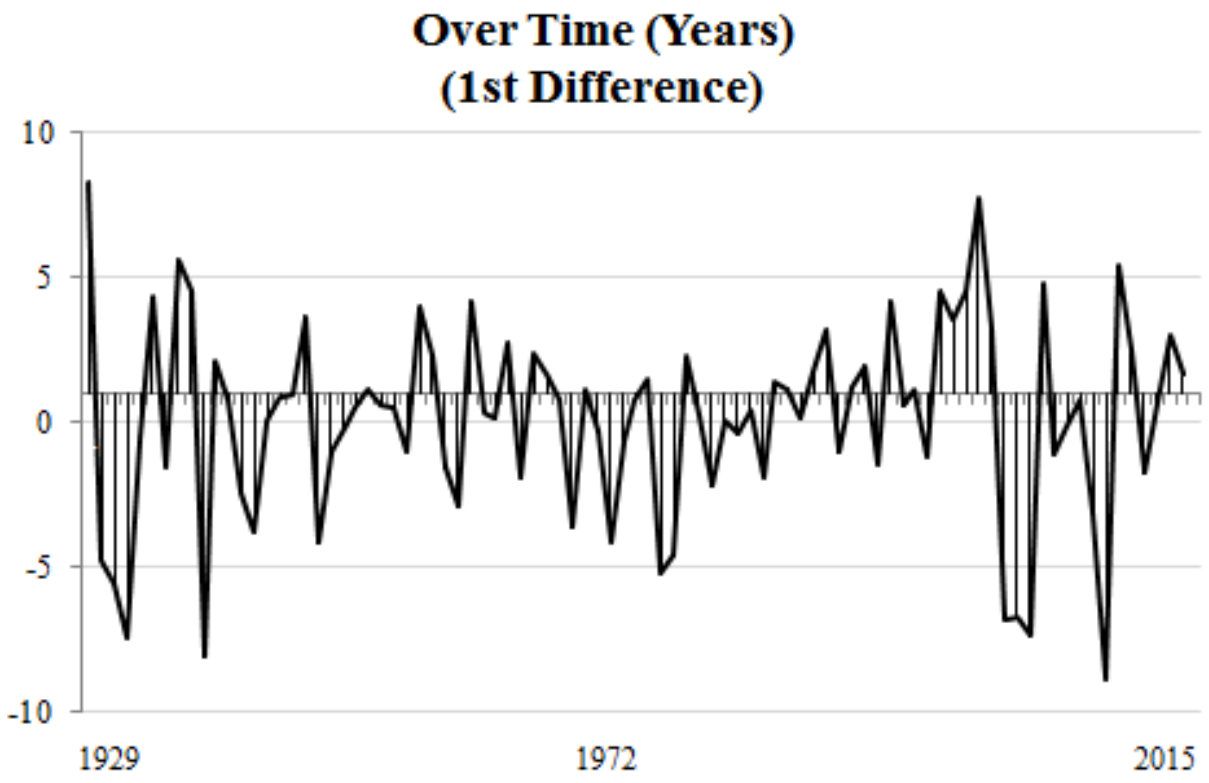

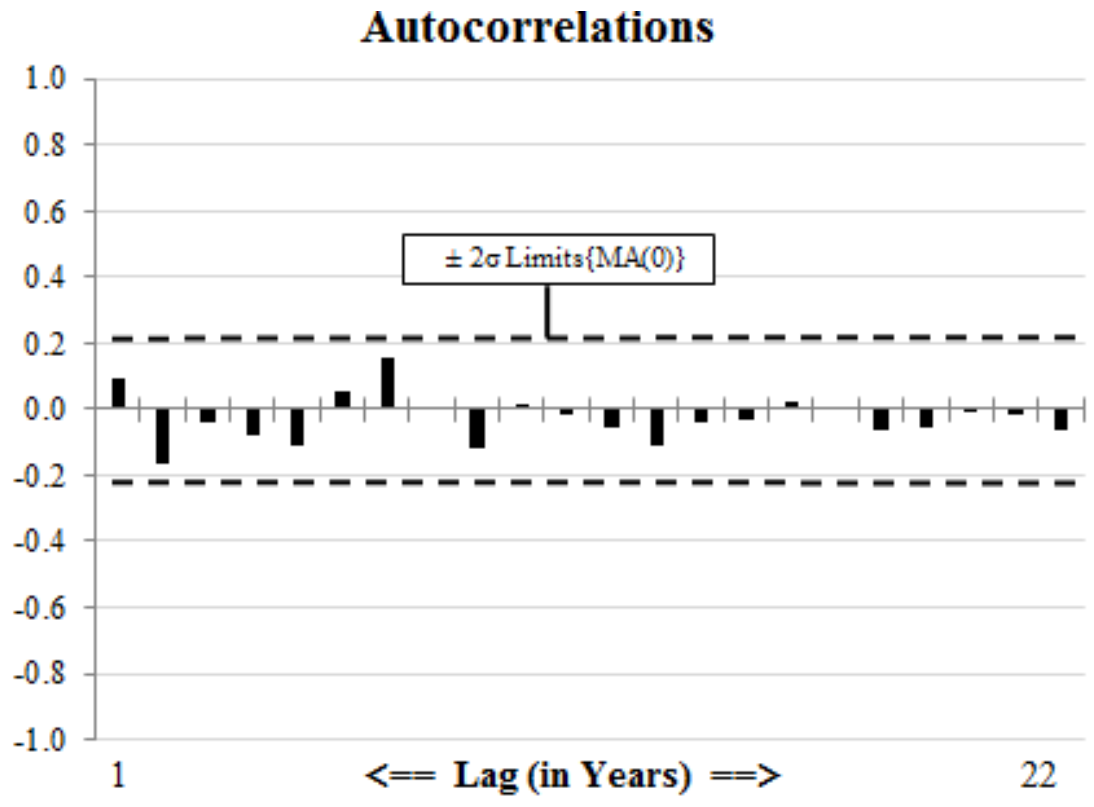

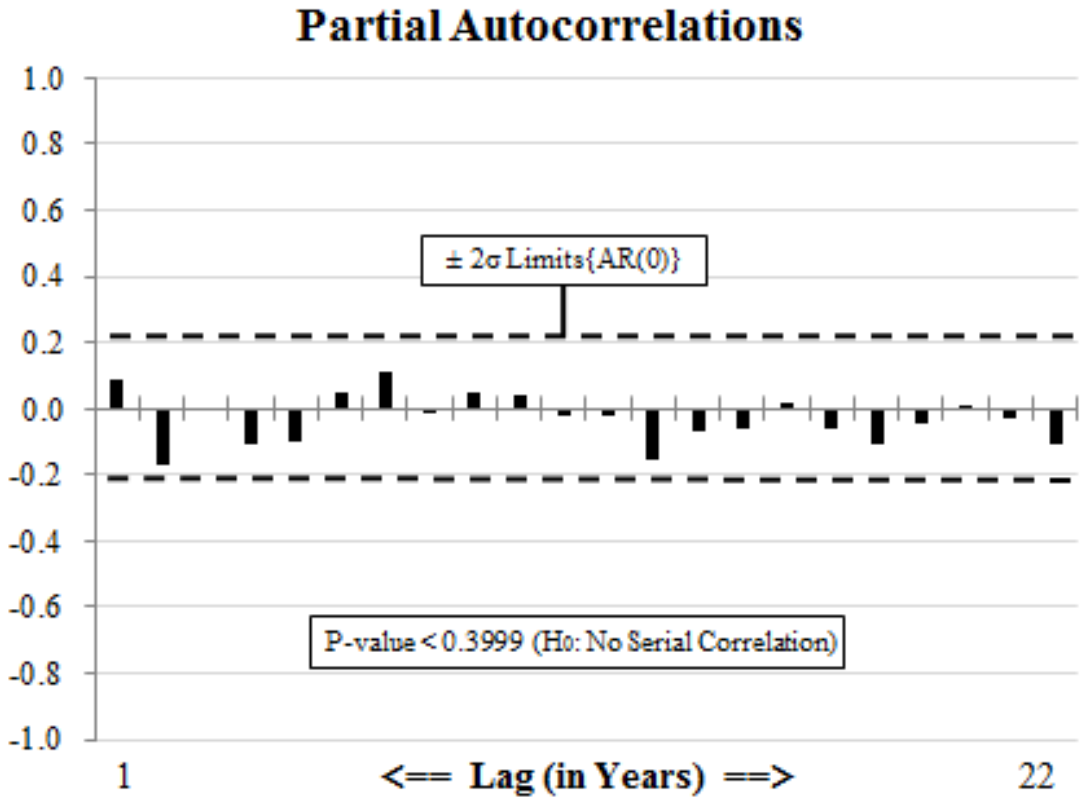

## *17.) 1st Diff. Log Shiller CAPE Ratio*

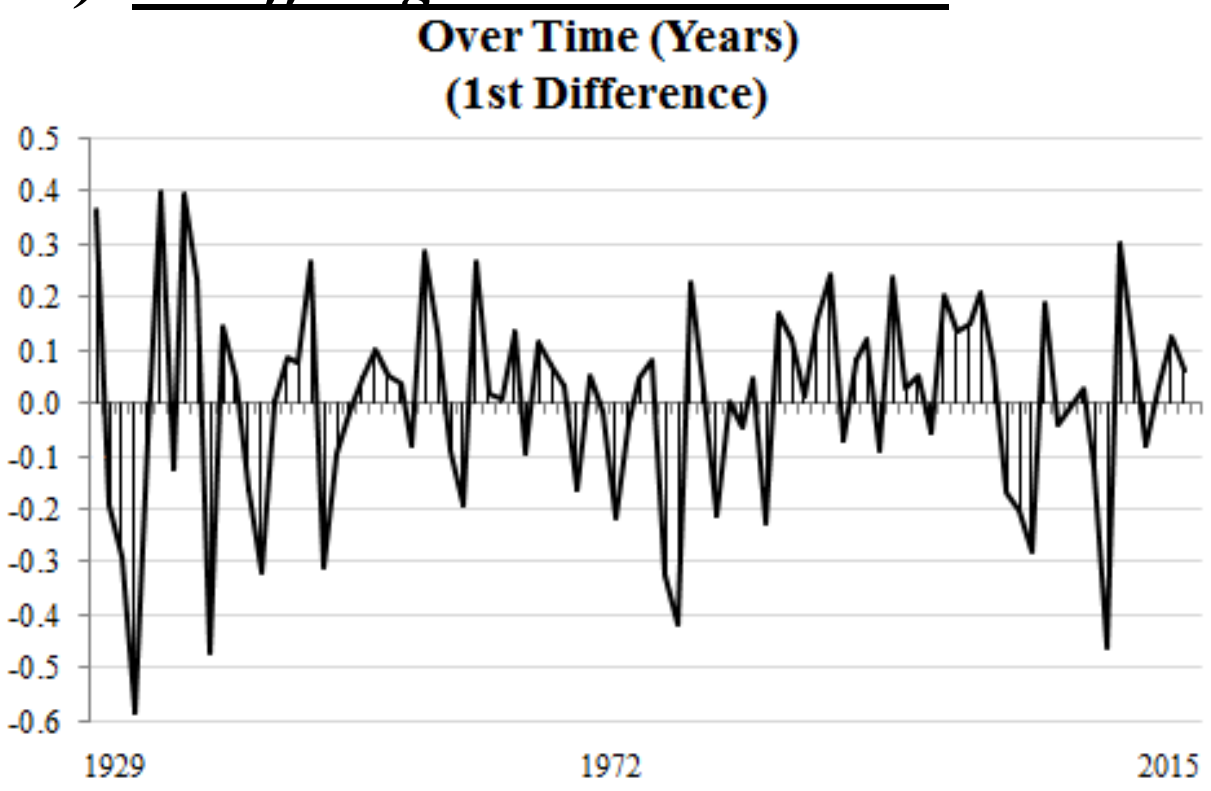

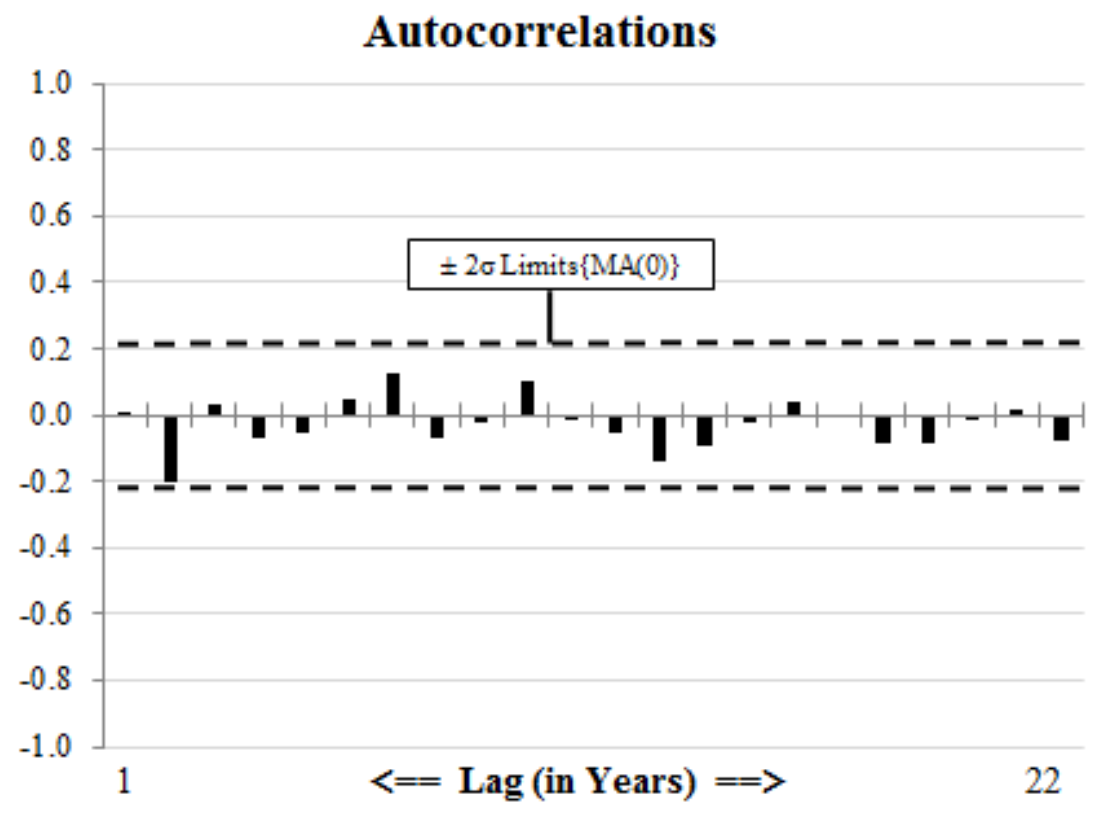

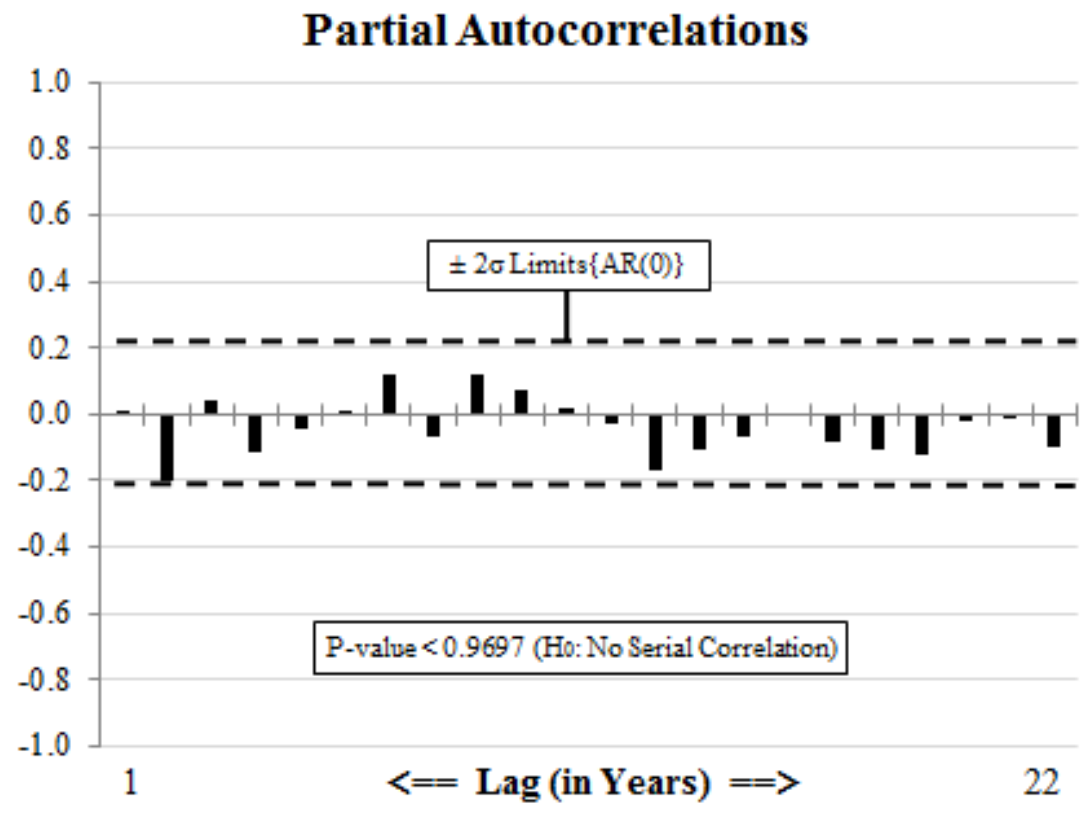

## *18.) Detrended S&P 10-Year Avg. Real Earnings*

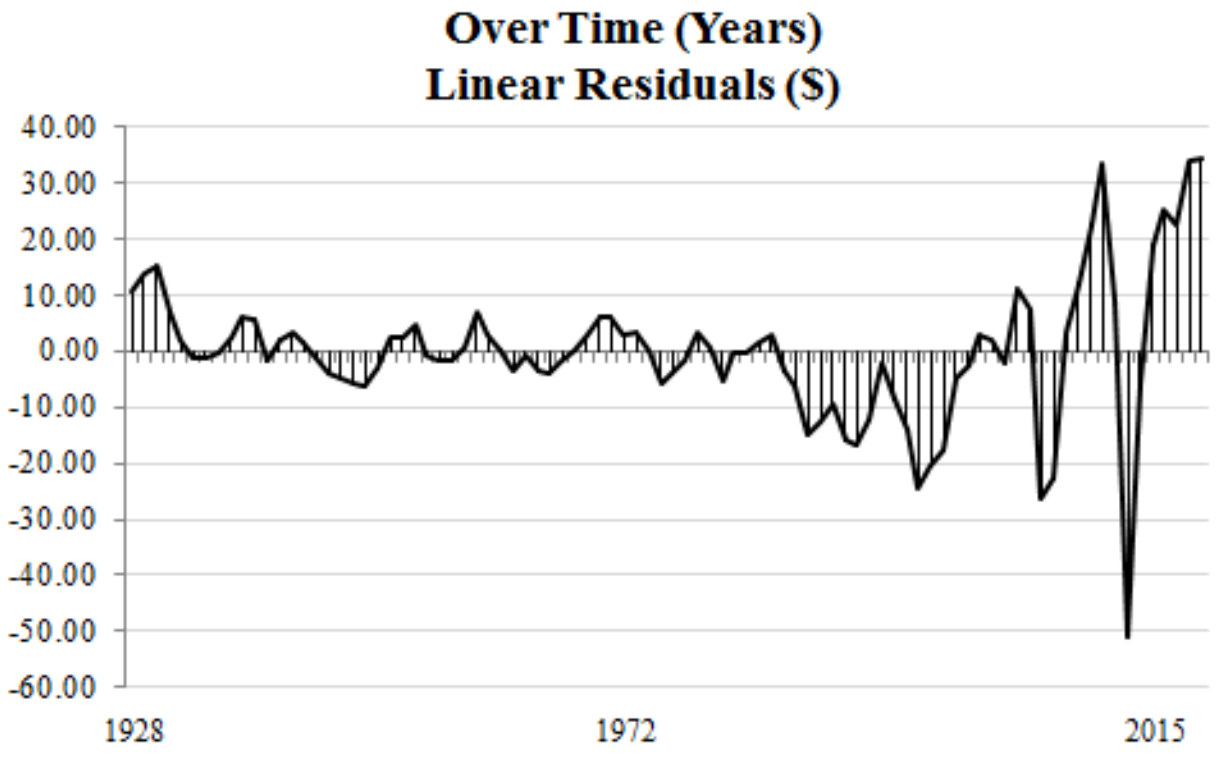

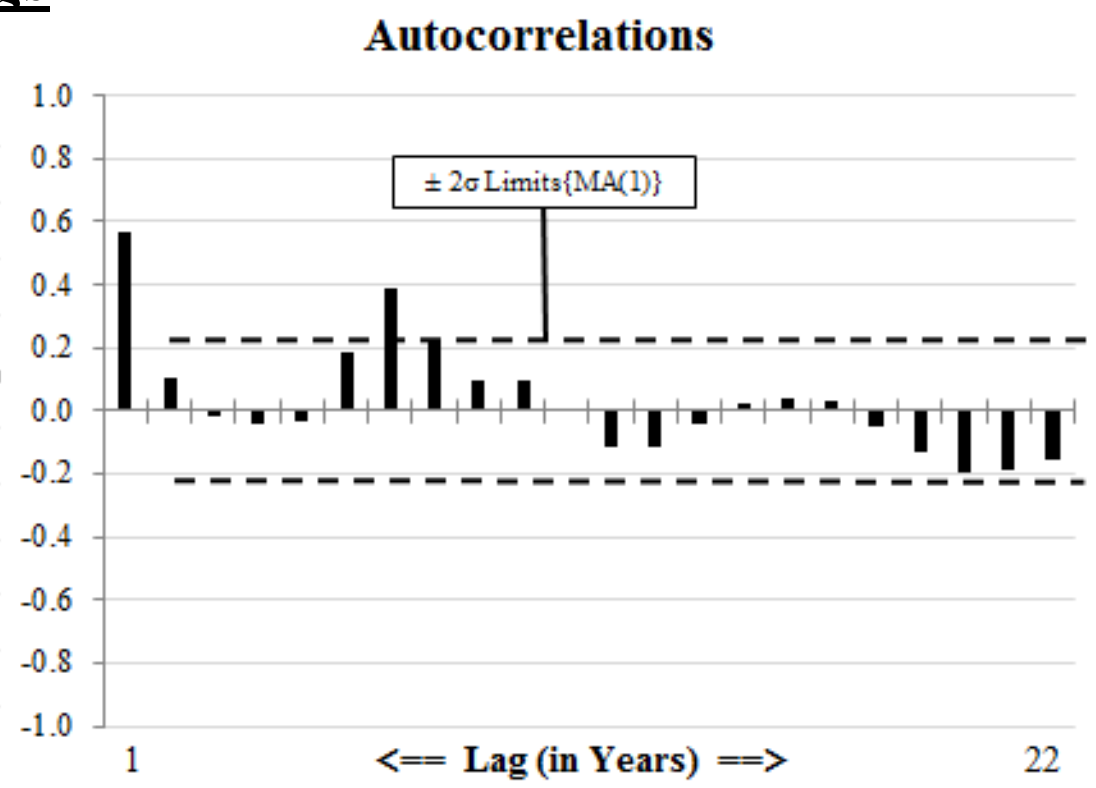

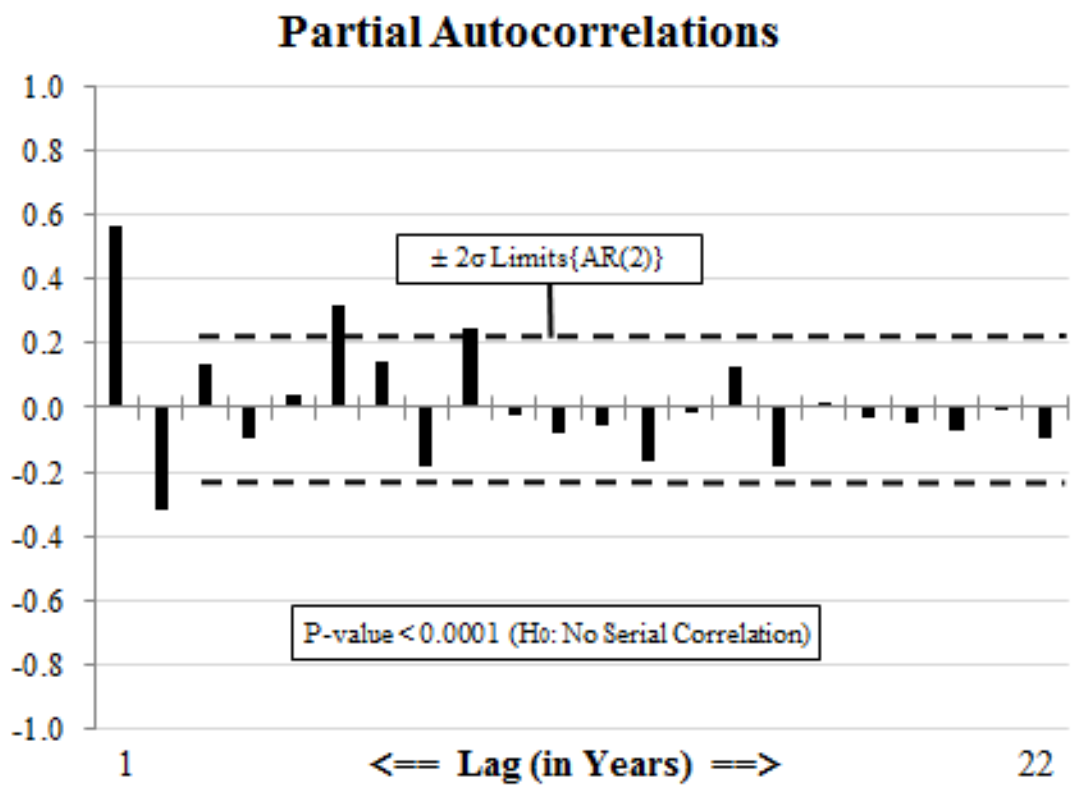

## Appendix C. C/C++ Source Code

The application accepts 2 input files: 1.) a control file with settings, and 2.) a text file of returns. Samples of each are shown below. The control file sets 7 parameters with the 1st and 2nd being the # of assets and # of time points, respectively. The 3rd parameter sets the # of random starts per CPU core for each execution of the EM algorithm when fitting univariate PDFs. The value is multiplied by 1 less than the # of components in the mixture being fit. For example, if fitting a 4-component univariate mixture the setting below results in 5·3 random starts/core. Since our machine contains 40 cores this mixture PDF is fit using 600 EM algorithm random starts. The 4th parameter is the maximum # of components for all univariate mixtures which we set to 5 in this implementation. The 5th parameter is the # of samples to use for each bootstrapped LRT within the proposed forward-backward framework that determines the optimal # of univariate mixture components for each asset. The final 2 values in "control.txt" are the forward-backward LRT p-values, respectively. The returns file contains a column of (compounding) returns for each asset. On our computer the control file shown can reproduce the multivariate PDF for (L,S,B) from (6.A.2) in a few minutes.

control.txt

```
3 88 5 5 100 0.25 0.25
```

returns.txt

```
1.45557 1.41387 1.02065
0.91700 0.48640 1.04200
0.76958 0.63566 1.07441
0.61647 0.55159 1.06959
```

Both univariate and multivariate parameter estimation results are written to the screen and to the file named "output.txt" in the folder supplied by the user upon application launch. The 2 files above must also exist within this directory. Lastly, a folder to contain error files for issues encountered during the optimization must exist and be specified by the user using the global constant "errFolder" which is also defined in the header file.

To save time, processing for both the univariate and multivariate PDF is multithreaded. Our implementation of the EM algorithm launches each random start in a separate thread (random starts end at local optimums). The ECME algorithm takes a small or large step in the general direction of steepest ascent while maximizing the multivariate log-likelihood function with respect to the covariance parameters, see §**V.A**. The line search and stepping parameters are simulated and values that yield the largest increase are used, which is also multithreaded. Note that our ECME implementation uses the actual or incomplete log-likelihood function from (2.E.2), conditioning first on the probabilities holding the covariances constant, and then on the covariances holding the probabilities constant. These are Steps 1 and 2 in (5.A.1) and (5.A.2).

This application uses the Boost (http://www.boost.org/), LPSolve (http://lpsolve.sourceforge.net/5.5/), and Eigen (http://eigen.tuxfamily.org/) external C/C++ libraries, which are freely available to download under terms and conditions described at the given sites. Our code consists of a header file and 20 functions and is being provided under the GNU General Public License, see http://www.gnu.org/licenses/ for full details.

```
/*
/ Copyright (C) 2016 Chris Rook
/
/ This program is free software: you can redistribute it and/or modify it under the terms of the GNU General Public License as published by the Free Software Foundation, either version 3 of the License,
/ or (at your option) any later version.  This program is distributed in the hope that it will be useful, but WITHOUT ANY WARRANTY; without even the implied warranty of MERCHANTABILITY or FITNESS FOR A
/ PARTICULAR PURPOSE.  See the GNU General Public License for more details: <http://www.gnu.org/licenses/>
/
/ Filename:  stdafx.h
/
/ Summary:
/
/    This is the header file which is included before each function call to define needed C/C++ libraries, global constants, inline functions, and function prototypes.  For function prototypes, see
/    headers attached to the code definition for a description of the purpose and parameters.
/
/ Inline Functions:
/
/    getNDens:      Compute/return the supplied normal density function evaluated at a given point, without the constant 1/sqrt(2*pi).
/    getMDens:      Compute/return the supplied normal mixture density function evaluated at a given point, without the constant 1/sqrt(2*pi).
/    getPost:       Compute/return the probability that a given value belongs to a given component of the supplied normal mixture density.
/    getUniMean:    Compute/return the mean of a given set of observations.
/    getUniStd:     Compute/return the standard deviation (MLE) of a given set of observations.
/    getLLVal:      Compute/return the log-likelihood value for a given observation set from the supplied univariate mixture density.
/    getVRatio:     Compute/return the ratio of the largest-to-smallest variance for a supplied set of standard deviations.
/    getEMMean:     Compute/return mean update for the EM algorithm along with a sum-of-squares quantity needed for the EM algorithm variance update.
/    getMVNDens:    Compute/return the supplied multivariate normal density function evaluated at a given vector of values for each random variable.
/    getMVNMDens:   Compute/return the supplied multivariate normal mixture density function evaluated at a given vector of values for each random variable.
/    getCovs:       Accept a multivariate normal mixture density and return the vector of covariances starting with VC element (1,2) of component #1, then element (1,3), etc...
/    setCovs:       Accept a vector of all covariances and insert the values into a single component's VC matrix as specified by the user.
/    getCofM:       Accept a matrix and element/position and convert it to one who's determinant is equal to the cofactor of the matrix with respect to that position/position.
/    getIDM:        Accept an empty square matrix and populate it with an identity matrix of the same size.
/    chkSum:        Accept a multivariate normal mixture density and check that no component has a likelihood of zero at all time points.
/    showVals:      Display the parameters (means, standard deviations, component probabilities) of a supplied multivariate mixture density to standard output.
/*********************************************************************************************************************************************************************************************************/
#pragma once

// Include libraries.
//=====================
#include "targetver.h"
#include <stdio.h>
#include <iostream>
#include <iomanip>
#include <string>
#include <fstream>
#include <random>
#include <boost/algorithm/string.hpp>
#include <boost/thread/thread.hpp>
#include <Eigen/Core>
#include <Eigen/Dense>
#include <Eigen/LU>
#include <Eigen/Eigenvalues>
#include "lp_lib.h"
using namespace std;

// Global constants.
//====================
const string cfile="control.txt";                     // Name of parameter control input file.
const string rfile="returns.txt";                     // Name of input data file.
const string ofile="output.txt";                      // Name of output file.
const string errFolder="c:\\mDensity\\Issues\\";      // Folder where issues/errors will be written to files (mandatory).
const long double pi=3.141592653589793;               // Constant representation of pi.
const long double logrecipsqrt2pi=-0.918938533204673; // ln(1.00/square root of 2*pi).
const long double stdRatio=16.0;                      // Square root of the variance ratio constraint threshold.
const long double epsilon=pow(0.1,15);                // Convergence criteria.
const long double LNegVal=-pow(9.0,10);               // Large negative value for use as an invalid value indicator.
const long double LPosVal=pow(9.0,10);                // Arbitrarily large positive value.
```

```
const long double rrMultMin=2.0;                    // Minimum multiple when performing a ridge repair on a broken variance-covariance matrix.
const long double rrMultMax=10.0;                   // Maximum multiple when performing a ridge repair on a broken variance-covariance matrix.
const long double pdmineval=pow(0.1,13);            // Threshold for minimum eigenvalue to ensure a positive definite variance-covariance matrix.
const long double detminval=pow(0.1,13);            // Threshold for determinant to ensure a positive definite variance-covariance matrix.
const int mIters=1000000000;                        // Maximum # of EM iterations allowed per single optimization.
const int K=1000000;                                // Large constant for use in LP objective function to enforce feasibility constraint.
const int dLvl=500;                                 // Discretization level for separable quadratic objective function in solveLP().
const long double minHessAdd=0.00;                  // Minimum additive factor to use for Hessian stepping.
const int mItersH=1000;                             // Maximum # Hessian steps per thread.  (Set to 1 for non-random ECME stepping.)
const int nCorMult=2;                               // Multiply by # cores to determine the total # threads used for Hessian stepping.  (Set to 1 for non-random ECME stepping.)
const int nBeats=80;                                // # of log-likelihood beats to randomly select from during the ECME algorithm 2nd Step.  (Set to 1 for non-random ECME stepping.)
const int dbug=0;                                   // Debug level (0,1,2) for output window details (higher # = more details).
const int cDown=2000;                               // Processor cool down time in between EM iterations in msec (value of 100000 = 100 seconds).
const int nECMEs=1;                                 // Number of outer loops for ECME processing.  (# of times to repeat the ECME procedure.)

// Inline functions.
//====================
inline long double getNDens(const long double val, const long double mn, const long double std)
                          {return (1.00/std)*exp(-0.5*pow((val-mn)/std,2));}

inline long double getMDens(const long double val, const int g, const long double **inMDist)
                          {long double dval=0.00; for (int c=0; c<g; ++c) dval=dval + inMDist[0][c]*getNDens(val,inMDist[1][c],inMDist[2][c]); return dval;}

inline long double getPost(const long double val, const int g, const long double **inMDist, const int cID)
                         {return inMDist[0][cID]*getNDens(val,inMDist[1][cID],inMDist[2][cID])/getMDens(val,g,inMDist);}

inline long double getUniMean(const int T, const long double *r)
                            {long double mn=0.00; for (int t=0; t<T; ++t) mn=mn + r[t]; return mn/T;}

inline long double getUniStd(const int T, const long double *r, const long double mn)
                           {long double var=0.00; for (int t=0; t<T; ++t) var=var + pow(r[t]-mn,2); return sqrt(var/T);}

inline long double getLLVal(const int T, const long double *r, const int g, const long double **inMDist)
                          {long double llval=((long double) T)*logrecipsqrt2pi; for (int t=0; t<T; ++t) llval=llval + log(getMDens(r[t],g,inMDist)); return llval;}

inline long double getVRatio(const int g, const long double *stds)
                           {long double minstd=LPosVal, maxstd=LNegVal; for (int c=0; c<g; ++c){if (stds[c] < minstd) minstd=stds[c]; if (stds[c] > maxstd) maxstd = stds[c];}
                            return pow(maxstd/minstd,2);}

inline long double getEMMean(const int T, const long double *r, const long double *pprbs, const long double cprbs, long double *ssq)
                  {long double mn=0.00, atrm; ssq[0]=0.00; for (int t=0; t<T; ++t) {atrm=pprbs[t]*r[t]; mn=mn+atrm; ssq[0]=ssq[0]+atrm*r[t];}; return mn/(cprbs);}

inline long double getMVNDens(const Eigen::VectorXd vals, const Eigen::VectorXd mn, const Eigen::MatrixXd vcmi, const long double sqrdet, const long double picnst)
                            {return (sqrdet/picnst)*exp(-0.5*((vals-mn).transpose())*vcmi*(vals-mn));}

inline long double getMVNMDens(const int uCells, const Eigen::VectorXd vals, const Eigen::VectorXd *mns, const Eigen::MatrixXd *vcmis, const Eigen::VectorXd prbs, const long double *sqrdets,
                                    const long double picnst){long double dval=0.00; for (int u=0; u<uCells; ++u) {dval=dval+prbs[u]*getMVNDens(vals,mns[u],vcmis[u],sqrdets[u],picnst);} return dval;}

inline void getCovs(const int uCells, const Eigen::MatrixXd *inVCs, Eigen::VectorXd *inDVarsM){int itr=0; for (int v=0; v<uCells; ++v) for (int r=0; r<(int) inVCs[v].rows(); ++r)
                                                                                                          for (int c=r+1; c<(int) inVCs[v].cols(); ++c)
                                                                                                              inDVarsM[0](itr++)=inVCs[v](r,c);}

inline void setCovs(const int uCell, Eigen::MatrixXd *inVCs, const Eigen::VectorXd *inDVarsM){int itr=uCell*((int) (((int) inVCs[uCell].rows())*((int) inVCs[uCell].rows()-1))/2);
                                                                                              for (int r=0; r<(int) inVCs[uCell].rows(); ++r)
                                                                                                  for (int c=r+1; c<(int) inVCs[uCell].cols(); ++c)
                                                                                                      {inVCs[uCell](r,c)=inDVarsM[0](itr++);
                                                                                                       inVCs[uCell](c,r)=inVCs[uCell](r,c);}}

inline void getCofM(const int numA, const int inr, const int inc, const Eigen::MatrixXd inE, Eigen::MatrixXd *inEjk){inEjk[0]=inE; for (int r=0; r<numA; ++r)
                                                                                                                      for (int c=0; c<numA; ++c)
                                                                                                                          {if (r==inr && c==inc) inEjk[0](r,c)=1.00;
                                                                                                                             else if (r==inr || c==inc) inEjk[0](r,c)=0.00;}}

inline void getIDM(Eigen::MatrixXd *ID){for (int r=0; r<(int) ID[0].rows(); ++r) for (int c=r; c<ID[0].cols(); ++c)
                                                                  {if (r==c) ID[0](r,c)=1.00; else ID[0](c,r)=ID[0](r,c)=0.00;}}
```

```
inline void chkSum(const int T, const int nUCmps, const long double **fvals){for (int v=0; v<nUCmps; ++v){long double tmpsum=0.00; for (int t=0; t<T; ++t) tmpsum=tmpsum + fvals[t][v];
             if (tmpsum <= 0.00){cout << endl << "ERROR:  Unique component # " << v << " has a likelihood that is zero for each time point." << endl
                                      << "        This will eliminate the corresponding component probability from the stage 2 objective function." << endl
                                      << "        The component probability should be treated as a constant and moved to the RHS constraint vector" << endl
                                      << "        and eliminated from the objective function.  The code for this has not yet been implemented." << endl
                                      << "EXITING...ECMEAlg()..." << endl; cin.get(); exit (EXIT_FAILURE);}}}

inline void showVals(const int T, const long double *r, const int g, const long double **inMDist, ostream& oVar=cout)
                    {oVar.setf(ios_base::fixed, ios_base::floatfield); oVar.precision(16); for (int x=0; x<T; ++x) oVar << r[x] << endl; for (int c=0; c<g; ++c){oVar << string(37,' ')
                                       << "Prob[" << c << "]=" << inMDist[0][c] << "  Mean[" << c << "]=" << inMDist[1][c] << "  Std.Dev.[" << c << "]=" << inMDist[2][c] << endl;}};

// Function prototypes.
//======================
int fitMixDist(const int a, const int T, const long double *r, const int maxcmps, const int nsmpls, const int nstrts, const long double sl[2], long double **fnlMDst, string rdir);
void getRVals(const int N, const int g, const long double **inMDist, long double *rvls);
void getRPrbsStds(const int T, const long double * r, const int g, long double *prbs, const long double *mns, long double *stds);
void EMAlg(const int T, const long double *r, const int g, long double **prbs, long double **mns, long double **stds, long double *llVal, const long double **inMDist, int *rprms);
int ECMEAlg(const int T, const long double **r, const int numA, const int nUCmps, const Eigen::MatrixXd cMtrx, const Eigen::VectorXd cVctr, long double *muPrbs, Eigen::VectorXd *muMns, Eigen::MatrixXd
*muVCs, int *uCellIDs, const string rdir);
long double ThrdEMAlg(const int T, const long double *r, const int rs, const int ing, const long double **inMDist, const int outg, long double **outMDist);
void mapCells(const int totCells, const int numA, int **inCellAry, const int curAst, const int *inCmps, int *cID, int *tmpAry);
int getCell(const int **inCellAry, const int totCells, const int *cmpLvls, const int numA);
void asgnObs(int *inAry, const int T, const long double *r, const int g, const long double **inMDst);
void getCor(const int T, const long double **r, const int **asgn, const int **inCellAry, const int cellID, Eigen::VectorXd Mn, Eigen::MatrixXd *VC, const int vCell);
int solveLP(const int totCells, const int numA, const int **inCellAry, const int *cmps, const long double **prbs, const int *nCellObs, const long double *cellProb, double *outPrbs, const int type);
void getCMtrx(const int totrows, const int nUCmps, const int sol, const int numA, const int *nCmps, const int *vCIDs, const int **inCellAry, const long double **prbs, Eigen::MatrixXd *fLHS,
Eigen::VectorXd *fRHS);
long double getHessE(const int T, const int inUCmps, const long double **infVals, const long double *inDNoms, const Eigen::MatrixXd inCMtrx, const Eigen::VectorXd inCVctr, Eigen::MatrixXd *inHess,
Eigen::MatrixXd *inLHS, Eigen::VectorXd *inRHS);
long double getHessM(const int T, Eigen::VectorXd *rs, const int inUCmps, const int numA, const long double **infVals, const long double *inDNoms, const long double *inPrbs, const Eigen::VectorXd *muMns,
const Eigen::MatrixXd *E, const Eigen::MatrixXd *Einv, const Eigen::MatrixXd *inA, Eigen::MatrixXd *inHess);
void getGradE(const int T, const int inUCmps, const long double **infVals, const long double *inDNoms, const Eigen::MatrixXd inLHS, const Eigen::VectorXd inRHS, const Eigen::VectorXd inDVars,
Eigen::VectorXd *inGrad);
void getGradM(const int T, Eigen::VectorXd *rs, const int inUCmps, const int numA, const long double **infVals, const long double *inDNoms, const long double *inPrbs, Eigen::VectorXd *muMns, const
Eigen::MatrixXd *E, const Eigen::MatrixXd *Einv, const Eigen::MatrixXd *inA, Eigen::VectorXd *inGrad, int chk=0);
long double getLFVals(const int T, const int numA, const int nUCmps, const Eigen::VectorXd *rs, const Eigen::VectorXd uPrbs, const Eigen::VectorXd *inMns, const Eigen::MatrixXd *inVCIs, const long double
*insqs, const long double inpicst, long double *denoms, long double **lfVals);
void wrtDens(const string typ, const int nUCmps, const int *uCells, const long double *muPrbs, const Eigen::VectorXd *muMns, const Eigen::MatrixXd *muVCs, ostream& oVar);
void stepHessM(int *inHess1, long double *inHess2, const Eigen::VectorXd *rs, Eigen::VectorXd *inDvars, const Eigen::VectorXd inGrad, const Eigen::MatrixXd inHess, const Eigen::VectorXd uPrbs, const
Eigen::VectorXd *inMns, const Eigen::MatrixXd *inVCs);
int ridgeRpr(const int uCell, Eigen::MatrixXd *E, long double mult);

/*
/ Copyright (C) 2016 Chris Rook
/
/ This program is free software: you can redistribute it and/or modify it under the terms of the GNU General Public License as published by the Free Software Foundation, either version 3 of the License,
/ or (at your option) any later version.  This program is distributed in the hope that it will be useful, but WITHOUT ANY WARRANTY; without even the implied warranty of MERCHANTABILITY or FITNESS FOR A
/ PARTICULAR PURPOSE.  See the GNU General Public License for more details: <http://www.gnu.org/licenses/>
/
/ Filename:  mDensity.cpp
/
/ Function:  main()
/
/ Summary:
/
/    This function defines the entry point for the console application and drives the analysis using 4 major sections.  Section #1.) The contents of the control file (control.txt: see global constant
/    cfile in the header program), and the data file (returns.txt: see global constant rfile in the header program) are read in and stored as variables.  Section #2.) Build the univariate mixture PDFs
/    for each asset using the EM algorithm with random starts and bootstrapped likelihood ratio test for determining the optimal # of components for each asset.  Section #3.) Combine the univariate
/    mixtures into a multivariate mixture PDF without disturbing the marginals and without correlations.  At this point, a multivariate mixture PDF is estimated with dependence but without regime
/    correlations.  (There are 2 multivariate mixture densities at this stage, one for each type of LP solved.  The 2 LP objectives for building the multivariate mixture PDF are minimax and minimum
/    squared distance.)  Section #4.) Use an ECME type algorithm to estimate the correlations and refine the component probabilities with maximum likelihood as the objective.  This 4th step is a 2-step
/    iterative procedure similar to the EM algorithm and random starts are used in the 2nd step where the covariances are estimated.  The 4th step is repeated nECMEs times and the optimal multivariate
/    mixture PDF for each is written to the output file.  Once all 4 steps have been completed there are 2*nECMEs multivariate mixture densities with fixed marginals and one is chosen based on some
/    criteria such as higher likelihood or higher information criteria.  This final step is left to the user (we use AIC) and the decision would account for the likelihood value as well as the total # of
/    parameters.  All PDFs and their log-likelihood values are written to the output file (output.txt: see global constant ofile in the header program), along with the univariate mixture PDF details.
```

```
/
/ Inputs:
/
/    No input arguments are processed by this function.  Critical inputs are supplied via the control.txt file (see global constant cfile in the header program), and data is supplied via returns.txt (see
/    global constant rfile in the header program).  In addition to control.txt, other inputs are set to global constants in the header file.
/
/ Outputs:
/
/    This function writes details of fitting univariate & multivariate mixture PDFs for the supplied assets to the screen and to the file output.txt (see ofile in the header program).
/*****************************************************************************************************************************************************************************************************/
#include "stdafx.h"
int main(int argc, char *argv[])
{
    // Declare/initialize local variables.
    //====================================
    string rootdir;
    int nBoots, nTPoints, nAssets, nRStarts, mComps, *nComps=nullptr, **asgnmnt=nullptr, uCell;
    long double **rtrn=nullptr, **prob=nullptr, **mean=nullptr, **stdev=nullptr, alpha[2], **optMDst=new long double *[3];
    ofstream fout;

    // Ensure that an error/issues folder has been provided.  (Set in the header file.)
    //================================================================================
    if ("" == errFolder)
    {
        cout << "ERROR:  An error/issues folder has not been provided.  Problems found during optimization will be written to files in this folder." << endl
             << "        Use global variable errFolder in the header file to set this destination." << endl << "EXITING...main()..." << endl; cin.get(); exit (EXIT_FAILURE);
    }

    // Retrieve directory location of setup files.
    //===========================================
    cout << "Enter the directory where the setup files reside (eg, c:\\mDensity\\): " << endl; cin >> rootdir; cin.get(); boost::algorithm::trim(rootdir); cout << endl;

    // Read in control file which contains:
    //   1.) # of asset classes.
    //   2.) # of timepoints with return data.
    //   3.) # of random starts for maximizing the likelihood of a g-component univariate mixture, as a multiple of # independent processing units and components (i.e., value of 2 with 20 independent
    //       processing units will use 2x20x1=40 random starts for a 2-component mixture, 2x20x2=80 random starts for a 3-component mixture, and 2x20x(g-1) random starts for a g-component mixture.)
    //   4.) Maximum # of components appropriate for this data set when fitting univariate mixture densities (i.e., the marginals).
    //   5.) Sample size to use when bootstrapping the LRT test statistic.
    //   6.) Alpha-level for LRTs for the # of components using a forward-backward selection algorithm, where the 1st alpha is used for forward selection and the 2nd alpha is used for backward selection.
    //===========================================================================================================================================================================================
    ifstream getparams(rootdir+cfile);
    if (getparams.is_open())
    {
          getparams >> nAssets >> nTPoints >> nRStarts >> mComps >> nBoots >> alpha[0] >> alpha[1];  getparams.close();
    }
    else
    {
        cout << "ERROR:  Could not open file: " << rootdir + cfile << endl << "EXITING...main()..." << endl; cin.get(); exit (EXIT_FAILURE);
    }

    // Instantiate 2-dimensional arrays to hold means, standard deviations, and proportion weights for mixtures.
    //=========================================================================================================
    nComps = new int [nAssets];                 // Array to hold # components for each asset.
    rtrn = new long double *[nAssets];          // One return per asset and time point up to time nTPoints.
    prob = new long double *[nAssets];          // One prob per asset and component for component sizes up to mComp.
    mean = new long double *[nAssets];          // One mean per asset and component for component sizes up to mComp.
    stdev = new long double *[nAssets];         // One stdev per asset and component for component sizes up to mComp.
    asgnmnt = new int *[nAssets];               // One assignment per asset and time point up to time nTPoints.
    for (int a=0; a<nAssets; ++a)
    {
          nComps[a] = 1;                          // Start with 1 component for each asset.
          rtrn[a] = new long double [nTPoints];   // Array of returns for each asset. Returns are then rtrns[a][0], rtrns[a][1], etc...
          asgnmnt[a] = new int [nTPoints];        // Array of component assignments for each asset after density has been determined.
    }
```

```
// Read in returns file which has a column of returns for each asset, and store in a 2-dimensional array.
//=========================================================================================================
ifstream getrtrns(rootdir+rfile);
if (getrtrns.is_open())
{
      int r=0;
      while (!getrtrns.eof() && r < nTPoints)
      {
          for (int a=0; a<nAssets; ++a)
              getrtrns >> rtrn[a][r];
          r=r+1;
      }
      getrtrns.close();
      if (r < nTPoints)
      {
          cout << "ERROR:  File " << rootdir + rfile << " should have " << nTPoints << " rows of returns, for " << nAssets << " assets, but it has fewer." << endl
               << "EXITING...main()..." << endl; cin.get(); exit (EXIT_FAILURE);
      }
}
else
{
      cout << "ERROR:  Could not open file: " << rootdir + rfile << endl << "EXITING...main()..." << endl; cin.get(); exit (EXIT_FAILURE);
}

// Build a large array to temporarily hold the optimal distribution for each asset.
//===================================================================================
for (int m=0; m<3; ++m)
      optMDst[m]=new long double [mComps];

// Calculate the component probability, mean and standard deviation (Normal MLE version).  (Note:  MLE standard deviation divides by n, not n-1 and is a biased estimator.)
//============================================================================================================================================================================
for (int a=0; a<nAssets; ++a)
{
      // Initialize all probabilities in the large array to zero.
      //===========================================================
      for (int c=0; c<mComps; ++c)
          optMDst[0][c]=0.00;

      // Find the best fitting mixture distribution for this asset.
      //=============================================================
      nComps[a]=fitMixDist(a,nTPoints,rtrn[a],mComps,nBoots,nRStarts,alpha,optMDst,rootdir);

      // Assign each observation to a component.  The most likely one using Bayes rule.
      //=================================================================================
      asgnObs(asgnmnt[a],nTPoints,rtrn[a],nComps[a],(const long double **) optMDst);

      // Build arrays to hold the optimal solution and transfer it to these arrays.
      //=============================================================================
      prob[a]=new long double [nComps[a]];
      mean[a]=new long double [nComps[a]];
      stdev[a]=new long double [nComps[a]];
      for (int c=0; c<nComps[a]; ++c)
      {
          prob[a][c]=optMDst[0][c];
          mean[a][c]=optMDst[1][c];
          stdev[a][c]=optMDst[2][c];
      }
}

// Write out the assignment of each observation time point to the corresponding components during debug mode.
//=============================================================================================================
if (dbug >= 2)
{
   cout << endl << string(97,'=') << endl << "Assignment of observations (at each time point) to a set of components using Bayes Decision Rule:" << endl << string(97,'=');
   for (int t=0; t<nTPoints; ++t)
```

```
   {
      cout << endl << "Time t = " << setfill('0') << setw(to_string((long long) nTPoints).size()) << t << ":";
      for (int a=0; a<nAssets; ++a)
          cout << "  Asset #" << a+1 << "/Component=" << asgnmnt[a][t];
   } cout << endl << endl;
}

// Mixture distribution has been fit for all assets.  Assemble the multivariate density.  Start by computing the total # of cells
// that need to be mapped (dealing with a k-dimensional cube).  Also compute the total # of components, summing across all assets.
//===================================================================================================================================
int nCells=1, totCmps=0;
for (int a=0; a<nAssets; ++a)
{
    nCells = nCells*nComps[a];
    totCmps = totCmps + nComps[a];
}

// Initialize variables.
//========================
int **allCells=new int *[nCells], cellID=0, *tmpVals=new int [nAssets];
for (int i=0; i<nCells; ++i)
    allCells[i] = new int [nAssets];

// Call function to map each cell of the k-dimensional cube to a single list value.
//===================================================================================
mapCells(nCells, nAssets, allCells, 0, nComps, &cellID, tmpVals);

// Derive the unique cell ID for each time point and count the # of obs per unique cell.
//========================================================================================
int *tmpCombo=new int [nAssets], *cellAsgn=new int [nTPoints], *cellCnt=new int [nCells];
long double *cellPrb=new long double [nCells];

for (int c=0; c<nCells; ++c)
    cellPrb[c]=(long double) (cellCnt[c]=0);
if (dbug>=2)
    cout << endl << endl << string(40,'=') << endl << "Assignment of each time point to a cell:" << endl << string(40,'=') << endl;
for (int t=0; t<nTPoints; ++t)
{
    if (dbug >= 2)
       cout << "Time t = " << setfill('0') << setw(to_string((long long) nTPoints).size()) << t << ": ";
    for (int a=0; a<nAssets; a++)
        tmpCombo[a]=asgnmnt[a][t];
    cellAsgn[t]=getCell((const int **) allCells, nCells, tmpCombo, nAssets);
    cellCnt[cellAsgn[t]]=cellCnt[cellAsgn[t]] + 1;
    cellPrb[cellAsgn[t]]=cellPrb[cellAsgn[t]] + 1.00/((long double) nTPoints);
}

// Build and solve the corresponding LP that determines the structure of the multivariate density using both a minimax and minimum squared distance objective.  Then build an array of cell IDs that
// have non-zero probabilities attached using each method.  These are the cells we must derive a correlation for, ensuring that the resulting VC matrix is positive-definite.  In solveLP(), type is
// defined as:  (These become the index values for all arrays that hold results from both methods to compare and select the better performer.)
//   0.) Minimax objective
//   1.) Minimum sum of squared distances objective
//======================================================================================================================================================================================================
string type[2]; type[0]="Minimax"; type[1]="Minimum Sum of Squared Distances (SSD)";
int numCmps [2], **valcIDs=new int *[2], ctr;
double **estPrbs=new double *[2];
Eigen::MatrixXd *mLHS=new Eigen::MatrixXd [2];
Eigen::VectorXd *mRHS=new Eigen::VectorXd [2];

cout << endl << string(120,'=') << endl << "Running " << type[0] << " and " << type[1] << " LPs to set the initial multivariate density structure." << endl << string(120,'=') << endl << endl;
for (int i=0; i<2; ++i)
{
      // Use an LP to approximate the structure of the multivariate density.
      //======================================================================
      estPrbs[i]=new double [nCells];
```

```
      numCmps[i]=solveLP(nCells,nAssets,(const int **) allCells,nComps,(const long double **) prob,cellCnt,cellPrb,estPrbs[i],i);
      valcIDs[i]=new int [numCmps[i]];

      // Identify and store the unique cells with non-zero probabilities.
      //===================================================================
      ctr=0;
      for (int c=0; c<nCells; ++c)
          if (estPrbs[i][c] > 0.00)
              valcIDs[i][ctr++]=c;

      // Retrieve the effective constraint matrix that applies to the LP solution and is full row rank.  Ensure it does not have more rows than non-zero components.
      //=====================================================================================================================================================
      getCMtrx(totCmps-nAssets+1,numCmps[i],i,nAssets,nComps,valcIDs[i],(const int **)allCells,(const long double **)prob,mLHS,mRHS);
      cout << "-----> " << type[i] << " Done.  There are " << numCmps[i] << " unique cells with non-zero probabilities using " << type[i] << "." << endl;
}

// Assemble multivariate densities using the minimax and minimum squared distance objectives as starting points and select the one with a better fit after
// iterating with the ECME algorithm until convergence.  (Where the 1st Step is replaced by a convex programming problem.)  Better fit means higher likelihood.
//=====================================================================================================================================================
int *numCmpsF=new int [2*nECMEs], **vCIDs=new int *[2*nECMEs];
long double **mProbs=new long double *[2*nECMEs];
Eigen::VectorXd **mMeans=new Eigen::VectorXd *[2*nECMEs];
Eigen::MatrixXd **mVCs=new Eigen::MatrixXd *[2*nECMEs];

for (int i=0; i<2*nECMEs; ++i)
{
      // Initialize the arrays to their proper size based on the # of components resulting from the above LPs.
      //=====================================================================================================
      mProbs[i]=new long double [numCmps[(i%2)]];
      mMeans[i]=new Eigen::VectorXd [numCmps[(i%2)]];
      mVCs[i]=new Eigen::MatrixXd [numCmps[(i%2)]];
      vCIDs[i]=new int [numCmps[(i%2)]];

      // Initialize each type and cell specific mean vector and VC matrix to the appropriate size, which is the # of assets.
      //===================================================================================================================
      for (int v=0; v<numCmps[(i%2)]; ++v)
      {
         mMeans[i][v]=Eigen::VectorXd(nAssets);
         mVCs[i][v]=Eigen::MatrixXd(nAssets,nAssets);
         vCIDs[i][v]=valcIDs[(i%2)][v];
      }

      // Populate probabilities, means, and the initial variance-covariance matrix for each valid cell.  Off-diagonal elements (i.e., covariances) are set to zero at this stage.
      //=====================================================================================================================================================
      for (int v=0; v<numCmps[(i%2)]; ++v)
      {
          uCell=vCIDs[i][v];
          mProbs[i][v]=estPrbs[(i%2)][uCell];
          for (int a1=0; a1<nAssets; ++a1)
          {
                mMeans[i][v](a1)=mean[a1][allCells[uCell][a1]];
                for (int a2=0; a2<nAssets; ++a2)
                {
                    if (a1 == a2)
                        mVCs[i][v](a1,a2)=pow(stdev[a1][allCells[uCell][a1]],2);
                    else
                        mVCs[i][v](a1,a2)=0.00;
                }
          }
      }
      // If debugging mode is on, write out the corresponding parameters of the distribution.  These are the initial LP values before improving with covariances.
      //=====================================================================================================================================================
      if (dbug >= 2)
         wrtDens(type[(i%2)],numCmps[(i%2)],vCIDs[i],mProbs[i],mMeans[i],mVCs[i],cout);
```

```
        // Improve the initial estimate using the ECME algorithm.  Stop when the multivariate likelihood is maximized.
        //====================================================================================================================
        cout << endl << string(85,'=') << endl << "ECME Algorithm #" << i+1 << " for " << type[(i%2)] << " initial solution." << endl << string(85,'=') << endl << endl;
        fout.open(rootdir+ofile, ios_base::out | ios_base::app);
        fout << endl << string(85,'=') << endl << "ECME Algorithm #" << i+1 << " for " << type[(i%2)] << " initial solution." << endl << string(85,'=') << endl << endl;
        fout.close();
        numCmpsF[i]=ECMEAlg(nTPoints,(const long double **) rtrn,nAssets,numCmps[(i%2)],mLHS[(i%2)],mRHS[(i%2)],mProbs[i],mMeans[i],mVCs[i],vCIDs[i],rootdir);

        // Write final density to the output file and display to user (which contains maximum likelihood estimates).
        //==================================================================================================================
        fout.open(rootdir+ofile, ios_base::out | ios_base::app);
        wrtDens(type[i],numCmpsF[i],vCIDs[i],mProbs[i],mMeans[i],mVCs[i],fout); fout.close();
        wrtDens(type[i],numCmpsF[i],vCIDs[i],mProbs[i],mMeans[i],mVCs[i],cout);
        cout << endl << "Done processing " << type[i] << " initial LP solution.  Final density shown above and written to file." << endl;
}

// Free temporary memory allocations.
//=====================================
for (int i=0; i<2*nECMEs; ++i)
{
    delete [] mProbs[i]; mProbs[i]=nullptr;
    delete [] mMeans[i]; mMeans[i]=nullptr;
    delete [] mVCs[i]; mVCs[i]=nullptr;
    delete [] vCIDs[i]; vCIDs[i]=nullptr;
}
for (int i=0; i<2; ++i)
{
    delete [] valcIDs[i]; valcIDs[i]=nullptr;
    delete [] estPrbs[i]; estPrbs[i]=nullptr;
}
delete [] mProbs; mProbs=nullptr;
delete [] mMeans; mMeans=nullptr;
delete [] mVCs; mVCs=nullptr;
delete [] valcIDs; valcIDs=nullptr;
delete [] mLHS; mLHS=nullptr;
delete [] mRHS; mRHS=nullptr;
delete [] estPrbs; estPrbs=nullptr;
for (int c=0; c<nCells; ++c)
{
    delete [] allCells[c]; allCells[c]=nullptr;
}

delete [] allCells; allCells=nullptr;
delete [] tmpVals; tmpVals=nullptr;
delete [] tmpCombo; tmpCombo=nullptr;
delete [] cellAsgn; cellAsgn=nullptr;
delete [] cellCnt; cellCnt=nullptr;
delete [] cellPrb; cellPrb=nullptr;
for (int m=0; m<3; ++m)
{
    delete [] optMDst[m]; optMDst[m]=nullptr;
}
for (int a=0; a<nAssets; ++a)
{
    delete [] rtrn[a]; rtrn[a]=nullptr;
    delete [] asgnmnt[a]; asgnmnt[a]=nullptr;
    delete [] prob[a]; prob[a]=nullptr;
    delete [] mean[a]; mean[a]=nullptr;
    delete [] stdev[a]; stdev[a]=nullptr;
}
delete [] optMDst; optMDst=nullptr;
delete [] rtrn; rtrn=nullptr;
delete [] asgnmnt; asgnmnt=nullptr;
delete [] prob; prob=nullptr;
delete [] mean; mean=nullptr;
```

```
    delete [] stdev; stdev=nullptr;
    delete [] nComps; nComps=nullptr;
    delete [] numCmpsF; numCmpsF=nullptr;

    // Exit.
    //========
    cout << endl << "Done.  (Hit return to exit.)" << endl; cin.get()
    return 0;
}

/*
/ Copyright (C) 2016 Chris Rook
/
/ This program is free software: you can redistribute it and/or modify it under the terms of the GNU General Public License as published by the Free Software Foundation, either version 3 of the License,
/ or (at your option) any later version.  This program is distributed in the hope that it will be useful, but WITHOUT ANY WARRANTY; without even the implied warranty of MERCHANTABILITY or FITNESS FOR A
/ PARTICULAR PURPOSE.  See the GNU General Public License for more details: <http://www.gnu.org/licenses/>
/
/ Filename:  fitMixDist.cpp
/
/ Function:  fitMixDist()
/
/ Summary:
/
/    This function fits a univariate normal mixture density to a set of observations using the EM algorithm within a forward-backward iterative procedure that tests for the optimal # of components.
/    First, a 1-component density is fit using standard maximum likelihood (ML) estimates and tested against a 2-component mixture density fit using the EM algorithm with a (user specified) large # of
/    random starts.  (Note:  The # of random starts is specified by the 3rd parameter in the input control file as a multiple of the # of cores/independent processing units on the computer running the
/    application.  In addition this value is multiplied by # components - 1, which uses an increasing # of random starts for densities with more components.)  When fitting a particular density, the one
/    with largest likelihood value from all random starts and obeying the variance ratio constraint (set in the header file using global constant: stdRatio) is selected as the ML estimator.  A hypothesis
/    test of Ho: 1-component vs Ha: 2-components is then conducted using the forward alpha (significance level) set by the user in the input control file (parameter #6).  The likelihood ratio test
/    statistic (LRT) is used and derived as -2Log(Lo/La), where Lo and La are the maximized likelihood values under Ho and Ha, respectively.  The observed LRT statistic value is then compared to the
/    critical value of the LRT's distribution under Ho that yields area to the right equal to the forward alpha.  The distribution of the LRT statistic under Ho is approximated via bootstrapping (see,
/    McLachlan (1987)).  Assuming the MLE's fit under Ho reflect the true distribution of the data (under Ho) random samples from the distribution under Ho are generated.  A 1-component and 2-component
/    univariate mixture density is fit to each sample which together generate a single null value from the true null distribution of the LRT statistic (again, assuming the true null distribution is that
/    fit by the MLEs). Using a large # of samples yields a set of values used to approximate the null distribution of the LRT statistic, with corresponding critical value for rejecting Ho based on the
/    forward alpha.  If Ho is rejected then a 2-component univariate mixture density becomes the new testing basis and the forward procedure continues with a test of 2-components vs 3 components.
/    Otherwise, if Ho is not rejected then the forward procedure testing basis remains a 1-component density, which is tested against 3-components in the exact same manner.  The key to bootstrapping the
/    LRT statistic's null distribution is to assume that the density with MLE parameters (under Ho) is the true null density and random samples are generated from it.  The forward procedure continues in
/    this manner testing Ho: go components vs Ha: ga components up to Ha = max # components specified in the control file (parameter #4).  If the maximum # of components is set to N, then the forward
/    procedure will conduct N-1 hypotheses tests sequentially, where the alternatives are Ha=2 components, 3 components, 4 components, ..., N components, respectively.  The forward procedure will end
/    with a certain # of mixture components being suggested as optimal.  It is followed by a backward procedure that tests a univariate mixture density with 1 less component, and repeats the test until
/    Ho is rejected, ending the forward-backward procedure.  If the forward procedure ends by suggesting that F components are needed to properly fit the data, then the backward procedure begins by
/    testing Ho: F-1 components vs Ha: F components using the backward alpha specified by the user in the control file (parameter #7).  Note that the forward procedure conducts a series of tests with Ha
/    having 2, 3, 4, ..., maxcmps (maxcmps is the maximum # of components to consider in the application, which is set by the 4th parameter in the control file) components, respectively.  Therefore, once
/    the forward procedure ends we have optimal univariate mixtures fit with ML estimates for all components from 1 to maxcmps.  During backward processing these are used to generate the random samples
/    needed under Ho to approximate the null distribution of the LRT.  If this hypothesis is rejected then the forward-backward procedure ends and F components are considered optimal for the given
/    observation set.  Else, if Ha is not rejected then the backward procedure basis becomes F-1 components (since it is not significantly different from F components, and applying the principle of
/    parsimony).  In this case, the backward procedure continues by testing Ho: F-2 components vs Ha: F-1 components.  If this test is not rejected, a mixture density with F-2 components becomes the new
/    basis and the procedure continues by testing Ho: F-3 components vs Ha: F-2 components.  If the test is rejected then F-1 components are considered optimal and the forward-backward procedure ends.
/    The backward portion continues in this manner until a test is rejected.  Consider an example with a maximum of N=10 components specified by the user in the control file (4th parameter) and a forward
/    alpha = 0.15 (6th parameter) and backward alpha = 0.20 (7th parameter).  The following details a possible sequence of testing events and illustrates how the procedure iterates:
/
/         Test #              Ho:                 Ha:                  Alpha               Type                Result (Ho)
/         ======              ============        =============        =======             ========            ===========
/            1                1-Component         2-Components          0.15               Forward               Accept
/            2                1-Component         3-Components          0.15               Forward               Accept
/            3                1-Component         4-Components          0.15               Forward               Reject
/            4                4-Components        5-Components          0.15               Forward               Accept
/            5                4-Components        6-Components          0.15               Forward               Accept
/            6                4-Components        7-Components          0.15               Forward               Reject
/            7                7-Components        8-Components          0.15               Forward               Accept
/            8                7-Components        9-Components          0.15               Forward               Accept
/            9                7-Components        10-Components         0.15               Forward               Accept
/           10                6-Components        7-Components          0.20               Backward              Accept
/           11                5-Components        6-Components          0.20               Backward              Accept
/           12                4-Components        5-Components          0.20               Backward              Reject
/         ======              ============        =============        =======             ========            ===========
```

```
/
/    Final Result:  Here, a 5-component mixture is considered to provide best fit based on the user-specified forward and backward alpha values (significance levels).  Forward processing ended with a 7-
/    component mixture producing the best fit and backward processing found no significant difference between 6 & 7 components (a test that was not performed during forward processing), and no difference
/    between 5 & 6 components (a test that was also not performed during forwared processing).  The backward test of 4 vs 5 components was performed during forward processing but with a smaller alpha and
/    was not rejected, but is rejected using the larger alpha for backward processing.  By tuning the forward/backward alphas the user customizes the procedure to favor a larger/smaller # of components.
/
/ Inputs:
/
/    1.) The asset # being processed (ranges from 0 thru (total # assets-1)).  The asset # is used for debugging and displaying results to the output window and output file. (a)
/
/    2.) The total # of observations (i.e., time points) in the current application.  Specified by the user via the control file (2nd parameter). (T)
/
/    3.) An array of size T holding the observations/returns at each time point for the asset currently being processed. (r)
/
/    4.) The maximum # of components to allow in the univariate mixture distribution fit by this function.  Specified by the user via the control file (4th parameter). (maxcmps)
/
/    5.) The # of bootstrap samples to use for each LRT when approximating the statistic's null distribution.  Specified by the user via the control file (5th parameter). (nsmpls)
/
/    6.) The # of random starts to use when finding the ML estimates for a specific mixture density.  Specified by the user via the control file (3rd parameter).  Note that this value is taken as a
/        multiple of the # of cores and is also increased by the multiple (# components - 1). (nstrts)
/
/    7.) An array of size 2 to hold the forward and backward alphas, respectively for the LRTs. (sl)
/
/    8.) An empty double array to hold the fitted univariate mixture distribution that results from applying the forward-backward procedure detailed here.  The array is indexed as [s][c], where s = 0, 1,
/        2, and c = component #.  Here, s=0 refers to the component's probability, s=1 refers to the component's mean, and s=2 refers to the component's standard deviation. (fnlMDst)
/
/    9.) A string for the output directory.  The final univariate mixture density for each # of components considered is written to the output file specified by the global constant ofile. (rdir)
/
/ Outputs:
/
/    This function populates the empty 2-dimensional array that is supplied via parameter #8 with the optimal univariate mixture density fit by the forward-backward procedure detailed above.  The number
/    of components in this density is returned at the function call.
/*******************************************************************************************************************************************************************************************************/
#include "stdafx.h"
int fitMixDist(const int a, const int T, const long double *r, const int maxcmps, const int nsmpls, const int nstrts, const long double sl[2], long double **fnlMDst, string rdir)
{
    // Declare/initialize local variables.
    //====================================
    int hnum=0, sas, strt, end, incr, rCmpsH0=1, cbfsol=1, cbbsol, curopt, tstid, nBootsAdj, pValCntr, fprms, cmpord;
    string yn, f_b, tmptxt;
    long double LRT[2], *StatLRT=new long double [nsmpls+1], **orProb=new long double *[maxcmps], **orMean=new long double *[maxcmps], **orStd=new long double *[maxcmps],
                *orLLVal=new long double [maxcmps], **orH0MDst=new long double *[3], *tmpRtrn=new long double[T], **outMDstH0=new long double *[3], **outMDstH1=new long double *[3],
                **rMDstH0=new long double *[3], **oriMDst=new long double *[3], **oroMDst=new long double *[3];
    vector<long double> pValue, alpha;
    vector<int> h0, h1;

    // Derive the 1-component solution for the incoming sample (i.e., using the MLE estimates).
    //=========================================================================================
    for (int m=0; m<3; ++m)
    {
        oriMDst[m]=new long double [1];
        rMDstH0[m]=new long double [1];              // 1-component mixture used for sample random starts.
    }
    oriMDst[0][0]=1.00;
    oriMDst[1][0]=getUniMean(T,r);
    oriMDst[2][0]=getUniStd(T,r,oriMDst[1][0]);

    // For each asset 2-dimensional arrays holds the optimal g-component solutions for the original sample.  This will avoid having to rebuild optimal solutions during forward-backward procedure.
    //===============================================================================================================================================================================================
    orProb[0]=new long double [1]; orProb[0][0]=oriMDst[0][0];
    orMean[0]=new long double [1]; orMean[0][0]=oriMDst[1][0];
    orStd[0]=new long double [1]; orStd[0][0]=oriMDst[2][0];
    orLLVal[0]=getLLVal(T,r,1,(const long double **) oriMDst);
    cout << string(20,'=') << endl << string(20,'=') << endl << "PROCESSING ASSET #" << a+1 << endl << string(20,'=') << endl << string(20,'=') << endl << endl;
    strt=2; end=maxcmps+1; incr=1; f_b="Forward";
```

```
for (int fb=0; fb<2; ++fb)
{
    // Iteration limits depend on the forward-backward procedure.
    //=============================================================
    if (1 == fb)
    {
        strt=cbfsol-1; end=0; incr=-1; f_b="Backward"; cbbsol=cbfsol;
    }
    for (int g=strt; g != end; g=g+incr)
    {
        // Set null and alternative hypotheses.
        //=======================================
        hnum=hnum+1;
        if (0 == fb)
        {
            h0.push_back(cbfsol); h1.push_back(g); sas=-1;
            alpha.push_back(sl[0]);
        }
        else if (1 == fb)
        {
            h0.push_back(g); h1.push_back(cbfsol);
            alpha.push_back(sl[1]);
        }

        // Write details of this iteration.
        //===================================
        cout << string(62,'-') << endl << string(62,'-') << endl << "Asset #" << a+1 << " - Hypothesis Test #" << hnum << " (" << h0[hnum-1] << " vs " << h1[hnum-1]
             << ") " << " [Direction: " << f_b << "]"  << endl << string(62,'-') << endl << string(62,'-') << endl << endl << "  " << string(62,'-') << endl << "  " << string(62,'-')
             << endl << "  H0: Asset #" << a+1 << " is best fit by a(n) " << h0[hnum-1] << " component Normal mixture." << endl << "      vs" << endl << "  H1: Asset #"
             << a+1 << " is best fit by a(n) " << h1[hnum-1] << " component Normal mixture." << endl << "  " << string(62,'-') << endl << "  " << string(62,'-') << endl;

        // If backward processing, check whether or not this hypothesis has already been tested.  If so, issue a warning then retrieve and use the existing solution.
        //===============================================================================================================================================================
        for (int h=0; h<hnum-1; ++h)
        {
            if (h0[h]==h0[hnum-1] && h1[h]==h1[hnum-1])
            {
                if (sl[0]==sl[1])
                   tmptxt=" the same ";
                else
                   tmptxt=" a different ";
                cout << endl << "  WARNING: This hypothesis has already been tested using" << tmptxt << "alpha. (See test #" << h+1 << ".)"; sas=h+1;
            }
        }

        // Rebuild input array under h0. (We need the null dist. to generate bootstrap samples.)
        // Note that h0[hnum-1] = # components under h0 for test # hnum and h0[hnum-1]-1 is the index of the optimal solution under h0 # of components.
        //====================================================================================================================================================
        for (int m=0; m<3; ++m)
            orH0MDst[m]=new long double [h0[hnum-1]];
        for (int c=0; c<h0[hnum-1]; ++c)
        {                                                //  In general, the g-component solution is stored in orXYZ[g-1][c].  For example:
            orH0MDst[0][c]=orProb[h0[hnum-1]-1][c];      //  -->  orProb[0] is a 1-dimensional array with single element orProb[0][0].
            orH0MDst[1][c]=orMean[h0[hnum-1]-1][c];      //  -->  orMean[1] is a 2-dimensional array with elements orMean[1][0], orMean[1][1].
            orH0MDst[2][c]=orStd[h0[hnum-1]-1][c];       //  -->  orStd[2] is a 3-dimensional array with elements orStd[2][0], orStd[2][1], orStd[2][2].
        }

        // Report the current optimal solution for the actual sample under H0 when forward processing.
        //==============================================================================================
        if (0 == fb && dbug >= 1)
        {
            cout.setf(ios_base::fixed, ios_base::floatfield); cout.precision(20);
            cout << endl << "  --> " << h0[hnum-1] << "-Component Optimal Solution:  Log-Likelihood: " << orLLVal[h0[hnum-1]-1] << endl << "                                      Variance Ratio: "
                 << getVRatio(h0[hnum-1], orH0MDst[2]) << endl;
            showVals(0,r,h0[hnum-1],(const long double **) orH0MDst);
```

```
} cout << endl;

// Simulation is used to approximate the null distribution of the LRT statistic.  If a sample is generated but no local optimum is found then reduce the simulation
// count by 1.  Also, reduce the simulation count by 1 when the LRT statistic is < 0.  This can happen if a local optimum is found under the full model that has a
// smaller likelihood than the optimum value under the reduced model (which may itself be a local optimum).  The adjusted count is held in nBootsAdj.
//=====================================================================================================================================================
nBootsAdj=nsmpls;

// LRT value for current solution.  Numerator uses the parameters already estimated.  We need to fit a new solution using h1 components on the incoming data for the
// denominator of the LRT.  This is done here.  Once fit, the estimates are transferred to the double arrays that hold optimal solutions for all component sizes.
// (Note: When forward processing, hnum == h1[hnum-1]-1 == g-1 so that hnum+1 == h1[hnum-1] == g.)
//=====================================================================================================================================================
if (0 == fb)
{
    for (int m=0; m<3; ++m)
        oroMDst[m]=new long double [h1[hnum-1]];  /* (2) Note that oriMDst has exactly hnum components during forward processing which is 1 less than the # of comps in H1. */
    orLLVal[hnum]=ThrdEMAlg(T,r,3*nstrts,hnum,(const long double **) oriMDst,h1[hnum-1],oroMDst);

    // Instantiate the arrays of size h1 and transfer the optimal solution for storage across component sizes.
    //=========================================================================================================
    orProb[hnum] = new long double [h1[hnum-1]]; orMean[hnum] = new long double [h1[hnum-1]]; orStd[hnum] = new long double [h1[hnum-1]];
    for (int c=0; c<h1[hnum-1]; ++c)
    {
        orProb[hnum][c]=oroMDst[0][c];
        orMean[hnum][c]=oroMDst[1][c];
        orStd[hnum][c]=oroMDst[2][c];
    }

    // If the algorithm fails to converge on the original data we must exit the program.
    //===================================================================================
    if (orLLVal[hnum] <= LNegVal)
    {
        cout << "ERROR:  Asset #" << a+1 << " did not converge to a local optimum when attempting to fit " << h1[hnum-1] << " components." << endl
             << "Try increasing the # of random starts, decreasing the # of components, or increasing the variance ratio constraint." << endl
             << "EXITING...fitMixDist()..." << endl; cin.get();
        exit (EXIT_FAILURE);
    }

    // Also exit if the algorithm finds an inferior optimum when compared to those with fewer #'s of components.
    //===========================================================================================================
    if (orLLVal[hnum-1] > orLLVal[hnum])
    {
        cout << "ERROR:  Likelihood for Asset #" << a+1 << " when fitting " << h1[hnum-1] << " components is less than the likelihood when " << endl
             << "fitting " << hnum << " components.  An inferior local optimum has been found.  (Increase the # of random starts to prevent this.)" << endl
             << "EXITING...fitMixDist()..." << endl; cin.get(); exit (EXIT_FAILURE);
    }
}
StatLRT[nsmpls] = -2.00*(orLLVal[h0[hnum-1]-1] - orLLVal[h1[hnum-1]-1]);

// Report the LRT statistic value for the actual sample.
//=======================================================
if (dbug >= 1)
{
    cout.setf(ios_base::fixed, ios_base::floatfield); cout.precision(20);
    cout << "  --> LRT Details for Original Sample:  H0 Log-L=" << orLLVal[h0[hnum-1]-1] << ", H1 Log-L=" << orLLVal[h1[hnum-1]-1] << ", LRTS for Asset #" << a+1 << "="
         << StatLRT[nsmpls] << endl << endl;
}

// Bootstrap the LRT statistic to determine its sampling dist.  The reduced model is the one with MLE's just derived for the given # of components.  This test is for # of components + 1.
//==============================================================================================================================================================================
if (-1==sas)
{
    if (0 == dbug)
       cout << "  Processing bootstrap samples ";
```

```
// Instantiate the arrays to hold the optimal solutions under H0 and H1.  These will be reused for each sample.
//===========================================================================================================
for (int m=0; m<3; ++m)
{
    outMDstH0[m]=new long double [h0[hnum-1]];
    outMDstH1[m]=new long double [h1[hnum-1]];
}

// Test the hypothesis.
//=======================
for (int b=0; b<nsmpls; ++b)
{
    if (0 == dbug)
    {
        cout << ".";
        if (b+1 != nsmpls && 0 == (b+1) % 100)
           cout << endl << string(31,' ');
    }
    else if (dbug >= 1)
    cout << string(162,'-') << endl << string(162,'-') << endl << string(40,' ') << "Asset #" << a+1 << ":  Hypothesis Test #" << hnum << " (" << h0[hnum-1]
         << " vs " << h1[hnum-1] << "): Start Processing Bootstrap Sample #" << b+1 << " of " << nsmpls << "." << endl << string(162,'-') << endl << string(162,'-') << endl << endl;

    // Generate sample of size nTPoints from the reduced model under the null hypothesis that the reduced model (using MLEs) is correct for this set of returns.
    // When bootstrapping the LRT statistic, use 1-component values (probs, means, stds) to generate the random starts under h0 (because we do not have access to
    // the solution using H0 # components less 1 on each sample) and use the model fitted under h0 to generate random starts when fitting the model specified by h1.
    //==========================================================================================================================================================
    getRVals(T,h0[hnum-1],(const long double **) orH0MDst,tmpRtrn);

    // Fit both H0 (# components) and H1 (# components) models and form the LRT statistic.  [Random start for H0 (always a 1-component mixture)].
    //=========================================================================================================================================
    rMDstH0[0][0]=1.00;
    rMDstH0[1][0]=getUniMean(T,tmpRtrn);
    rMDstH0[2][0]=getUniStd(T,tmpRtrn,rMDstH0[1][0]);

    // H0.
    //=====
    LRT[0]=ThrdEMAlg(T,tmpRtrn,nstrts,rCmpsH0,(const long double **) rMDstH0,h0[hnum-1],outMDstH0);

    // H1
    //=====
    if (LRT[0] > LNegVal)
       LRT[1]=ThrdEMAlg(T,tmpRtrn,nstrts,h0[hnum-1],(const long double **) outMDstH0,h1[hnum-1],outMDstH1);

    // Random observation from the null distribution of the following LRT:  (Reduce the simulation size count by 1 if the LRT statistic is negative.)
    // H0:  Data originate from mixture with h0[hnum-1] components vs. H1:  Data originate from mixture with h1[hnum-1] components
    //============================================================================================================================================
    if (LRT[0] > LNegVal && LRT[1] > LNegVal)
       StatLRT[b] = -2.00*(LRT[0] - LRT[1]);
    else
       StatLRT[b] = -1.00;

    // The denominator gets decremented in 2 ways:
    //   1.) The LRT statistic is negative which means that an inferior local optimum was found.
    //   2.) No local optimum was found when fitting either the null component or the alternative component distributions, all were spurious (i.e., leading to unboundedness).
    //=====================================================================================================================================================================
    if (StatLRT[b] < 0.00)
       nBootsAdj=nBootsAdj-1;

    // Report the LRT statistic value for the bootstrap sample.
    //==========================================================
    if (dbug >= 1)
    {
        cout.setf(ios_base::fixed, ios_base::floatfield); cout.precision(20);
        if (LRT[0] > LNegVal && LRT[1] > LNegVal && StatLRT[b] >= 0.00)
           cout << "  --> LRT Details for Sample #" << b+1 << ":  LRT[0]=" << LRT[0] << ", LRT[1]=" << LRT[1] << ", LRTS for bootstrap sample "
```

```
                    << b+1 << " of " << nsmpls << "=" << StatLRT[b] << endl << endl;
            else
            {
                if (LRT[0] <= LNegVal || LRT[1] <= LNegVal)
                {
                    tstid=1;
                    if (LRT[0] <= LNegVal)
                       tstid=0;
                    cout << "  --> The EM algorithm did not find a local optimum under H" << tstid << ".  (Sample " << b+1 << " will be discarded.)" << endl << endl;
                }
                else
                    cout << "  --> Negative LRT test statistic: " << StatLRT[b] <<".  (Sample " << b+1 << " will be discarded.)" << endl
                         << "      [An inferior (not the largest) local optimum was found under H1.]" << endl << endl;
            }
        }
    }

    // Clear out sample solution arrays.
    //==================================
    for (int m=0; m<3; ++m)
    {
        delete [] outMDstH0[m]; outMDstH0[m]=nullptr;
        delete [] outMDstH1[m]; outMDstH1[m]=nullptr;
    }
}


// Clear the array that holds the optimal solution for the original sample under H0.
//==================================================================================
for (int m=0; m<3; ++m)
{
    delete [] orH0MDst[m]; orH0MDst[m]=nullptr;
}

if (-1==sas && 0 == dbug)
   cout << " (" << nsmpls << ")" << endl;

// Determine p-value & test result.  If H0 is rejected then move the temporary values into their permanant placeholders.  If null hypothesis is not rejected then set variable stop to 1.
//=====================================================================================================================================================================================
if (-1 == sas)
{
    pValCntr=1;
    for (int n=0; n<nsmpls; ++n)
        if (StatLRT[n] > 0.00 && StatLRT[n] >= StatLRT[nsmpls])
           pValCntr=pValCntr+1;
    pValue.push_back((long double) pValCntr/(nBootsAdj+1));

    // Uncomment to write the LRT[] array to a file in the error folder.  These values form the null distribution of the test statistic.
    //=================================================================================================================================
    // ofstream fout;
    //  fout.open(errFolder + "Asset_" + to_string((long long) a+1) + "_LRT_" + to_string((long long) h0[hnum-1]) + "_vs_" + to_string((long long) h1[hnum-1]) + ".txt");
    //  fout.setf(ios_base::fixed, ios_base::floatfield); fout.precision(16);
    //  fout << "--> LRT statistic for original data:  " << StatLRT[nsmpls] << endl << endl;
    //  fout << "--> Null distribution LRT statistic values for " << nsmpls << " bootstrap samples (set values of -1.00 to missing):" << endl << endl;
    //  for (int b=0; b<nsmpls; ++b)
    //    fout << "StatLRT[" << b << "] = " << StatLRT[b] << endl;
    // fout.close();
}
else
    pValue.push_back(pValue[sas-1]);

// Write out the hypothesis test result.  If this is a retest, report the prior result.  (Note that it is a retest when sas >= 0, and sas holds the prior test number.)
//====================================================================================================================================================================
if (-1 == sas)
{
```

```
            cout << endl << "  " << string(37,'=') << endl << "  Hypothesis test #" << hnum << " (" << h0[hnum-1] << " vs " << h1[hnum-1] << ") result:" << endl << "  " << string(37,'=')
                 << endl << "  --> The hypothesis test uses " << nBootsAdj << " valid LRTS values from resampling along with the 1 LRTS value from the original sample. (Total="
                 << nBootsAdj+1 << ")" << endl << "  --> Of these, there are " << pValCntr << " values >= the sample LRT statistic." << endl;
            cout.setf(ios_base::fixed, ios_base::floatfield); cout.precision(20);
            cout << "  --> The resulting p-value for testing H0 vs H1 is: " << pValue[hnum-1] << " vs. " << f_b << " alpha: " << sl[fb] << "." << endl;
        }
        else
        {
            cout << endl << "  " << string(67,'=') << endl << "  Hypothesis test #" << hnum << " (" << h0[hnum-1] << " vs " << h1[hnum-1] << ") is a retest of Hypothesis test # " << sas
                 << "." << endl << "  " << string(67,'=') << endl;
            cout.setf(ios_base::fixed, ios_base::floatfield); cout.precision(20); cout << "  --> The resulting p-value for testing H0 vs H1 is: " << pValue[hnum-1] << " vs. " << f_b
                 << " alpha: " << sl[fb] << "." << endl;
        }

        // Accept or reject the hypothesis.  If Ho is rejected then Ha is considered a better fit.  When forward testing, Ha has additional components (i.e., the integer value of h1).
        // When backward testing Ho is considered an acceptable fit when not rejected (i.e., integer value of h0).  Here we keep track of the best solutions for both forward and backward
        // testing.  The index of the optimal solution when accessing the orXYZ[][] double arrays is (cbbsol-1) once finished processing.  That is, orXYZ[cbbsol-1][0] - orXYZ[cbbsol-1][cbbsol-1].
        //==============================================================================================================================================================================
        yn=" NOT ";
        if (pValue[hnum-1] <= sl[fb])
        {
            yn=" ";
            if (0 == fb)
               cbfsol=h1[hnum-1];       // H0 was rejected, set CBFSOL to H1 # components.
            else                        // Stop backward processing after first H0 is rejected.
               end=g+incr;
        }
        else if (1 == fb)
                cbbsol=h0[hnum-1];
        curopt = (int) (0==fb)*cbfsol + (int) (1==fb)*cbbsol;     // For displaying current solution to user.

        // If forward processing and not at the max # components we should reset oriMDst[][] and clear out oroMDst[][].
        //==================================================================================================================
        if (0 == fb && g+incr != end)
        {
            // Clear/update the input and output arrays.  The oriMDst[][] densities will generate the random starts when fitting the original
            // sample to one additional component.  (Note that the original double mixture arrays are only needed during forward processing.)
            //====================================================================================================================
            for (int m=0; m<3; ++m)
            {
                delete [] oriMDst[m];
                oriMDst[m]=new long double [h1[hnum-1]];
                for (int c=0; c<h1[hnum-1]; ++c)
                    oriMDst[m][c]=oroMDst[m][c];
                delete [] oroMDst[m]; oroMDst[m]=nullptr;
            }
        }

        // Report the current best solution after testing this hypothesis.
        //==================================================================
        cout << "  --> The null hypothesis is" << yn << "rejected in favor of the alternative hypothesis." << endl << "  --> At this stage a(n) " << curopt
             << "-component Normal mixture provides the best fit for Asset #" << a+1 << "." << endl << endl << string(162,'-') << endl << string(162,'-') << endl << endl;

        // Processor cool down.
        //======================
        if (dbug >= 2)
        {
            cout.setf(ios_base::fixed, ios_base::floatfield); cout.precision(4);
            cout << endl << "***** Processor cool Down: " << ((double) cDown/1000.00)/60.00 << " minutes. *****" << endl;
        } Sleep(cDown);
    }
}

// Display p-values then clear out the containers.
//===================================================
```

```
cout << "  All p-values for Asset #" << a+1 << " by test number:" << endl << "  " << string(41,'=') << endl;
for (int m=0; m<pValue.size(); ++m)
{
    cout.setf(ios_base::fixed, ios_base::floatfield); cout.precision(10);
    cout << "  P-value for hypothesis test #" << setfill('0') << setw(2) << m+1 << " (" << setw(2) << h0[m] << " vs " << setw(2) << h1[m] << "): " << pValue[m];  cout.precision(5);
    cout << " (alpha = " << alpha[m] << ")"; cout.precision(10); cout << " (Log-Likelihood Values:  LL[H0]=" << orLLVal[h0[m]-1] << ", LL[H1]=" << orLLVal[h1[m]-1] << ")" << endl;
}

// Insert the optimal solution into the double array passed to this function. The # components is returned by the function.  1st order the mixture returned by the respective means.
//==========================================================================================================================================================================
for (int c1=0; c1<cbbsol; ++c1)
{
    cmpord=0;
    for (int c2=0; (c2<cbbsol); ++c2)
        if (orMean[cbbsol-1][c2] < orMean[cbbsol-1][c1])
           cmpord++;

    // Account for ties.
    //===================
    for (int c3=cmpord; (c3<cbbsol && fnlMDst[0][cmpord] > 0.00); ++c3)
        cmpord++;

    // Populate the outgoing array.
    //==============================
    fnlMDst[0][cmpord]=orProb[cbbsol-1][c1];
    fnlMDst[1][cmpord]=orMean[cbbsol-1][c1];
    fnlMDst[2][cmpord]=orStd[cbbsol-1][c1];
}

// Display the final result for each asset.
//=========================================
fprms = (3*(cbbsol)-1);  // Free parameters.
cout << endl << "  --> Distribution for Asset #" << a+1 << " is a(n) " << cbbsol << "-Component Normal Mixture (full details below)." << endl
     << endl << "      Log-Likelihood: " << orLLVal[cbbsol-1]
     << endl << "      Variance Ratio: " << getVRatio(cbbsol, orStd[cbbsol-1])
     << endl << "      AIC:  " << (long double) 2.00*fprms - 2.00*orLLVal[cbbsol-1]
     << endl << "      BIC:  " << -2.00*orLLVal[cbbsol-1] + (long double) fprms*log((long double) T)
     << endl << "      AICC: " << (long double) 2.00*fprms - 2.00*orLLVal[cbbsol-1] + ((long double) 2.00*(fprms + 1)*(fprms + 2))/((long double) (T - fprms - 2))
     << endl << "      Density Parameters: " << endl;
showVals(0,r,cbbsol,(const long double **) fnlMDst);
cout << endl << string(162,'=') << endl << string(162,'=') << endl;

// Write all densities to output file for this asset. Along with all p-values.
//============================================================================
ofstream fout;
if (0 == a)
    fout.open(rdir+ofile, ios_base::out);
else
    fout.open(rdir+ofile, ios_base::out | ios_base::app);
fout << string(28,'=') << endl << "Asset #" << a+1 << " Optimal Densities:" << endl << string(28,'=') << endl;
for (int i=0; i<maxcmps; ++i)
{
    fout << endl << "  "  << i+1 << "-Component Optimal Solution: " << endl << "  " << string(43,'=');
    fout.setf(ios_base::fixed, ios_base::floatfield); fout.precision(16);
    fprms = (3*(i+1)-1);  // Free parameters.
    fout << endl << "  --> Log-Likelihood: " << orLLVal[i]
         << endl << "      Variance Ratio: " << getVRatio(i+1, orStd[i])
         << endl << "      AIC:  " <<  (long double) 2.00*fprms - 2.00*orLLVal[i]
         << endl << "      BIC:  " << -2.00*orLLVal[i] + (long double) fprms*log((long double) T)
         << endl << "      AICC: " << (long double) 2.00*fprms - 2.00*orLLVal[i] + ((long double) 2.00*(fprms + 1)*(fprms + 2))/((long double) (T - fprms - 2))
         << endl << "      Density Parameters: " << endl;
    for (int c=0; c<i+1; ++c)
        fout << string(37,' ') << "Prob[" << c << "]=" << orProb[i][c] << "  Mean[" << c << "]=" << orMean[i][c] << "  Std.Dev.[" << c << "]=" << orStd[i][c] << endl;
}
fout << endl << endl << "  All p-values for Asset #" << a+1 << " by test number:" << endl << "  " << string(41,'=') << endl;
```

```
    for (int m=0; m<pValue.size(); ++m)
    {
        fout.setf(ios_base::fixed, ios_base::floatfield); fout.precision(10);
        fout << "  P-value for hypothesis test #" << setfill('0') << setw(2) << m+1 << " (" << setw(2) << h0[m] << " vs " << setw(2) << h1[m] << "): " << pValue[m];  fout.precision(5);
        fout << " (alpha = " << alpha[m] << ")"; fout.precision(10); fout << " (Log-Likelihood Values:  LL[H0]=" << orLLVal[h0[m]-1] << ", LL[H1]=" << orLLVal[h1[m]-1] << ")" << endl;
    }
    fout << endl << endl << "  Optimal Density for Asset #" << a+1 << endl << "  " << string(31,'=') << endl;
    showVals(0,r,cbbsol,(const long double **) fnlMDst, fout); fout << endl << endl;
    fout.close();

    // Clear vectors and free temporary memory allocations.
    //======================================================
    pValue.clear(); pValue.shrink_to_fit(); alpha.clear(); alpha.shrink_to_fit();
    h0.clear(); h0.shrink_to_fit(); h1.clear(); h1.shrink_to_fit();
    for (int m=0; m<3; ++m)
    {
        delete [] oriMDst[m]; oriMDst[m]=nullptr;
        delete [] oroMDst[m]; oroMDst[m]=nullptr;
        delete [] rMDstH0[m]; rMDstH0[m]=nullptr;
    }
    delete [] oriMDst; oriMDst=nullptr;
    delete [] oroMDst; oroMDst=nullptr;
    delete [] rMDstH0; rMDstH0=nullptr;
    for (int c=0; c<maxcmps; ++c)
    {
        delete [] orProb[c]; orProb[c]=nullptr;
        delete [] orMean[c]; orMean[c]=nullptr;
        delete [] orStd[c]; orStd[c]=nullptr;
    }
    delete [] orProb; orProb=nullptr;
    delete [] orMean; orMean=nullptr;
    delete [] orStd; orStd=nullptr;
    delete [] tmpRtrn; tmpRtrn=nullptr;
    delete [] StatLRT; StatLRT=nullptr;
    delete [] orLLVal; orLLVal=nullptr;
    delete [] orH0MDst; orH0MDst=nullptr;
    delete [] outMDstH0; outMDstH0=nullptr;
    delete [] outMDstH1; outMDstH1=nullptr;

    // Return the optimal # of components.
    //=====================================
    return cbbsol;
}

/*
/ Copyright (C) 2016 Chris Rook
/
/ This program is free software: you can redistribute it and/or modify it under the terms of the GNU General Public License as published by the Free Software Foundation, either version 3 of the License,
/ or (at your option) any later version.  This program is distributed in the hope that it will be useful, but WITHOUT ANY WARRANTY; without even the implied warranty of MERCHANTABILITY or FITNESS FOR A
/ PARTICULAR PURPOSE.  See the GNU General Public License for more details: <http://www.gnu.org/licenses/>
/
/ Filename:  ThrdEMAlg.cpp
/
/ Function:  ThrdEMAlg()
/
/ Summary:
/
/    This function threads the EM algorithm.  An individual thread is used to maximize the likelihood function for 1 random start.  Likelihood functions for mixture PDFs have multiple local optimums, and
/    they are unbounded.  The MLEs for the parameters are those that yield the largest of the local maximums, after removing spurious optimizers.  Spurious optimizers are those where a single component
/    is used to fit one or a small number of closely clustered observations.  In such cases the variance of the corresponding component becomes very small or approaches zero, which drives the likelihood
/    value to infinity.  Spurious optimizers can be eliminated by imposing a variance ratio constraint which does not allow the ratio of the largest to smallest variance (across components) to exceed
/    some given constant.  This will prevent a variance from approaching zero or a small number.  The user must set the variance ratio constraint to a value that is appropriate for the data in their
/    application.  If the intent is to construct a density function that memorizes the training data then this value can be set very high.  If the intent is to build a density that extends well to out-
/    of-sample data then this value can be set very low.  The global constant stdRatio is set in the header file as the square root of the desired variance ratio constraint.  The # of threads is set by
/    the # of independent processing units on the computer running the application as well as the # of components in the mixture being fit and the user-specified # of random starts parameter setting in
```

```
/    the control file (3rd parameter).  The # of random starts is equal to: rs*(# cores)*(g-1), where rs = # of random starts specified by the 3rd parameter in control.txt, # cores = # of indpendent
/    processing units on the computer running the application, and g = # of components in the univariate mixture density being fit.  The # of random starts therefore increases with the size of the
/    density being fit.  Each random start is assigned its own thread and launches the EMAlg() function to find the likelihood maximizer based on that random start (this is a unique value and reflects
/    the local maximum nearest to the parameter settings in the random start).  Once all threads finish, the parameter settings from the random start which yields the largest likelihood value (and
/    obeying the variance ratio constraint) are taken as the MLEs.  An empty double array is populated with these values and the maximum log-likelihood function value is returned at the call.
/
/ Inputs:
/
/    1.) The total # of observations (i.e., time points) in the current application.  Specified by the user via the control file (2nd parameter). (T)
/
/    2.) An array of size T holding the observations/returns at each time point for the asset currently being processed. (r)
/
/    3.) The # of random starts to use when finding the ML estimates for a specific mixture density.  Specified by the user via the control file (3rd parameter).  Note that this value is taken as a
/        multiple of the # of cores and is also increased by the multiple (# components - 1), so that an increasing # of random starts is used as the # of components increases. (nstrts)
/
/    4.) The # of components in the univariate mixture distribution used to generate random starts (i.e., means) for the EM algorithm.  For example if fitting a 5-component univariate mixture
/        distribution to an asset and an optimal 4-component univariate mixture distribution is available for that asset, then a 4-component mixture distribution is used to generate outg (argument #6 to
/        this function) means as the starting point for constructing the random starts.  If no such 4-component mixture distribution is available (for example during bootstrapping of the LRT with Ho
/        having 4 components), then a 1-component mixture will be used to generate the means for an outg-component random start. (Note: Recall that a g-component random start is built by generating g-
/        random values which serve as the component means.  Each observation is then attached to the nearest mean (component).  The probability for that component is the # attached divided by the total #
/        of observations, and the standard deviation for that component is the sample standard deviation for all observations assigned to it.) (ing)
/
/    5.) A 2-dimensional array to hold the univariate mixture distribution used to generate means for a random start.  This density function has ing components as mentioned above and the double array is
/        indexed as [s][c], where s = 0, 1, 2, and c = component #.  Here, s=0 refers to the component's probability, s=1 refers to the component's mean, and s=2 refers to the component's standard
/        deviation.  As noted above, when generating random starts for the EM algorithm a density function that is most similar to the # of components being fit is desirable.  Therefore, if fitting a 5-
/        component mixture density and an optimal 4-component mixture density is available then it will be used to generate the random starts.  If an optimal 4-component mixture density is not available
/        then a univariate normal distribution (i.e., 1-component mixture) can be used to generate the means for a random start. (inMDst)
/
/    6.) The # of components in the outgoing univariate mixture density that is being fit using the EM algorithm. (outg)
/
/    7.) An empty double array to hold the fitted univariate mixture distribution fit using the EM algorithm.  The double array is indexed as [s][c], where s = 0, 1, 2, and c = component #.  Here, s=0
/        refers to the component's probability, s=1 refers to the component's mean, and s=2 refers to the component's standard deviation. (outMDist)
/
/ Outputs:
/
/    This function returns the log-likelihood for the optimal univariate mixture fit using the EM algorithm at the function call.  It also populates the supplied empty double array (outMDist) with the
/    corresponding optimal univariate mixture distribution.
/************************************************************************************************************************************************************************************************/
#include "stdafx.h"
long double ThrdEMAlg(const int T, const long double *r, const int rs, const int ing, const long double **inMDist, const int outg, long double **outMDist)
{
    // Local variables.
    //===================
    long double llval;

    // Handle trivial case first (i.e., 1-component), then non-trivial case (i.e., > 1-component).
    //==========================================================================================
    if (outg == 1)
    {
        for (int m=0; m<3; ++m)
            outMDist[m][0]=inMDist[m][0];                              // Transfer probabilities, means, and standard deviations.
        llval = getLLVal(T,r,outg,(const long double **) outMDist);     // Compute likelihood.
    }
    else
    {
        // Declare/initialize local variables.  Find # of independent processing units and create array of thread objects using the random start multiplication factor specified in the control file.
        // The # of random starts specified in the control file is a multiple of the # of independent processing units.
        //====================================================================================================================================================================================
        int p = (int) (outg-1)*rs*boost::thread::hardware_concurrency(), fstgrp=1, **runprms=new int *[p];
        long double **rPrbs=new long double *[p], **rMns=new long double *[p], **rStds=new long double *[p], *llVals=new long double [p];
        boost::thread *t=new boost::thread[p];
        string cmt, spc;

        // Display message with details of the optimization process.
        //============================================================
```

```
if (dbug >= 1)
{
    cout << "  Threading EM Algorithm will use " << p << " random starts for the current optimization." << endl;
    if (dbug >=2)
    {
        cout << "  [Note:  When the variance ratio constraint is violated, or the maximum # of iterations has been reached, or an invalid probability detected,"  << endl
             << "          the log-likelihood is set to the arbitrarily large negative #: ";  cout.setf(ios_base::fixed, ios_base::floatfield); cout.precision(7); cout << LNegVal << "]" << endl;
    }   cout << endl;
}

// Create arrays to hold the random starts and return values from the EM algorithm.  A random start must specify all unknown parameters for
// a ng-component mixture. (i.e., component probabilities, means, standard deviations)  These p arrays will be reused within each run group.
//==================================================================================================================================
for (int j=0; j<p; ++j)
{
    rPrbs[j] = new long double [outg];
    rMns[j] = new long double [outg];
    rStds[j] = new long double [outg];
    runprms[j] = new int [4]; runprms[j][0]=ing; runprms[j][1]=j;
}

// Iterate over # of random starts and determine the start, then launch optimization calls.
//=====================================================================================
for (int j=0; j<p; ++j)
{
    // Launch call to EMAlg() for each thread.  Once all threads finish, we scan the likelihood values across the p solutions (inner loop) and select the largest local  optimum as the optimum.
    // The outer loop then repeats this process rs*(outg-1) number of times.
    //===================================================================================================================================================================================
    t[j] = boost::thread(EMAlg,T,boost::cref(r),outg,boost::ref(rPrbs),boost::ref(rMns),boost::ref(rStds),boost::ref(llVals), boost::cref(inMDist),boost::ref(runprms[j]));
}

// Pause until all sub-threads finish, and save the optimal solution from this group of runs.
//=======================================================================================
for (int j=0; j<p; ++j)
{
    t[j].join();

    // Report results when requested.
    //================================
    if (dbug >= 2)
    {
        spc=""; cmt="";
        if (0 == runprms[j][2])
           cmt="[Variance ratio constraint violated, solution not used.]";
        else if (1 == runprms[j][2])
                cmt="[Maximum # of iterations reached, solution not used.]";
        else if (2 == runprms[j][2])
        {
            if (llVals[j] > 0.00 && llVals[j] < 10.00) spc="  ";
            else if (llVals[j] >= 10.00 && llVals[j] < 100.00) spc=" ";
        }
        else if (3 == runprms[j][2])
                cmt="[Invalid probability encountered, solution not used.]";
        cout << "  --> " << outg << "-Component Log-Likelihood for random start # " << setfill('0') << setw(to_string((long long) p).size()) << j+1 << " = ";
        cout.setf(ios_base::fixed, ios_base::floatfield); cout.precision(8*(1+(2 == runprms[j][2]))); cout << spc << llVals[j]; cout.precision(0); cout << " (Iterations = " << setfill('0')
             << setw(to_string((long long) mIters).size()) << runprms[j][3] << ") " << cmt << endl;
    }

    // Retrieve optimal solution:  Scan random starts and locate the one associated with the highest likelihood value.  The optimal probabilities, means, standard deviations
    // are transferred to the placeholders passed to this function.  If no local optimum has been found then set the return code to 0, otherwise use a return code of 1.
    //=========================================================================================================================================================
    if (0 == j && 1 == fstgrp)
    {
        fstgrp=0;
        llval=llVals[0];
```

```
                for (int c=0; c<outg; ++c)
                {
                    outMDist[0][c]=rPrbs[0][c];
                    outMDist[1][c]=rMns[0][c];
                    outMDist[2][c]=rStds[0][c];
                }
            }
            else if (llVals[j] > llval)
            {
                llval=llVals[j];
                for (int c=0; c<outg; ++c)
                {
                    outMDist[0][c]=rPrbs[j][c];
                    outMDist[1][c]=rMns[j][c];
                    outMDist[2][c]=rStds[j][c];
                }
            }
        }
        if (dbug >= 1)
           cout << endl;

        // Free temporary memory allocations.
        //=====================================
        for (int j=0; j<p; ++j)
        {
            delete [] rPrbs[j]; rPrbs[j]=nullptr;
            delete [] rMns[j]; rMns[j]=nullptr;
            delete [] rStds[j]; rStds[j]=nullptr;
            delete [] runprms[j]; runprms[j]=nullptr;
        }
        delete [] rPrbs; rPrbs=nullptr;
        delete [] rMns; rMns=nullptr;
        delete [] rStds; rStds=nullptr;
        delete [] runprms; runprms=nullptr;
        delete [] llVals; llVals=nullptr;
        delete [] t; t=nullptr;
    }

    // Report the optimal solution.
    //===============================
    if (dbug >= 1)
    {
        cout.setf(ios_base::fixed, ios_base::floatfield); cout.precision(16);
        cout << "  --> " << outg << "-Component Optimal Solution:  Log-Likelihood: " << llval << endl << "                                     Variance Ratio: " << getVRatio(outg, outMDist[2]) << endl;
        if (llval > LNegVal)
           showVals(0,r,outg,(const long double **) outMDist); cout << endl;
    }

    // Return the log-likelihood value corresponding to the optimal solution.
    //=========================================================================
    return llval;
}

/*
/ Copyright (C) 2016 Chris Rook
/
/ This program is free software: you can redistribute it and/or modify it under the terms of the GNU General Public License as published by the Free Software Foundation, either version 3 of the License,
/ or (at your option) any later version. This program is distributed in the hope that it will be useful, but WITHOUT ANY WARRANTY; without even the implied warranty of MERCHANTABILITY or FITNESS FOR A
/ PARTICULAR PURPOSE.  See the GNU General Public License for more details: <http://www.gnu.org/licenses/>
/
/ Filename:  EMAlg.cpp
/
/ Function:  EMAlg()
/
/ Summary:
```

```
/
/    This function implements the EM algorithm for estimating parameters of a univariate mixture PDF.  The observations from a univariate mixture PDF can be viewed as an incomplete data problem, where
/    the component # for each observation/time point is missing or unobserved.  That is, at each time point 2 random variables produce a value:  1.) the component #, and 2.) the value from it.  Note that
/    both sets of RVs will have parameters to be estimated.  If viewed this way, the likelihood function can be expressed using both the missing and non-missing random variables and their corresponding
/    parameters.  For example, the parameters for the unobserved component random variable that each observation originates from are the component probabilities, and the parameters for the observed
/    density values are means and standard deviations from each component distribution.  The EM algorithm then estimates parameters for both random variables iteratively as follows:
/
/    Step 1:  Select starting values for all parameters (for both the missing/unobserved and non-missing/observed random variables).
/    Step 2:  Compute the expected values for all missing/unobserved random variables using the most recent parameter estimates.
/    Step 3:  Replace all instances of the missing random variables with their expected values in the likelihood or log-likelihood function.
/    Step 4:  Optimize the resulting likelihood or log-likelihood function with respect to the parameters for the non-missing random variables.
/    Step 5:  Check the %-change in likelihood or log-likelihood value maximized in step 4 for convergence (i.e., small %-change) or error/exit condition.
/    Step 6:  If no convergence or error/exit condition is met in Step 5 then return to Step 2.
/
/    Starting values for Step 1 are computed as random starts for all parameter values.  This is acheived by first generating g values from the closest distribution to the one being fit (i.e., if fitting
/    a 5-component univariate mixture and an optimal 4-component univariate mixture is available, then use the 4-component mixture to generate random starts, and if no other density is available then use
/    a 1-component mixture).  These g values are taken as the means, and each observation is attached to the closest mean.  The standard deviation of the set of obserations attached to each mean is
/    computed as the component standard deviation, and the proportion of observations attached to each mean is the component probability for the random start.  Convergence in Step 5 is checked using the
/    global constant epsilon set in the header file.  In Step 5 we also check for error/exit conditions and exit the optimization if any of the following error conditions are met.  These are:
/
/    Error/Exit 1.) Variance ratio constraint is violated (control with global constant stdRatio set in the header file)
/    Error/Exit 2.) Maximum # of iterations reached (control with global constant mIters set in the header file)
/    Error/Exit 3.) Any component probability becomes zero or negative while iterating
/
/ Inputs:
/
/    1.) The total # of observations (i.e., time points) in the current application.  Specified by the user via the control file (2nd parameter). (T)
/
/    2.) An array of size T holding the observations/returns at each time point for the asset currently being processed. (r)
/
/    3.) The # of components in the univariate mixture density being fit. (g)
/
/    4.) An empty array of size g to hold the probability for each component in the univariate mixture.  This array is updated during each iteration of the EM algorithm and therefore holds the optimal
/        probabilities upon convergence, which are returned in this array to the calling function.  (Note that this parameter is a double array indexed as [t][g], where t is the thread # assigned by
/        ThrdEMAlg(), which invokes this function.  The EM algorithm is implemented within threaded calls where a random start is assigned to its own thread.) (prbs)
/
/    5.) An empty array of size g to hold the mean for each component in the univariate mixture.  This array is updated during each iteration of the EM algorithm and therefore holds the optimal means
/        upon convergence, which are returned in this array to the calling function.  (Note that this parameter is a double array indexed as [t][g], where t is the thread # assigned by ThrdEMAlg(), which
/        invokes this function.  The EM algorithm is implemented within threaded calls where a random start is assigned to its own thread.) (mns)
/
/    6.) An empty array of size g to hold the standard deviation for each component in the univariate mixture.  This array is updated during each iteration of the EM algorithm and therefore holds the
/        optimal standard deviations upon convergence, which are returned in this array to the calling function.  (Note that this parameter is a double array indexed as [t][g], where t is the thread #
/        assigned by ThrdEMAlg(), which invokes this function.  The EM algorithm is implemented within threaded calls where a random start is assigned to its own thread.) (stds)
/
/    7.) An array to hold the log-likelihood value that the EM algorithm converges to.  It is the optimal value for the given random start.  The array is indexed by thread #, therefore a single call to
/        this function generates 1 optimal log-likelihood value which is returned in the element # for the current thread. (llVal)
/
/    8.) A double array to hold the univariate mixture density that is used to generate the random starts.  This function begins by generating a random start, then optimizes the log-likelihood function
/        based on that start (i.e., finds the nearest local maximum).  This double array is indexed as [s][c], where s = 0, 1, 2, and c = component #.  Here, s=0 refers to the component's probability,
/        s=1 refers to the component's mean, and s=2 refers to the component's standard deviation. (inMDist)
/
/    9.) An array of integer parameters holding values passed to and returned by this function.  Element [0] holds the # of components in the mixture density that is used to generate the random starts,
/        which is parameterized in inMDist (parameter #8 to this function); element [1] is the thread # of the current call; element [2] is a return code which takes the following values: 0=variance
/        ratio constraint is violated, 1=maximum # of iterations reached, 2=convergence based on change to log-likelihood value, 3=A component probability is not between 0.0 and 1.0; element [3] is the
/        iteration # of convergence or algorithm termination for one of the reasons decoded in element [2]. (rprms)
/
/ Outputs:
/
/    This function does not return a value at the call but updates several empty arrays that are supplied by the user.  The prbs array is updated with the final estimated component probabilities, the mns
/    array is updated with the final estimated component means, the stds array is updated with the final estimated standard deviations, the llval array is updated with the final log-likelihood value, and
/    the rprms array is updated at element [2] with the functions return code, and element [3] with the # iterations before convergence or termination.
/**************************************************************************************************************************************************************************************************/
#include "stdafx.h"
void EMAlg(const int T, const long double *r, const int g, long double **prbs, long double **mns, long double **stds, long double *llVal, const long double **inMDist, int *rprms)
{
```

```
// Declare/initialize local variables.
//=====================================
const int ing=rprms[0], thrd=rprms[1];
int stop;
long double oldllval, newllval, minstd, maxstd, cprbs, ssqrs[1], var, **pdens=nullptr, *postprbs=new long double [T], *mdens=nullptr, psum, **orMDst=new long double *[3],
            **uMDst=new long double *[3];

// Generate g random obs from solution with rprms[0] components which are specified in inMDist[][].  Use these means to generate probabilities and standard deviations for each random start.
//=====================================================================================================================================================================================================
do
{
      stop=1; minstd=LPosVal; maxstd=LNegVal;
      getRVals(g,ing,(const long double **) inMDist,mns[thrd]);
      getRPrbsStds(T,r,g,prbs[thrd],mns[thrd],stds[thrd]);

      // Check that none of the probabilities are zero in the random sample and that the variance ratio constraint is not violated in the random sample.  Note that any
      // component having standard deviation = 0 will disqualify that sample.  Generate a new sample if a probability is zero or the variance ratio constraint is violated.
      //===========================================================================================================================================================================
      for (int c=0; c<g; ++c)
      {
          if (prbs[thrd][c] <= 0.00)
             stop=0;
          if (stds[thrd][c] < minstd)
             minstd=stds[thrd][c];
          if (stds[thrd][c] > maxstd)
             maxstd=stds[thrd][c];
      }
      if (maxstd > stdRatio*minstd)
         stop=0;
} while (stop == 0);

// Store random start and updated mixture distribution as 2-dimensional array for debugging and compliance with other functions.
//=================================================================================================================================
for (int m=0; m<3; ++m)
{
    orMDst[m]=new long double [g];
    uMDst[m]=new long double [g];
}
for (int c=0; c<g; ++c)
{
    orMDst[0][c]=prbs[thrd][c];
    orMDst[1][c]=mns[thrd][c];
    orMDst[2][c]=stds[thrd][c];
}

// Store (component likelihood)x(component probability) (in pdens[][]) for each observation/component using the current solution.  Store the mixture likelihood value (in mdens[]) for each observation
// using the current solution.  These are needed to implement the updating equations.  The log-likelihood for the initial random start parameters is also computed here and used below.
//=========================================================================================================================================================================================================
pdens = new long double *[T];
mdens = new long double [T];
oldllval = ((long double) T)*logrecipsqrt2pi;
for (int t=0; t<T; ++t)
{
    pdens[t] = new long double [g];
    mdens[t] = 0.00;
    for (int c=0; c<g; ++c)
    {
        pdens[t][c]=prbs[thrd][c]*getNDens(r[t],mns[thrd][c],stds[thrd][c]);
        mdens[t]=mdens[t] + pdens[t][c];
    }
    oldllval = oldllval+log(mdens[t]);
}

// Iterate using the EM algorithm.  Component probabilities are updated first and independently of the means/variances.
//========================================================================================================================
```

```
stop=0;
int itcntr=1;
do
{
    // Get updated component probabilities, means, and standard deviations.  Store in temporary placeholders.  The mean and standard deviation update formulas will not
    // work (as written) when any component probability is zero because there will be a division by zero.  If this happens end the optimization with an error.
    //===================================================================================================================================================================
    psum=0.00;
    for (int c=0; (c<g && 0==stop); ++c)
    {
        // Derive the posterior probabilities for this component along with the component probabilities.
        //================================================================================================
        uMDst[0][c]=0.00;
        if (c < g-1)                                // Derive component manually if not at last component.
        {
             for (int t=0; t<T; ++t)
             {
                postprbs[t]=(pdens[t][c])/mdens[t];
                uMDst[0][c] = uMDst[0][c] + postprbs[t];
             }
             uMDst[0][c] = uMDst[0][c]/T;
             psum = psum + uMDst[0][c];
        }
        else                                        // Final component probability is:  1.00 - (sum of all others).
        {
            for (int t=0; t<T; ++t)
                postprbs[t]=(pdens[t][c])/mdens[t];
            uMDst[0][c] = 1.00 - psum;
        }

        // Exit the optimization with appropriate code if any single probability is not between 0.0 and 1.0.
        //====================================================================================================
        if (uMDst[0][c] <= 0.00 || uMDst[0][c] >= 1.00)
        {
            stop=1; llVal[thrd]=LNegVal;
            rprms[2]=3; rprms[3]=itcntr;
        }
        else
        {
            // New updating equations (faster processing).
            //==============================================
            cprbs=uMDst[0][c]*T;
            uMDst[1][c]=getEMMean(T,r,postprbs,cprbs,ssqrs);
            var=(ssqrs[0] - uMDst[1][c]*uMDst[1][c]*cprbs)/cprbs;

            // Updating equations can result in zero (but stored value is negative) variance.  If it happens the variance ratio constraint is automatically violated.
            //=========================================================================================================================================================
            if (var > 0.00)
               uMDst[2][c]=sqrt(var);
            else
            {
                stop=1; llVal[thrd]=LNegVal;
                rprms[2]=0; rprms[3]=itcntr;
            }
        }
    }

    // Find maximum and minimum StDev values.  The EM algorithm will stop when the ratio of largest to smallest variance exceeds a constant (variable is stdRatio set in the header file).  This
    // prevents an unbounded likelihood value and once the constraint is violated we conclude that the solution is spurious.  (The likelihood will be set to a large negative # to ensure it is never
    // the maximum across random starts.)  We also stop when the maximum # of iterations exceeds some value (mIters set in the header file) or when any component probability is not between 0 and 1.
    //=================================================================================================================================================================================================
    if (stop == 0)
    {
        minstd=LPosVal; maxstd=LNegVal;
        for (int c=0; c<g; ++c)
```

```
        {
            if (uMDst[2][c] < minstd)
               minstd = uMDst[2][c];
            if (uMDst[2][c] > maxstd)
               maxstd = uMDst[2][c];
        }
        if (maxstd > stdRatio*minstd)
        {
            stop=1; llVal[thrd]=LNegVal;
            rprms[2]=0; rprms[3]=itcntr;
        }
        else if (itcntr > mIters)
        {
            stop=1; llVal[thrd]=LNegVal;
            rprms[2]=1; rprms[3]=itcntr;
        }
    }

    // If the variance ratio constraint is not violated then proceed as usual.
    //============================================================================
    if (stop == 0)
    {
        // Transfer values to permanant placeholders (variance ratio constraint has not been violated).
        //=================================================================================================
        for (int c=0; c<g; ++c)
        {
            prbs[thrd][c]=uMDst[0][c];
            mns[thrd][c]=uMDst[1][c];
            stds[thrd][c]=uMDst[2][c];
        }

        // Store the (component likelihood)x(component probability) (in pdens[][]) for each component/observation using the updated solution.  Store the mixture likelihood value (in mdens[]) for each
        // observation using the updated solution.  These are to implement the updating equations.  The log-likelihood for the initial random start parameters is also computed here and used below.
        //===============================================================================================================================================================================================
        newllval = ((long double) T)*logrecipsqrt2pi;
        for (int t=0; t<T; ++t)
        {
            mdens[t] = 0;
            for (int c=0; c<g; ++c)
            {
                pdens[t][c]=prbs[thrd][c]*getNDens(r[t],mns[thrd][c],stds[thrd][c]);
                mdens[t]=mdens[t] + pdens[t][c];
            }
            newllval = newllval+log(mdens[t]);
        }

        // Terminate the algorithm when %-change criteria is met.
        //==========================================================
        if ((newllval-oldllval) < epsilon*abs(oldllval))
        {
            stop=1; llVal[thrd]=newllval;
            rprms[2]=2; rprms[3]=itcntr;
        }
        else
        {
            oldllval=newllval;
            ++itcntr;
        }
    }
} while (stop == 0);

// Write error/issue files for debugging.
//===========================================
if (1 == rprms[2])
{
```

```
        ofstream fout;
        fout.open(errFolder + "Max_iterations_reached_" + to_string((long long) g) + "_" + to_string((long long) thrd) + ".txt");
        fout << "--> Maximum # of iterations (" << mIters << ") has been reached for this optimization. (Check the solution below.)" << endl << endl;
        fout.setf(ios_base::fixed, ios_base::floatfield); fout.precision(25);
        fout << "--> New log-likelihood value: " << newllval << endl;
        fout << "--> Epsilon = " << epsilon << endl << endl;
        fout << endl << "--> Original observation vector and parameter starting values for this EMAlg() call: " << endl << endl;
        for (int x=0; x<T; ++x) fout << r[x] << endl; fout << endl;
        for (int cc=0; cc<g; ++cc)
        {
          fout << "Prob[" << cc << "]=" << orMDst[0][cc] << "  Mean[" << cc << "]=" << orMDst[1][cc] << "  Std.Dev.[" << cc << "]=" << orMDst[2][cc] << endl;
        }
        fout.close();
    }
    else if (3 == rprms[2])
    {
        ofstream fout;
        fout.open(errFolder + "Prob_of_0_or_1_in_" + to_string((long long) g) + "_" + to_string((long long) thrd) + ".txt");
        fout << "--> Probability of 0.00 or 1.00 has been encountered and updating equations will not work. (Check the solution below.)" << endl << endl;
        fout.setf(ios_base::fixed, ios_base::floatfield); fout.precision(25);
        fout << "--> Original observation vector and parameter starting values for this EMAlg() call: " << endl << endl;
        for (int x=0; x<T; ++x) fout << r[x] << endl; fout << endl;
        for (int cc=0; cc<g; ++cc)
        {
            fout << "Prob[" << cc << "]=" << orMDst[0][cc] << "  Mean[" << cc << "]=" << orMDst[1][cc] << "  Std.Dev.[" << cc << "]=" << orMDst[2][cc] << endl;
        }
        fout.close();
    }

    // Delete temporary memory allocations.
    //=====================================
    for (int t=0; t<T; ++t)
    {
        delete [] pdens[t]; pdens[t]=nullptr;
    }
    delete [] pdens; pdens=nullptr;
    delete [] mdens; mdens=nullptr;
    for (int m=0; m<3; ++m)
    {
        delete [] orMDst[m]; orMDst[m]=nullptr;
        delete [] uMDst[m]; uMDst[m]=nullptr;
    }
    delete [] orMDst; orMDst=nullptr;
    delete [] uMDst; uMDst=nullptr;
    delete [] postprbs; postprbs=nullptr;
}

/*
/ Copyright (C) 2016 Chris Rook
/
/ This program is free software: you can redistribute it and/or modify it under the terms of the GNU General Public License as published by the Free Software Foundation, either version 3 of the License,
/ or (at your option) any later version.  This program is distributed in the hope that it will be useful, but WITHOUT ANY WARRANTY; without even the implied warranty of MERCHANTABILITY or FITNESS FOR A
/ PARTICULAR PURPOSE.  See the GNU General Public License for more details: <http://www.gnu.org/licenses/>
/
/ Filename:  getRVals.cpp
/
/ Function:  getRVals()
/
/ Summary:
/
/    This function accepts a mixture distribution as input and generates a random sample of observations from that distribution.  The # of observations to generate is specified by the user in the call
/    and the sample is placed into an empty array supplied to this function by the user.  Generating a random sample from a mixture distribution is a 2-step process.  (1) Generate a uniform random value
/    and compare it to the component probabilities to determine the component, and once the component is selected (2) generate an observation from the corresponding component density.  Let p(i) be the
/    probability for component i and let there be a total of C components, then i=0,1,2,...,C-1.  If the uniform random value U <= p(0) then component 0 is selected, else if U <= p(0)+p(1) then component
/    1 is selected, else if U <= p(0)+p(1)+p(2) then component 2 is selected, etc...  Once a component is selected an observation is generated from the corresponding density function.  The result is an
```

```
/    observation generated from the supplied mixture distribution.  Here, there are 2 reasons for generating random observations from a univariate mixture distribution as described below:
/
/    1.) In this application, univariate mixture distributions are fit using the EM algorithm with random starts.  A random start must specify a value for all parameters in the given univariate mixture
/        density which is of known size, but with unknown parameters.  The EM algorithm continues to increase the likelihood function until a local optimum is found based on the given parameter settings
/        from the random start.  To optimize a g-component mixture density when a g-1 component mixture density is available from the same data set, we generate a single random start by producing g
/        random observations from the g-1 component density.  These g values are taken as means for the g-component mixture.  The observations are then assigned to the component with nearest mean by
/        using a standard distance computation.  Once all observations are assigned to the closest mean, the standard deviation for each component (i.e., mean) is computed by taking the standard
/        deviation of the corresponding set of observations assigned to that mean.  The probability assigned to each component is the # of observations assigned to the component divided by the total # of
/        observations in the data set.  At this point, values for all parameters (i.e., g means, g standard deviations, g component probabilities) have been derived and the EM algorithm can be applied
/        starting at the given point in the parameter space.  The EM algorithm then climbs to the top of the nearest hill and declares a local optimum.  The maximum likelihood estimators would be the set
/        of parameters that yields the largest value for the likelihood function amongst the set of local optimums found via a large # of random starts.
/
/    2.) The likelihood ratio test (LRT) is used to select the optimal # of components for a univariate mixture density fit to a given observation set.  The null hypothesis is that g0 components are
/        optimal vs the alternative of ga components being optimal (here, ga > g0).  By fitting univariate mixtures of both sizes (g0 and ga) to the data we can generate a single value for the LRT
/        statistic.  The value is compared to the null distribution of the LRT statistic, which is the distribution of the LRT under the assumption that a g0 component mixture is the appropriate size.
/        The distribution of the LRT under H0 is not known and is not asymptotically chi-square due to the relevant regularity conditions not being satisfied.  We can estimate the null distribution of
/        the LRT by bootstrapping (see, McLachlan (1987)).  To bootstrap the LRT distribution we generate a random sample from the distribution specified by H0 of the same size as our data set.  (Note
/        that H0 does not specify a particular univariate mixture, but rather just a # of components, namely g0.  We take the ML Estimates of our data under g0 components as the distribution governed by
/        the null hypothesis, and this distribution is used to generate the sample of observations.)  This sample is then fit to a univariate mixture under both g0 and ga components by applying the EM
/        algorithm using random starts, and a single value of the LRT statistic is produced.  By repeating the process a large # of times we can approximate the null distribution of the LRT statistic.
/        We then compare the value derived from our data set and reject the null hypothesis when the LRT is large, where the critical point is determined by the user's choice of alpha = type 1 error for
/        the test.  (Type 1 error probability = probability that the null hypothesis is rejected when it is true.)  This function is used to generate the random samples used for the LRT just described.
/
/ Inputs:
/
/    1.) The # of random observations to generate from the supplied univariate mixture density.  The observations are inserted into an empty array of the same size that is supplied by the user as the
/        last parameter to this function. (N)
/
/    2.) The # of components in the univariate mixture density that a sample will be generated from. (g)
/
/    3.) A double array holding the univariate mixture distribution definition from which a sample of size N will be generated.  The array is indexed as [s][c], where s = 0, 1, 2, and c = component #.
/        Here, s=0 refers to the component's probability, s=1 refers to the component's mean, and s=2 refers to the component's standard deviation. (inMDst)
/
/    4.) An empty array of size N to be populated by this function as a random sample from the supplied univariate mixture distribution. (rvls)
/
/ Outputs:
/
/    This function populates an empty array with a random sample of user-specified size from the supplied univariate mixture distribution.  No value is returned at the function call.
/*************************************************************************************************************************************************************************************************************/
#include "stdafx.h"
void getRVals(const int N, const int g, const long double **inMDist, long double *rvls)
{
    // Generate N random observations from the current optimal solution that uses g components.  These can be the means to use as a random start for fitting a specific model or observations used to
    // approximate the null distribution of the LRT statistic.
    //=================================================================================================================================================================================================
    random_device rd;
    default_random_engine gen(rd());
    normal_distribution<long double> *ndist = new normal_distribution<long double> [g];
    uniform_real_distribution<long double> udist(0.0,1.0);

    // Define the array of normal distribution objects, one for each component.
    //=========================================================================
    for (int c=0; c<g; ++c)
    ndist[c] = normal_distribution<long double> (inMDist[1][c],inMDist[2][c]);

    // Generate N observations from the existing g-component mixture distribution.
    //============================================================================
    int cid;
    long double uval, psum;
    for (int i=0; i<N; ++i)
    {
        // Initialize variables.
        //========================
        cid=-1;
        psum=inMDist[0][0];
```

```
        // Generate uniform random value.
        //================================
        uval = udist(gen);

        // Find the corresponding component.
        //===================================
        for (int c=0; (c<g && cid<0); ++c)
        {
            if (uval <= psum || c == g-1)
               cid=c;
            else
               psum = psum + inMDist[0][c+1];
        }

        // Generate a single obs from that component and store in array provided.
        //======================================================================
        rvls[i] = ndist[cid](gen);
    }

    // Free up temporary memory allocations.
    //=======================================
    delete [] ndist; ndist=nullptr;
}

/*
/ Copyright (C) 2016 Chris Rook
/
/ This program is free software: you can redistribute it and/or modify it under the terms of the GNU General Public License as published by the Free Software Foundation, either version 3 of the License,
/ or (at your option) any later version.  This program is distributed in the hope that it will be useful, but WITHOUT ANY WARRANTY; without even the implied warranty of MERCHANTABILITY or FITNESS FOR A
/ PARTICULAR PURPOSE.  See the GNU General Public License for more details: <http://www.gnu.org/licenses/>
/
/ Filename:  getRPrbsStds.cpp
/
/ Function:  getRPrbsStds()
/
/ Summary:
/
/    This function accepts a set of T observations along with a set of g means for a g-component mixture distribution and it derives corresponding values for the component probabilities and standard
/    deviations.  Each observation is assigned to a single component (i.e., mean) and the # of observations assigned to a component divided by the total # of observations is the corresponding estimate
/    for that component's probability.  Observations are assigned to components using a simple distance function, and specifically each observation is assigned to the component with nearest mean.  The
/    standard deviation for each component is then estimated as the sample standard deviation of the set assigned to that component assuming the mean is known.  Empty arrays of size g are supplied to
/    this function to hold the set of component probabilities and set of standard deviations, respectively.  Combined with the existing set of g means (derived as a random sample via the function
/    getRVals()) these 3 arrays (i.e., mean, standard deviation, and component probability) completely define a g-component mixture distribution.  If the means were generated as a random sample, then it
/    defines a single random start for fitting a g-component mixture distribution using the EM algorithm.
/
/ Inputs:
/
/    1.) The total # of time points with data collected.  (T)
/
/    2.) An array holding the set of observations (i.e., returns) for the asset being processed. (r)
/
/    3.) The # of means (i.e., components) in the array provided (by parameter #5) used to generate a random start for a mixture density. (g)
/
/    4.) An empty array of size g to be populated by this function as component probabilities for the univariate mixture density being constructed. (prbs)
/
/    5.) An array of means (i.e., components) to use as the basis for generating a g-component mixture distribution.  (mns)
/
/    6.) An empty array of size g to be populated by this function as standard deviations for the univariate mixture density being constructed. (stds)
/
/ Outputs:
/
/    This function populates two empty arrays of size g supplied (prbs & stds) with component probability estimates and standard deviation estimates, for each of the g components defined by the means
/    supplied in mns.  No value is returned at the function call.
/*************************************************************************************************************************************************************************************************************/
#include "stdafx.h"
```

```
void getRPrbsStds(const int T, const long double *r, const int g, long double *prbs, const long double *mns, long double *stds)
{
    // Local variables.
    //===================
    long double *cssqrs=new long double [g], mindist;
    int cid, *ccntr=new int [g];

    // Initialize component counter and sum-of-squares arrays to all zeros.
    //======================================================================
    for (int c=0; c<g; ++c)
    {
        ccntr[c]=0;
        cssqrs[c]=0.0;
    }

    // Iterate over all observations and classify each into the component whose mean it is closest to.
    //================================================================================================
    for (int n=0; n<T; ++n)
    {
        mindist=abs(r[n]-mns[0]);
        cid=0;
        for (int c=1; c<g; ++c)
            if (abs(r[n]-mns[c]) < mindist)
            {
                cid=c;
                mindist=abs(r[n]-mns[c]);
            }
        ++ccntr[cid];
        cssqrs[cid]=cssqrs[cid]+pow(mindist,2);
    }

    // Compute the estimated probabilities and standard deviations (MLEs).
    //=====================================================================
    for (int c=0; c<g; ++c)
    {
        prbs[c] = ((long double) ccntr[c]/(long double) T);
        stds[c] = sqrt((long double) (cssqrs[c]/ccntr[c]));
    }

    // Delete temporary memory allocations.
    //=====================================
    delete [] ccntr; ccntr=nullptr;
    delete [] cssqrs; cssqrs=nullptr;
}

/*
/ Copyright (C) 2016 Chris Rook
/
/ This program is free software: you can redistribute it and/or modify it under the terms of the GNU General Public License as published by the Free Software Foundation, either version 3 of the License,
/ or (at your option) any later version.  This program is distributed in the hope that it will be useful, but WITHOUT ANY WARRANTY; without even the implied warranty of MERCHANTABILITY or FITNESS FOR A
/ PARTICULAR PURPOSE.  See the GNU General Public License for more details: <http://www.gnu.org/licenses/>
/
/ Filename:  asgnObs.cpp
/
/ Function:  asgnObs()
/
/ Summary:
/
/    This function assigns each observation (which is a time point in this application) to one of the univariate mixture components for a given asset.  Each asset has been fit to a univariate mixture
/    distribution containing a certain # of components.  Each component can be viewed as a generator of observations for that asset, with the corresponding component probability that the observation
/    originates from any single component's density.  Bayes' decision rule is used to assign each observation to a corresponding component by computing the posterior probability that each observation
/    originates from each component.  The observation is then assigned to the component with the highest posterior probability.  This function performs that task.  The user supplies an empty array for
/    the given asset of size equal to the # of time points.  Inserted into the array at each position is the component that the observation most likely originates from, and therefore is assigned to.
/
/ Inputs:
```

```
/
/    1.) An empty array of size T that will hold the component that the observation is assigned to. (Determined by this function.) (inAry)
/
/    2.) The total number of time points with data collected. (T)
/
/    3.) An array of size T holding the returns for the asset being processed. (r)
/
/    4.) The number of univariate components for the asset being processed. (g)
/
/    5.) The (optimal) univariate mixture distribution fit for the given asset as a 2-dimensional array indexed as [c][s], where c = component #, and s = 0, 1, 2.  Here, s=0 refers to the component's
/        probability, s=1 refers to the component's mean, and s=2 refers to the component's standard deviation. (inMDst)
/
/ Outputs:
/
/    This function populates an empty array of size T that is supplied with the component that the given observation most likely originates from.  Bayes decision rule is used and the observation is
/    assigned to the component with highest posterior probability.  This function returns no value at the call.
/*******************************************************************************************************************************************************************************************************/
#include "stdafx.h"
void asgnObs(int *inAry, const int T, const long double *r, const int g, const long double **inMDst)
{
    // Declare local variables.
    //==========================
    long double maxProb, tmpProb;

    // Assign each observation per time point to the component with highest posterior probability (Bayes' rule).
    //==========================================================================================================
    for (int t=0; t<T; ++t)
    {
       maxProb=-1.00;
       for (int c=0; c<g; ++c)
       {
          tmpProb=getPost(r[t], g, (const long double **) inMDst, c);
          if (tmpProb > maxProb)
          {
             inAry[t]=c;
             maxProb=tmpProb;
          }
       }
    }
}

/*
/ Copyright (C) 2016 Chris Rook
/
/ This program is free software: you can redistribute it and/or modify it under the terms of the GNU General Public License as published by the Free Software Foundation, either version 3 of the License,
/ or (at your option) any later version.  This program is distributed in the hope that it will be useful, but WITHOUT ANY WARRANTY; without even the implied warranty of MERCHANTABILITY or FITNESS FOR A
/ PARTICULAR PURPOSE.  See the GNU General Public License for more details: <http://www.gnu.org/licenses/>
/
/ Filename:  mapCells.cpp
/
/ Function:  mapCells()
/
/ Summary:
/
/    A multidimensional grid is formed using the components from the univariate marginal mixture densities for all assets.  If there are "a" assets with asset i having c[i] univariate mixture components,
/    for i=0,1,2,...,a-1, then the multidimensional grid used as the basis for the multivariate density will have a total of c[0]xc[1]xc[2]x...xc[a-1] cells.  This function converts the multidimensional
/    grid to a single array/list holding all cells.  Each value in this array/list represents one cell in the grid and therefore contains a set of components, one per asset.  For example, the first
/    element of this list contains all assets at their first component, the 2nd element of this list contains all assets at their first component, but the final asset at it's 2nd component.  The 3rd
/    element of this list contains all assets at their first component but the last asset at its 3rd component, etc...   The list is ordered as a design matrix for a full factorial experiment, from left
/    to right, with values to the right repeatedly cycling through all of their components for each set of values to the left.  The function populates an empty list that is supplied.  To keep track of
/    which component levels are used in any given cell, a 2nd term is added to the list making it a 2-dimensional array indexed as [c][a], where c = unique cell # and a = asset #.  That is, element
/    [0][0] would indicate the component of the first asset within the first cell of the multidimensional grid, and element [0][1] would indicate the component level for the 2nd asset within the first
/    cell of the multidimensional grid.  If there are a total of 4 assets, then elements at positions [5][0], [5][1], [5][2], [5][3] would indicate the set of components for each of the 4 assets within
/    the 6th cell of the multidimensional grid.  Essentially, this function converts the multidimensional grid to a list which is easier to manage and each cell in the list contains a 2nd term to
/    identify the contents of the list item (which is a single cell).  Example:  Suppose there are 3 assets with c[0]=2, c[1]=3, c[2]=2 components in the respective univariate mixture densities.  The
```

```
/    multidimensional grid is formed by crossing all univariate components across assets and will contain 2x3x2=12 total cells.  This grid forms the basis for building the multivariate mixture density.
/    A list of length 12 will be used to represent each cell and the components that are contained within each cell as follows:
/
/                        =======     ========    ========    ========
/                        Cell ID     Asset #0    Asset #1    Asset #2
/                        =======     ========    ========    ========
/                           0           0           0           0
/                           1           0           0           1
/                           2           0           1           0
/                           3           0           1           1
/                           4           0           2           0
/                           5           0           2           1
/                           6           1           0           0
/                           7           1           0           1
/                           8           1           1           0
/                           9           1           1           1
/                          10           1           2           0
/                          11           1           2           1
/                        =======     ================================
/           Array Index:   [ ]                    [ ]
/
/    For example, last row is defined as:  inCellAry[11][0]=1, inCellAry[11][1]=2, inCellAry[11][2]=1.
/
/    Note that the list can be derived by starting with the last asset and repeatedly cycling through all component levels, then proceeding to the 2nd last asset and cycling through all component levels
/    for each set just defined, then proceeding to the 3rd last asset and cycling through all component levels for the sets defined to the right.  This is the strategy used to convert the
/    multidimensional grid to a single list.  This function is recursive with a single call for each asset.  That call cycles through all levels of the given asset, invoking a call for the asset to the
/    right at each level.  When at the final asset, no additional recursive calls are made and all levels of that asset are posted.  In this manner, the function begins at asset #0, burrows inward to
/    asset #(a-1), then expands outward back to the 1st asset.  When the debug level is set to a value >= 2 all details of the mapping are printed for review, similar to the table shown above.
/
/ Inputs:
/
/    1.) The total # of cells in the multidimensional grid, which is:  c[0]xc[1]xc[2]x...xc[a-1], where there are "a" assets with asset i having c[i] components in the corresponding univariate mixture
/        density. (totCells)
/
/    2.) The number of assets in the current application. (numA)
/
/    3.) A 2-dimensional array that is populated by this function indexed as [c][a], where c = unique cell #, and a = asset #.  Here, both "c" and "a" begin at 0.  The value contained is the component
/        level of asset a within unique cell c. (inCellAry)
/
/    4.) The current asset that is being processed.  This function processes each asset separately and iterates over all of its component levels.  At each component level the function recursively invokes
/        itself for the next asset, which similarly processes each component in order. (curAst)
/
/    5.) An array to hold the # of components for each asset in their respective univariate mixture density.  This array will hold the values for c[i] as described above. (inCmps)
/
/    6.) An integer value to hold the current unique cell #.  Value begins at 0 and ends at totCells-1.  After each cell is defined and written to the inCellAry[][] array this value is incremented. (cID)
/
/    7.) An array to hold the component level being processed for each asset.  This function starts at the first asset (i.e., a=0) and iterates over all c[0] levels of the corresponding univariate
/        mixture.  At each level, the function is invoked recursively to process the next asset (i.e., a=1).  The function then iterates over all c[1] levels of the 2nd asset and recursively invokes
/        itself to process the next asset (i.e., a=2).  When at the final asset a cell is completely defined and the result is appended to the list.  This array is persistant and of size numA to hold the
/        current component level being processed for each asset.  When at the final asset (i.e., final recursive call) the numA components held in this array defines the given unique cell. (tmpAry)
/
/ Outputs:
/
/    This function converts the multidimensional grid formed by crossing all assets and their univariate component levels to a single list where each element in the list defines one cell of the
/    multidimensional grid.  It is more straightforward to navigate the grid in this manner.  The list has 2 elements, first is the unique cell # and 2nd is the array of component levels that define the
/    given cell.  This function is recursive (invokes itself) and returns no value at the call.
/*********************************************************************************************************************************************************************************************************/
#include "stdafx.h"
void mapCells(const int totCells, const int numA, int **inCellAry, const int curAst, const int *inCmps, int *cID, int *tmpAry)
{
    // Output cell mappings when debugging mode is on.
    //=================================================
    if (0 == curAst && dbug >= 2)
       cout << string(14,'=') << endl << "Cell Mappings: " << endl << string(14,'=') << endl;
```

```
    // Iterate over all components of the current asset.  At each component recursively invoke this function to process the next asset.
    //====================================================================================================================================
    for (int i=0; i<inCmps[curAst]; ++i)
    {
        // Store the component level for the current asset.
        //===================================================
        tmpAry[curAst]=i;

        // Recursive call, if not processing the final asset.  If processing final asset, populate the array and increment the cell counter.
        //=====================================================================================================================================
        if (curAst < numA-1)
           mapCells(totCells, numA, inCellAry, curAst+1, inCmps, cID, tmpAry);
        else
        {
            if (dbug >= 2)
               cout << "Cell #" << setfill('0') << setw(to_string((long long) totCells).size()) << *cID;
            for (int j=0; j<numA; ++j)
            {
                if (dbug >= 2)
                   cout << ", Asset #" << j+1 << " = " << tmpAry[j];
                inCellAry[*cID][j] = tmpAry[j];
            }
            if (dbug >= 2)
                cout << endl;
            *cID = *cID + 1;
        }
    }
}

/*
/ Copyright (C) 2016 Chris Rook
/
/ This program is free software: you can redistribute it and/or modify it under the terms of the GNU General Public License as published by the Free Software Foundation, either version 3 of the License,
/ or (at your option) any later version.  This program is distributed in the hope that it will be useful, but WITHOUT ANY WARRANTY; without even the implied warranty of MERCHANTABILITY or FITNESS FOR A
/ PARTICULAR PURPOSE.  See the GNU General Public License for more details: <http://www.gnu.org/licenses/>
/
/ Filename:  getCell.cpp
/
/ Function:  getCell()
/
/ Summary:
/
/    This function accepts a set of component levels (one per asset) and returns the unique cell ID # from the multidimensional grid that contains this set.  The multi-dimensional grid contains a cell
/    for each combination of univariate mixture components across all assets.  If no match is found a value of -1 is returned.  This function iterates over the 2-dimensional array that is supplied until
/    a match is found, then returns the cell ID # and exits.  The 2-dimensional array supplied is indexed as [c][a], where c = unique cell #, and a = asset #.
/
/ Inputs:
/
/    1.) 2-dimensional array as [c][a], where c=unique cell #, a=asset #.  Here, both "c" & "a" begin at 0.  The value contained is the component level of asset "a" within unique cell "c". (inCellAry)
/
/    2.) The total # cells in the multidimensional grid: c[0]xc[1]xc[2]x...xc[a-1], there are "a" assets with asset i having c[i] components in the corresponding univariate mixture PDF. (totCells)
/
/    3.) An array of size numA containing the set of asset component levels we are attempting to match.  The unique cell # for the match is returned at the call. (cmpLvls)
/
/    4.) The number of assets in the current application. (numA)
/
/ Outputs:
/
/    This function searches for a match on a set of numA asset component levels and returns the unique cell ID #.  The cell ID # ranges from 0 - (totCells-1).  Return -1 if no match is found.
/***********************************************************************************************************************************************************************************************************/
#include "stdafx.h"
int getCell(const int **inCellAry, const int totCells, const int *cmpLvls, const int numA)
{
    // Iterate over all mapped values and find the match.
    //=====================================================
```

```
    int cntr;
    for (int i=0; i<totCells; ++i)
    {
        cntr=0;
        for (int j=0; j<numA; ++j)
            if (inCellAry[i][j] == cmpLvls[j])
               ++cntr;

        // Check for match, then return the cell position of the match.
        //=================================================================
        if (cntr == numA)
        {
            if (dbug >= 2)
               cout << "Cell = " << i << endl;
            return i;  // Match found.
        }
    }
    return -1;  // No match found.
}

/*
/ Copyright (C) 2016 Chris Rook
/
/ This program is free software: you can redistribute it and/or modify it under the terms of the GNU General Public License as published by the Free Software Foundation, either version 3 of the License,
/ or (at your option) any later version.  This program is distributed in the hope that it will be useful, but WITHOUT ANY WARRANTY; without even the implied warranty of MERCHANTABILITY or FITNESS FOR A
/ PARTICULAR PURPOSE.  See the GNU General Public License for more details: <http://www.gnu.org/licenses/>
/
/ Filename:  solveLP.cpp
/
/ Function:  solveLP()
/
/ Summary:
/
/    This function solves 2 linear programs (LPs) which become (feasible) initial solutions for maximizing the multivariate mixture distribution likelihood.  The univariate mixture densities for each
/    asset have already been fit using the EM algorithm and then combined into a multi-dimensional grid where the grid levels in each dimension are the corresponding asset/mixture component
/    combinations.  This multidimensional grid forms the basis for the multivariate density function, which is also a mixture PDF.  Each cell in the multivariate grid defines a unique combination of
/    assets and their components.  Using Bayes' decision rule we can assign the observation for a given asset at each time point to a corresponding univariate component based on the component with
/    highest probability of membership.  Using these individual component memberships we combine them and assign each multivariate observation to a single cell in the multidimensional grid.  The
/    probability of an observation originating from that grid cell is then the # of observations in the given cell divided by the total # of observations (or time points).  We now have an estimated
/    probability that a new observations originates from each multi-dimensional grid cell.  (Refer to this estimate as ek, applicable to cell k.)  Note that each grid cell defines a multivariate density
/    function using the corresponding means and variances for the assets/components that define the cell, but at this point all covariances are undefined (i.e., zero).  An important aspect of this
/    research is that we must maintain the univariate marginals that have already been fit.  To accomplish this, the sum of probabilities for each cell containing a given asset/component must equal the
/    corresponding probability for that component in the univariate density.  This implies that we can use linear constraints on the grid cell probabilities to maintain the marginals.  Since the sum of
/    all grid cell probabilites must equal 1.00, there will be (# univariate components) - 1 constraints needed to maintain the univariate marginal for each asset.  (Once these constraints are enforced
/    the final (sum of probabilities) constraint for each component is automatically enforced by the fact that the sum of all probabilities equals 1.00.)  This means that there will be a total of (Total
/    # of components across all assets - Total # of assets + 1) constraints required to maintain the marginals.  A set of linear equality constraints can be formulated in matrix notation as Ax = b. Here,
/    A is a matrix with one row holding the coefficients (0 or 1) for a single constraint and the matrix A has a column for each decision variable (i.e., single grid cell probability which is a
/    probability for the multivariate mixture density).  We will refer to A as the LHS constraint matrix and b as the RHS constraint vector.  The vector b will hold the component probabilities from the
/    univariate marginals, with only the first (# components) - 1 needed since the probability for the final component within each asset is automatically enforced by the last row of A (i.e., final
/    constraint) that the sum of all probabilities equals 1.00.  Note that the LHS constraint matrix should be of full row rank, meaning that we include the minimum # of constraints needed to enforce the
/    marginals.  The rows of A must be linearly independent.  (This requirement is needed for a future optimization that uses this matrix.)  The cell probabilities that have been estimated using the data
/    (i.e., the ek via Bayes' decision rule, see above) will generally not satisfy the marginal constraints in Ax = b, therefore these estimates will not (in general) maintain the marginals.  The purpose
/    of this function is to find cell probabilities that do maintain the marginals, and that are in some way as close as possible to the estimated probabilities.  We offer 2 methods here.  First, to
/    formulate the problem we assign an unknown decision variable to each unique cell in the multidimensional grid and that represents the true probability of membership in that cell.  These decision
/    variables are then estimated using the following 2 LP objectives:  1.) Minimize the maximum distance between all estimated unique cell probabilties and the decision variable that represents each
/    cell, and 2.) Minimize the sum of squared distances between the decision variables and the estimated unique cell probabilities.  LP #1 is a classic minimax objective of the form:
/    Min{Max[|p1-e1|,|p2-e2|,|p3-e3|, ..., |pu-eu|]}, where there are u unique cells in the multidimensional grid and pk is the decision variable (true cell probability) for cell k, and ek is the
/    corresponding estimated cell probability.  The distance between the two is: |pk-ek|.  In LP #1 our objective is to select the pk, k=1,2,...,u that minimizes the maximum of these distance values such
/    that the constraints Ax = b are satisfied (i.e., the marginal densities are maintained).  As written, the objective in LP #1 contains absolute values and therefore is not linear, however, it can be
/    rewritten as an equivalent linear program.  For example, note that the objective can be rewritten as:  Min{Max[p1-e1,e1-p1, p2-e2,e2-p2, p3-e3,e3-p3, ..., pu-eu,eu-pu]} since |pk-ek| = pk-ek or
/    ek-pk, for k=1,2,...,u.  The absolute values have now been removed.  Next, let Z = Max[p1-e1,e1-p1, p2-e2,e2-p2, p3-e3,e3-p3, ..., pu-eu,eu-pu], then the objective becomes Min{Z}, and note that the
/    following inequalities must hold:  Z >= p1-e1, Z >= e1-p1, Z >= p2-e2, Z >= e2-p2, ... Z >= pu-eu, Z >= eu-pu.  This is because Z is the maximum of a set of values therefore it must be >= all
/    members of the set.  Further, since the objective is to minimize Z, it must take one of the values that bounds it below at an optimal solution.  Using the new objective and added constraints, the
/    problem is now an equivalent linear program.  Lastly, it will be important to keep the multivariate density as parsimonious as possible, meaning carrying fewer unique cells into the multivariate
/    mixture density is desirable.  Fewer components translates to fewer parameters and also we will run into problems when optimizing the multivariate mixture likelihood when a component is included
```

```
/    that generates a zero likelihood value for the set of observations at all time points.  A future gradient/Hessian computation will see the component probabilities prefixed by the corresponding
/    likelihood value in the objective.  If the likelihood of this multivariate component density is zero for all time points then the decision variable is effectively removed from the problem and the
/    corresponding Hessian will not be of full rank.  This will be a problem when applying Newton's method, for example.  To prevent components with zero likelihood from being carried merely to satisfy
/    the constraints, we will add a 2nd decision variable to each unique cell in the multidimensional grid that penalizes the objective function when a cell with zero observations is included in the LP
/    solution.  This will help to guarantee that we only keep cells that contain actual data points and it prevents likelihoods from being zero at all time points.  LP #2 is similar to LP #1 but the
/    objective is to minimize the sum of squared distances between the actual and estimated probabilities, subject to the marginal constraints.  That is, in LP #2 the objective takes the form:
/    Min {(p1-e1)^2 + (p2-e2)^2 + (p3-e3)^2 + ... + (pu-eu)^2}.  Note that each term in this sum (pk-ek)^2, k=1,2,...,u is quadratic and concave (U-shaped) in pk and centered at ek.  Furthermore, no
/    decision variable exists in more than 1 term of the sum.  This objective is known as a quadratic program that is separable, which by definition means it is not linear.  It can, however, be
/    approximated arbitrarily close by a linear program.  Since 0 <= pk <= 1, for k=1,2,...,u, we will define a set of decision variables to approximate each concave quadratic term in the sum over this
/    pk range.  The approximation will consist of line segments that trace out the U-shaped term.  For the range 0 <= pk <= 1, we first set the number of line segments to use when tracing out the curve,
/    and this is done via the global constant dLvl defined in the header file (current value=500).  Next we compute the pk values on the horizontal axis (i.e., the probability axis) that are equidistant
/    and cover the region between 0 and 1.  Note that there will be dLvl+1 such points.  At the current setting of 500 line segments, these points will be:  0, 1/500, 2/500, 3/500, ..., 1.  They are
/    fixed once dLvl is known and they also do not change per term in the sum.  That is, these points on the probability axis are used to trace out each concave term in the objective.  Once these points
/    are known we can compute the function evaluated at each point.  For example, considering the first term in the sum, the function evaluated at each point is:  (0-e1)^2, (1/500-e1)^2, (2/500-e1)^2,
/    ..., (1-e1)^2.  These values are fixed and constant once dLvl is determined and the estimates ek are known, but they do change per term in the sum.  If there are a total of U components then there
/    will be U*(dLvl+1) such quantities defined.  (Connecting the dots of these function values will trace out the curve.)  In LP #1 the decision variables were the true probabilities, for each unique
/    cell but in LP #2 the decision variables are dLvl+1 alpha values for each term in the sum that are >= 0 and sum to 1.  Each such alpha variable is attached to a segment boundary on the horizontal
/    axis (i.e., 0, 1/500, 2/500, ..., 1).  A probability value pk is then defined by using the 2 alpha variables that bound the pk value.  For example, we can create the value (1.3)/500 by using
/    (alpha1)*(1/500) + (alpha2)*(2/500), where alpha1=0.7 and alpha2=0.3, and all other alpha values = 0.0.   That is, in general, we define pk=(alpha_k_0)*(0) + (alpha_k_1)*(1/500) +
/    (alpha_k_2)*(2/500) + (alpha_k_3)*(3/500) + ... + (alpha_k_500)*(1).  Note that the alpha variables are specific to a component of the sum.  Lastly, the quadratic objective component (pk-ek)^2 is
/    estimated arbitrarily close by (alpha_k_0)*(0-ek)^2 + (alpha_k_1)*(1/500-ek)^2 + (alpha_k_2)*(2/500-ek)^2 + (alpha_k_3)*(3/500-ek)^2 + ... + (alpha_k_500)*(1-ek)^2, where ek is the known cell
/    estimate.  These constant multipliers were defined above and are stored in variables.  Now, both the objective function and constraints are linear in the U*(dLvl+1) alpha decision variables.  Within
/    a component, all alpha variables must sum to 1.00, that is:  alpha_k_0 + alpha_k_1 + alpha_k_2 + ... + alpha_k_500 = 1.00.  Once the pk are defined, we use these variables to build the constraints
/    that maintain the marginals, that is Ax = b.  We now have a linear objective, with linear constraints that approximates the quadratic separable program.  Linear programs can be solved fast for a
/    global solution using the simplex algorithm.  Here, we use the lp_solve free library of functions to solve both LPs.
/
/ Variable Summary:
/
/    Minimax Objective:
/    ==================
/
/      Total # of decision variables defined in the array vbl[]:  2*totCells + 1
/
/      vbl[0] - vbl[totCells-1] will hold the (totCells - 1 - 0 + 1) = totCells "probability" decision variables (pk, for k=1,2,...,u) as detailed above. These are referenced in the LP as column #'s:  1,
/      2, 3, ..., totCells.
/
/      vbl[totCells] - vbl[2*totCells-1] will hold the (2*totCells - 1 - totCells + 1) = totCells "feasibility factor" decision variables (fk, for k=1,2,...,u).  These are referenced in the LP as column
/      #'s:  totCells+1, totCells+2, totCells+3, ..., 2*totCells.  One constraint is that the true cell probability (i.e., value of the decision variable) must be <= the total # of observations that fall
/      in a cell plus the cell's corresponding feasiblity factor.  Thus, when a cell with zero actual observations assigned to it needs a non-zero probability for an optimal solution this feasibility
/      factor must be set to a positive value to satisfy the constraint.  In the objective we then add each of these feasibility factors multiplied by the large constant K (set in the header file). This
/      applies a penalty to the objective when a cell with no observations is included in the optimal solution, and it makes the occurence rare.  As noted above we want to avoid including grid cells that
/      have a zero likelihood across all time points since it will lead to the Hessian of a future optimization being non-singular.
/
/      vbl[2*totCells] holds the objective function decision variable.  This is referenced in the LP as column #:  2*totCells+1.
/
/    Minimum Sum of Squared Distances Objective:
/    ===========================================
/
/      Total # of decision variables defined in the array vbl[]:  totCells*(dLvl+2)
/
/      vbl[0] - vbl[dLvl] will hold the alpha values (which sum to 1.00) within the 1st unique cell (index=0) of the multidimensional grid, say, alpha_0_0, alpha_0_1, alpha_0_2, ..., alpha_0_500.  The
/      corresponding optimal probability value p0 is then derived using the 2 non-zero values combined with the corresponding 2 segment endpoints.
/
/      vbl[dlvl+1] - vbl[2dLvl+1] will hold the alpha values (which sum to 1.00) within the 2nd unique cell (index=1) of the multidimensional grid, say, alpha_1_0, alpha_1_1, alpha_1_2, ...,
/      alpha_1_500.  The corresponding optimal probability value p1 is then derived using the 2 non-zero values combined with the corresponding 2 segment endpoints.   ... Etc ... (cont. below)
/
/      vbl[(totCells-1)*dlvl+(totCells-1)] - vbl[(totCells)*dLvl+(totCells-1)] will hold the alpha values (which sum to 1.00) within the last unique cell (index=totCells-1) of the multidimensional grid,
/      say, alpha_(totCells-1)_0, alpha_(totCells-1)_1, alpha_(totCells-1)_2, ..., alpha_(totCells-1)_500.  The corresponding optimal probability value p2 is then derived using the 2 non-zero values
/      combined with the corresponding 2 segment endpoints.
/
/      vbl[(totCells)*dLvl+totCells] - vbl[(totCells)*(dLvl+2)-1] will hold the single feasiblity factor assigned per cell.  The corresponding cell probability must be <= the # observations which are
/      assigned to that cell + this variable.  When there are 0 observations in a cell and it is required to be non-zero in an optimal solution then the feasibility factor must be forced to a non-zero
/      (positive) value.  The objective function contains a term for each feasibility factor multiplied by a large constant (see, K=1000000 in the header file).  This serves as a penalty on the objective
/      when a grid cell with no observations from the actual data is included in the optimal solution.  We want to avoid such solutions whenever possible since it can lead to the Hessian matrix of an
/      upcoming optimization being non full rank.
```

```
/
/ Inputs:
/
/    1.) The total # of unique cells in the multidimensional grid formed by combining all components across the estimated univariate mixture density functions for all assets.  For example, if there are a
/        total of 4 assets being considered and the univariate mixture densities have 2, 3, 4, 2 levels, respectively, the complete multidimensional grid will have 2x3x4x2 = 48 unique cells.  This
/        function provides 2 methods for determining which cells are important and needed in the multivariate mixture density. (totCells)
/
/    2.) The total number of assets being considered in the problem.  (numA)
/
/    3.) A double array indexed as [c][a], with "c" being a unique cell ID (values are from 0 - totCells-1), and "a" being an asset ID (values are from 0 - numA-1).  The value held at this position is
/        the univariate mixture component level for asset "a" within unique cell "c".  Recall that the multidimensional grid is formed by crossing all assets/components with all other assets/components.
/        (Here component refers back the the univariate mixture density for the asset.)  (inCellAry)
/
/    4.) An array of size numA holding the # of univariate mixture components for each asset.  Index "a" of this array will return the number of components that were needed to fit the univariate mixture
/        for component "a".  (Range is a=0 to numA-1.)  (cmps)
/
/    5.) A double array indexed as [a][g], with "a" being an asset ID (from 0 - numA-1), and "g" being a component ID for asset "a" ("g" ranges from 0 - cmps[a]-1).  The corresponding univariate mixture
/        component probability is stored at the indexed position.  (prbs)
/
/    6.) An array of size totCells that holds the number of observations (i.e., time points) that are assigned to the given unique cell of the multidimensional grid.  Observations are assigned to
/        specific components of an asset using Bayes' decision rule, that is they are assigned to the component with highest probability of membership.  Once a time point has been processed, it is
/        assigned to a component for each asset and this defines the cell of the multidimensional grid that it is assigned to.  For example, nCellObs[4] = 11 implies that there are 11 observations which
/        fall into unique cell ID #4.  (nCellObs)
/
/    7.) An array of estimated cell probabilities, ek, k=1,2,...,totCells.  These are derived as nCellObs[c]/T and drive both LP optimizations.  (cellProb)
/
/    8.) An empty array of true probabilities estimated by the 2 LPs and populated by this function.  This array will be of size totCells.  (outPrbs)
/
/    9.) The type of LP to use for a given function call, where 0 = Minimax objective, 1 = Minimum Sum of Squared Distances (SSD) objective.  (type)
/
/ Outputs:
/
/    This function counts the number of non-zero probabilities in the array outPrbs and returns this value at the call.  This function also populates the empty array outPrbs with the estimated true
/    probability for each unique cell which is "close" to the estimated ek values (using Bayes' decision rule), but that satisfies the marginal constraints.
/*****************************************************************************************************************************************************************************************/
#include "stdafx.h"
int solveLP(const int totCells, const int numA, const int **inCellAry, const int *cmps, const long double **prbs, const int *nCellObs, const long double *cellProb, double *outPrbs, const int type)
{
    // Initialize local variables.  (Variable dVars=total # of decision variables for the given LP and it depends on type.)
    //==========================================================================================================
    lprec *lp=NULL;
    int dVars=(int) (2*totCells+1)*(0==type) + (totCells*(dLvl+2))*(1==type), *vnum=new int [dVars], j, non0cmps=0;
    double *vbl=new double [dVars], objval, *epnt=new double [dLvl+1], *fpnt=new double [dVars-totCells];
    string strlabel;
    char **vlabels=new char *[dVars];

    // Build LP model to derive the joint density with zero covariances.
    //==================================================================
    lp = make_lp(0,dVars);
    if (lp == NULL)
    {
        cout << "ERROR:  LP Model did not build.  Something is wrong with the lp_solve setup (type=" << type << ")." << endl << "EXITING...solveLP()..." << endl; cin.get(); exit (EXIT_FAILURE);
    }

    // Add labels to the decision variables.  Include the cell index and asset/component values for minimax objective.  Include the cell index and alpha index for minimum squared distance objective.
    //===============================================================================================================================================================
    if (0 == type)     // Minimax objective.
    {
        for (int c=0; c<totCells; ++c)
        {
            strlabel="Cell_" + to_string((long long) c) + "_";
            for (int a=0; a<numA; ++a)
            {
                strlabel=strlabel + "A" + to_string((long long) a+1) + "C" + to_string((long long) inCellAry[c][a]+1);
                if (a < (numA-1))
```

```
                strlabel=strlabel + "_";
            }
            vlabels[c]=new char [strlabel.size()+1];
            strcpy_s(vlabels[c], strlabel.size()+1, strlabel.c_str());
            set_col_name(lp, c+1, vlabels[c]);

            // Now label the corresponding feasibility factor.  The decision variables come in (probability, feasibility factor) pairs.
            //==========================================================================================================================
            strlabel="F_" + strlabel;
            vlabels[totCells+c]=new char [strlabel.size()+1];
            strcpy_s(vlabels[totCells+c], strlabel.size()+1, strlabel.c_str());
            set_col_name(lp, totCells+c+1, vlabels[totCells+c]);
        }
        strlabel="Z";
        vlabels[dVars-1]=new char [strlabel.size()+1];
        strcpy_s(vlabels[dVars-1], strlabel.size()+1, strlabel.c_str());
        set_col_name(lp, (int) dVars, vlabels[dVars-1]);
    }
    else if (1 == type)     // Sum of squared distances objective.
        for (int c=0; c<totCells; ++c)
        {
            for (int s=0; s<=dLvl; ++s)
            {
                // Build array of constants for the piecewise linear function endpoints. There are dLvl segments, therefore dLvl+1 endpoints and these
                // do not change by cell.  Build an array of corresponding function endpoints, which do change by cell since the est. prob changes.
                //======================================================================================================================================
                if (0 == c)
                   epnt[s]=(double) (s*(1.00/dLvl));
                fpnt[c*(dLvl+1)+s]=pow((epnt[s] - (double) cellProb[c]),2);

                // Labels for alpha decision-variables using minimum squared distance objective.
                //================================================================================
                strlabel="Cell_" + to_string((long long) c) + "_Alpha_" + to_string((long long) s);
                vlabels[c*(dLvl+1)+s]=new char [strlabel.size()+1];
                strcpy_s(vlabels[c*(dLvl+1)+s], strlabel.size()+1, strlabel.c_str());
                set_col_name(lp, c*(dLvl+1)+s+1, vlabels[c*(dLvl+1)+s]);
            }

            // Label feasibility factor decision-variables using minimum squared distance objective.
            //======================================================================================
            strlabel="F_Cell_" + to_string((long long) c) + "_";
            for (int a=0; a<numA; ++a)
            {
                strlabel=strlabel + "A" + to_string((long long) a+1) + "C" + to_string((long long) inCellAry[c][a]+1);
                if (a < (numA-1))
                   strlabel=strlabel + "_";
            }
            vlabels[totCells*(dLvl+1) + c]=new char [strlabel.size()+1];
            strcpy_s(vlabels[totCells*(dLvl+1) + c], strlabel.size()+1, strlabel.c_str());
            set_col_name(lp, totCells*(dLvl+1) + (c+1), vlabels[totCells*(dLvl+1) + c]);
        }

    // Add marginal constraints on the cell probabilities. Sum of all probabilities attached to an Asset/Component must equal that Asset/Component probability.  For each asset with G components, once the
    // first G-1 constraints have been satisfied the G-th constraint is set since probabilities sum to 1.00.  We have not added the sum to 1.00 constraint thus will keep all G for each component.
    //====================================================================================================================================================================================================
    set_add_rowmode(lp, TRUE);
    for (int a=0; a<numA; ++a)
        for (int g=0; g<cmps[a]; ++g)
        {
            j=0;
            for (int c=0; c<totCells; ++c)
                if (inCellAry[c][a] == g)
                {
                    if (0 == type)                    // Minimax objective.
                    {
```

```
                    vnum[j]=c+1;
                    vbl[j++]=1.00;
                }
                else if (1 == type)               // Sum of squared distances objective.
                    for (int s=0; s<=dLvl; ++s)
                    {
                        vnum[j]=(c*(dLvl+1)+s)+1;
                        vbl[j++]=epnt[s];
                    }
            }
            if (!add_constraintex(lp, j, vbl, vnum, EQ, (double) prbs[a][g]))
            {
                cout << "ERROR: [LP Issue] Marginal probability constraint for Asset=" << a << "/Component=" << g << " failed to load (type=" << type << ")." << endl
                     << "EXITING...solveLP()..." << endl; cin.get(); exit (EXIT_FAILURE);
            }
    }

// Add feasibility constraints on the cell probabilities when using a minimax or minimum squared distance objective.  When an individual cell has zero observations assigned to it, force the cell
// probability to zero (relax if there is no feasible solution).
//=====================================================================================================================================================================================================
for (int c=0; c<(int) totCells; ++c)
{
    j=0;
    if (0 == type)
    {
        vnum[j]=c+1; vbl[j++]=1.00;
        vnum[j]=totCells+c+1; vbl[j++]=-1.00;
    }
    else if (1 == type)
    {
        for (int s=0; s<=dLvl; ++s)
        {
            vnum[j]=c*(dLvl+1)+s+1;
            vbl[j++]=epnt[s];
        }
        vnum[j]=totCells*(dLvl+1) + (c+1);
        vbl[j++]=-1.00;
    }
    if (!add_constraintex(lp, j, vbl, vnum, LE, (double) nCellObs[c]))
    {
        cout << "ERROR: [LP Issue] Feasibility factor constraint for Cell=" << c << " failed to load (type=" << type << ")." << endl << "EXITING...solveLP()..." << endl; cin.get();
        exit (EXIT_FAILURE);
    }
}

// Add inequality constraints on the minimax objective function.  Objective is transformed into an LP using appropriate inequality constraints.
//==================================================================================================================================================
if (0 == type)
   for (int c=0; c<(int) totCells; ++c)
   {
       // First absolute value constraint for the objective function pertaining to this cell probability.
       //====================================================================================================
       j=0;
       vnum[j]=(int) dVars; vbl[j++]=1.00;
       vnum[j]=c+1; vbl[j++]=-1.00;

       if (!add_constraintex(lp, j, vbl, vnum, GE, -1.00*cellProb[c]))
       {
           cout << "ERROR: [LP Issue] Minimax objective function absolute value constraint (#1) for Cell=" << c << " failed to load." << endl << "EXITING...solveLP()..." << endl; cin.get();
           exit (EXIT_FAILURE);
       }

       // Second absolute value constraint for the objective function pertaining to this cell probability.
       //=====================================================================================================
       j=0;
```

```
        vnum[j]=(int) dVars; vbl[j++]=1.00;
        vnum[j]=c+1; vbl[j++]=1.00;
        if (!add_constraintex(lp, j, vbl, vnum, GE, cellProb[c]))
        {
            cout << "ERROR: [LP Issue] Minimax objective function absolute value constraint (#2) for Cell=" << c << " failed to load." << endl << "EXITING...solveLP()..." << endl; cin.get();
            exit (EXIT_FAILURE);
        }
    }

// The sum of squared distances objective requires a constraint that the decision variables sum to 1 within each cell.
//===================================================================================================================
if (1 == type)
    for (int c=0; c<(int) totCells; ++c)
    {
        j=0;
        for (int s=0; s<=dLvl; ++s)
        {
            vnum[j]=(c*(dLvl+1)+s)+1;
            vbl[j++]=1.00;
        }

        if (!add_constraintex(lp, j, vbl, vnum, EQ, 1.00))
        {
            cout << "ERROR: [LP Issue] Sum of squared distances constraint that sum of " << dLvl+1 << " decision variables = 1.00 within Cell=" << c << "failed to load."
                 << endl << "EXITING...solveLP()..." << endl; cin.get(); exit (EXIT_FAILURE);
        }
    }

// Add minimization objective.  Output the entire LP formulation when requested.
//==============================================================================
set_add_rowmode(lp, FALSE);
j=0;
if (0 == type)                              // Minimax objective.
{
    vnum[j]=(int) dVars; vbl[j++]=1.00;
    for (int c=0; c<totCells; ++c)
    {
        vnum[j]=totCells+c+1;
        vbl[j++]=K;
    }
}
else if (1 == type)                         // Sum of squared distances objective.
    for (int c=0; c<totCells; ++c)
    {
        for (int s=0; s<=dLvl; ++s)
        {
            vnum[j]=c*(dLvl+1)+s+1;
            vbl[j++]=fpnt[c*(dLvl+1)+s];
        }
        vnum[j]=totCells*(dLvl+1) + (c+1);
        vbl[j++]=K;
    }

if (!set_obj_fnex(lp, j, vbl, vnum))
{
    cout << "ERROR: [LP Issue] Objective failed to load.  (Type=" << type << ")." << endl << "EXITING...solveLP()..." << endl; cin.get(); exit (EXIT_FAILURE);
}
set_minim(lp);
if (dbug >= 2)
{
    // Note:  Uncomment to write out full LP details when debugging.  (Note: Output can be large.)
    //===============================================================================================
    /* cout << string(11,'=') << endl << "LP Details:" << endl << string(11,'=') << endl;
       write_LP(lp, stdout); */
}
```

```
// Solve the LP and retrieve the results.
//=========================================
set_verbose(lp, IMPORTANT);
if (solve(lp) != OPTIMAL)
{
    cout << "ERROR: [LP Issue] No solution found, something has gone wrong.  (Type=" << type << ")." << endl << "EXITING...solveLP()..." << endl; cin.get(); exit (EXIT_FAILURE);
}

// Get the objective function value as well as the value of the unknown probabilities that define the multivariate density.  (Output the values.)
//==================================================================================================================================================
objval=get_objective(lp);
get_variables(lp, vbl);

// Compute the probabilities.  For Minimax these are the values from the first (Total # Cells) decision variables. For Min SSD objective, these are the weighted sum of the dLvl+1 decision variables.
//======================================================================================================================================================================================================
for (int c=0; c<totCells; ++c)
{
    if (0 == type)
    outPrbs[c] = vbl[c];
    else if (1 == type)
    {
        outPrbs[c] = 0.00;
        for (int s=0; s<=dLvl; ++s)
            outPrbs[c] = outPrbs[c] + vbl[c*(dLvl+1)+s]*epnt[s];
    }
}

// Write out the LP estimated probabilities along with the empirical values and corresponding distance between the estimated probabilities LP solution probabilities when debugging is requested.
//==================================================================================================================================================================================================
if (dbug >= 2)
{
    // Add correct labels if using sum of squared distances objective.
    //==================================================================
    if (1 == type )
       for (int c=0; c<totCells; ++c)
       {
           strlabel="Cell_" + to_string((long long) c) + "_";
           for (int a=0; a<numA; ++a)
           {
               strlabel=strlabel + "A" + to_string((long long) a+1) + "C" + to_string((long long) inCellAry[c][a]+1);
               if (a < (numA-1))
                  strlabel=strlabel + "_";
           }
           vlabels[c]=new char [strlabel.size()+1];
           strcpy_s(vlabels[c], strlabel.size()+1, strlabel.c_str());
           set_col_name(lp, c+1, vlabels[c]);
       }

    // Output the estimated probabilities along with the actual and absolute distance between the values.
    //=========================================================================================================
    cout.setf(ios_base::fixed, ios_base::floatfield); cout.precision(20);
    cout << endl << "Objective function (Type=" << type << ") value = " << objval << endl;
    cout << "LP Solution:" << endl << endl;
    for (int c=0; c<totCells; ++c)
        cout << get_col_name(lp, c+1) << " = " << outPrbs[c] << " vs. (actual) " << cellProb[c] << " and abs(diff) = " << abs(outPrbs[c]-cellProb[c]) << endl;
    if (0 == type)
       for (int c=totCells; c<(int) 2*totCells+1; ++c)
           cout << get_col_name(lp, c+1) << " = " << vbl[c] << endl;
    else if (1 == type)
           for (int c=totCells*(dLvl+1); c<totCells*(dLvl+2); ++c)
    cout << get_col_name(lp, c+1) << " = " << vbl[c] << endl;
}

// Issue a warning if a cell with zero observations is assigned a non-zero probability.
//=======================================================================================
```

```
    for (int c=0; c<totCells; ++c)
        if (cellProb[c] <= 0.00 && outPrbs[c] > 0.00)
           cout << endl << "WARNING: [LP Issue] Unique Cell # " << c << " has no observations but is assigned a non-zero probability.  (Type=" << type << ")." << endl
                        << "         The danger is that the likelihood function using this cell density could be zero at all time points.  If it occurs then the stage 2" << endl
                        << "         optimization will eliminate this decision variable (i.e., the unique cell probability) in the objective function (of the 1st Step)" << endl
                        << "         causing the corresponding Hessian to be singular because the upper left block is singular (it is a border matrix).  A solution is" << endl
                        << "         to change the forward-backward alphas so that a simpler solution is used as there may be too many unique cells.  Also, the variance" << endl
                        << "         ratio constraint may be too large resulting in spurious solutions being combined across assets resulting in cells with no obs." << endl << endl;

    // Free the memory allocated for the LP and the labels array.
    //===============================================================
    delete_lp(lp);
    delete [] vnum; vnum=nullptr;
    delete [] vbl; vbl=nullptr;
    delete [] epnt; epnt=nullptr;
    delete [] fpnt; fpnt=nullptr;
    for (int c=0; c<dVars; ++c)
    {
        delete [] vlabels[c]; vlabels[c]=nullptr;
    }
    delete [] vlabels; vlabels=nullptr;

    // Count the # of non-zero cell probability decision variables.  This is the # of components in the multivariate density and is returned by this function.
    //=================================================================================================================================================
    for (int c=0; c<totCells; ++c)
        if (outPrbs[c] > 0.00)
           non0cmps++;
    return non0cmps;
}

/*
/ Copyright (C) 2016 Chris Rook
/
/ This program is free software: you can redistribute it and/or modify it under the terms of the GNU General Public License as published by the Free Software Foundation, either version 3 of the License,
/ or (at your option) any later version.  This program is distributed in the hope that it will be useful, but WITHOUT ANY WARRANTY; without even the implied warranty of MERCHANTABILITY or FITNESS FOR A
/ PARTICULAR PURPOSE.  See the GNU General Public License for more details: <http://www.gnu.org/licenses/>
/
/ Filename:  getCMtrx.cpp
/
/ Function:  getCMtrx()
/
/ Summary:
/
/    This function ensures that the constraint matrix used to maintain the marginal mixture densities is of full row rank using only the non-zero component probabilities from the given LP solution.  A
/    full row rank component matrix is required for a future optimization.  The marginal probability densities are maintained via a set of linear constraints on the multidimensional grid cell
/    probabilities.  Since the sum of all probabilities must equal 1.00, each asset will require: (# univariate mixture components for that asset) - 1 constraints to maintain the marginals.  Note that
/    the constraint on the last univariate component for each asset will be automatically satisfied by the final constraint that the sum of all cell probabilities equals 1.00.  If there are "a" assets,
/    and C[i] components for asset i in the univariate mixture density, i=1,2,...,a, then there will be: (C[0]-1) + (C[1]-1) + (C[2]-1) + ... + (C[a-1]-1) + 1 = C[0] + C[1] + C[2] + ... + C[a-1] - a + 1
/    total rows in the original constraint matrix used to solve the LPs.  The constraints can be written in matrix form as Ax = b, where the vector b contains the corresponding probabilities for each row
/    of A, and x is a vector that holds the unique cell probabilities (i.e., decision variables) for the multidimensional grid.  Before the LP has been solved rows(A) <= elements(x), since elements(x) =
/    C[0]xC[1]xC[2]x...xC[a-1].  Therefore, the matrix A will be of full row rank.  After each LP has been fit many elements of x will be zero.  The effective constraint matrix that applies to the LP
/    solution will be the constraint matrix A with all columns that correspond to zeros of the vector x removed.  This effective constraint matrix is not necessarily of full row rank after the LP has
/    been solved.  Consider for example the case of 2 assets with 3 components in their corresponding univariate mixture densities.  In this scenario, the multidimensional grid is 3x3.  Suppose the
/    optimal LP solution contains non-zero probabilities on the diagonal of this grid and zeros in all off-diagonal positions.  The matrix A will have (3-1)+(3-1)+1 = 5 rows initially, however, after the
/    LP has been fit only 3 decision variables (i.e., columns of A) remain.  Therefore, the effective constraint matrix A will be of dimension 5x3, and not of full row rank.  That is, not all 5 rows of A
/    are needed to enforce the marginal constraints given the current LP solution and 2 rows may be dropped.  This function determines the rows that can be dropped and removes them from A producing a
/    full row rank effective constraint matrix, which is needed for an upcoming optimization.  To promote parsimony in the fitted multivariate mixture density, components with zero probabilities are
/    always permanantly eliminated at any point in any optimization.  Once dropped, a component is not permitted to return to the multivariate density.
/
/ Inputs:
/
/    1.) The number of rows in the original constraint matrix A before solving the LP (either Minimax or Minimum Sum of Squared Distances (SSD)).  (totrows)
/
/    2.) The number of non-zero probabilities (i.e., decision variables) after solving the LP (either Minimax or Minimum Sum of Squared Distances (SSD)).  (nUCmps)
/
```

```
/    3.) Type of LP objective, 0=Minimax or 1=Minimum Sum of Squared Distances (SSD).  (type)
/
/    4.) The total number of assets under consideration.  (numA)
/
/    5.) Array to hold the optimal/chosen # of univariate mixture components for each asset.  (nCmps)
/
/    6.) Array of unique cell IDs for the multidimensional grid w/non-0 probs in the LP solution, i.e., cells that are used to structure the initial multivariate mixture PDF (w/out covariances).  (vCIDs)
/
/    7.) A double array indexed as [c][a], with "c" being a unique cell ID (values are from 0 - totCells-1), and "a" being an asset ID (values are from 0 - numA-1).  The value held at this position is
/        the univariate mixture component level for asset "a" within unique cell "c".  Recall that the multidimensional grid is formed by crossing all univariate assets/components with all other
/        univariate assets/components.  (Here component refers back the univariate mixture density for the asset.)  (inCellAry)
/
/    8.) A double array indexed as [a][r], with "a" being the asset indicator (values 0 thru numA-1), and "r" being the component of the optimal univariate mixture distribution for that asset (values 0
/        thru nCmps[a]-1).  The corresponding univariate mixture component probability is held by the array.  Note that these probabilities are used to construct all but the last element of the vector b,
/        where the marginals are maintained via:  Ax = b.  (prbs)
/
/    9.) An empty matrix to hold the full row rank version of A for the given LP solution.  This function derives and returns the corresponding matrix.  (inLHS)
/
/   10.) An empty vector to hold the corresponding probabilities for the new full row rank version of A.  This is the b vector and it is derived here as the original b vector without the corresponding
/        rows that were dropped from A to make it full row rank.  (inRHS)
/
/ Outputs:
/
/   This function returns no value at the call, but it derives and populates the empty LHS matrix and RHS vector for the full row rank version of the constraint set that maintains the marginals.  The
/   constraints are linear of the form Ax = b.
/***********************************************************************************************************************************************************************************************/
#include "stdafx.h"
void getCMtrx(const int totrows, const int nUCmps, const int type, const int numA, const int *nCmps, const int *vCIDs, const int **inCellAry, const long double **prbs, Eigen::MatrixXd *inLHS,
              Eigen::VectorXd *inRHS)
{
    // Declare/initialize local variables.
    //====================================
    int rnk, trnk, *remrows=nullptr, rr, kr, rw;
    Eigen::MatrixXd oLHS(totrows,nUCmps);
    Eigen::VectorXd oRHS(totrows), tVec(nUCmps);

    // Build modified constraint matrix that applies to the LP solution.
    //==================================================================
    rw=0;
    for (int a=0; a<numA; ++a)
        for (int r=0; r<(nCmps[a]-1); ++r)
        {
            for (int c=0; c<nUCmps; ++c)
                oLHS(rw,c) = (int) (inCellAry[vCIDs[c]][a] == r);
            oRHS(rw++) = prbs[a][r];
        }
    for (int c=0; c<nUCmps; ++c)
    oLHS(rw,c) = 1;
    oRHS(rw) = 1.00;

    // Rank of constraint matrix with non-basic and zero-basic columns removed.
    //=========================================================================
    rnk = (int) Eigen::FullPivLU<Eigen::MatrixXd>(oLHS).rank();

    // Full row rank modified constraint matrix must have # rows equal to its rank.
    //=============================================================================
    inLHS[type] = Eigen::MatrixXd(rnk,nUCmps);
    inRHS[type] = Eigen::VectorXd(rnk);

    // If the constraint matrix is not of full row rank, identify (totrows - rnk) rows that can be removed from the constraint matrix to make it full row rank.
    //=========================================================================================================================================================
    if (rnk < totrows)
    {
        rr=0;
        remrows = new int [totrows-rnk];
```

```
        for (int r=0; r<totrows; ++r)
        {
            // Store the values at this row and then set the row to all zeros.
            //==================================================================
            for (int c=0; c<nUCmps; ++c)
            {
                tVec[c]=oLHS(r,c);
                oLHS(r,c)=0.00;
            }

            // Recheck the rank.  If it changes replace the row with its original values.  Otherwise, leave it as all zeros and store the row # that can be dropped.
            //===========================================================================================================================================================
            trnk = (int) Eigen::FullPivLU<Eigen::MatrixXd>(oLHS).rank();
            if (trnk < rnk)
               for (int c=0; c<nUCmps; ++c)
                   oLHS(r,c)=tVec(c);
            else
               remrows[rr++]=r;
        }
    }

    // Populate the full row rank modified constraint matrix.
    //=========================================================
    rw=0;
    for (int r=0; r<totrows; ++r)
    {
        kr=1;
        for (int k=0; k<(totrows-rnk); ++k)
            if (r == remrows[k])
               kr=0;

        if (1 == kr)
        {
            for (int c=0; c<nUCmps; ++c)
            {
                inLHS[type](rw,c)=oLHS(r,c);
                inRHS[type](rw)=oRHS(r);
            }
            rw++;
        }
    }

    // Print out the modified constraint matrix when debugging is on.
    //=================================================================
    if (dbug >= 2)
    {
        cout.setf(ios_base::fixed, ios_base::floatfield); cout.precision(0);
        cout << endl << "Modified Full Row-Rank LHS Constraint Matrix: " << endl << inLHS[type] << endl;
        cout.setf(ios_base::fixed, ios_base::floatfield); cout.precision(10);
        cout << endl << "Modified RHS Constraint Vector: " << endl << inRHS[type] << endl;
        cout.setf(ios_base::fixed, ios_base::floatfield); cout.precision(0);
        int chkrnk; chkrnk = (int) Eigen::FullPivLU<Eigen::MatrixXd>(inLHS[type]).rank();
        cout << endl << "The rank of this modified constraint matrix is: " << chkrnk << endl;
    }

    // Free up temporary memory allocations.
    //========================================
    delete [] remrows; remrows=nullptr;
}

/*
/ Copyright (C) 2016 Chris Rook
/
/ This program is free software: you can redistribute it and/or modify it under the terms of the GNU General Public License as published by the Free Software Foundation, either version 3 of the License,
/ or (at your option) any later version.  This program is distributed in the hope that it will be useful, but WITHOUT ANY WARRANTY; without even the implied warranty of MERCHANTABILITY or FITNESS FOR A
```

```
/ PARTICULAR PURPOSE.  See the GNU General Public License for more details: <http://www.gnu.org/licenses/>
/
/ Filename:  ECMEAlg.cpp
/
/ Function:  ECMEAlg()
/
/ Summary:
/
/    An extension of the ECME algorithm (Liu & Rubin, 1994) is implemented by this function.  The multivariate mixture likelihood is optimized with respect to the mixing proportions and the covariances.
/    The means and variances are held fixed to maintain the mixture marginals.  The 1st Step optimization is convex in the mixing proportions and constrained to maintain the mixture marginals.  The
/    corresponding Lagrangian is formed (1st Step optimization) and its zeros are determined iteratively using Newton's method.  The resulting mixture proportions are unique global optimizers of the 1st
/    Step likelihood (all other parameters are fixed at this stage).  Once 1st Step convergence is achieved, processing is passed to the 2nd Step where the likelihoood is maximized with respect to only the
/    covariances.  This optimization is constrained by the positive-definiteness of the corresponding estimated VC matrices (there is 1 VC matrix per multivariate density component).  It is also non-
/    convex as there may be multiple local optimums of the likelihood function with only the covariances unknown, and the positive-definite constraint is not convex.  The goal of the 2nd Step is to find
/    the largest local optimum given that the means, variances, and mixing proportions are fixed and only the covariances are unknown.  We will attempt to climb to the top of the current hill (local
/    optimum) using both the gradient and Hessian while also searching for larger hills in the general direction of steepest ascent (see Marquardt, 1963).  This is considered a compromise between
/    strictly applying Newton's Method and Gradient Ascent and is useful when a single-step Newton's Method overshoots or a single step lands in an infeasible region.  Once no further progress is made
/    during the 2nd Step, we return to the 1st Step with the newly estimated covariances and repeat the optimization over the mixing proportions.  Convergence is achieved when the 2nd Step fails to improve
/    the likelihood function returned from the 1st Step.  Both the 1st and 2nd Steps are iterative with the corresponding gradients and Hessians updated repeatedly during a single corresponding iteration
/    of the extended ECME algorithm.  Solutions found here will not be considered spurious, which differs from the search for a solution when dealing with univariate mixtures.  This is due to the fact
/    that the variances have already been fixed and do not change.  The corresponding variance ratio constraint specified by the user remains in force.  We can justify this approach by noting that
/    commercial software packages such as SAS (R) use a line-search based quasi-Newton algorithm to find the MLE for a univariate mixture density (instead of the EM algorithm).  We have not proven that
/    the extended ECME method used here will guarantee convergence to the largest local optimum, only that we have located the nearest local optimum in the vicinity of the informed start.
/
/    Important Note:  The 1st Step imposes equality constraints that maintain the mixture marginals.  Greater-than-zero constraints on the mixing proportions (i.e., probabilities) are not explicitly
/    imposed, therefore negative probabilities may maximize the 1st Step likelihood function.  The 1st Step likelihood function treats the mixing probabilities as unknowns and all other parameters (means,
/    variances, covariances) as known constants.  Any components that require a negative probability to maximize the likelihood function are dropped during the 1st Step and the entire problem is resized
/    accordingly (fewer components).  This may result in a likelihood that decreases, however the overall objective is to balance parsimony with maximizing the likelihood function.
/
/ Inputs:
/
/    1.) The total number of time points with data collected. (T)
/
/    2.) The double array of returns for each asset and at each time point, indexed as r[a][t].
/
/    3.) The number of assets with returns collected. (numA)
/
/    4.) The number of unique components in the multivariate mixture that results from either the Minimax or Minimum SSD LP optimizations.  Each multivariate mixture has fixed means and variances and
/        this function will optimize the mixing probabilities and covariance terms.  All covariance terms begin the optimization at zero. (nUCmps)
/
/    5.) The LHS matrix required to enforce the constraint that the marginal density for each asset equals its fixed univariate mixture.  This matrix is built during the LP optimization and resized there
/        accordingly to ensure it is of full row rank.  The LHS matrix has a column for each component in the multivariate density. (cMtrx)
/
/    6.) The RHS vector required to enforce the constraint that the marginal density for each asset equals its fixed univariate mixture.  This vector is built during the LP optimization and contains the
/        marginal mixture component probabilities for each asset (less the last probability for each asset, which is fixed once all others have been fixed for that asset). (cVctr)
/
/    7.) The array of multivariate mixture probabilities returned from the corresponding LP optimization (either Minimax of Minimum SSD).  Note that these probabilities are passed as an array but
/        converted to a vector within this program as other functions require the values to be stored in a vector.  (muPrbs)
/
/    8.) The array of mean vectors.  Each component of the multivariate mixture density is a multivariate density function which has its own set of means.  The first element of this array is the vector
/        of means for the first multivariate component, etc...  All means supplied to this function are fixed and do not change which is required to maintain the marginals, with the exception that
/        components may be dropped.  When a component is dropped the corresponding mean vector for that component is dropped.  (muMns)
/
/    9.) The array of VC matrices.  Each component in the multivariate mixture density is a multivariate density function with a corresponding VC matrix.  Each VC matrix is of dimension (numA)x(numA).
/        The diagonals of each VC matrix are the corresponding variances for that asset within that component.  All variances supplied to this function and are fixed and do not change which is required
/        to maintain the marginals, with the exception that components may be dropped.  When a component is dropped the corresponding VC matrix for that component is dropped.  (muVCs)
/
/   10.) The array of unique cell IDs that link each component of the multivariate density back to the full factorial of components.  The full factorial of components represents each cell in the
/        multidimensional grid formed by considering all combinations of assets and their levels.  Note that the full factorial would be required to build a multivariate mixture density with given
/        marginals under the assumption that the assets were all mutually independent random variables.  (uCellIDs)
/
/   11.) A string to hold the directory where the output file resides.  (rdir)
/
/ Outputs:
/
```

```
/    This function updates the arrays of multivariate mixture probabilities (muPrbs), mean vectors (muMns), and VC matrices (muVCs).  Note that muMns is updated only when components are dropped, and
/    muVCs is updated when components are dropped and when covariances are estimated.  This function returns the total number of unique multivariate mixture components in the final density.
/*************************************************************************************************************************************************************************************/
#include "stdafx.h"
int ECMEAlg(const int T, const long double **r, const int numA, const int nUCmps, const Eigen::MatrixXd cMtrx, const Eigen::VectorXd cVctr, long double *muPrbs, Eigen::VectorXd *muMns, Eigen::MatrixXd
            *muVCs, int *uCellIDs, const string rdir)
{
    // Local variables.
    //===================
    long long mHessMag;
    int uCmps=nUCmps, ecnvg, mcnvg, cnvg, n0vals, nLMs=(int) cVctr.size(), ld=uCmps+nLMs, itr, nCovs, nCores=(int) boost::thread::hardware_concurrency(), nUpdts, nThrds=(int) nCorMult*nCores,
        **sHess1=new int *[nThrds], ecmeItr=1, eItrs, mItrs, sumval, n_beats=min(nBeats,nThrds), *beat=new int [n_beats], mtch, npos;
    long double **fVals=new long double *[T], *dnom=new long double [T], picst=exp(((double) (numA/2.00))*log(2.00*pi)), *sqdets=new long double [uCmps], LL, curmlt, oldLL, **sHess2=new long double
                *[nThrds], *maxHLL=new long double [n_beats], eLL, mLL, lBound, sumLL, uval, max1, max2, cnum;
    Eigen::VectorXd *gradE=new Eigen::VectorXd [1], *dvarsE=new Eigen::VectorXd [2], *rts=new Eigen::VectorXd [T], *RHS=new Eigen::VectorXd [1], *tmpCVctr=new Eigen::VectorXd [1], *tmpDv=new
                    Eigen::VectorXd [1], *gradM=new Eigen::VectorXd [1], *dvarsM=new Eigen::VectorXd [4], **tmpDvarsM=new Eigen::VectorXd *[nThrds];
    Eigen::MatrixXd *hessE=new Eigen::MatrixXd [1], *VCMInv=new Eigen::MatrixXd [nUCmps], *LHS=new Eigen::MatrixXd [1], *tmpCMtrx=new Eigen::MatrixXd [1], *hessM=new Eigen::MatrixXd [1], *A=new
                    Eigen::MatrixXd [(int) numA*(numA-1)/2], *p1Opt=new Eigen::MatrixXd [1], *P=new Eigen::MatrixXd [1], *Pinv=new Eigen::MatrixXd [1];
    string lblvar, ndef;
    boost::thread *t1=new boost::thread[nThrds];
    random_device rd;  default_random_engine gen(rd());
    uniform_real_distribution<long double> udist(0.0,1.0);
    Eigen::EigenSolver<Eigen::MatrixXd> egnslvr, normslvr, normslvrt;
    ofstream fout;

    // Populate an array of vectors with the returns at each time point indexed as [t][a] to ease computations.
    //==========================================================================================================
    for (int t=0; t<T; ++t)
    {
        fVals[t] = new long double [nUCmps];
        rts[t]=Eigen::VectorXd(numA);
        for (int a=0; a<numA; ++a)
            rts[t](a)=r[a][t];
    }

    // Derive VC inverses and corresponding determinants.
    //====================================================
    for (int v=0; v<uCmps; ++v)
    {
        VCMInv[v]=Eigen::MatrixXd(numA,numA);
        VCMInv[v]=muVCs[v].inverse();
        sqdets[v]=sqrt(VCMInv[v].determinant());
    }

    // Populate the covariance identifier matrices for use in the 2nd Step.
    //=====================================================================
    itr=0;
    for (int a1=0; a1<numA; ++a1)
        for (int a2=a1+1; a2<numA; ++a2)
        {
            A[itr]=Eigen::MatrixXd(numA,numA);
            for (int r=0; r<numA; ++r)
                for (int c=0; c<numA; ++c)
                {
                    if (r==a1 && c==a2 || r==a2 && c==a1)
                       A[itr](r,c)=1.0;
                    else
                       A[itr](r,c)=0.0;
                }
            itr++;
        }

    // Initialize decision variables for 1st step.  Probabilities will be set to their values as determined by solving the corresponding LP and the Lagrange multipliers will be initialized to zeros.
    //================================================================================================================================================================================================
    for (int i=0; i<2; ++i)
```

```
    dvarsE[i]=Eigen::VectorXd(ld);
for (int d=0; d<uCmps; ++d)
    dvarsE[0](d)=muPrbs[d];
for (int d=uCmps; d<ld; ++d)
    dvarsE[0](d)=0.00;

// Populate a double array with the likelihood values for each timepoint and component.  The multivariate likelihood of each observation is also stored in an array (dnom[]).
// Any component with zero likelihood for all time points is a variable that does not exist in the objective function.  It should be treated as a constant and moved to the right
// hand side of each constraint and the problem needs to be resized accordingly.  This check is made via the function call chkSum() below.
//=================================================================================================================================================================================
LL=getLFVals(T,numA,uCmps,rts,dvarsE[0],muMns,VCMInv,sqdets,picst,dnom,fVals);
cout.setf(ios_base::fixed, ios_base::floatfield); cout.precision(15); cout << "Initial Log-Likelihood value is: " << LL << endl << endl;
chkSum(T,uCmps,(const long double **) fVals);

// Build the corresponding Hessian (of the Lagrangian).  The matrix is stored in hessE[0](.,.).  When building the Hessian, iterate until it is invertible by multiplying the constraint LHS and RHS by
// a constant value (multiple of 10) until full rank.  (Note:  A matrix with large and small eigenvalues may be ill-conditioned and there may be computational issues when attempting to invert it.)
//======================================================================================================================================================================================================
hessE[0]=Eigen::MatrixXd(ld,ld);
curmlt=getHessE(T,uCmps,(const long double **) fVals,dnom,cMtrx,cVctr,hessE,LHS,RHS);

// Build the gradient (of the Lagrangian) (using the modifed constraint matrix/vector as required above).
//=========================================================================================================
gradE[0]=Eigen::VectorXd(ld);
getGradE(T,uCmps,(const long double **) fVals,dnom,LHS[0],RHS[0],dvarsE[0],gradE);

// Initialize arrays used and reused during the 2nd Step.
//========================================================
for (int h=0; h<nThrds; ++h)
{
    sHess1[h]=new int [8];
    sHess2[h]=new long double [7];
    tmpDvarsM[h]=new Eigen::VectorXd [4];
}

// Iterate using the ECME algorithm until convergence.
// Iterate and update the probabilities (the 1st Step is a maximization problem with concave objective and convex constraints).  Stationary points for the Lagrangian will therefore be taken as global
// optimizers and these are determined using Newton's method.  (The Hessian here is a bordered matrix which is invertible under certain met conditions on the non-zero sections.)  The 2nd Step is a
// constrained maximization problem having multiple local optimums.  We attempt to find the largest local optimum using an iterative technique that steps in the general direction of steepest ascent.
//======================================================================================================================================================================================================
cnvg=0;
do
{
    //==================
    // (ECME) 1st STEP:
    //==================
    oldLL=LL; ecnvg=0; n0vals=0; eItrs=0;
    cout << endl << "1st Step Start:  ECME Algorithm (Iteration=" << ecmeItr << ") (Beginning LL = ";
    cout.setf(ios_base::fixed, ios_base::floatfield); cout.precision(15); cout << oldLL << ")." << endl << endl << string(15,' ') << "-----> Iterating:  ";
    do
    {
        // 1st Step iteration counter.
        //=============================
        eItrs++;
        cout << ".";

        // Solve for new component probabilities, which are the decision variables in the 1st Step.
        //============================================================================================
        dvarsE[1]=hessE[0].colPivHouseholderQr().solve(-1.00*gradE[0]);
        dvarsE[0]=(dvarsE[1]+dvarsE[0]);

        // Check for zero or negative probabilities and prepare for next iteration.
        //=============================================================================
        n0vals=0;
        for (int v=0; v<uCmps; ++v)
            if (dvarsE[0](v) <= 0.00)
```

```
       n0vals++;
// Resize the problem (if needed) and perform another iteration.
//================================================================
uCmps=uCmps-n0vals;
nLMs=(int) RHS[0].size();
ld=uCmps+nLMs;

// Update constraint matrix/vector.  Undo the multiplier and adjust size if multivariate component probabilities have been set to zero.
//========================================================================================================================================
tmpCMtrx[0]=Eigen::MatrixXd(nLMs,uCmps);
for (int r=0; r<(int) LHS[0].rows(); ++r)
{
    itr=0;
    for (int c=0; c<(int) LHS[0].cols(); ++c)
        if (dvarsE[0](c) > 0.00)
           tmpCMtrx[0](r,itr++)=(1.00/curmlt)*LHS[0](r,c);
}
tmpCVctr[0]=Eigen::VectorXd(nLMs);
tmpCVctr[0]=(1.00/curmlt)*RHS[0];

// If a component is dropped then update mean vectors, VC matrices, and unique cell IDs.
//========================================================================================
if (n0vals > 0)
{
    itr=0;
    for (int v=0; v<(int) LHS[0].cols(); ++v)
    if (dvarsE[0](v) > 0.00)
    {
        muMns[itr]=muMns[v];
        muVCs[itr]=muVCs[v];
        VCMInv[itr]=muVCs[itr].inverse();
        sqdets[itr]=sqrt(VCMInv[itr].determinant());
        uCellIDs[itr++]=uCellIDs[v];
    }

    // Resize the internal array that holds the likelihood values for each timepoint and component when the # of components changes.
    //=================================================================================================================================
    for (int t=0; t<T; ++t)
    {
        delete [] fVals[t]; fVals[t]=new long double [uCmps];
    }

    // Update decision variable vectors, dropping relevant rows.
    //============================================================
    tmpDv[0]=Eigen::VectorXd((int) dvarsE[0].size());
    tmpDv[0]=dvarsE[0];
    delete [] dvarsE; dvarsE=new Eigen::VectorXd [2];
    for (int i=0; i<2; ++i)
        dvarsE[i]=Eigen::VectorXd(ld);
    itr=0;
    for (int v=0; v<(int) LHS[0].cols(); ++v)
        if (tmpDv[0](v) > 0.00)
           dvarsE[0](itr++)=tmpDv[0](v);
    for (int v=(int) LHS[0].cols(); v<(int) tmpDv[0].size(); ++v)
        dvarsE[0](itr++)=tmpDv[0](v);
    delete [] tmpDv; tmpDv=new Eigen::VectorXd [1];
}

// Update density function values grid timepoint-specific likelihood function values.
//=====================================================================================
LL=getLFVals(T,numA,uCmps,rts,dvarsE[0],muMns,VCMInv,sqdets,picst,dnom,fVals);

// Check for 1st Step convergence.  Need no negative/zero component probabilities and unchanged LL.  Otherwise, rebuild Hessian/gradient and iterate again.
//===========================================================================================================================================================
if (0 == n0vals && (LL - oldLL) < (epsilon)*oldLL)
```

```
    {
        ecnvg=1;
        itr=0;
        oldLL=LL;
        for (int v=0; v<uCmps; ++v)
            muPrbs[itr++]=dvarsE[0](v);
    }

    // No 1st Step convergence, iterate again.
    //=========================================
    if (0 == ecnvg)
    {
        // Reset Log-Likelihood, which is only valid when there are no negative probabilities.
        //=====================================================================================
        if (0 == n0vals)
           oldLL=LL;

        // Rebuild Hessian.
        //===================
        delete [] LHS;  LHS=new Eigen::MatrixXd [1];
        delete [] RHS;  RHS=new Eigen::VectorXd [1];
        delete [] hessE; hessE=new Eigen::MatrixXd [1]; hessE[0]=Eigen::MatrixXd(ld,ld);
        curmlt=getHessE(T,uCmps,(const long double **) fVals,dnom,tmpCMtrx[0],tmpCVctr[0],hessE,LHS,RHS);

        // Rebuild Gradient then delete temporary memory allocations.
        //=============================================================
        delete [] gradE; gradE=new Eigen::VectorXd [1]; gradE[0]=Eigen::VectorXd(ld);
        getGradE(T,uCmps,(const long double **) fVals,dnom,LHS[0],RHS[0],dvarsE[0],gradE);
        delete [] tmpCMtrx; tmpCMtrx=new Eigen::MatrixXd [1];
        delete [] tmpCVctr; tmpCVctr=new Eigen::VectorXd [1];
    }
} while (0 == ecnvg);

eLL=oldLL;
cout << "  (Done.)" << endl << endl << "1st Step Done:   ECME Algorithm (Iteration=" << ecmeItr << ") Converged in " << eItrs << " iterations.  (New LL = ";
cout.setf(ios_base::fixed, ios_base::floatfield); cout.precision(15); cout << eLL << ")." << endl << endl;

//==================
// (ECME) 2nd STEP:
//==================
nCovs=(int) uCmps*numA*(numA-1)/2;
p1Opt[0]=Eigen::VectorXd(nCovs);
for (int i=0; i<4; ++i)
    dvarsM[i]=Eigen::VectorXd(nCovs);
getCovs(uCmps,muVCs,dvarsM);
dvarsM[2]=dvarsM[0];
gradM[0]=Eigen::VectorXd(nCovs);
hessM[0]=Eigen::MatrixXd(nCovs,nCovs);
P[0]=Eigen::MatrixXd(nCovs,nCovs);
Pinv[0]=Eigen::MatrixXd(nCovs,nCovs);
for (int h=0; h<nThrds; ++h)
    for (int i=0; i<4; ++i)
        tmpDvarsM[h][i]=Eigen::VectorXd(nCovs);
mcnvg=0; mItrs=0;
cout << endl << "2nd Step Start:  ECME Algorithm (Iteration=" << ecmeItr << ") (Beginning LL = ";
cout.setf(ios_base::fixed, ios_base::floatfield); cout.precision(15); cout << oldLL << ")." << endl << endl;
do
{
    // 2nd Step iteration counter.
    //=============================
    mItrs++;

    // Build gradient for 2nd step.
    //==============================
    getGradM(T,rts,uCmps,numA,(const long double **) fVals,dnom,muPrbs,muMns,muVCs,VCMInv,A,gradM);
```

```
// Build Hessian for 2nd step.  Find the length (# digits) of the element with largest magnitude.
//================================================================================================
mHessMag=(long long) getHessM(T,rts,uCmps,numA,(const long double **) fVals,dnom,muPrbs,muMns,muVCs,VCMInv,A,hessM);

// Write out the max eigenvalue, condition # and # of negative/positive eigenvalues of the Hessian just derived.
//================================================================================================================
egnslvr.compute(hessM[0], true);
P[0]=egnslvr.eigenvectors().real();
Pinv[0]=P[0].inverse();
normslvr.compute(P[0].transpose()*P[0],false);
normslvrt.compute(Pinv[0].transpose()*Pinv[0],false);
max1=LNegVal; max2=LNegVal; npos=0;
for (int a=0; a<(int) hessM[0].rows(); ++a)
{
    if (egnslvr.eigenvalues()[a].real() > 0)
       npos++;
    if (abs(normslvr.eigenvalues()[a].real()) > max1)
       max1=abs(normslvr.eigenvalues()[a].real());
    if (abs(normslvrt.eigenvalues()[a].real()) > max2)
       max2=abs(normslvrt.eigenvalues()[a].real());
}
cnum = sqrt(max1)*sqrt(max2);
cout << endl << string(15,' ')  << "Total # of (-,+) eigenvalues: ("<< nCovs-npos << "," << npos << ") (Hessian condition # = " << cnum << ")" << endl;

// The function stepHessM() uses the Hessian to step in the direction of the gradient.  To step we add a (random) constant to the diagonal
// with larger (random) constants translating to smaller steps, and smaller (random) constants translating to larger steps.
//==========================================================================================================================================
Eigen::initParallel();
lblvar=string(15,' ') + "-----> Run #" + to_string((long long) mItrs) + " - Threads Launched: ";
cout << lblvar;
for (int h=0; h<nThrds; ++h)
{
     sHess1[h][0]=T; sHess1[h][1]=uCmps; sHess1[h][2]=h; sHess1[h][3]=0; sHess1[h][4]=(int) to_string(mHessMag).size() + 2; sHess1[h][6]=0; sHess1[h][7]=0;
     sHess2[h][0]=oldLL; sHess2[h][1]=0.00; sHess2[h][2]=(long double) (minHessAdd + h*mItersH*sHess2[h][1]); sHess2[h][5]=0.00;
     tmpDvarsM[h][0]=dvarsM[0];
     tmpDvarsM[h][2]=dvarsM[2];
     t1[h]=boost::thread(stepHessM,boost::ref(sHess1[h]),boost::ref(sHess2[h]),boost::cref(rts),boost::ref(tmpDvarsM[h]),gradM[0],hessM[0],dvarsE[0],boost::cref(muMns),boost::cref(muVCs));
}

// Conditionally output a line feed and alignment spaces once all threads have successfully launched.
//=====================================================================================================
if (dbug <= 1)
{
    do
    {
        Sleep(1000); sumval=0;
        for (int h=0; h<nThrds; ++h)
        sumval=sumval+sHess1[h][3];
    } while (sumval < nThrds); cout << string(10,' ') << "(Pr LL = " << oldLL << ")" << endl << string(lblvar.size()-18,' ') << "Threads Finished: ";
    sumval=0;
    do
    {
        Sleep(100);
        for (int h=0; h<nThrds; ++h)
            if (1 == sHess1[h][7])
            {
                cout << ".";
                sHess1[h][7]=0;
                sumval++;
            }
    } while (sumval < nThrds);
}

// Pause until all threads finish.
//==================================
```

```
for (int j=0; j<nThrds; ++j)
    t1[j].join();

// Randomly select one of the top "nBeats" performers to begin the next iteration.  (Weight values by their LL to favor higher values.)
// The value of "nBeats" is set in the header file and only LL values that exceed the current LL are considered beats.
//==============================================================================================================================
nUpdts=0; sumLL=0.00;
for (int b=0; b<n_beats; ++b)
    maxHLL[b]=oldLL;
for (int h1=0; h1<nThrds; ++h1)
{
    // No LL values returned should be less than the existing maximum.
    //================================================================
    if (sHess2[h1][3] < oldLL)
    {
        cout.setf(ios_base::fixed, ios_base::floatfield); cout.precision(15);
        cout << "ERROR:  ECME Algorithm 2nd Step stepping function has returned a LL value inferior to the current maximum, which should not happen." << endl
             << "        Must inspect and fix.  Thread # = " << h1 << endl
             << "        The current maximum LL = " << oldLL << endl
             << "        Maximum LL value returned from stepping function = " << sHess2[h1][3] << endl
             << "EXITING...ECMEAlg()..." << endl; cin.get(); exit (EXIT_FAILURE);
    }

    // Find and process the improvements.
    //====================================
    itr=0;
    for (int h2=0; h2<nThrds; ++h2)
        if (sHess2[h1][3] < sHess2[h2][3])
           itr++;
    for (int b=0; b<n_beats; ++b)              // Deal with LL ties.
        if (sHess2[h1][3] > oldLL && sHess2[h1][3] == maxHLL[b])
           itr++;
    if (sHess2[h1][3] > oldLL && itr < n_beats)
    {
        maxHLL[itr]=sHess2[h1][3];             // Store the log-likelihood for the given beat.
        beat[itr] = h1;                        // Store the thread index for the given beat.
        sumLL = sumLL + (maxHLL[itr]-oldLL);   // Sum the magnitude of log-likelihood improvements.
        nUpdts++;                              // Record the # of improvements.
    }
}

// Randomly select one of the LL beats to begin the next 2nd Step iteration.
//=========================================================================
if (nUpdts > 0)
{
    uval = udist(gen)*sumLL; mtch=0;
    do
    {
        lBound = sumLL - (maxHLL[nUpdts-1]-oldLL);
        if (uval >= lBound)
           mtch = 1;
        else
           sumLL = sumLL - (maxHLL[(nUpdts--)-1]-oldLL);
    } while (0 == mtch);
    dvarsM[0]=tmpDvarsM[beat[nUpdts-1]][3];
}
else
    nUpdts=1;

if (0 == dbug)
{
    cout.setf(ios_base::fixed, ios_base::floatfield); cout.precision(15);
    cout << " (Done.)  (RC LL = " << maxHLL[nUpdts-1] << ", #(RRs)=" << sHess1[beat[nUpdts-1]][6] << ")" << endl << endl; cout.precision(15); /* RC = Randomly Chosen */
}
```

```
    // Update the VC and inverse VC matrices along with the vector of corresponding determinants.
    //===============================================================================================
    for (int v=0; v<uCmps; ++v)
    {
        setCovs(v,muVCs,dvarsM);
        VCMInv[v]=muVCs[v].inverse();
        sqdets[v]=sqrt(VCMInv[v].determinant());
    }
    LL=getLFVals(T,numA,uCmps,rts,dvarsE[0],muMns,VCMInv,sqdets,picst,dnom,fVals);

    // QC check that Max LL equals the beat value chosen above.
    //============================================================
    if (abs(LL - maxHLL[nUpdts-1]) > epsilon*maxHLL[nUpdts-1])
    {
        cout.setf(ios_base::fixed, ios_base::floatfield); cout.precision(15);
        cout << "ERROR:  ECME Algorithm has derived LL not equal to the beat LL chosen randomly, which should not happen.  Must inspect and fix." << endl
             << "        Value of maxHLL[nUpdts-1] = " << maxHLL[nUpdts-1] << endl
             << "        Value of LL = " << LL << endl << "EXITING...ECMEAlg()..." << endl; cin.get(); exit (EXIT_FAILURE);
    }

    // Check for convergence of the 2nd step.
    //========================================
    if ((LL - oldLL) < pow(10.0,6)*(epsilon)*oldLL)
       mcnvg=1;
    oldLL=LL; dvarsM[2]=dvarsM[0];

} while (0 == mcnvg);

mLL=oldLL;
cout << endl << "2nd Step Done:   ECME Algorithm (Iteration=" << ecmeItr << ") Converged in " << mItrs << " iterations. (New LL=";
cout.setf(ios_base::fixed, ios_base::floatfield); cout.precision(15); cout << mLL << ")." << endl << endl;

// Check ECME convergence. (2nd Step did not improve 1st Step likelihood.)  If non-convergence, prepare another 1st Step iteration.  If convergence, find/display eigenvalues to check concavity.
//===================================================================================================================================================================================================
if ((mLL - eLL) < pow(10.0,6)*(epsilon)*eLL)
   cnvg=1;

// Free temporary memory allocations.
//=====================================
delete [] gradM; gradM=new Eigen::VectorXd [1];
delete [] hessM; hessM=new Eigen::MatrixXd [1];
delete [] P; P=new Eigen::MatrixXd [1];
delete [] Pinv; Pinv=new Eigen::MatrixXd [1];
delete [] dvarsM; dvarsM=new Eigen::VectorXd [4];
delete [] p1Opt; p1Opt=new Eigen::MatrixXd [1];
for (int h=0; h<nThrds; ++h)
{
    delete [] tmpDvarsM[h]; tmpDvarsM[h]=new Eigen::VectorXd [4];
}

if (0 == cnvg)
{
    // Populate the dnom and fVals arrays.
    //======================================
    LL=getLFVals(T,numA,uCmps,rts,dvarsE[0],muMns,VCMInv,sqdets,picst,dnom,fVals);
    chkSum(T,uCmps,(const long double **) fVals);

    // Build Hessian and Gradient.
    //==============================
    delete [] LHS;  LHS=new Eigen::MatrixXd [1];
    delete [] RHS;  RHS=new Eigen::VectorXd [1];
    delete [] hessE; hessE=new Eigen::MatrixXd [1]; hessE[0]=Eigen::MatrixXd(ld,ld);
    curmlt=getHessE(T,uCmps,(const long double **) fVals,dnom,tmpCMtrx[0],tmpCVctr[0],hessE,LHS,RHS);
    delete [] gradE; gradE=new Eigen::VectorXd [1]; gradE[0]=Eigen::VectorXd(ld);
    getGradE(T,uCmps,(const long double **) fVals,dnom,LHS[0],RHS[0],dvarsE[0],gradE);
```

```
        // Delete temporary memory allocations.
        //======================================
        delete [] tmpCMtrx; tmpCMtrx=new Eigen::MatrixXd [1];
        delete [] tmpCVctr; tmpCVctr=new Eigen::VectorXd [1];
        ecmeItr++;
    }

    // Processor cool down.
    //======================
    if (dbug >= 2)
    {
        cout.setf(ios_base::fixed, ios_base::floatfield); cout.precision(4);
        cout << endl << "***** Processor cool Down: " << ((double) cDown/1000.00)/60.00 << " minutes. *****" << endl;
    } Sleep(cDown);

} while (0 == cnvg);

cout << "ECME Algorithm Converged in " << ecmeItr << " Iterations (Maximum LL=" << LL << ")." << endl;
fout.open(rdir+ofile, ios_base::out | ios_base::app);
   fout << "ECME Algorithm Converged in " << ecmeItr << " Iterations (Maximum LL=" << LL << ")." << endl;
fout.close();

// Delete temporary memory allocations.
//======================================
delete [] dnom; dnom=nullptr;
delete [] sqdets; sqdets=nullptr;
delete [] gradE; gradE=nullptr;
delete [] gradM; gradM=nullptr;
delete [] rts; rts=nullptr;
delete [] VCMInv; VCMInv=nullptr;
delete [] hessE; hessE=nullptr;
delete [] hessM; hessM=nullptr;
delete [] P; P=nullptr;
delete [] Pinv; Pinv=nullptr;
delete [] dvarsE; dvarsE=nullptr;
delete [] dvarsM; dvarsM=nullptr;
delete [] LHS; LHS=nullptr;
delete [] RHS; RHS=nullptr;
delete [] tmpCMtrx; tmpCMtrx=nullptr;
delete [] tmpCVctr; tmpCVctr=nullptr;
delete [] tmpDv; tmpDv=nullptr;
delete [] A; A=nullptr;
delete [] beat; beat=nullptr;
delete [] maxHLL; maxHLL=nullptr;
for (int t=0; t<T; ++t)
{
    delete [] fVals[t]; fVals[t]=nullptr;
}
delete [] fVals; fVals=nullptr;
for (int h=0; h<nThrds; ++h)
{
    delete [] sHess1[h]; sHess1[h]=nullptr;
    delete [] sHess2[h]; sHess2[h]=nullptr;
    delete [] tmpDvarsM[h]; tmpDvarsM[h]=nullptr;
}
delete [] sHess1; sHess1=nullptr;
delete [] sHess2; sHess2=nullptr;
delete [] tmpDvarsM; tmpDvarsM=nullptr;
delete [] t1; t1=nullptr;

// Count and return the final number of non-zero unique cell probabilities for this solution.
//==============================================================================================
return uCmps;
}
```

```
/*
/ Copyright (C) 2016 Chris Rook
/
/ This program is free software: you can redistribute it and/or modify it under the terms of the GNU General Public License as published by the Free Software Foundation, either version 3 of the License,
/ or (at your option) any later version.  This program is distributed in the hope that it will be useful, but WITHOUT ANY WARRANTY; without even the implied warranty of MERCHANTABILITY or FITNESS FOR A
/ PARTICULAR PURPOSE.  See the GNU General Public License for more details: <http://www.gnu.org/licenses/>
/
/ Filename:  getLFVals.cpp
/
/ Function:  getLFVals()
/
/ Summary:
/
/    This function decomposes the multivariate mixture likelihood as a 2-dimensional grid of values with time (T) on the vertical axis and component (U) on the horizontal axis.  Each cell in the 2-
/    dimensional grid is a likelihood value for the data at that timepoint using the corresponding multivariate density function for that component.  This function computes each value in the grid and
/    stores the value in the 2-dimensional array supplied by parameter #11.  In addition, if the values in each row are summed using the component probabilities as weights, the value is the multivariate
/    mixture likelihood at the given time point.  These values are derived and stored in the single array supplied by parameter #10.  Lastly, if the log of the values computed for parameter #10 are taken
/    and summed across all time points then this value is the log-likelihood for the data using the supplied multivariate mixture density.  This log-likelihood value is computed/returned at the call.
/
/ Inputs:
/
/    1.) The total number of time points with data collected.  (T)
/
/    2.) The number of assets with returns.  (numA)
/
/    3.) The number of unique components in the multivariate mixture.  (Initial value is from either the Minimax or Minimum SSD LP optimizations.)  (inUCmps)
/
/    4.) The array of vector returns at each time point indexed as rs[t](a).  There are T vectors of returns and each is of size numA, where numA=# assets.  (rs)
/
/    5.) The current vector of multivariate mixture probabilities.  (Initial values are from either the Minimax or Minimum SSD LP optimizations.)  (uPrbs)
/
/    6.) The array of mean vectors.  Each component of the multivariate mixture density is a multivariate density function which has its own set of means.  The first element of this array is the vector
/        of means for the first multivariate component, etc...  (inMns)
/
/    7.) The array of VC inverse matrices.  Each component in the multivariate mixture density is a multivariate density function with a corresponding VC matrix.  Each VC matrix is invertible and of
/        dimension (numA)x(numA).  The diagonals of each VC matrix are the corresponding variances for that asset within that component.  (inVCIs)
/
/    8.) The array of square roots of the determinants of the VC inverse matrices from parmeter #7.  This term is required to construct the multivariate normal density.  (insqs)
/
/    9.) This parameter equals (2*pi)^(numA/2), where numA = total # of assets in the application.  (inpicst)
/
/   10.) A single array of T values that sum the double array in parameter #11 across the components at each time point, weighting each component by its corresponding estimated probability.  Each value
/        in this array is the multivariate mixture likelihood value for the data at each individual time point.  This parameter is supplied empty and populated/returned by this function.  (denoms)
/
/   11.) A double array of likelihood values indexed by time and component.  At each time point the likelihood for each component is computed and stored in this double array for reuse.  This forms a 2-
/        dimensional grid of values of size TxU, where T=# of time points, and U=# components.  This parameter is supplied empty and populated/returned by this function.  (lfVals)
/
/ Outputs:
/
/    This function returns the log-likelihood value at the call and also populates the 2 incoming arrays denoms (see parameter #10) and lfVals (parameter #11).
/*****************************************************************************************************************************************************************************************************/
#include "stdafx.h"
long double getLFVals(const int T, const int numA, const int inUCmps, const Eigen::VectorXd *rs, const Eigen::VectorXd uPrbs, const Eigen::VectorXd *inMns, const Eigen::MatrixXd *inVCIs,
                      const long double *insqs, const long double inpicst, long double *denoms, long double **lfVals)
{
    // Local variables.
    //=================
    long double LL=0.00;

    // Populate 2 containers:  1.) 2-dimensional grid of all component likelihoods evaluated at each time point.  2.) 1-dimensional array of all full likelihood values evaluated at each time point.
    //==================================================================================================================================================================================================
    for (int t=0; t<T; ++t)
    {
        denoms[t]=0.00;
        for (int v=0; v<inUCmps; ++v)
```

```
        {
            lfVals[t][v]=getMVNDens(rs[t],inMns[v],inVCIs[v],insqs[v],inpicst);
            denoms[t]=denoms[t] + uPrbs(v)*lfVals[t][v];
        }
        if (denoms[t] > 0.00)
           LL=LL+log(denoms[t]);
        else
           LL=LL+LNegVal;
    }
    // Return the log-likelihood value.
    //==================================
    return LL;
}

/*
/ Copyright (C) 2016 Chris Rook
/
/ This program is free software: you can redistribute it and/or modify it under the terms of the GNU General Public License as published by the Free Software Foundation, either version 3 of the License,
/ or (at your option) any later version.  This program is distributed in the hope that it will be useful, but WITHOUT ANY WARRANTY; without even the implied warranty of MERCHANTABILITY or FITNESS FOR A
/ PARTICULAR PURPOSE.  See the GNU General Public License for more details: <http://www.gnu.org/licenses/>
/
/ Filename:  getGradE.cpp
/
/ Function:  getGradE()
/
/ Summary:
/
/    This function derives the Gradient for the ECME algorithm 1st Step optimization, which is convex in the decision variables (multivariate mixture component probabilities).  All means, variances and
/    covariances are treated as constants.  The objective is to maximize the corresponding log-likelihood function subject to constraints that the marginal densities are fixed and known univariate
/    mixtures.  The marginal constraints can be enforced via linear functions on the decision variables (component probabilities).  By incorporating the constraints into the objective we form the
/    Lagrangian.  Stationary points of the Lagrangian will be unique global optimizers of the constrainted convex optimization problem.  These points are found by applying Newton's method to the
/    Lagrangian.  Newton's method requires that the gradient and Hessian of the Lagrangian be constructed during each iteration.  The optimization problem converges when the log-likelihood fails to
/    improve.  In this function we compute the Gradient of the Lagrangian for the 1st Step.  The Gradient is the vector of 1st partial derivatives of the Lagrangian.  The number of elements is the sum of
/    the # of unique components (multivariate mixture probabilities) and the # of constraints (Lagrange multipliers).
/
/ Inputs:
/
/    1.) The total number of time points with data collected.  (T)
/
/    2.) The number of unique components in the multivariate mixture.  (Initial value is from either the Minimax or Minimum SSD LP optimizations.)  (inUCmps)
/
/    3.) A double array of likelihood values indexed by time and component.  At each time point the likelihood for each component is computed and stored for reuse.  This forms a 2-dimensional grid of
/        values of size TxU, where T=# of time points, and U=# components.  (infVals)
/
/    4.) A single array of T values that sum the double array in parameter #3 across the components at each time point, weighting each likelihood value by its corresponding estimated component
/        probability.  Therefore, each value in this array is the multivariate mixture likelihood value for the data at each individual time point.  (inDNoms)
/
/    5.) The LHS matrix needed to enforce the marginal mixture density constraints.  These constraints are linear in the component probabilities therefore can be represented using a LHS matrix and RHS
/        vector.  This matrix has 1's & 0's.  It may be necessary to multiply both sides of each constraint by a constant factor to make the corresponding Hessian full rank (computationally).  The # of
/        columns is equal to the # of components and the # of rows is equal to the # of constraints (i.e., Lagrange multipliers) needed to maintain the marginal univariate mixtures (inLHS).
/
/    6.) The RHS constraint vector required to enforce the fixed marginal density constraints using actual univariate marginal mixture probabilities.  The constraints needed to ensure that given fixed
/        mixture marginals add multivariate component probabilities up to equal the given marginal mixture probabilities for each component of each asset.  This vector is scaled when the corresponding
/        LHS matrix above is scaled to ensure the Hessian of the Lagrangian is full rank (computationally).  (inRHS)
/
/    7.) Current values of the decision variables.  Values for the probabilities are not needed given that we have the double array infVals[t][u] above.  However we do need the current values of the
/        Lagrange multipliers as these change during each iteration.  Therefore we will pull these from the vector of all decision variables passed via this parameter.  (inDVars)
/
/    8.) The empty Gradient vector to be filled by this function.  The vector is of dimension equal to the sum of the # of unique components and the # of Lagrange multipliers.  The # of Lagrange
/        multipliers equals the # of rows in the LHS constraint matrix, which equals the # of constraints.  (inGrad)
/
/ Outputs:
/
/    This function populates the empty gradient vector supplied to it but does not return any other output at the function call.
/**********************************************************************************************************************************************************************************************************/
```

```
#include "stdafx.h"
void getGradE(const int T, const int inUCmps, const long double **infVals, const long double *inDNoms, const Eigen::MatrixXd inLHS, const Eigen::VectorXd inRHS, const Eigen::VectorXd inDVars,
              Eigen::VectorXd *inGrad)
{
    // Local variables.
    //===================
    int ld=inUCmps + (int) inRHS.size();

    // Gradient:  Partials wrt Probabilities.
    //========================================
    for (int v=0; v<inUCmps; ++v)
    {
        inGrad[0](v)=0.00;
        for (int t=0; t<T; ++t)
            inGrad[0](v)=inGrad[0](v) + infVals[t][v]/inDNoms[t];
        for (int r=0; r<(int) inRHS.size(); ++r)
            inGrad[0](v)=inGrad[0](v) - inDVars(inUCmps+r)*inLHS(r,v);
    }

    // Gradient:  Partials wrt Multipliers.
    //======================================
    for (int m=inUCmps; m<ld; ++m)
    {
        inGrad[0](m)=0.00;
        for (int v=0; v<inUCmps; ++v)
            inGrad[0](m)=inGrad[0](m) + ((double) inLHS(m-inUCmps,v))*inDVars(v);
        inGrad[0](m)=-1.00*(inGrad[0](m) - inRHS(m-inUCmps));
    }
}

/*
/ Copyright (C) 2016 Chris Rook
/
/ This program is free software: you can redistribute it and/or modify it under the terms of the GNU General Public License as published by the Free Software Foundation, either version 3 of the License,
/ or (at your option) any later version.  This program is distributed in the hope that it will be useful, but WITHOUT ANY WARRANTY; without even the implied warranty of MERCHANTABILITY or FITNESS FOR A
/ PARTICULAR PURPOSE.  See the GNU General Public License for more details: <http://www.gnu.org/licenses/>
/
/ Filename:  getHessE.cpp
/
/ Function:  getHessE()
/
/ Summary:
/
/    This function derives the Hessian for the ECME algorithm 1st Step optimization, which is convex in the decision variables (multivariate mixture component probabilities).  All means, variances and
/    covariances are treated as constants during this optimization (i.e., ECME 1st Step).  The objective is to maximize the corresponding log-likelihood function subject to constraints that the marginal
/    densities are fixed and known univariate mixtures.  The marginal constraints can be enforced via linear functions on the decision variables (component probabilities).  By incorporating the
/    constraints into the objective we form the Lagrangian.  Stationary points of the Lagrangian will be unique global optimizers of the constrained convex optimization problem.  These points are found
/    by applying Newton's method to the Lagrangian.  Newton's method requires that the gradient and Hessian of the Lagrangian be constructed during each iteration.  The optimization problem converges
/    when the log-likelihood fails to improve (i.e., at a zero of the Lagrangian).  In this function we compute the Hessian of the Lagrangian for the 1st Step.  The Hessian is a border matrix since the 2nd
/    derivative WRT the Lagrange multipliers is always zero, therefore there will be a block matrix of zeros in the lower right corner.  A border matrix is invertible under certain conditions on the 3
/    block matrices that border the zero block.  These conditions will be met for this optimization, however it may be necessary to inflate the constraint matrix by using a constant larger than 1.  This
/    would be needed when the Hessian is ill-conditioned using 0/1 indicator variables to enforce the constraints.
/
/ Inputs:
/
/    1.) The total number of time points with data collected.  (T)
/
/    2.) The number of unique components in the multivariate mixture.  (Initial values are from either the Minimax or Minimum SSD LP optimizations.)  (nUCmps)
/
/    3.) A double array of likelihood values indexed by time and component.  At each time point the likelihood for each component is computed and stored for reuse.  This forms a 2-dimensional grid of
/        values of size TxU, where T=# of time points, and U=# components.  (infVals)
/
/    4.) A single array of T values that sum the double array in parameter #3 across the components at each time point, weighting each component by its corresponding estimated probability.  Therefore,
/        each value in this array is the multivariate mixture likelihood value for the data at each individual time point.  (inDNoms)
/
```

```
/    5.) The LHS matrix (of 1's and 0's) needed to enforce the marginal mixture density constraints.  These constraints are linear in the decision variables (component probabilities) therefore can be
/        represented using a LHS matrix and RHS vector.  This matrix has 1's and 0's but these may be multiplied by a constant to ensure the Hessian returned is full rank (computationally). The # of
/        columns is equal to the # of components and the # of rows is equal to the # of constraints (Lagrange multipliers) that are required to ensure that the marginals match their fixed mixtures as
/        found earlier.  The matrix must be of full rank, therefore if multivariate components are set to zero we should check that it remains full rank, if not force it to be full rank by removing rows
/        one at a time until it is.  (This code is yet to be implemented.  Problem has not been encountered.  The function getCMtrx() can be used to perform the task.)  (inCMtrx)
/
/    6.) The RHS constraint vector required to enforce the fixed marginal density constraints using actual marginal probabilities.  The constraints needed to ensure given fixed mixture marginals add
/        multivariate component probabilities up to equal the given marginal mixture probabilities for each component of each asset.  This vector is scaled when the corresponding LHS matrix above is
/        scaled to ensure the Hessian of the Lagrangian is invertible.  (inCVctr)
/
/    7.) The empty Hessian matrix to be filled by this function.  The matrix is square with dimension equal to the # of unique components plus the # of Lagrange multipliers.  The # of Lagrange
/        multipliers equals the # of rows in the LHS constraint matrix, which equals the # of constraints.  (inHess)
/
/    8.) An empty matrix to be filled with the updated LHS constraint matrix once scaled to ensure the resulting Hessian is invertible.  As noted, the Hessian is for the Lagrangian which is a border
/        matrix with a block of zeros in the lower right corner.  The upper right corner is the constraint matrix of 1's and 0's, while the upper left corner is the Hessian of the original objective
/        function (i.e., without the Lagrange multipliers).  In rare cases, large values in the upper left matrix coupled with 1's and 0's in the upper right matrix can cause the matrix to be ill-
/        conditioned, therefore not invertible.  We have found that a solution is to scale the constraint matrix up by a large constant.  That is, we multiple the LHS and RHS of each constraint by a
/        given large constant.  This fixes the singularity of the Hessian.  Note that the gradient also uses the constraints therefore when the constraint matrix is scaled we must use the same scaled
/        version when constructing the gradient.  This parameter returns the scaled LHS constraint matrix.  Note that the scale factor is returned by the function.  (inLHS)
/
/    9.) An empty vector to be filled with the updated RHS constraint values when the constraint matrix is scaled to be invertible.  (inRHS)
/
/ Outputs:
/
/    This function returns the the multiplier used to scale the constraint matrix and vector to ensure that the resulting Hessian is computationally invertible.  It also populates the empty Hessian
/    matrix supplied to it along with the (scaled) LHS matrix and RHS vector.
/*************************************************************************************************************************************************************************************/
#include "stdafx.h"
long double getHessE(const int T, const int inUCmps, const long double **infVals, const long double *inDNoms, const Eigen::MatrixXd inCMtrx, const Eigen::VectorXd inCVctr, Eigen::MatrixXd *inHess,
                     Eigen::MatrixXd *inLHS, Eigen::VectorXd *inRHS)
{
    // Local variables.
    //===================
    int rnk, ld=inUCmps + (int) inCVctr.size();
    long double mult=1.00;
    Eigen::MatrixXd *ulHess=new Eigen::MatrixXd [1];
    ulHess[0]=Eigen::MatrixXd(inUCmps,inUCmps);

    // Hessian:  Upper left.
    //========================
    for (int r=0; r<inUCmps; ++r)
        for (int c=r; c<inUCmps; ++c)
        {
            inHess[0](r,c)=0.00;
            for (int t=0; t<T; ++t)
                inHess[0](r,c) = inHess[0](r,c) + infVals[t][r]*infVals[t][c]/pow(inDNoms[t],2);
            inHess[0](r,c)=-1.00*inHess[0](r,c);
            ulHess[0](r,c)=inHess[0](r,c);
            if (c > r)
            {
                inHess[0](c,r)=inHess[0](r,c);
                ulHess[0](c,r)=ulHess[0](r,c);
            }
        }

    // Before proceeding with the UR, LL, and LR sections, check that the UL Hessian is full rank and put out a warning if it is not.
    // (This may or may not prevent the optimization from working.  Often it does not prevent the optimization from working.)
    //==============================================================================================================
    rnk=(int) Eigen::FullPivLU<Eigen::MatrixXd>(ulHess[0]).rank();
    if (rnk < inUCmps && dbug >=2)
       cout << endl << "WARNING:  The UL Hessian matrix is singular which may prevent the component probabilities from being optimized." << endl
                    << "          This can happen for various reasons, two of which are:" << endl
                    << "             1.) The likelihood of a single component is zero at all time points, which eliminates the decision variable." << endl
                    << "             2.) The Upper Left matrix is ill-conditioned having large and small elements at different diagonal positions." << endl
                    << "Message from ... getHessE() ..." << endl;
```

```
// Hessian:  Build upper right, lower left, and lower right sections of the Hessian.  Iterate until the entire Hessian is full rank so that Newton's method may be applied.  This may require
//    multiplying all constraints by a constant (both LHS and RHS). First, make sure that the constraint matrix is of full rank since components may be dropped during the 1st Step.
//==================================================================================================================================================================================
rnk=(int) Eigen::FullPivLU<Eigen::MatrixXd>(inCMtrx).rank();
if (rnk < (int) inCVctr.size())
{
    cout << "ERROR:  Detection of a non-full rank constraint matrix rank during the ECME 1st Step, which is likely due to" << endl
         << "        components with < 0 probabilities being dropped.  The function getCMtrx() can be used to fix this" << endl
         << "        by sequentially removing linearly dependent rows until the constraint matrix becomes full rank." << endl
         << "EXITING...getHessE()..." << endl; cin.get(); exit (EXIT_FAILURE);
}
inLHS[0]=Eigen::MatrixXd((int) inCVctr.size(),inUCmps); inLHS[0]=mult*inCMtrx;
inRHS[0]=Eigen::VectorXd((int) inCVctr.size());         inRHS[0]=mult*inCVctr;
do
{
    // Hessian:  Upper right & Lower left.
    //=====================================
    for (int r=0; r<inUCmps; ++r)
        for (int c=inUCmps; c<ld; ++c)
        {
            inHess[0](r,c)=-1.00*inLHS[0](c-inUCmps,r);
            inHess[0](c,r)=inHess[0](r,c);
        }

    // Hessian:  Lower right.
    //========================
    for (int r=inUCmps; r<ld; ++r)
        for (int c=r; c<ld; ++c)
        {
            inHess[0](r,c)=0.00;
            if (c > r)
               inHess[0](c,r)=inHess[0](r,c);
        }

    // Before proceeding, ensure that the entire Hessian is full rank.
    //=================================================================
    rnk=(int) Eigen::FullPivLU<Eigen::MatrixXd>(inHess[0]).rank();

    // If another iteration is needed, update the multiplier and the LHS/RHS constraints.
    //===================================================================================
    if (rnk < ld)
    {
        mult=mult*10.00;
        inRHS[0]=mult*inRHS[0];
        inLHS[0]=mult*inLHS[0];
    }

} while (rnk < ld && mult < LPosVal);

// If the entire Hessian is not full rank then put out a warning but do not exit, the optimization may still succeed.
//===================================================================================================================
if (rnk < ld && dbug >=2)
{
    cout << endl << "WARNING:  The 1st Step Hessian matrix is singular which may prevent the multivariate component probabilities from being optimized." << endl
                 << "          This can happen for various reasons, including:" << endl
                 << "             1.) There are very large and very small eigenvalues, resulting in an ill-conditioned matrix." << endl
                 << "             2.) Components have been dropped and the constraint matrix is no longer of full row rank.  (Note:  Already checked above.)" << endl
                 << "Message from ... getHessE() ..." << endl;
}

// Free temporary memory allocations.
//====================================
delete [] ulHess; ulHess=nullptr;

// Return the multiplier used to correct an ill-conditioned Hessian.
//===================================================================
```

```
    return mult;
}

/*
/ Copyright (C) 2016 Chris Rook
/
/ This program is free software: you can redistribute it and/or modify it under the terms of the GNU General Public License as published by the Free Software Foundation, either version 3 of the License,
/ or (at your option) any later version.  This program is distributed in the hope that it will be useful, but WITHOUT ANY WARRANTY; without even the implied warranty of MERCHANTABILITY or FITNESS FOR A
/ PARTICULAR PURPOSE.  See the GNU General Public License for more details: <http://www.gnu.org/licenses/>
/
/ Filename:  getGradM.cpp
/
/ Function:  getGradM()
/
/ Summary:
/
/    This function derives the Gradient for the ECME algorithm 2nd Step optimization, which is (evidently) NOT convex in the decision variables (covariances).  In general mixture density likelihoods are
/    not concave functions and have many local optimums.  We are dealing with a multivariate mixture density here.  All means, variances and component probabilities are treated as constants during this
/    ECME 2nd Step optimization.  The objective is to maximize the corresponding log-likelihood function (only covariances are unknown) subject to constraints that all variance-covariance matrices are
/    positive definite.  That is, we seek the covariances that maximize the multivariate mixture log-likelihood function with the variance-covariance matrix at each component being positive definite
/    (and all means, variances, component probabilities being fixed and known).  These points are found by applying a modified Newton's method to the log-likelihood (in which only the covariances are
/    unknown).  Newton's method requires that the gradient and Hessian of the log-likelihood be constructed during each iteration.  The gradient is the vector of first order partial derivatives WRT each
/    covariance term, and is derived in this function.  The Hessian is the matrix of second order partial derivatives WRT all covariance terms.  If the problem has "A" total assets with returns measured,
/    and "U" components in the multivariate mixture density then there will be a total of U*A*(A-1)/2 unique covariance terms that require estimation.  Clearly this problem suffers from the curse of
/    dimensionality and will work best with a limited number of assets relative to the number of observations (i.e., time points).  The optimization problem converges when the log-likelihood fails to
/    improve.  The method used to find the largest local optimum is due to Marquardt (1963), which uses the Hessian to step in the general direction of the gradient by adding a constant to the Hessian
/    diagonals prior to solving the updating equation.  In practice we will iterate over a large number of random step sizes searching for the local/global log-likelihood function maximizer.  Once the
/    maximizer is found we recompute the gradient and Hessian and iterate again.  Note that large additive quantities (i.e., added to the Hessian diagonal) translate into small steps and small quantities
/    translate into large steps.  An additive factor of zero translates into using Newton's method without modification, which is assumed here to overshoot the local optimizer.  This method is approprate
/    when strictly applying Newton's method overshoots.  Here, the goal is to find the largest local optimum in the vicinity of the carefully constructed starting point (i.e., LP solution), but also to
/    search outside the current region/hill in an attempt to find a better solution.  The constraints on the resulting variance-covariance matrices are enforced implictly.  At each step, the resulting
/    matrix is decomposed and the eigenvalues are inspected.  If none are <= 0 the resulting variance-covariance matrix is positive definite.  Otherwise, it is not and a ridge repair is immediately
/    performed.  Stepping continues using the repaired variance-covariance matrix.  In general, we find that there are large regions of the covariance set where stepping proceeds without the need for
/    repairs, and other large regions of the covariance set where repairs are needed after each step.  The variance-covariance matrix of each multivariate component is examined and repaired (if
/    necessary) by the function ridgeRpr().  The feasible region is any covariance set that results in all component variance-covariance matrices being valid (i.e., positive definite).
/
/ Inputs:
/
/    1.) The total number of time points with data collected.  (T)
/
/    2.) The array of vector returns at each time point indexed as rs[t](a).  There are T vectors of returns and each is of size numA, where numA=# assets.  (rs)
/
/    3.) The number of unique components in the multivariate mixture.  (Initial value is from either the Minimax or Minimum SSD LP optimizations.)  (inUCmps)
/
/    4.) The number of assets with returns collected.  (numA)
/
/    5.) A double array of likelihood values indexed by time and component.  At each time point the likelihood for each component is computed and stored for reuse.  This forms a 2-dimensional grid of
/        values of size TxU, where T=# of time points, and U=# components.  (infVals)
/
/    6.) A single array of T values that sum the double array in parameter #5 across the components at each time point, weighting each likelihood value by its corresponding estimated component
/        probability.  Therefore, each value in this array is the multivariate mixture likelihood value for the data at each individual time point.  (inDNoms)
/
/    7.) The current array of multivariate mixture probabilities.  (Initial values from either the Minimax or Minimum SSD LP optimizations.)  (inPrbs)
/
/    8.) The array of mean vectors.  Each component of the multivariate mixture density is a multivariate density function which has its own set of means.  The first element of this array is the vector
/        of means for the first multivariate component, etc...  (muMns)
/
/    9.) The array of VC matrices.  Each component in the multivariate mixture density is a multivariate density function with a corresponding VC matrix.  Each VC matrix is of dimension (numA)x(numA).
/        The diagonals of each VC matrix are the corresponding variances for that asset within that component.  (E)
/
/   10.) The array of VC matrix inverses.  Each component in the multivariate mixture density is a multivariate density function with a corresponding VC matrix.  Each VC matrix is of dimension
/        (numA)x(numA).  The diagonals of each VC matrix are the corresponding variances for that asset within that component.  This parameter holds the corresponding array of inverses of the VC
/        matrices.  (Einv)
/
/   11.) The array of identifier matrices for each covariance term.  A VC matrix can be decomposed into the sum of a matrix of diagonal elements and a term for each unique covariance.  The constant
```

```
/        matrix multiplied by the covariance term has a 1/0 in each element where 1 is in the location of the corresponding covariance term.  The constant matrices are contained in this array.  The
/        constant matrices are identical across components and can be reused.  The # of these matrices is the number of unique covariance terms for a single multivariate mixture: numA*(numA-1)/2.  (inA)
/
/   12.) The empty Gradient vector to be filled by this function.  The vector is of dimension equal to the total # of covariances in the problem.  If the problem has "A" total assets with returns
/        measured, and "U" components in the multivariate mixture density then there are a total of U*A*(A-1)/2 covariance terms.  (inGrad)
/
/ Outputs:
/
/    This function populates the empty gradient vector supplied to it but does not return any other output at the function call.
/*********************************************************************************************************************************************************************************************************/
#include "stdafx.h"
void getGradM(const int T, Eigen::VectorXd *rs, const int inUCmps, const int numA, const long double **infVals, const long double *inDNoms, const long double *inPrbs, Eigen::VectorXd *muMns, const
              Eigen::MatrixXd *E, const Eigen::MatrixXd *Einv, const Eigen::MatrixXd *inA, Eigen::VectorXd *inGrad, int chk)
{
    // Local variables.
    //===================
    int itr=0, itrA;
    long double Qtijk;
    Eigen::MatrixXd *Ejk=new Eigen::MatrixXd [1];
    Ejk[0]=Eigen::MatrixXd(numA,numA);

    // Populate gradient vector.
    //============================
    for (int i=0; i<inUCmps; ++i)
    {
        itrA=0;
        for (int j=0; j<numA; ++j)
            for (int k=j+1; k<numA; ++k)
            {
                getCofM(numA,j,k,E[i],Ejk);
                inGrad[0](itr)=0.00;
                for (int t=0; t<T; ++t)
                {
                    Qtijk = (0.50)*((rs[t]-muMns[i]).transpose())*Einv[i]*inA[itrA]*Einv[i]*(rs[t]-muMns[i]) - Ejk[0].determinant()/E[i].determinant();
                    inGrad[0](itr)=inGrad[0](itr) + (inPrbs[i]*infVals[t][i]*Qtijk)/inDNoms[t];
                }
                itr++;
                itrA++;
            }
    }

    // Delete temporary memory allocations.
    //=========================================
    delete [] Ejk; Ejk=nullptr;
}

/*
/ Copyright (C) 2016 Chris Rook
/
/ This program is free software: you can redistribute it and/or modify it under the terms of the GNU General Public License as published by the Free Software Foundation, either version 3 of the License,
/ or (at your option) any later version.  This program is distributed in the hope that it will be useful, but WITHOUT ANY WARRANTY; without even the implied warranty of MERCHANTABILITY or FITNESS FOR A
/ PARTICULAR PURPOSE.  See the GNU General Public License for more details: <http://www.gnu.org/licenses/>
/
/ Filename:  getHessM.cpp
/
/ Function:  getHessM()
/
/ Summary:
/
/    This function derives the Hessian matrix for the ECME algorithm 2nd Step optimization, which is (evidently) NOT convex in the decision variables (covariances).  In general a univariate/multivariate
/    mixture density likelihood function is not strictly concave and has many local optimums.  We are dealing with a multivariate mixture density here where all means, variances and component
/    probabilities fixed during the ECME 2nd Step optimization.  The objective is to maximize the corresponding log-likelihood function (only covariances are unknown) subject to constraints that all
/    resulting variance-covariance matrices are positive definite.  That is, we seek the covariances that maximize the multivariate mixture log-likelihood function with the variance-covariance matrix at
/    each component being positive definite (and all means, variances, component probabilities are fixed).  These points are found by applying a modified Newton's method to the log-likelihood (in which
/    only the covariances are unknown).  Newton's method requires that the gradient and Hessian of the log-likelihood be constructed during each iteration.  The gradient is the vector of first order
```

```
/    partial derivatives WRT each covariance term, and is derived in the function getGradM().  The Hessian is the matrix of second order partial derivatives WRT all covariance terms and is derived in
/    this function.  If the problem has "A" total assets with returns measured, and "U" components in the multivariate mixture density then there will be a total of U*A*(A-1)/2 unique covariance terms
/    that require estimation.  This problem suffers from the curse of dimensionality and will work best with a limited number of assets relative to the # of observations (i.e., time points).  The
/    optimization problem converges when the log-likelihood fails to improve.  The method used to find the largest local optimum is due to Marquardt (1963), which uses the Hessian to step in the general
/    direction of the gradient by adding a constant to the Hessian diagonals prior to solving the updating equation.  In practice we will iterate over a large number of random step sizes searching for
/    the best local and global log-likelihood function maximizer.  Once the maximizer is found we recompute the gradient and Hessian and iterate again.  Note that large additive quantities translate into
/    small steps and small quantities translate into large steps.  An additive factor of zero translates into using Newton's method without modification. This method is appropriate when strictly applying
/    Newton's method overshoots.  Here, the goal is to find the largest local optimum in the vicinity of the carefully constructed starting point (LP solution) using small step sizes, but also to search
/    outside the current region/hill in an attempt to find a better solution (using large step sizes).  The constraints on the resulting variance-covariance matrices are enforced implictly.  At each
/    step, the resulting matrix is decomposed and the eigenvalues are inspected.  If none are <= 0 the resulting variance-covariance matrix is positive definite.  Otherwise, it is not and a ridge repair
/    is immediately performed and stepping continues using the repaired variance-covariance matrix.  In general, we find that there are large regions of the covariance set where stepping proceeds without
/    the need for repairs, and other large regions of the covariance set where repairs are needed after each step.  The variance-covariance matrix of each multivariate component is examined and repaired
/    (if necessary) by the function ridgeRpr().  The feasible region is any covariance set that results in all component variance-covariance matrices being valid (i.e., positive definite).
/
/ Inputs:
/
/    1.) The total number of time points with data collected.  (T)
/
/    2.) The array of vector returns at each time point indexed as rs[t](a).  There are T vectors of returns and each is of size numA, where numA=# assets.  (rs)
/
/    3.) The number of unique components in the multivariate mixture.  (Initial value is from either the Minimax or Minimum SSD LP optimizations.)  (inUCmps)
/
/    4.) The number of assets with returns collected.  (numA)
/
/    5.) A double array of likelihood values indexed by time and component.  At each time point the likelihood for each component is computed and stored for reuse.  This forms a 2-dimensional grid of
/        values of size TxU, where T=# of time points, and U=# components.  (infVals)
/
/    6.) A single array of T values that sum the double array in parameter #5 across the components at each time point, weighting each likelihood value by its corresponding estimated component
/        probability.  Therefore, each value in this array is the multivariate mixture likelihood value for the data at each individual time point.  (inDNoms)
/
/    7.) The current array of multivariate mixture probabilities.  (Initial values from either the Minimax or Minimum SSD LP optimizations.)  (inPrbs)
/
/    8.) The array of mean vectors.  Each component of the multivariate mixture density is a multivariate density function which has its own set of means.  The first element of this array is the vector
/        of means for the first multivariate component, etc...  (muMns)
/
/    9.) The array of VC matrices.  Each component in the multivariate mixture density is a multivariate density function with a corresponding VC matrix.  Each VC matrix is of dimension (numA)x(numA).
/        The diagonals of each VC matrix are the corresponding variances for that asset within that component.  (E)
/
/   10.) The array of VC matrix inverses.  Each component in the multivariate mixture density is a multivariate density function with a corresponding VC matrix.  Each VC matrix is of dimension
/        (numA)x(numA).  The diagonals of each VC matrix are the corresponding variances for that asset within that component.  This parameter holds the inverses of the VC matrices.  (Einv)
/
/   11.) The array of identifier matrices for each covariance term.  A VC matrix can be decomposed into the sum of a matrix of diagonal elements and a term for each unique covariance.  The constant
/        matrix multiplied by the covariance term has a 1/0 in each element where 1 is in the location of the corresponding covariance term.  The constant matrices are contained in this array.  The
/        constant matrices are identical across components and can be reused.  The # of these matrices is the number of unique covariance terms for a single multivariate mixture: numA*(numA-1)/2.  (inA)
/
/   12.) The empty Hessian matrix to be filled by this function.  The matrix is square and of dimension equal to the total # of covariances in the problem.  If the problem has "A" total assets with
/        returns measured, and "U" components in the multivariate mixture density then there are a total of U*A*(A-1)/2 covariance terms.  (inHess)
/
/ Internal Variables:
/
/   Note:  Each Hessian element is a partial WRT Sigma(i,j,k), then another partial WRT Sigma(p,r,s).  Where i is the component index for the 1st partial derivative, and p is the component index for the
/          2nd partial derivative.  The paired index (j,k) identifies the covariance term from component i, whereas the paired index (r,s) identifies the covariance term from component p.  Note that the
/          covariance term at (j,k) is equivalent to the covariance term at (k,j) therefore we will also assume that j < k, and r < s.
/
/          itrAjk:     Index of the indicator matrix array A[] for the 1st covariance term in the partial.
/          itrArs:     Index of the indicator matrix array A[] for the 2nd covariance term in the partial.
/          fti:        Product of the likelihood value for the observations at time t using density for component i (of C) and the corresponding component probability.
/          fti_p:      Partial derivative of fti (defined above) WRT a covariance term from component p (of C) where (p == i).  [Note that fti is just a constant when (p != i) since it does not contain
/                      the 2nd covariance term in its function.]
/          ftp:        Product of the likelihood value for the observations at time t using density for component p (of C) and the corresponding component probability.
/          Qtijk:      This is the extra term that arises in the numerator when differentiating the density of component i WRT the covariance term (j,k).
/          Qtprs:      This is the extra term that arises in the numerator when differentiating the density of component p WRT the covariance term (r,s).
/          Qtijk_prs:  Partial derivative of Qtijk WRT Sigma(p,r,s).
/          gt:         Overall likelihood of all data points using the full multivariate mixture density.
/          gt_prs:     Partial derivative of gt WRT Sigma(p,r,s).
```

```
/   ---->  Note:       If there are n assets then there are n(n-1)/2 distinct covariance terms within each component and C*n*(n-1)/2 total distinct covariance terms.
/
/ Outputs:
/
/   This function populates the empty Hessian matrix supplied.  The magnitude of the largest element is returned at the call and used to help determine the best step size.
/*************************************************************************************************************************************************************************************/
#include "stdafx.h"
long double getHessM(const int T, Eigen::VectorXd *rs, const int inUCmps, const int numA, const long double **infVals, const long double *inDNoms, const long double *inPrbs, const Eigen::VectorXd *muMns,
                     const Eigen::MatrixXd *E, const Eigen::MatrixXd *Einv, const Eigen::MatrixXd *inA, Eigen::MatrixXd *inHess)
{
    // Local variables.
    //===================
    int hr=0, hc, itrAjk, itrArs;
    long double fti, fti_p, ftp, Qtijk, Qtprs, Qtijk_prs, gt, gt_prs, q1, q2, q3, maxmag=0.00;
    Eigen::MatrixXd *Eijk=new Eigen::MatrixXd [1], *Eprs=new Eigen::MatrixXd [1], *Ejkrs=new Eigen::MatrixXd [1], *Ejksr=new Eigen::MatrixXd [1];
    Eijk[0]=Eigen::MatrixXd(numA,numA); Eprs[0]=Eigen::MatrixXd(numA,numA); Ejkrs[0]=Eigen::MatrixXd(numA,numA); Ejksr[0]=Eigen::MatrixXd(numA,numA);
    Eigen::MatrixXd q4=Eigen::MatrixXd(numA,numA);

    // Populate Hessian matrix.
    //===========================
    for (int i=0; i<inUCmps; ++i)
    {
        itrAjk=0;
        for (int j=0; j<numA; ++j)
            for (int k=j+1; k<numA; ++k)                // 1st covariance term fixed:  Cov(i,j,k)
            {
                hc=0;                                   // Hessian column entry indicator (0 thru nCovs-1).
                getCofM(numA,j,k,E[i],Eijk);            // Build Eijk.
                for (int p=0; p<inUCmps; ++p)
                {
                    itrArs=0;
                    for (int r=0; r<numA; ++r)
                        for (int s=r+1; s<numA; ++s)    // 2nd covariance term fixed:  Cov(p,r,s)
                        {
                            if (hr <= hc)
                            {
                                // Unconditional quantities that are not functions of time.
                                //==========================================================
                                getCofM(numA,r,s,E[p],Eprs);                        // Build Ers.

                                // Conditional quantities that are not functions of time.
                                //========================================================
                                if (i == p)    /* 2nd covariance term is from the same component. */
                                {
                                    getCofM(numA,r,s,Eijk[0],Ejkrs);                // Build Ejkrs, conditionally.
                                    getCofM(numA,s,r,Eijk[0],Ejksr);                // Build Ejksr, conditionally.
                                    q1=2.00*(Eijk[0].determinant())*(Eprs[0].determinant())/pow(E[i].determinant(),2);
                                    q2=(Ejksr[0].determinant())/(E[i].determinant());
                                    q3=(Ejkrs[0].determinant())/(E[i].determinant());
                                    q4=Einv[i]*inA[itrAjk]*Einv[i]*inA[itrArs]*Einv[i] + Einv[i]*inA[itrArs]*Einv[i]*inA[itrAjk]*Einv[i];
                                }

                                // Initialize the element.
                                //==========================
                                inHess[0](hr,hc)=0.00;

                                // Iterate over the time dimension.
                                //===================================
                                for (int t=0; t<T; ++t)
                                {
                                    // Derive unconditional quantities that are functions of time.
                                    //==============================================================
                                    fti=inPrbs[i]*infVals[t][i];
                                    Qtijk=(0.50)*((rs[t]-muMns[i]).transpose())*Einv[i]*inA[itrAjk]*Einv[i]*(rs[t]-muMns[i]) - Eijk[0].determinant()/E[i].determinant();
                                    ftp=inPrbs[p]*infVals[t][p];
```

```
                                        Qtprs=(0.50)*((rs[t]-muMns[p]).transpose())*Einv[p]*inA[itrArs]*Einv[p]*(rs[t]-muMns[p]) - Eprs[0].determinant()/E[p].determinant();
                                        gt=inDNoms[t];
                                        // Derive conditional quantities that are functions of time, including the Hessian value itself.
                                        //==================================================================================================
                                        if (i == p)     // 2nd correlation term is from same component.
                                        {
                                            fti_p=fti*Qtprs;
                                            gt_prs=ftp*Qtprs;
                                            if (j == r || k == s)     // 2nd partial WRT term below diagonal only.
                                               Qtijk_prs=q1 - q2 - 0.50*((rs[t]-muMns[i]).transpose())*(q4)*(rs[t]-muMns[i]);
                                            else if (k == r)          // 2nd partial WRT term above diagonal only.
                                               Qtijk_prs=q1 - q3 - 0.50*((rs[t]-muMns[i]).transpose())*(q4)*(rs[t]-muMns[i]);
                                            else                      // 2nd partial WRT term above and below diagonal.
                                               Qtijk_prs=q1 - (q2+q3) - 0.50*((rs[t]-muMns[i]).transpose())*(q4)*(rs[t]-muMns[i]);
                                            inHess[0](hr,hc)=inHess[0](hr,hc) + (gt*(fti*Qtijk_prs + fti_p*Qtijk) - (fti*Qtijk)*gt_prs)/pow(gt,2);
                                        }
                                        else
                                            inHess[0](hr,hc)=inHess[0](hr,hc) - (fti*Qtijk*ftp*Qtprs)/pow(gt,2);
                                    }

                                    // Populate corresponding element below the diagonal.
                                    //====================================================
                                    if (hr < hc)
                                       inHess[0](hc,hr)=inHess[0](hr,hc);
                                }
                                hc++;          // Hessian column entry indicator (0 thru nCovs-1).
                                itrArs++;
                            }
                    }
                    hr++;
                    itrAjk++;
                }
        }

        // Delete temporary memory allocations.
        //========================================
        delete [] Eijk; Eijk=nullptr;
        delete [] Eprs; Eprs=nullptr;
        delete [] Ejkrs; Ejkrs=nullptr;
        delete [] Ejksr; Ejksr=nullptr;

        // Return the magnitude of the largest element.
        //================================================
        for (int i=0; i<(int) inHess[0].rows(); ++i)
            for (int j=0; j<(int) inHess[0].cols(); ++j)
                if (abs(inHess[0](i,j)) > maxmag)
                   maxmag=abs(inHess[0](i,j));

        return maxmag;
}
/*
/ Copyright (C) 2016 Chris Rook
/
/ This program is free software: you can redistribute it and/or modify it under the terms of the GNU General Public License as published by the Free Software Foundation, either version 3 of the License,
/ or (at your option) any later version.  This program is distributed in the hope that it will be useful, but WITHOUT ANY WARRANTY; without even the implied warranty of MERCHANTABILITY or FITNESS FOR A
/ PARTICULAR PURPOSE.  See the GNU General Public License for more details: <http://www.gnu.org/licenses/>
/
/ Filename:  stepHessM.cpp
/
/ Function:  stepHessM()
/
/ Summary:
/
/    This function steps in the general direction of steepest ascent for the multivariate mixture likelihood that maintains the marginal mixture densities, using Marquardt (1963).  The multivariate
/    mixture likelihood has fixed means, variances, and component probabilities but unknown covariances.  The multivariate mixture here is therefore a function of only the unknown covariances, and we
```

```
/    seek to maximize it.  A problem is that the function can have multiple local optimums.  Using Marquardt (1963) a constant term of random size is added/subtracted to the diagonals of the
/    correpsonding Hessian matrix prior to solving the updating equations when implementing Newton's method.  Adding a large constant results in taking a small step, and adding a small constant results
/    in taking a large step.  This method will allow us to take small steps towards the local optimum of the current hill and simultaneously search for larger hills in the general direction of steepest
/    ascent.  The minimum step size is set (to zero) in the header file (see minHessAdd) and the maximum step size is set to 10^X where X = # digits in the maximum Hessian element + 2.  Within the
/    min/max range a step size is randomly generated for each iteration.  The stepping is multi-threaded (see the wrapper that invokes this function) to save time and the total # of threads used is equal
/    to the # of independent processing units on the PC running the application multiplied by the global constant nCorMult (also set in the header file).  Each thread will take a total of mItersH random
/    sized steps, which is also a global constant set in the header file.  To cover the step size range from min-to-max using a total of nCorMult*(# of independent processing units) threads we first
/    generate a random value of X, between 1 and the maximum # of digits in the largest Hessian element + 2, then generate a step size randomly between 0 and 10^X.  At each iteration, a step is taken by
/    adding/subtracting the random step size to the Hessian diagonals and solving the Newton's method updating equations.  The decision variables are the U*A*(A-1)/2 unique covariances, where U = total #
/    multivariate mixture components and A = total # assets.  After obtaining the new solution, all corresponding variance-covariance matrices are confirmed to be positive definite, and if not a ridge
/    repair is immediately performed using a random multiplier between the 2 values (rrMultMin and rrMultMax) specified as global constants in the header file.  The decision variables that maximize the
/    likelihood function are returned, as is the maximum log-likelihood value along with the random step size that generates the maximum.  The best solution across all threaded calls is then used for the
/    current iteration of the ECME 2nd Step.  After stepping finishes, processing returns to the top of the 2nd Step and the 2nd Step gradient and Hessian are rebuilt for another iteration (in ECMEAlg()).
/
/ Inputs:
/
/    1.) Input supplied to this function as a 4-element integer array.  The element at position [0] is the # of time points with data.  The element at position [1] is the # of unique componenents in the
/        current multivariate mixture solution.  The element at position [2] is the current thread # determined by the function that generates threaded calls.  The thread # is only used within this
/        function for reporting results to the output window, for example during debugging.  The element at position [3] is an indicator that the current thread has launched.  (inHess1)
/
/    2.) Input/output both supplied to this function and generated by this function as a 5-element long double array.  Elements at positions [0], [1], [2] are input parameters and elements at positions
/        [3] and [4] are output generated by this function and returned to the calling function.  The input at position [0] is the current optimal log-likelihood value we are attempting to improve upon
/        during this ECME 2nd Step iteration.  Input at position [1] is the step size, and the input at position [2] is the starting value for stepping in this threaded call.  Elements at positions [3]
/        and [4] are placeholders for return values.  The maximum log-likelihood value found during this stepping iteration is returned in element [3], and the step size multiplier that generates this
/        maximum is returned in element [4].  (inHess2)
/
/    3.) The array of vector returns at each time point indexed as rs[t](a).  There are T vectors of returns and each is of size numA, where numA=# assets.  (rs)
/
/    4.) The current vector of 2nd Step decision variables (i.e., all covariances), as an array of 4 elements.  The vector at indices [0] and [2] hold the current covariance estimates.  The vector at
/        index [3] holds the returned estimates that maximize the log-likelihood for this function call.  The vector at position [1] holds the updated covariance estimates derived at each step. (inDvars)
/
/    5.) The current gradient vector evaluated at the current values of the covariance estimates (i.e., decision variables).  (inGrad)
/
/    6.) The current Hessian matrix evaluated at the current values of the covariance estimates (i.e., decision variables).  (inHess)
/
/    7.) The current vector of multivariate mixture probabilities.  (Initial values are from either the Minimax or Minimum SSD LP optimizations.)  (uPrbs)
/
/    8.) The array of mean vectors.  Each component of the multivariate mixture density is a multivariate density function which has its own set of means.  The first element of this array is the vector
/        of means for the first multivariate component, etc...  (muMns)
/
/    9.) The array of VC matrices.  Each component in the multivariate mixture density is a multivariate density function with a corresponding VC matrix.  Each VC matrix is of dimension (numA)x(numA).
/        The diagonals of each VC matrix are the corresponding variances for that asset within that component.  (inVCs)
/
/ Outputs:
/
/    This function updates element [3] of incoming parameter inDvars[] array with the covariance estimates that maximize the log-likelihood during this stepping iteration.  In addition, elements at
/    positions [3] and [4] of incoming array inHess2 are updated with the maximum value of the log-likelihood, and the step size multiplier that maximizes the log-likelihood, respectively.
/************************************************************************************************************************************************************************************************/
#include "stdafx.h"
void stepHessM(int *inHess1, long double *inHess2, const Eigen::VectorXd *rs, Eigen::VectorXd *inDvars, const Eigen::VectorXd inGrad, const Eigen::MatrixXd inHess, const Eigen::VectorXd uPrbs, const
               Eigen::VectorXd *inMns, const Eigen::MatrixXd *inVCs)
{
    // Declare/initialize local variables.
    //=====================================
    Eigen::initParallel();
    int rpr, nA=(int) inMns[0].size(), inT=inHess1[0], inUCmps=inHess1[1], hMaxLen=inHess1[4], expon, curExp;
    long double hIdx=1.00, hmult, curMaxLL=inHess2[0], LL, picst=exp(((double) (nA/2.00))*log(2.00*pi)), *sq_dets=new long double [inUCmps], *d_nom=new long double [inT], **f_vals=new long double
                *[inT], strt=inHess2[2], jmp, curMaxMult=strt, mult;
    random_device rd; default_random_engine gen(rd());
    uniform_real_distribution<long double> udist(0.0,1.0);
    Eigen::MatrixXd *Idm=new Eigen::MatrixXd [1], tHessM(inHess.rows(),inHess.cols()), *VC_Inv=new Eigen::MatrixXd [inUCmps], *V_C=new Eigen::MatrixXd [inUCmps];
    Idm[0]=Eigen::MatrixXd(inHess.rows(),inHess.cols()); getIDM(Idm);

    // Size the array holding all component likelihood values at all time points.
    //============================================================================
```

```
for (int t=0; t<inT; ++t)
    f_vals[t]=new long double [inUCmps];

// Size the local VC and inverse matrices.
//==========================================
for (int v=0; v<inUCmps; ++v)
{
    V_C[v]=Eigen::MatrixXd(nA,nA);
    V_C[v]=inVCs[v];
    VC_Inv[v]=Eigen::MatrixXd(nA,nA);
}

// Initialize the covariance maximizers to the starting values.  (Write details when debugging.)
//================================================================================================
inDvars[3]=inDvars[0];
if (dbug <= 1)
{
    cout << ".";
    inHess1[3]=1;
}
else if (dbug >=2)
{
    if (0 == inHess1[2])
       cout << endl;
    cout << "Launching thread # " << inHess1[2] << endl;
}

// Set the multiple for variance-covariance matrix repairs.  This applies to all steps.
//=======================================================================================
mult = rrMultMin + udist(gen)*(rrMultMax - rrMultMin);

// Iterate using the Hessian and solve for new covariances.
//===========================================================
for (int i=0; i<mItersH; ++i)
{
    // Hessian has element with maximum length equal to hMaxLen digits.  Randomly select value between 1 and this # to use as the max for stepping.
    //================================================================================================================================================
    expon = 1 + (int) (udist(gen)*((double) hMaxLen));
    jmp = pow(10.0,expon);
    hmult = minHessAdd + udist(gen)*jmp;

    if (udist(gen) <= 0.50)
    {
        hmult=-1.00*hmult;
        expon=0-expon;     // Indicate the sign/direction. (Value of -10 means exponent of 10 used for backward stepping.)
    }

    tHessM=inHess + hmult*Idm[0];
    inDvars[1]=tHessM.colPivHouseholderQr().solve(-1.00*udist(gen)*inGrad);
    inDvars[0]=(inDvars[1]+inDvars[2]);

    // Check that the new covariance estimates yield PD VC matrices.  (Repair if broken, and update the corresponding vector of current decision variables.)
    //=========================================================================================================================================================
    rpr=0;
    for (int v=0; v<inUCmps; ++v)
    {
        setCovs(v,V_C,inDvars);
        rpr=rpr + ridgeRpr(v,V_C,mult);
        VC_Inv[v]=V_C[v].inverse();
        sq_dets[v]=sqrt(VC_Inv[v].determinant());
    }
    LL=getLFVals(inT,nA,inUCmps,rs,uPrbs,inMns,VC_Inv,sq_dets,picst,d_nom,f_vals);
    if (rpr > 0)
       getCovs(inUCmps,V_C,inDvars);
```

```
        // Check vs existing maximum LL.  If larger, then update the current maximum and covariance array.
        //=================================================================================================
        if (LL > curMaxLL)
        {
            curMaxLL=LL;
            curMaxMult=hmult;
            inDvars[3]=inDvars[0];
            curExp=expon;
            inHess1[6]=rpr;
        }
    }

    // Return the maximum log-likelihood value along with the mutliplier that generated it.  (Write details when debugging.)
    //=====================================================================================================================
    inHess2[3]=curMaxLL;
    inHess2[4]=curMaxMult;
    inHess2[6]=mult;
    inHess1[5]=curExp;
    if (dbug <= 1)
       inHess1[7]=1;
    else if (dbug >= 2)
       cout << "Done with thread # " << inHess1[2] << " which started at: " << strt << " maximum LL is: " << curMaxLL << endl;

    // Delete temporary memory allocations.
    //======================================
    delete [] Idm; Idm=nullptr;
    delete [] d_nom; d_nom=nullptr;
    delete [] sq_dets; sq_dets=nullptr;
    delete [] V_C; V_C=nullptr;
    delete [] VC_Inv; VC_Inv=nullptr;
    for (int t=0; t<inT; ++t)
    {
        delete [] f_vals[t]; f_vals[t]=nullptr;
    }
    delete [] f_vals; f_vals=nullptr;
}

/*
/ Copyright (C) 2016 Chris Rook
/
/ This program is free software: you can redistribute it and/or modify it under the terms of the GNU General Public License as published by the Free Software Foundation, either version 3 of the License,
/ or (at your option) any later version.  This program is distributed in the hope that it will be useful, but WITHOUT ANY WARRANTY; without even the implied warranty of MERCHANTABILITY or FITNESS FOR A
/ PARTICULAR PURPOSE.  See the GNU General Public License for more details: <http://www.gnu.org/licenses/>
/
/ Filename:  ridgeRpr.cpp
/
/ Function:  ridgeRpr()
/
/ Summary:
/
/    This function accepts a variance-covariance matrix as input and determines whether or not it is positive-definite.  If not, it performs a ridge repair to make the matrix positive definite.  To
/    determine whether or not the matrix is positive definite, an eigenvalue decomposition is performed.  If all eigenvalues are > 0 then the matrix is positive definite.  A necessary and sufficient
/    condition for a variance-covariance matrix to be valid is that it is positive definite.  To perform a ridge repair, the diagonal elements are all multiplied by the same constant value (> 1).  Since
/    the diagonals are the variances, this implies that we increase the variances.  Doing so will automatically reduce the size of the covariances relative to the variances.  It will also reduce the
/    magnitude of the correlations.  The covariances are the correlations multiplied by the standard deviations.  That is, Cov(X,Y) = rho*std(X)*std(Y).  When std(X) and std(Y) increase and Cov(X,Y)
/    remains constant the correlations decrease in magnitude.  If the constant multiplier is large enough we will drive the correlations to near zero, and at this point the covariances will be extremely
/    small relative to the variances.  The resulting matrix approaches a diagonal matrix which is positive definite.  The point is to use a small multiplier and increase it iteratively until the repaired
/    matrix becomes positive definite and then scale it back so that the variances are undisturbed, but the off-diagonal elements are smaller relative to the diagonal elements.  If a matrix is positive
/    definite, then that matrix multiplied by a constant is also positive definite (easily proven with the definition of positive definite).  The initial multiplier is randomly generated in the calling
/    function and bounded by the global constants rrMultMin and rrMultMax, which are set in the header file.  This function iterates, multiplying the diagonals by (1.00 + i*rrmult) and checking for
/    positive definiteness after each iteration ("i" is the iteration index).  A variance-covariance matrix that is badly broken (for example with a correlation term > 100, when all such quantities
/    should be between -1.00 and +1.00) may require a large number of iterations to repair.  Therefore, to speed up processing we increase the multiplier by a factor of 10 after each 100 iterations.
/    That is, after 100 iterations rrmult is multiplied by 10, and again after 200 iterations, etc...  Note that The variance-covariance matrix supplied to this function is a modifiable value and is
/    updated in place with care taken to ensure that the diagonals are not disturbed when it is returned.
/
```

```
/ Inputs:
/
/    A single variance-covariance matrix is supplied to this function using 2 inputs:  (1) An array of variance-covariance matrices and (2) an index to identify the one we are checking for positive-
/    definiteness, and repairing if necessary.
/
/    1.) The index that identifies the variance-covariance matrix to be checked/repaired.  (uCell)
/
/    2.) The array of variance-covariance matrices for the current multivariate mixture solution.  (E)
/
/    3.) The multiple used to add a ridge.  (mult)
/
/ Outputs:
/
/    This function returns an integer value of 1 or 0.  A 1 is returned if the variance-covariance matrix is repaired, and a 0 is returned if it is not in need of repair.
/*****************************************************************************************************************************************************************************************************/
#include "stdafx.h"
int ridgeRpr(const int uCell, Eigen::MatrixXd *E, long double mult)
{
    // Local variables.
    //==================
    Eigen::initParallel();
    int pd, retvar=0;
    long double sclfctr, rr_mult=mult, det;
    Eigen::EigenSolver<Eigen::MatrixXd> egnslvr;
    Eigen::MatrixXd newVCm((int) E[uCell].rows(),(int) E[uCell].cols());

    // Is the VC matrix Positive-definite as required?  If not, repair it.
    //=====================================================================
    egnslvr.compute(E[uCell], false);
    pd=1; det=1.0;
    for (int a=0; a<(int) E[uCell].rows(); ++a)
    {
        if (egnslvr.eigenvalues()[a].real() <= pdmineval)
           pd=0;
        det=det*egnslvr.eigenvalues()[a].real();
    }
    if (det <= detminval)
       pd=0;

    // Repair is necessary.
    //======================
    if (0 == pd)
    {
        // VC matrix is not positive definite and a ridge repair will be performed.
        //=========================================================================
        newVCm=E[uCell];

        // Repair the matrix.
        //====================
        for (int i=1; 0==pd; ++i)
        {
            // Increase each variance by a factor > 1 and rescale.
            //====================================================
            sclfctr=(1 + ((double) i)*rr_mult);
            for (int r=0; r<(int) E[uCell].rows(); ++r)
                for (int c=0; c<(int) E[uCell].cols(); ++c)
                    if (r != c)
                       newVCm(r,c)=E[uCell](r,c)/sclfctr;

            // Check if the updated VC matrix is positive-definite.
            //=====================================================
            egnslvr.compute(newVCm, false);
            pd=1; det=1.0;
            for (int a=0; a<(int) E[uCell].rows(); ++a)
            {
```

```
                if (egnslvr.eigenvalues()[a].real() <= pdmineval)
                   pd=0;
                det=det*egnslvr.eigenvalues()[a].real();
            }
            if (det <= detminval)
               pd=0;

            // Increase the scale factor by a multiple of 10 each 100 iterations.
            //==================================================================
            if (0 == (int) (i % 100))
               rr_mult=10*rr_mult;
        }

        // Replace the original VC matrix with the repaired version.
        //============================================================
        E[uCell]=newVCm;
        retvar=1;
    }

    // If the VC matrix has been repaired return a 1, otherwise return a 0.
    //======================================================================
    return retvar;
}

/*
/ Copyright (C) 2016 Chris Rook
/
/ This program is free software: you can redistribute it and/or modify it under the terms of the GNU General Public License as published by the Free Software Foundation, either version 3 of the License,
/ or (at your option) any later version. This program is distributed in the hope that it will be useful, but WITHOUT ANY WARRANTY; without even the implied warranty of MERCHANTABILITY or FITNESS FOR A
/ PARTICULAR PURPOSE.  See the GNU General Public License for more details: <http://www.gnu.org/licenses/>
/
/ Filename:  wrtDens.cpp
/
/ Function:  wrtDens()
/
/ Summary:
/
/    This function writes out the structure of the final multivariate density along with the actual values contained in that structure.  The output has 2 parts: 1.) the structure, which is a sum of
/    multivariate normal densities weighted by the corresponding component probability using parameters, and 2.) details of the parameter values.  For each component, the definitions include the actual
/    component probability, the mean vector, the variance-covariance matrix, and the unique cell that this component originates from with respect to the full factorial of combinations defined after
/    fitting the univariate mixture densities.  Finally, the inverse of the variance-covariance matrix is printed as is the rank and the determinant.  This function can be used to write the supplied
/    density to either standard output (using cout in the last parameter), or a file (using an ofstream fout definition in the last parameter).
/
/ Inputs:
/
/    1.) Specify the type of starting point when transitioning between univariate marginal densities and the multivariate mixture PDF.  In this application we offer 2 transition methods:  Minimax or
/        Minimum Sum of Squared Distances (SSD).  Each are solved as constrained linear programs (LPs).  The constraints maintain the univariate marginal mixture densities.  This string variable takes
/        one of two values:  "Minimax" or "Minimum Sum of Squared Distances (SSD)".  This is for display purposes only and informs the user which transition method was used for the given multivariate
/        density function being written.  (typ)
/
/    2.) The number of unique components in the multivariate mixture that results from either the Minimax or Minimum SSD LP optimization starting points.  (nUCmps)
/
/    3.) The array of unique cell IDs that link each component of the multivariate density back to the full factorial of components.  The full factorial of components represents each cell in the
/        multidimensional grid formed by considering all combinations of assets and their levels.  (Note that the full factorial would be required to build a multivariate mixture density with given
/        marginals under the assumption that the assets were all mutuallly independent random variables (RVs).)  (uCellIDs)
/
/    4.) The array of final multivariate mixture component probabilities.  (muPrbs)
/
/    5.) The array of mean vectors.  Each component of the multivariate mixture density is a multivariate density function which has its own set of means. The first element of this array is the vector of
/        means for the first multivariate component, etc...  (muMns)
/
/    6.) The array of VC matrices.  Each component in the multivariate mixture density is a multivariate density function with a corresponding VC matrix.  Each VC matrix is of dimension (numA)x(numA),
/        where numA = total # of assets.  The diagonals of each VC matrix are the corresponding variances for that asset within that component.  (muVCs)
/
/    7.) The output destination as a variable reference.  Use either "cout" for display to the screen or a valid ofstream output object.  (oVar)
```

```
/
/ Outputs:
/
/    The density supplied to this function is written to the output desination supplied in 2 parts.  First, the structure of the density is written using parameters followed by a detailed definition of
/    those parameters.  No value is returned at the call.
/**********************************************************************************************************************************************************************************************************/
#include "stdafx.h"
void wrtDens(const string typ, const int nUCmps, const int *uCells, const long double *muPrbs, const Eigen::VectorXd *muMns, const Eigen::MatrixXd *muVCs, ostream& oVar)
{
    // Local variables.
    //===================
    long long w=to_string((long long) nUCmps).size();
    string ps="";
    int ln=135;
    if ("Minimum Sum of Squared Distances (SSD)" == typ)
       ln=ln+20;

    // Start by writing the structure of the multivariate density, without actual details of the values for the means, variances, covariances, and component probabilities.
    //========================================================================================================================================================================
    oVar << endl << string(ln,'=') << endl << "The structure of the multivariate density function for the supplied assets using an initial " << typ << " LP objective is given by: "
         << endl << string(ln,'=') << endl << endl;

    for (int v=0; v<nUCmps; ++v)
    {
        oVar << ps << "p" << setfill('0') << setw(w) << v << "*" << "f(m" << setfill('0') << setw(w) << v << ",V" << setfill('0') << setw(w) << v << ")";
        ps=" + ";
        if (0 == ((1 + v) % 5))
           oVar << endl;
    }

    // Write out the details for each component.  That is, the means, variances, covariances, and component probabilities.
    //========================================================================================================================
    oVar << endl << endl << string(112,'=') << endl << "Where values for the multivariate normal pdfs f(m,V) with mean vector m and " << "variance-covariance matrix V, are:"
         << endl << string(112,'=') << endl;

    oVar.setf(ios_base::fixed, ios_base::floatfield); oVar.precision(30);
    for (int v=0; v<nUCmps; ++v)
        oVar << endl << string(35,'=') << endl << "Component #" << v << " is unique cell #" << uCells[v] << endl << string(35,'=') << endl << "p" << setfill('0') << setw(w)
             << v << " = " << muPrbs[v] << endl << "m" << setfill('0') << setw(w) << v << " = " << endl << muMns[v] << endl << "V" << v << " = " << endl << muVCs[v] << endl
             << "V" << v << setfill('0') << setw(w) << "^(-1) = " << endl << muVCs[v].inverse() << endl << "--> The rank of V" << v << " is: "
             << (int) Eigen::FullPivLU<Eigen::MatrixXd>(muVCs[v]).rank() << " and the determinant is: " << muVCs[v].determinant() << endl;
}
```